\documentclass[11pt]{article}
\usepackage[a4paper,margin=26mm]{geometry}
\usepackage[T1]{fontenc}
\usepackage{lmodern}
\usepackage{amsmath,amssymb,amsthm,mathrsfs}
\usepackage{xcolor}
\usepackage{tikz}
\usepackage{microtype}
\usepackage{booktabs}
\usepackage[numbers,sort&compress]{natbib}
\usepackage{hyperref}
\hypersetup{hidelinks,pdftitle={Propagation and horizon formation in Einstein--Gauss--Bonnet scalar collapse. I. Characteristic conditions, bounds and formation}}
\numberwithin{equation}{section}
\allowdisplaybreaks[1]
\theoremstyle{definition}
\newtheorem{theorem}{Theorem}[section]
\newtheorem{proposition}[theorem]{Proposition}
\newtheorem{corollary}[theorem]{Corollary}
\newcommand{\dd}{\mathrm d}
\newcommand{\R}{\mathbb R}
\title{Propagation and horizon formation in\\Einstein--Gauss--Bonnet scalar collapse.\\
{\large I. Characteristic conditions, bounds and formation}}
\author{
Cendikiawan Suryaatmadja\thanks{\texttt{cendikiawan.suryaatmadja@uwaterloo.ca}}\\
{\small Department of Physics and Astronomy, University of Waterloo}\\
{\small Waterloo, ON N2L 3G1, Canada}
}
\date{}
\begin{document}
\maketitle
\begin{abstract}
We study scalar collapse in five- and six-dimensional Einstein--Gauss--Bonnet gravity with positive coupling as an exact classical theory. On spherical scalar backgrounds, gravitational characteristic conditions reduce to inequalities in two local invariants. In five dimensions, they imply $R_h^2>\tilde\alpha$ for regular marginal spheres on the GR branch and the sharp Kretschmann bound $\tilde\alpha^2\mathcal R_{abcd}\mathcal R^{abcd}<28$ for nonnegative mass. Under the stated finite-domain hypotheses, spherical continuation places the first characteristic degeneration on a smooth background. Within this setting, a future marginal sphere formed while the characteristic conditions hold throughout the preceding evolution forces their later failure. For every allowed radius, separately chosen complete smooth data form marginal spheres with mass arbitrarily close to the sharp lower bound; in six dimensions both radius and mass can be arbitrarily small. At sufficiently large scale, complete nonspherical data with weak curvature contain marginal surfaces with prescribed $S^3$ metrics under a spectral condition. These surfaces are outermost and strictly stable. A separate Berger family has arbitrarily small area at one common ADM mass; formation from untrapped data remains open. Exact nonspherical developments have smoothly convergent initial data but tidal tensors that fail to converge to the reference values at a chosen time after the reference core loses hyperbolicity. Under local geometric uniqueness for developments of the same data on overlaps connecting the initial slice to the observers, classical existence on a common region and continuous dependence of the tidal tensor cannot both hold. The constructed solutions themselves extend smoothly. Fixed time averages converge on the controlled interval, but this does not establish a predictive effective-field-theory continuation.
\end{abstract}

\section{Introduction}

{Einstein--Gauss--Bonnet (EGB) gravity provides a higher-curvature model whose metric equations remain second order, while the propagation of gravitational disturbances depends on the background curvature. The Gauss--Bonnet combination also appears in curvature-squared corrections to string effective actions \cite{Zwiebach1985,MetsaevTseytlin1987}. This supplies a motivation for studying its effects, but does not identify the constant-coupling EGB action with a complete string effective action: additional fields and interactions are generally present, and the minimally coupled scalar used here is not a string dilaton.

We use a canonical massless scalar as a simple dynamical matter model with positive energy and a well-defined ingoing flux. Spherical symmetry makes the collapse background tractable without restricting the gravitational perturbations to be spherical. The questions are whether the characteristic conditions for those perturbations restrict the formation of marginal surfaces, whether the background can remain smooth when those conditions fail, and what this failure implies for dependence on the initial data.

Throughout, we treat the displayed EGB--scalar equations as an exact classical model. This permits an assessment of their local initial-value problem and geometric restrictions without assuming that the Gauss--Bonnet contribution is perturbatively small. It does not imply that the model remains a controlled low-energy effective field theory (EFT) at every scale reached by its solutions. In an EFT interpretation, the size of omitted operators and the proper wavelengths of perturbations must be assessed separately. In particular, a sharp characteristic boundary of the exact equations need not persist at the same location after those operators are included. Reall discusses the distinction between mathematical characteristics and their interpretation within an EFT \cite{Reall2021}.}

Numerical studies of scalar collapse in EGB gravity found a breakdown of the discrete self-similarity seen in general relativity \cite{Golod2012}. Deppe et al.\ found evidence for a minimum apparent-horizon radius at formation in five dimensions, while no radius gap was evident in six dimensions \cite{Deppe2012}. This paper and its companion, Paper~II~\cite{PaperII}, study the classical EGB--scalar equations analytically to understand this contrast and its relation to gravitational propagation.

Gravitational characteristics are the wavefronts of high-frequency gravitational disturbances. In Lovelock gravity, their propagation cones depend on curvature and can differ from the matter null cone \cite{Reall2014,Izumi2014}. Small static EGB black holes in five and six dimensions have a gravitational instability \cite{Beroiz2007}, which Reall, Tanahashi and Way related to loss of hyperbolicity \cite{Reall2014}. Konoplya and Zhidenko studied the related eikonal instability of Gauss--Bonnet black holes with a cosmological constant \cite{Konoplya2017}, and Papallo and Reall analyzed the local well-posedness of Lovelock and Horndeski theories on weakly curved backgrounds \cite{Papallo2017}. Cao and Wu derived the tensor perturbation equation on time-dependent warped-product backgrounds \cite{Cao2021}.

{Related numerical work in four-dimensional scalar--Gauss--Bonnet gravity illustrates why smooth background evolution and hyperbolic evolution must be distinguished. Ripley and Pretorius found a change of PDE type during collapse \cite{RipleyPretorius2019}; East and Ripley evolved nonlinear solutions using a modified harmonic formulation \cite{EastRipley2021}; and Hegade, Ripley and Yunes developed a covariant diagnostic for loss of hyperbolicity in spherical evolution \cite{HegadeRipleyYunes2023}. These theories couple the scalar to the Gauss--Bonnet invariant. Their results provide relevant comparisons, but their characteristic thresholds do not directly apply to the higher-dimensional theory with constant Gauss--Bonnet coupling and a minimally coupled scalar studied here.

The relation to EFT validity also depends on the theory, background and region considered. In the spin-induced scalarization evolutions studied by Doneva et al., the weak-coupling diagnostic fails before the observed loss of gravitational hyperbolicity \cite{Doneva2023}. This is not a universal equivalence between the two failures. Corman et al.\ compare evolution of the full equations, perturbative evolution and equation fixing, and show why the behavior of a regulated system must be checked against the scales introduced by the regulator \cite{Corman2024}.

Five-dimensional black-string evolutions provide a closer comparison within EGB gravity. Figueras, Kov{\'a}cs and Yao study the nonlinear instability and its dependence on the coupling sign in an EFT treatment \cite{FiguerasKovacsYao2026EFT}, and analyze the associated hyperbolicity conditions using an effective metric for their symmetry-reduced degree of freedom \cite{FiguerasKovacsYao2026EGB}. For positive coupling, their simulations encounter hyperbolicity failure inside the apparent horizon while their diagnostic for the neglected six-derivative terms remains small on the horizon. That effective metric remains Lorentzian on and outside the horizon at that stage. These regional diagnostics should not be identified with a universal equivalence between hyperbolicity loss and EFT failure, or with control of every omitted operator. Our scalar-collapse results instead give explicit spherical characteristic inequalities, sharp conditional bounds, complete data that form marginal spheres, and an exact nonlinear comparison of tidal curvature.}

Kov{\'a}cs and Reall established strong hyperbolicity of the modified harmonic formulation at weak coupling \cite{Kovacs2020}. For the strictly admissible spherical backgrounds considered here, Appendix~\ref{app:symbol} constructs a suitable modified harmonic formulation and a local symmetrizer without assuming weak coupling. This gives local well-posedness for nearby initial data, which may be nonspherical. The data must satisfy the constraints and the initial gauge conditions, with the regularity and foliation specified in Appendix~\ref{app:hyperbolic}. Real characteristic frequencies alone would not establish this result; the symmetrizer argument is essential. At characteristic degeneration this argument no longer applies, whereas the separate spherical continuation theorem can still apply.

We obtain all three gravitational sectors from the curvature derivative of the field equations. On spherical scalar backgrounds, the characteristic conditions reduce to inequalities for two invariants, $K=1+\tilde\alpha F/R^2$, which measures the mass inside a sphere relative to its radius, and $j$, which measures the sign and size of the squared norm of the scalar gradient. This reduction, derived in Section~\ref{sec:characteristics}, makes it possible to compare characteristic degeneration with geometric restrictions and to follow both along explicit collapse solutions.

In five dimensions the admissible region of Fig.~\ref{fig:region} requires $R_h^2>\tilde\alpha$ for a regular marginal sphere and bounds the Kretschmann invariant for nonnegative mass (Section~\ref{sec:radius}). Near the radius bound the minimum rest-frame gravitational speed tends to zero; with the stated exterior hypotheses and ADM mass interval, the scalar-field stress-energy cannot vanish at the marginal sphere. The separate spherical PDE argument gives smooth continuation on the finite domains specified in Section~\ref{sec:radius}. Within that setting, a future marginal sphere formed while the characteristic inequalities hold throughout the preceding evolution forces their later failure. Complete smooth data can be chosen to form marginal spheres at each larger radius with arbitrarily small mass excess; in six dimensions there is no corresponding positive radius or mass gap (Section~\ref{sec:formation}). These are existence constructions, not statements about generic collapse or thresholds for a prescribed family.

Without spherical symmetry, we construct complete data with weak curvature containing marginal surfaces whose intrinsic metrics are prescribed on $S^3$, at sufficiently large scale, under a spectral condition analogous to that of Mantoulidis and Schoen \cite{MantoulidisSchoen2015}. A separate explicit construction gives Berger surfaces of arbitrarily small area with one common exterior and ADM mass and uniformly weak curvature (Section~\ref{sec:nonspherical-area}). The common mass is supplied by the construction. Neither formation from untrapped data nor the presence of an event horizon is established.

Section~\ref{sec:tidal-prediction} compares exact nonspherical solutions near a contracting homogeneous core. Their constrained initial data converge with every fixed number of derivatives, but their tidal matrices do not converge at a selected time $T$ after the reference core loses hyperbolicity. For that same sequence, smooth convergence holds at every fixed earlier time, and fixed time averages over the common interval converge. The construction is normalized separately for each chosen $T$. For classical developments that are geometrically unique on overlaps connecting the initial slice to the observers, a continuation cannot both exist on a common observation region and make the tidal tensor depend continuously on the initial data. It establishes neither generic inextendibility nor failure of smooth extension for the constructed solutions, which do extend. The convergent averages do not themselves supply an EFT evolution law.

Paper~II applies these results to narrow incoming scalar pulses. It determines which gravitational characteristic condition fails first, whether this happens before trapping, and the amplitude at which characteristic degeneration begins. For a symmetric pulse whose first degeneration is in the scalar gravitational sector, it also studies persistence under small nonspherical perturbations. It approaches the radius bound of Section~\ref{sec:radius} from initial data with arbitrarily weak curvature at finite mass.
\section{Field equations and characteristics}
\label{sec:characteristics}

We use the dimension, coupling and mass conventions of
Ref.~\cite{Deppe2012}. With signature $(-+\cdots+)$ and zero
cosmological constant, the action has the form of its Eq.~(2.1),
with the canonical sign for the scalar kinetic term:\footnote{The plus sign printed
in Eq.~(2.1) of Ref.~\cite{Deppe2012} would give negative scalar energy
for this signature. We use the minus sign in Eq.~(2.4) of
Ref.~\cite{Taves2012}, which is the sign implied by the positive-energy evolution
equations of Ref.~\cite{Deppe2012}.}
\begin{equation}
 \begin{split}
 \mathcal I=\frac1{2\kappa_n^2}\int\dd^n x\,\sqrt{-g}\Big(&\mathcal R
 +\alpha_{(2)}\bigl[\mathcal R^2-4\mathcal R_{\mu\nu}\mathcal R^{\mu\nu}
                  +\mathcal R_{\mu\nu\rho\sigma}\mathcal R^{\mu\nu\rho\sigma}\bigr]-\kappa_n^2(\nabla\psi)^2\Big)
 \end{split}
 \label{def:action}
\end{equation}
We are mainly interested in $n=5$ and $n=6$; identities stated for general $n$ hold for
$n\ge5$. We assume $G_n>0$, set $\kappa_n^2=8\pi G_n$, and use
$\tilde\alpha=2(n-3)(n-4)\alpha_{(2)}>0$, as defined below Eq.~(2.7) of
Ref.~\cite{Deppe2012}.
Following that reference, $k=8(n-3)/(n-2)^2$ and
$kG=\kappa_n^2/[(n-2)A_{n-2}]$, where $A_{n-2}$ is the area of
the unit $(n-2)$-sphere. We keep $G$ explicit and use the
rescaled scalar $\Psi=\sqrt{A_{n-2}}\,\psi$. The field equations are
\begin{equation}
 G_{ab}+\alpha_{(2)}H^{\rm GB}_{ab}
 =(n-2)kG\left(\Psi_a\Psi_b-\frac12g_{ab}(\nabla\Psi)^2\right),
 \qquad \Box\Psi=0
 \label{eq:field}
\end{equation}
where $\Psi_a=\nabla_a\Psi$ and $H^{\rm GB}_{ab}$ is the
Gauss--Bonnet tensor. We write
$T_{ab}[\Psi]=\Psi_a\Psi_b-\tfrac12g_{ab}(\nabla\Psi)^2$ for the
stress-energy tensor of the rescaled scalar.

{All results concern the exact equations~\eqref{eq:field}, with a
canonical, minimally coupled massless scalar. Reference solutions and
initial data are smooth unless a Sobolev regularity class is specified.
The dimension and additional assumptions, including symmetry,
nonnegative mass, completeness and asymptotic flatness, are stated
separately for each result.}

For a spherical metric $\dd s^2=g_{AB}\dd x^A\dd x^B+R^2\dd\Omega_{n-2}^2$, set
\begin{equation}
 F=1-(\nabla R)^2,\qquad A=1-F,\qquad
 K=1+\frac{\tilde\alpha F}{R^2}
 \label{sym:d1}
\end{equation}
The generalized Misner--Sharp mass is \cite{Maeda2008,Deppe2012}
\begin{equation}
 \mathcal M=\frac1{2kG}\left[
 R^{n-3}F+\frac{\tilde\alpha}{2}R^{n-5}F^2\right]
 \label{def:mass}
\end{equation}
We work on the GR branch $K>0$, where
$K^2=1+4kG\tilde\alpha\mathcal M/R^{n-1}$; our $K$ is the quantity called $q$
in Eq.~(5.15) of Ref.~\cite{Reall2014}. A marginal sphere has $F=1$,
or $A=0$; $A$ is the horizon function $\mathrm{AH}$ of
Ref.~\cite{Deppe2012}. In an orthonormal frame with future timelike
$e_0$ and outward radial $e_1$, the null expansions are
$\theta_\pm=(n-2)(e_0\pm e_1)R/R$, so that
$\theta_+\theta_-=-(n-2)^2A/R^2$. At a standard asymptotically flat
end, $M_{\rm ADM}=\mathcal M_\infty$.

The quantity $K$ measures the size of the Gauss--Bonnet correction
at a sphere. In Einstein gravity $K=1$. With $\tilde\alpha>0$, $K$ is
close to one wherever $kG\tilde\alpha|\mathcal M|\ll R^{n-1}$, in
particular on every untrapped sphere of nonnegative mass with
$R\gg\sqrt{\tilde\alpha}$, where $0\le K-1<\tilde\alpha/R^2$. At a
marginal sphere $F=1$, so
$K-1=\tilde\alpha/R^2$ there: $K$ compares the Gauss--Bonnet length
$\sqrt{\tilde\alpha}$ with the size of the sphere, and the departure
from Einstein gravity is large when the sphere is not
much larger than $\sqrt{\tilde\alpha}$. In terms of the mass, $K^2-1$
is proportional to $\mathcal M/R^{n-1}$, so $K$ grows when mass is
concentrated within a small radius. On spherical scalar
backgrounds the characteristic conditions are controlled by $K$
together with one invariant of the scalar gradient, defined next.

Set
\begin{equation}
 \begin{gathered}
 P=e_0\Psi,\qquad Q=e_1\Psi\\
 j=\tilde\alpha kG(P^2-Q^2),\qquad
 b_0=n-3+\frac{n-5}{2}(K^2-1)
 \end{gathered}
 \label{def:jb0}
\end{equation}
The invariant $j$ distinguishes a timelike scalar gradient ($j>0$)
from a spacelike one ($j<0$). It is not the local energy density,
which is proportional to $P^2+Q^2$ in this frame. Two configurations
with the same energy density can have different values of $j$, and the
propagation conditions below depend on the scalar gradient only
through $j$. Physically, $j>0$ describes a field that changes
mainly in time at the point in question, as for a spatially
homogeneous field, and $j<0$ one that changes mainly in space, as for
a static profile; for a purely incoming or outgoing null wave,
$P=\pm Q$ and $j=0$. Because $j$ is a Lorentz invariant of the
gradient, its value does not depend on the observer, whereas the
energy density does.

Characteristics are the surfaces along which short-wavelength
disturbances propagate, and their structure determines whether the
field equations can be solved forward in time. In Einstein gravity every polarization
of the gravitational field propagates on the null cone of the metric.
In Einstein--Gauss--Bonnet gravity the coefficients of the highest
derivatives depend on the curvature, so each of the tensor, vector
and scalar sectors of the gravitational field acquires its own
propagation cone, which can open wider or narrower than the light
cone of matter. Near the spherical backgrounds considered here, strict positivity
of the characteristic coefficients permits a strongly hyperbolic
modified harmonic formulation. Besides a common time direction for
the propagation cones, this requires the complete, uniformly bounded
set of modes established in Appendix~\ref{app:hyperbolic}. As a cone degenerates,
the phase speed of the corresponding sector tends to zero in some
direction, and the equations cease to be strongly hyperbolic. Locating where this happens
along a collapse, and what it implies for horizons, is the main
subject of this paper.

To determine the gravitational characteristics, linearize about the
spherical solution and retain the terms with two derivatives of the
metric perturbation $\delta g_{ab}$. In the short-wavelength limit these dominate
the lower-derivative terms and determine the possible wavefronts.
Unlike in Einstein gravity, their coefficients depend on the
background curvature. Curvature itself contains two metric
derivatives, and its variation is
\begin{equation}
 \delta\mathcal R_{abcd}\simeq\frac12\bigl(
 \partial_b\partial_c\,\delta g_{ad}+\partial_a\partial_d\,\delta g_{bc}
 -\partial_a\partial_c\,\delta g_{bd}-\partial_b\partial_d\,\delta g_{ac}\bigr)
 \label{sym:curvature-variation}
\end{equation}
where $\simeq$ keeps only second derivatives of the perturbation. The
field equations contain second derivatives of the metric only through
the curvature (Eq.~\eqref{app:deltaform}), so differentiating them with
respect to curvature extracts these coefficients.

The resulting coefficients are governed by one tensor $\mathcal W$:
\begin{equation}
 \delta\bigl(G^a{}_b+\alpha_{(2)}H^{{\rm GB}\,a}{}_b\bigr)
 \simeq-\frac14\,\mathcal W^{acd}{}_{bef}\,
        \delta\mathcal R^{ef}{}_{cd},
 \qquad
 \mathcal W^{acd}{}_{bef}
 =\delta^{acd}_{bef}
  +\alpha_{(2)}\,\delta^{acdgh}_{befpq}\,\mathcal R^{pq}{}_{gh}
 \label{sym:lovelock}
\end{equation}
where $\delta^{a_1\cdots a_r}_{b_1\cdots b_r}=\det(\delta^{a_\mu}_{b_\nu})$
are generalized Kronecker deltas. Up to the factor $-1/4$, $\mathcal W$
is the derivative of the metric field equations with respect to
curvature. The principal symbol of the linearized equations is the
matrix of their highest-derivative coefficients, with each derivative
$\partial_a$ replaced by the wave covector $\xi_a$. Making this
replacement in Eq.~\eqref{sym:curvature-variation} and inserting the
result in Eq.~\eqref{sym:lovelock} gives the principal symbol,
written out in Eqs.~\eqref{app:d1} and \eqref{app:d2}.

The components of $\mathcal W$ are easiest to read in an
orthonormal frame in which the curvature is diagonal on the planes
spanned by pairs of frame vectors. Since $\delta^{acd}_{bef}$, the
Einstein part of $\mathcal W$, vanishes unless $\{b,e,f\}=\{a,c,d\}$,
the Einstein principal symbol is a sum of contributions, one for each
set $\mathcal T$ of three frame directions. In the diagonal frame,
adding the Gauss--Bonnet term multiplies the contribution of
$\mathcal T$ by
\begin{equation}
 w(\mathcal T)=1+4\alpha_{(2)}\Sigma_1(\mathcal T^\perp)
 \label{sym:weights}
\end{equation}
where $\Sigma_1(\mathcal T^\perp)=\sum_{p<q}\mathcal R^{pq}{}_{pq}$ is the sum
of the sectional curvatures of the planes spanned by pairs of frame
vectors orthogonal to $\mathcal T$ (Appendix~\ref{app:symbol}). In five
dimensions $\mathcal T^\perp$ is a single plane and
$\tilde\alpha=4\alpha_{(2)}$, so
\begin{equation}
 w(\mathcal T)=1+\tilde\alpha\operatorname{sec}(\mathcal T^\perp)
 \label{sym:weights-five}
\end{equation}
where $\operatorname{sec}(\mathcal T^\perp)$ is the sectional curvature
of that plane.

On a spherical background, a diagonal frame has four distinct
types of weight. With $i,k,l$
distinct angular indices, the triples $\{0,1,i\}$, $\{0,i,k\}$,
$\{1,i,k\}$ and $\{i,k,l\}$ have weights $K$, $X$, $Y$ and $C_T$
(Table~\ref{tab:weights}), whose
values on solutions are given in Eqs.~\eqref{sym:base} and
\eqref{sym:general} below. Every
angular plane has sectional curvature $F/R^2$, which is why the
weight of the triples containing both orbit directions is the $K$
defined above.
\begin{table}[t]
\centering
\setlength{\tabcolsep}{5pt}
\begin{tabular}{@{}llcl@{}}
\toprule
triple $\mathcal T$ & complementary plane $\mathcal T^\perp$ & number of triples & weight $w(\mathcal T)$\\
\midrule
$\{0,1,i\}$ & two angular directions, sectional curvature $F/R^2$ & 3 & $K$\\
$\{0,i,k\}$ & radial and angular & 3 & $X$\\
$\{1,i,k\}$ & time and angular & 3 & $Y$\\
$\{i,k,l\}$ & time and radial & 1 & $C_T$\\
\bottomrule
\end{tabular}
\caption{The ten triples of frame directions on a five-dimensional
spherical background, in an orthonormal frame diagonalizing the
curvature, and their weights. Here $0$ and $1$ are the time and radial
directions and $i,k,l$ are distinct angular directions. By
Eq.~\eqref{sym:weights-five}, the weight of a triple is one plus
$\tilde\alpha$ times the sectional curvature of the complementary
plane; in general dimension, Eq.~\eqref{sym:weights}, it is one plus
$4\alpha_{(2)}$ times the sum of the sectional curvatures of the planes
spanned by the remaining $n-3$ directions. The counts
$3,3,3,1$ are the multiplicities in the curvature identity of
Eq.~\eqref{cont:deviation}.}
\label{tab:weights}
\end{table}In a general spherical frame the time--radial second-derivative
matrix $\nabla_A\nabla_BR$ need not be diagonal. Its contribution
then includes an off-diagonal coefficient $H$. The block formulation
below applies also when no diagonal orthonormal frame exists. The values
$X,Y,H,C_T$ are the tensor coefficients of Ref.~\cite{Cao2021}. Using
the background curvature relations in Eq.~\eqref{app:background},
\begin{equation}
 \begin{aligned}
 X&=\frac{b_0+\tilde\alpha kG(P^2+Q^2)}{(n-3)K},&
 Y&=\frac{b_0-\tilde\alpha kG(P^2+Q^2)}{(n-3)K}\\
 H&=\frac{2\tilde\alpha kG PQ}{(n-3)K},&
 \Delta&=XY+H^2=\frac{b_0^2-j^2}{(n-3)^2K^2}
 \end{aligned}
 \label{sym:base}
\end{equation}
These weights form the time and radial block common to all
gravitational sectors below. Covariantly,
\begin{equation}
 G_{\rm orb}^{AB}
 =\frac{(b_0-j)g^{AB}-2\tilde\alpha kG\,\nabla^A\Psi\nabla^B\Psi}{(n-3)K},
 \qquad
 K(\lambda_++\lambda_-)=\frac{2b_0}{n-3}
 \label{sym:orbit}
\end{equation}
where $\lambda_\pm=(b_0\pm j)/((n-3)K)$ are the eigenvalues of the mixed
block $G_{\rm orb}{}^A{}_B$, on $\nabla\Psi$ and on the orthogonal
orbit direction when $\nabla\Psi$ is not null; thus
$\Delta=\lambda_+\lambda_-$. The trace relation holds because the orbit
stress of a massless scalar is trace-free.

High-frequency gravitational perturbations of a spherical background
separate into tensor, vector and scalar sectors, labeled $I=T,V,S$,
according to their transformation under rotations of the sphere.
After removing gauge and constraint components, the physical
characteristic equation of each sector can be written with an inverse effective metric $G_I^{ab}$
\cite{Reall2014}. Their null cones govern gravitational propagation;
the matter scalar follows the null cone of $g_{ab}$. Let $\xi_a$ be
normal to a wavefront. In the orthonormal frame, rotate the angular
axes so that its components are $(\omega,\xi_R,\xi_\perp,0,\ldots)$,
where $\omega=\xi(e_0)$ is the signed temporal component,
$\xi_R=\xi(e_1)$ and $\xi_\perp$ is the magnitude of the angular part.
Then
\begin{equation}
 G_I^{ab}\xi_a\xi_b
 =-X\omega^2+2H\omega\xi_R+Y\xi_R^2+C_I\xi_\perp^2=0,
 \qquad I=T,V,S
 \label{sym:factors}
\end{equation}
with angular coefficients $C_T$, $C_V$ and $C_S$ given below.
The tensor, vector and scalar gravitational sectors have
$(n-1)(n-4)/2$, $n-3$ and one polarization, respectively; all of
them include angularly dependent perturbations of the spherical
background. The
scalar gravitational sector is distinct from the matter scalar.
For $\xi_R=0$, their squared angular phase speeds in this frame are
$v_I^2=C_I/X$. In static vacuum $H=0$ and $X=Y$; dividing by $X$
gives the normalization $c_I=C_I/X$ of Eq.~(5.7) in
Ref.~\cite{Reall2014}. Overall normalization does not affect a
characteristic cone. Distances, areas and matter stress remain
defined by $g_{ab}$.

Appendix~\ref{app:symbol} derives all three sectors from the weighted
symbol. The tensor coefficient $C_T$ is the weight of three angular
directions. The vector block has the algebraic form of the Maxwell
symbol in one time and two spatial directions and gives
$C_V=\Delta/K$. Eliminating the constrained scalar amplitudes gives
\begin{equation}
 (n-4)C_T+(n-2)C_S
 =2(n-3)C_V=\frac{2(n-3)\Delta}{K}
 \label{sym:dependence}
\end{equation}
This relation holds for every spherically symmetric metric with
$K\ne0$; the field equations are not used. The calculation thus
separates into a geometric identity and a restriction to actual
scalar-collapse solutions. For the latter, the background equations
reduce the angular coefficients to functions of $K$ and the
boost-invariant $j$. These two local quantities suffice to test all
three sectors. In every dimension $n\ge5$,
\begin{equation}
 \begin{aligned}
 4(n-3)(n-4)K^3C_T&=4j^2-4(n-2)K^2j+(n-5)(n-7)K^4\\
   &\quad+2(n-1)(2n-7)K^2-(n-1)^2\\
 4(n-2)(n-3)K^3C_S&=-12j^2+4(n-2)K^2j+(n-3)(n-5)K^4\\
   &\quad-6(n-1)K^2+3(n-1)^2
 \end{aligned}
 \label{sym:general}
\end{equation}
The Cao--Wu tensor equation \cite{Cao2021} gives the same tensor
coefficient. In five dimensions,
\begin{equation}
 \begin{aligned}
 \Delta&=\frac{4-j^2}{4K^2},&
 C_T&=\frac{(2-j)(3K^2-j-2)}{2K^3}\\
 C_V&=\frac{4-j^2}{4K^3},&
 C_S&=\frac{(2-j)(j+2-K^2)}{2K^3}
 \end{aligned}
 \label{sym:five}
\end{equation}
Written in the orbit eigenvalues, the part of $C_T$ not proportional
to $\lambda_-$ carries the factor $n-5$ in Eq.~\eqref{app:lambdaform},
so in five dimensions each of $\Delta,C_T,C_V,C_S$ contains the factor
$2-j=2K\lambda_-$. In six dimensions,
\begin{equation}
 \begin{aligned}
 \Delta_6&=\frac{(K^2+5)^2}{4}-j^2\\
 T_6&=4j^2-16K^2j-K^4+50K^2-25\\
 S_6&=3K^4+16K^2j-30K^2-12j^2+75,
 \end{aligned}
 \qquad
 \begin{aligned}
 \Delta&=\frac{\Delta_6}{9K^2},&
 C_V&=\frac{\Delta_6}{9K^3}\\
 C_T&=\frac{T_6}{24K^3},&
 C_S&=\frac{S_6}{48K^3}
 \end{aligned}
 \label{sym:six}
\end{equation}

Since $K>0$, the effective metrics are nondegenerate and Lorentzian,
with a timelike covector in common with matter, exactly when
\begin{equation}
 C_T>0,\qquad C_S>0
 \label{sym:conditions}
\end{equation}
Equation~\eqref{sym:dependence} then gives $C_V,\Delta>0$.
Appendix~\ref{app:symbol} derives these conditions by separating
the trace-free angular amplitudes from their trace. We call
Eq.~\eqref{sym:conditions} the characteristic conditions, a
local state $(K,j)$ admissible when they hold, and the set of
admissible states the admissible region (Fig.~\ref{fig:region} shows
it for $n=5$ and nonnegative mass).

Since $C_I$ multiplies the angular part of the wave
covector, $C_I=0$ means that waves of sector $I$ have zero phase speed
in the angular directions. The common block $(X,Y,H)$ describes propagation in the
time and radial directions; on the closure of the admissible region
its determinant $\Delta$ can only vanish together with all three
angular coefficients, as shown below. The conditions depend on the
scalar field only through $j$ because $C_T$, $C_S$ and $\Delta$ are
invariant under boosts in the time--radial plane, and $j$ is the only
boost invariant of the scalar gradient; the individual weights $X$,
$Y$ and $H$ do depend on the frame. Timelike and spacelike gradients
restrict propagation in different ways. In five dimensions a timelike
gradient reaching $j=2$ degenerates the common time--radial block and
with it all three sectors; the scalar gravitational sector degenerates at $j=K^2-2$, where
$C_S=0$. For $j<2$,
this coefficient is negative below that boundary (for $K^2<2$ the boundary requires a spacelike
gradient, $j<0$). For $1\le K^2<4/3$, increasing a timelike gradient
instead reaches $C_T=0$ at $j=3K^2-2<2$
(Fig.~\ref{fig:region}).

On the closure of the admissible region, $\Delta=0$ forces
$C_T=C_V=C_S=0$. By continuity, $C_T$ and $C_S$ are nonnegative on
the closure, and so are the terms $(n-4)C_T$ and $(n-2)C_S$ in
Eq.~\eqref{sym:dependence}. Their sum is $2(n-3)\Delta/K$, so where
$\Delta=0$ both vanish, and $C_V=\Delta/K$ vanishes with them. A
degeneration of the common time--radial block also makes all three
effective metrics degenerate.

By Eq.~\eqref{sym:base}, $\Delta$ vanishes exactly at $j=b_0$ and at
$j=-b_0$, with $b_0=[n-1+(n-5)K^2]/2$ from Eq.~\eqref{def:jb0}. At
$j=b_0$, Eq.~\eqref{sym:general} gives
\begin{equation}
 (n-3)(n-4)KC_T=(n-5)(n-1-2K^2)
 \label{sym:ct-plus}
\end{equation}
In five dimensions $b_0=2$ and the right side vanishes, so all
sectors degenerate along $j=2$. In six dimensions the right side is
$5-2K^2$, and this simultaneous degeneration occurs only at
$K^2=5/2$, $j=b_0=15/4$. At the other six-dimensional states with
$j=b_0$, $C_T\ne0$ while $\Delta=0$, so by the preceding paragraph
they lie outside the closure.

The other zero of $\Delta$, $j=-b_0$, lies outside the closure. There
Eq.~\eqref{sym:general} gives
\begin{equation}
 (n-3)(n-4)KC_T=(n-4)(n-5)K^2+(n-1)(2n-7)>0
 \label{sym:ct-minus}
\end{equation}
and Eq.~\eqref{sym:dependence} with $\Delta=0$ then gives
$(n-2)C_S=-(n-4)C_T<0$.

Along a regular evolution on the GR branch that starts with strict
inequalities, the first zero of $C_T$ or $C_S$ marks the degeneration
of an effective metric. We refer to this event as characteristic loss.

When the inequalities hold, the covector dual to $e_0$ is timelike
for matter and all gravitational sectors, since $X>0$ and the form in
Eq.~\eqref{sym:factors} reduces to $-X\omega^2$ on this covector. A
negative $C_I$ can preserve Lorentzian signature while having no
timelike covector in common with matter, so signature alone is not
enough. For $\Delta>0$ and $C_I<0$ the form has signature
$(+-\cdots-)$. Its timelike covectors are then those on which it is
positive, while it is negative on every covector that is timelike for
matter (see Eq.~\eqref{app:commoncone}).

\paragraph{Regular centers.}
At a regular center the scalar gradient has no
radial part, $Q=0$, and the curvature is invariant under spatial
rotations. The field equations then force
\begin{equation}
 \begin{gathered}
 j=\frac{n-1}{2}(K^2-1),\qquad X=\lambda_+=K,\qquad H=0\\
 Y=C_T=C_V=C_S=\lambda_-=\frac{n-1-2K^2}{(n-3)K}
 \end{gathered}
 \label{sym:center}
\end{equation}
(Appendix~\ref{app:symbol}). All sectors propagate isotropically, with
the lower orbit eigenvalue as their common coefficient, and a regular
center is admissible exactly when $K^2<(n-1)/2$. A characteristic loss
at a regular center therefore makes every gravitational effective
metric degenerate at the same event, at $K^2=(n-1)/2$ and $j=b_0$. Flat FLRW
solutions satisfy Eq.~\eqref{sym:center} everywhere in the comoving
frame. The common cone can also be read directly from the weighted
symbol. At an isotropic
point every spatial plane has the same sectional curvature and every
time--space plane has the same sectional curvature, so only two
distinct weights survive in Eq.~\eqref{sym:weights}: $X=K$ on the
triples containing $e_0$ and $C_T=Y$ on the purely spatial triples
(Table~\ref{tab:weights}).
The sector coefficients of Appendix~\ref{app:symbol} then assign the
same angular coefficient $Y$ to all three sectors. The relation
between $j$ and $K$ holds because
at a regular center $Q=0$, so $j$ is proportional to the local energy
density, and $K^2-1$, which measures the mass inside a small sphere
relative to its radius, is determined by the same energy density.

Completing the square in Eq.~\eqref{sym:factors} gives positive
radial and angular coefficients $\Delta/X$ and $C_I$, hence real
frequencies in every spatial direction. Real frequencies alone do
not give a well-posed Cauchy problem: the corresponding modes must
also span the perturbations without becoming arbitrarily ill
conditioned. Otherwise a small change in the fields need not remain
small under evolution.

{\begin{proposition}[Characteristics and local evolution]
\label{prop:local-evolution}
Let $(g,\Psi)$ be a smooth spherical solution of
Eq.~\eqref{eq:field} in dimension $n\ge5$, on the GR branch $K>0$.
Its physical gravitational characteristics are the three effective
metrics in Eq.~\eqref{sym:factors}; the matter scalar has the
characteristics of $g$. The gravitational effective metrics are
nondegenerate and Lorentzian, with a timelike covector in common with
matter, precisely when $C_T>0$ and $C_S>0$.

At any point satisfying these strict inequalities, there is a
local foliation spacelike for matter and all gravitational sectors,
and a modified harmonic gauge with separated auxiliary cones, for
which the full equations are strongly hyperbolic near the reference
solution.

Consider initial data for the metric and scalar on a relatively
compact part of such a slice. They must satisfy the EGB constraints
and the initial conditions of the chosen modified harmonic gauge.
If they are sufficiently close to the reference data in the Sobolev
norms specified in Appendix~\ref{app:hyperbolic}, they admit a unique
local solution on a smaller patch, with the regularity stated there.
The solution depends continuously on the data in those norms and is
locally unique up to diffeomorphism. The data need not be spherical,
and weak coupling is not assumed.
\end{proposition}

Appendix~\ref{app:symbol} gives the characteristic calculation;
Appendix~\ref{app:hyperbolic} constructs the symmetrizer, fixes the
foliation and gauge, and proves the evolution statement. The
neighborhood and existence time depend on the reference solution;
no uniform time is asserted as $C_T$ or $C_S$ tends to zero. This is
a local result for the specified formulation, not a claim that every
gauge or slicing is strongly hyperbolic, or that general nonspherical
backgrounds admit the same effective-metric factorization.}

\section{Five-dimensional bounds and continuation}
\label{sec:radius}

This section draws the consequences of the characteristic conditions
in five dimensions, the lowest dimension in which the Gauss--Bonnet
term is dynamical and, of the two dimensions studied here, the one
in which the characteristic conditions are more restrictive. The
conditions are inequalities between $K$ and $j$, and $K$ is fixed
by the mass inside a sphere and its radius. Requiring that
gravitational waves still propagate at a marginal sphere therefore
restricts the size and mass of that sphere; requiring it at every
point restricts the Kretschmann invariant. Section~\ref{sec:marginal} derives the
restrictions on marginal spheres. Section~\ref{sec:curvature-continuation}
derives the curvature bound and gives a separate continuation theorem
on finite conformal domains. Within those domains, the first
characteristic loss occurs on a smooth spherical background.

\subsection{Marginal spheres}
\label{sec:marginal}

In five dimensions the three gravitational effective metrics have a
single form. Set
\begin{equation}
 \varepsilon=1-\frac j2,\qquad
 \kappa_T=\frac{3K^2-j-2}{K^2},\qquad
 \kappa_V=\frac{2+j}{2K^2},\qquad
 \kappa_S=\frac{j+2-K^2}{K^2}
 \label{form:kappas}
\end{equation}
Since $b_0=2$ for $n=5$, the orbit eigenvalues of
Eq.~\eqref{sym:orbit} are $\lambda_\pm=(2\pm j)/(2K)$, so that
\begin{equation}
 K\lambda_-=\varepsilon,\qquad K\lambda_+=2-\varepsilon
 \label{form:orbit-eigenvalues}
\end{equation}
Equations~\eqref{sym:orbit} and \eqref{sym:five} then combine into
\begin{equation}
 K\,G_I^{ab}=\varepsilon\bigl(g_{\rm orb}^{ab}+\kappa_I\,g_{\rm sph}^{ab}\bigr)
 -\tilde\alpha kG\,\Psi^a\Psi^b,\qquad I=T,V,S
 \label{form:five-metrics}
\end{equation}
where $g_{\rm orb}^{ab}$ and $g_{\rm sph}^{ab}$ are the orbit and sphere
parts of $g^{ab}$, and $j=-\tilde\alpha kG\,\Psi_a\Psi^a$.

The three sectors share the orbit part and the gradient term, and
differ only by the factor $\kappa_I$ on the spheres. The orbit block
of Eq.~\eqref{sym:orbit}, multiplied by $K$, supplies the shared
terms, since in $KG_{\rm orb}^{AB}$ the coefficient of $g^{AB}$ is
$K\lambda_-=\varepsilon$ and that of $\nabla^A\Psi\nabla^B\Psi$ is
$-\tilde\alpha kG$. The angular term $C_I\xi_\perp^2$ of
Eq.~\eqref{sym:factors} gives the sphere part $KC_I\,g_{\rm sph}^{ab}$
of $KG_I^{ab}$, and by Eqs.~\eqref{sym:five} and \eqref{form:kappas},
\begin{equation}
 KC_I=\varepsilon\kappa_I,\qquad
 \kappa_T+\kappa_S=2,\qquad
 \kappa_V=\frac{1+\kappa_S}{2}
 \label{form:kappa-relations}
\end{equation}

Since $K>0$, Eq.~\eqref{form:kappa-relations} turns the conditions in
Eq.~\eqref{sym:conditions} into $\varepsilon\kappa_T>0$ and
$\varepsilon\kappa_S>0$. The factors $\kappa_T$ and $\kappa_S$ add up
to $2$ and cannot both be negative, so these
inequalities hold exactly when $\varepsilon$, $\kappa_T$ and
$\kappa_S$ are all positive. Positivity of $\kappa_S$ means
$K^2<j+2$ and implies $j>-2$. In terms of $K$ and $j$, the
characteristic conditions reduce to
\begin{equation}
 |j|<2,\qquad K^2<j+2,\qquad 3K^2>j+2
 \label{form:five-conditions}
\end{equation}
that is, to
\begin{equation}
 0<\varepsilon<2,\qquad 0<\kappa_S<2
 \label{form:open-square}
\end{equation}

The two variables $(\varepsilon,\kappa_S)$ therefore give coordinates
on an open square. They determine $K$ and $j$ uniquely through
\begin{equation}
 j=2(1-\varepsilon),\qquad K^2=\frac{2(2-\varepsilon)}{1+\kappa_S}
 \label{form:square}
\end{equation}
and every point of the square corresponds to an admissible state,
since the second relation gives $K^2>0$ there. By
Eqs.~\eqref{form:orbit-eigenvalues} and \eqref{form:kappa-relations},
the edges $\varepsilon=0$, $\kappa_S=0$ and $\kappa_S=2$ correspond
to simultaneous degeneration at $j=2$, degeneration of the scalar gravitational sector and degeneration of the tensor sector;
the fourth edge, $\varepsilon=2$, is the singular limit $K\to0$.

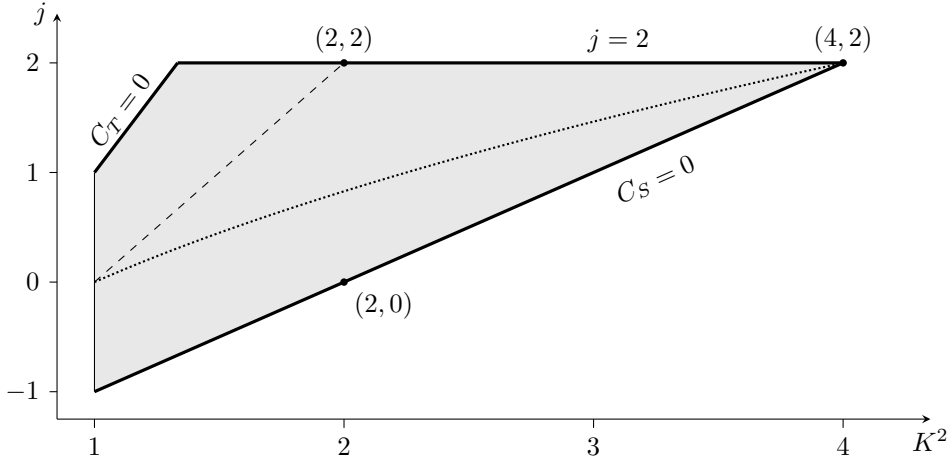
\begin{figure}[htbp]
 \centering
 \begin{tikzpicture}[x=3.3cm,y=1.45cm,>=stealth,font=\small]
  \fill[black!9] (1,-1) -- (4,2) -- ({4/3},2) -- (1,1) -- cycle;
  \draw[->] (0.85,-1.25) -- (4.35,-1.25) node[below] {$K^2$};
  \draw[->] (0.85,-1.25) -- (0.85,2.45) node[left] {$j$};
  \foreach \k in {1,2,3,4} \draw (\k,-1.25) -- (\k,-1.33) node[below] {$\k$};
  \foreach \jj in {-1,0,1,2} \draw (0.85,\jj) -- (0.82,\jj) node[left] {$\jj$};
  \draw[very thick] (1,-1) -- (4,2);
  \draw[very thick] (1,1) -- ({4/3},2);
  \draw[very thick] ({4/3},2) -- (4,2);
  \draw (1,-1) -- (1,1);
  \draw[dashed] (1,0) -- (2,2);
  \draw[densely dotted,thick] plot[domain=1:2,samples=60] ({\x*\x},{2*\x-2});
  \fill (2,0) circle (1.4pt);
  \fill (2,2) circle (1.4pt);
  \fill (4,2) circle (1.4pt);
  \node[below right] at (2,0) {$(2,0)$};
  \node[above] at (2,2) {$(2,2)$};
  \node[above] at (4,2) {$(4,2)$};
  \node[rotate=24] at (3.25,0.93) {$C_S=0$};
  \node[rotate=53,above,inner sep=2pt] at (1.16,1.47) {$C_T=0$};
  \node[above] at (3.1,2) {$j=2$};
 \end{tikzpicture}
 \caption{The five-dimensional admissible region for nonnegative
 mass, shown in the $(K^2,j)$ plane. Its boundaries are degeneration
 of the scalar gravitational sector ($C_S=0$, $\kappa_S=0$),
 degeneration of the tensor sector ($C_T=0$, $\kappa_S=2$),
 simultaneous degeneration at $j=2$ ($\varepsilon=0$), and the
 zero-mass segment $K=1$. Regular central values lie on the dashed
 segment $j=2(K^2-1)$, which leaves the region at $(2,2)$.
 The dotted curve $j=2(K-1)$ maximizes the angular scalar-sector
 speed in the scalar rest frame at fixed $K$. It ends at $(4,2)$, which marginal spheres approach as
 $R_h^2\to\tilde\alpha^+$. The dimensionless Kretschmann scalar
 then tends to $28$.
 On the vacuum line $j=0$, the scalar gravitational coefficient $C_S$ vanishes at $(2,0)$.}
 \label{fig:region}
\end{figure}

The conditions are linear in $(K^2,j)$, and $K^2$ is affine in the
mass at fixed radius. Figure~\ref{fig:region} shows the admissible
region for $K\ge1$, the case of nonnegative mass. Each plotted point
represents the local fields at a spacetime event. Moving to larger
$K^2$ means increasing $\mathcal M/R^4$, whereas changing $j$ changes
the balance between the scalar's temporal and radial derivatives.
The diagram is therefore a space of local states, not a spacetime
diagram; an evolution determines its own path through it.

The closure is the quadrilateral with vertices $(K^2,j)=(1,-1)$,
$(1,1)$, $(4/3,2)$ and $(4,2)$. On its left edge, $K=1$, the local
mass vanishes; this is not a boundary at which a characteristic metric degenerates. On either sloping edge the
angular coefficient of one sector vanishes, while at $j=2$ the common
orbit block and all three angular coefficients degenerate. Lines of
constant $\kappa_S$ are the rays $j+2=(1+\kappa_S)K^2$.

On the line $j=0$ the conditions reduce, for $K\ge1$, to the static vacuum
condition $K^2<2$ \cite{Reall2014}; a timelike scalar gradient,
$j>0$, raises the allowed $K$. At fixed $K$ the admissible values of
$j$ form the interval
\begin{equation}
 K^2-2<j<\min\bigl(2,\,3K^2-2\bigr)
 \label{form:d1}
\end{equation}
which is empty for $K^2\ge4$. Thus $K^2<4$ throughout the admissible
region, as also follows from Eq.~\eqref{form:square}, and $K\to2$ only
at the corner $(K^2,j)=(4,2)$, where $\varepsilon=\kappa_S=0$.

\begin{samepage}
This upper limit on $K$ gives the following local geometric bound.

\begin{proposition}[Radius and mass of a marginal sphere]
\label{prop:marginal-bounds}
Consider a smooth five-dimensional spherical solution of
Eq.~\eqref{eq:field} on the GR branch. At every regular marginal
sphere with $F=1$ and $C_T,C_S>0$, the areal radius and generalized
Misner--Sharp mass satisfy
\begin{equation}
 K_h=1+\frac{\tilde\alpha}{R_h^2}<2,\qquad
 R_h^2>\tilde\alpha,\qquad
 2kG\mathcal M_h=R_h^2+\frac{\tilde\alpha}{2}>\frac{3\tilde\alpha}{2}
 \label{form:radius-bound}
\end{equation}
Equality is excluded by characteristic degeneration. No
assumption on the data away from the sphere is required.
\end{proposition}
\end{samepage}

Indeed, $F=1$ fixes $K_h=1+\tilde\alpha/R_h^2$, while
Eq.~\eqref{form:d1} requires $K_h<2$; the mass follows from
Eq.~\eqref{def:mass}. This is a bound on regular marginal spheres,
including any that form later in an evolution, and does not by itself
identify an event horizon. A marginal sphere with $R_h\le\sqrt{\tilde\alpha}$ can
therefore exist only where at least one of the strict
characteristic inequalities fails.

Without the characteristic conditions, the mass relation alone gives
only the weaker spherical mass bound $\mathcal M_h>\tilde\alpha/(4kG)$
\cite{Deppe2012,Nozawa2008}; the radius limit $R_h^2=\tilde\alpha$ in
Ref.~\cite{Nozawa2008} concerns negatively curved horizons. The bound
in Eq.~\eqref{form:radius-bound} is only a necessary condition: an
allowed pair $(K,j)$ at one sphere need not extend to complete
constrained data, still less arise from initially untrapped data.
Section~\ref{sec:formation} addresses these separate requirements and
shows that the bound can be approached.

In six dimensions the admissible values of $j$ form a nonempty
interval for every $K\ge1$; for example,
$j=(K-1)(6-K)/8$ satisfies Eq.~\eqref{sym:conditions} for all $K\ge1$
(Appendix~\ref{app:admissibility}). No radius bound arises there.

The approach to the bound also restricts propagation speeds. At a
marginal sphere satisfying Proposition~\ref{prop:marginal-bounds},
assume in addition that the scalar gradient is timelike, $j_h>0$, and let $v_{\min}$ be the smallest phase speed
over all directions and polarizations in the scalar rest frame. There
the squared radial phase speed is $v_R^2=\varepsilon/(2-\varepsilon)$ and the
squared angular phase speeds are $v_I^2=\kappa_Iv_R^2$, by
Eq.~\eqref{form:five-metrics}. Appendix~\ref{app:speed} gives the
sharp bound
\begin{equation}
 v_{\min,h}\le
 \frac{R_h^2-\tilde\alpha}{R_h^2+\tilde\alpha}
 \label{form:speed-bound}
\end{equation}
The timelike condition follows from Eq.~\eqref{form:five-conditions}
when $\tilde\alpha<R_h^2\le(1+\sqrt2)\tilde\alpha$.
Thus at least one speed tends to zero as $R_h\to\sqrt{\tilde\alpha}^+$.

In a fixed mass interval, the same characteristic conditions require
nonzero stress-energy of the scalar field at the marginal sphere.
Assume the hypotheses of Proposition~\ref{prop:marginal-bounds} and,
in addition, a spacelike exterior slice extending from that sphere to
a standard asymptotically flat end, with $F<1$ and $R$ increasing
outward throughout the exterior. The canonical scalar obeys the dominant
energy condition, so mass monotonicity in the untrapped exterior gives
$\mathcal M_h\le\mathcal M_\infty$ \cite{Maeda2008,Nozawa2008}. Consequently
\begin{equation}
 j_h>
 \left(1+\frac{\tilde\alpha}{2kG\mathcal M_\infty-\tilde\alpha/2}\right)^2-2>0,
 \qquad
 \frac{3\tilde\alpha}{2}<2kG\mathcal M_\infty<
           \left(\frac32+\sqrt2\right)\tilde\alpha
 \label{form:vacuum-obstruction}
\end{equation}
For each mass in this interval, $j_h$ has a positive lower bound.
The static vacuum restriction is
$R_h^2>(1+\sqrt2)\tilde\alpha$
\cite{Reall2014,Beroiz2007}. The scalar stress-energy tensor at the marginal sphere therefore cannot
tend to zero in this mass range while the characteristic conditions hold.
Marginal spheres with
$\tilde\alpha<R_h^2<(1+\sqrt2)\tilde\alpha$, between the bound with
matter and the static vacuum bound, require a timelike scalar gradient
at the sphere, since $j_h>K_h^2-2>0$ there.

\subsection{Curvature and continuation}
\label{sec:curvature-continuation}

The characteristic conditions also bound the Kretschmann
invariant in five dimensions. The bound uses the weights in
Eq.~\eqref{sym:weights-five}; smooth continuation requires the
additional spherical PDE estimates stated below. The invariant bound
alone does not bound every orthonormal curvature component, since
Lorentzian contractions can contain cancellations.

In a frame diagonalizing the curvature, the Kretschmann scalar is
four times the sum of the squared sectional curvatures of the ten
planes spanned by pairs of frame vectors. By Eq.~\eqref{sym:weights-five},
$\tilde\alpha^2$ times it is therefore four times the sum of the
squared deviations of the ten triple weights from one. On a spherical
background three triples carry $K$, three pairs of triples carry the
orbit block, and one triple carries $C_T$ (Table~\ref{tab:weights}). In a general frame this
becomes
\begin{equation}
 \tilde\alpha^2\mathcal R_{abcd}\mathcal R^{abcd}
 =4(1-C_T)^2+12\,{\rm tr}\bigl[(G_{\rm orb}-\mathsf 1)^2\bigr]+12(K-1)^2
 \label{cont:deviation}
\end{equation}
where $G_{\rm orb}$ is the mixed block $G_{\rm orb}{}^A{}_B$,
$\mathsf 1$ is its identity matrix, and
\begin{equation}
 {\rm tr}\bigl[(G_{\rm orb}-\mathsf 1)^2\bigr]=(X-1)^2+(Y-1)^2-2H^2
 \label{form:d2}
\end{equation}
is its boost-invariant square. This trace is indefinite in
components, and it does not control the individual deviations: a null
scalar gradient at $K=1$ moves $X$, $Y$ and $H$ away from their
Einstein values while the trace vanishes.

On solutions the trace is fixed by $K$ and $j$. It equals
$(\lambda_+-1)^2+(\lambda_--1)^2=[2(K-1)^2+j^2/2]/K^2$, so on a
five-dimensional spherical scalar background
\begin{equation}
 \tilde\alpha^2\mathcal R_{abcd}\mathcal R^{abcd}
 =4(1-C_T)^2+
   \frac{24(K-1)^2+6j^2}{K^2}+12(K-1)^2
 \label{cont:curvature-identity}
\end{equation}
\begin{proposition}[Kretschmann invariant]
\label{prop:kretschmann}
At a point of a smooth five-dimensional spherical solution of
Eq.~\eqref{eq:field}, assume nonnegative generalized Misner--Sharp
mass, equivalently $K\ge1$ on the GR branch, and strict
characteristic inequalities $C_T,C_S>0$. Then
\begin{equation}
 0\le\tilde\alpha^2\mathcal R_{abcd}\mathcal R^{abcd}<28
 \label{cont:curvature-bounds}
\end{equation}
The upper bound is sharp as a supremum.
\end{proposition}

The estimate in Eq.~\eqref{cont:d7} proves the upper bound and
quantifies its strictness.

The number $28$ is the supremum of the right-hand side of
Eq.~\eqref{cont:deviation} over the admissible states with $K\ge1$,
approached
only at the corner $(K^2,j)=(4,2)$. There the weights
$(K,\lambda_+,\lambda_-,C_T)=(2,1,0,0)$ have multiplicities
$(3,3,3,1)$, the triple counts of Table~\ref{tab:weights}, giving
$4(3+3+1)=28$ in Eq.~\eqref{cont:deviation}. In
physical units, the square root of the Kretschmann scalar,
$(\mathcal R_{abcd}\mathcal R^{abcd})^{1/2}$, stays below
$2\sqrt7/\tilde\alpha$, a fixed multiple
of the inverse square of the Gauss--Bonnet length $\sqrt{\tilde\alpha}$,
while every sector propagates. Every admissible sequence of marginal spheres with
$R_h^2\to\tilde\alpha^+$ approaches the corner. The data constructed in
Section~\ref{sec:formation} form marginal spheres at every radius
above the bound, so the constant $28$ is sharp.

The second consequence concerns the spherical evolution itself. In
spherical symmetry there are no propagating gravitational degrees of
freedom, so the degeneration of a gravitational propagation cone need
not stop the spherical solution: the spherical equations for the radius, conformal factor and scalar
field in Eq.~\eqref{cont:conformal-equations} form a closed system, and the coefficients that
vanish at a characteristic loss never appear in them as divisors.
{\begin{theorem}[Spherical continuation on a finite domain]
\label{thm:spherical-continuation}
Consider smooth spherically symmetric initial data for the
five-dimensional EGB--scalar equations, satisfying the constraints
on the GR branch. Assume a regular center and $F<1$ on the initial
slice, with areal radius increasing outward. In the conformal coordinates of Eq.~\eqref{evol:metric}, choose
$R(0,x)=x$ and reflect across the center, with signed $R$ odd and the
scalar and conformal factor even. For a fixed finite $v_b>0$, set
\begin{equation}
 \mathscr D_{v_b}=\{0\le\tau\le v_b,\ |x|\le v_b-\tau\},
 \qquad u=\tau-x,\quad v=\tau+x.
 \label{cont:main-domain}
\end{equation}
Assume that the development is smooth in a neighborhood of the
closed initial interval $[-v_b,v_b]$. If $K$ remains bounded on the
smooth part of $\mathscr D_{v_b}$, the spherical solution extends
smoothly throughout this domain, including the axis and terminal
vertex. A bound on the Kretschmann invariant suffices in place of
the bound on $K$. No separate bound on $j$ or positivity of
$C_T,C_S$ is required.
\end{theorem}

Nonnegative mass, and hence $K\ge1$, follows from the initial
assumptions, the mass constraint in Eq.~\eqref{evol:generalconstraints},
and the propagation identities in Eq.~\eqref{cont:mass-monotonicity}. Completeness and asymptotic
flatness are not hypotheses of this finite-domain result. The proof
is in Appendix~\ref{app:curvature-continuation}, beginning with
Eq.~\eqref{cont:domain}; Eq.~\eqref{cont:curvature-identity} gives the
invariant criterion. This is
more than an algebraic curvature bound. The spherical equations must
also control the individual scalar derivatives, for which the
invariant $j$ alone would allow cancellations. Null integration away
from the center and a maximum principle for the normalized scalar
derivatives near the regular center supply that control; the regular auxiliary system in Eq.~\eqref{cont:auxiliary-equations}
then gives the higher-derivative bounds needed for extension.}

Since Eq.~\eqref{form:five-conditions} gives $K<2$ while the
characteristic conditions hold, a first characteristic loss in the stated domains, at
a finite value of the conformal time $\tau$ of Eq.~\eqref{evol:metric},
occurs on a smooth spherical solution which continues briefly beyond
it. At the center, Eq.~\eqref{cont:central-first} bounds the normalized
scalar derivatives near the endpoint in terms of the supremum of $j$
along the axis, defined in Eq.~\eqref{cont:d2}. Failure of
smooth extension through a central endpoint therefore requires
unbounded $j$ along the axis, where
$j=2(K^2-1)$. If $j<2$ at an earlier axis point, such a breakdown must
therefore be preceded by simultaneous degeneration at $(K^2,j)=(2,2)$.
The spherical equations control individual scalar derivatives and
higher derivatives even when the strict characteristic conditions fail.
This extension concerns exactly spherical solutions. It does not
restore the well-posed nonspherical evolution established under
strict positivity in Section~\ref{sec:characteristics}; those
perturbations retain the angular modes absent from the spherical
system. Section~\ref{sec:tidal-prediction} examines what such modes
can do after characteristic degeneration.

\begin{corollary}[Characteristic degeneration after trapping]
\label{cor:later-degeneration}
In the setting of Theorem~\ref{thm:spherical-continuation},
suppose a future-trapped sphere occurs at
$(u_{\rm tr},v_{\rm tr})$, and the initial regularity assumptions
cover the closed base of $\mathscr D_{v_{\rm tr}}$.
If Eq.~\eqref{form:five-conditions} holds throughout the evolution in
this domain up to and including the slice containing that sphere,
a characteristic condition must fail at a later finite conformal
time somewhere in $\mathscr D_{v_{\rm tr}}$. The first failure occurs
on a smooth spherical background. The same conclusion applies to a
future marginal sphere by passing to a nearby future-trapped point
on its ingoing null ray, with the corresponding initial base covered
by the same regularity assumptions.
\end{corollary}

The marginal sphere is followed by a future-trapped sphere.
In the conformal null coordinates $u,v$ of
Appendix~\ref{app:evolution}, an ingoing ray has constant $v$.
The five-dimensional radius equation, Eq.~\eqref{cont:conformal-equations}, gives $R_{,uv}\le0$,
with strict inequality where $K>1$, as on the marginal sphere, so
$R_{,v}$ remains negative along the future ingoing ray from the trapped sphere. If the characteristic
conditions persisted, continuation would extend this ray to a regular
center, where $R_{,v}>0$, a contradiction. This is the argument of Appendix~\ref{app:curvature-continuation}.
It does not specify where the condition first fails or provide a
universal bound in proper time. For narrow incoming pulses,
Paper~II determines the location. The conclusion concerns the
specified spherical domain; it does not assert continuation for
general nonspherical data.
\section{Formation from complete smooth data}
\label{sec:formation}

The bounds of Section~\ref{sec:radius} are necessary conditions: they
say what a marginal sphere cannot be while gravitational waves
propagate, but not whether spheres near the bound occur in the
collapse of regular data. We show that marginal spheres forming
dynamically from untrapped data come arbitrarily close to the bound, so that the
constants in it are the best possible. The construction first
solves the constraints on a complete slice containing a marginal
sphere, then shows that this slice lies in a smooth evolution with no
marginal sphere immediately beforehand. The second step is needed to
obtain formation rather than merely an initial apparent horizon.

\begin{theorem}[Formation at every allowed radius]
\label{thm:formation}
Fix $\tilde\alpha>0$ and $n\in\{5,6\}$. For each radius
parameter and mass excess specified below, there are smooth spherical
initial data on $\R^{n-1}$ satisfying the constraints on the GR branch.
Their initial slice is complete, with a regular center and one
asymptotically flat end. A short evolution
from an untrapped initial slice first forms a future marginal sphere
in the chosen spherical foliation. The strict inequalities
$C_T,C_S>0$ hold throughout a time interval containing formation.

For $n=5$, every $\rho_h>1$ and $\eta_{\mathcal M}>0$ can be
realized with
\begin{equation}
 R_h=\rho_h\sqrt{\tilde\alpha},\qquad
 \tilde\alpha\left(\rho_h^2+\frac12\right)<2kG\mathcal M_\infty
 <\tilde\alpha\left(\rho_h^2+\frac12+\eta_{\mathcal M}\right)
 \label{form:five-existence}
\end{equation}
At formation, the minimum gravitational phase speed in the
scalar rest frame equals the upper bound in
Eq.~\eqref{form:speed-bound}. For $n=6$, every $\rho_h>0$ and
$\eta_{\mathcal M}>0$ can be realized with
\begin{equation}
 R_h=\rho_h\sqrt{\tilde\alpha},\quad \rho_h>0,\qquad
 \tilde\alpha^{3/2}\left(\rho_h^3+\frac{\rho_h}2\right)<2kG\mathcal M_\infty
 <\tilde\alpha^{3/2}\left(\rho_h^3+\frac{\rho_h}2+\eta_{\mathcal M}\right)
 \label{form:six-existence}
\end{equation}
\end{theorem}

{Appendices~\ref{app:evolution}, \ref{app:admissibility} and
\ref{app:formation-construction} give the local evolution, the
crossing of the marginal condition and the complete data,
respectively. The data are chosen separately for each radius and mass excess.
The result concerns a short interval beginning just before formation
in the chosen foliation. It establishes existence without determining
generic collapse or thresholds for a prescribed family. An event
horizon and an existence time uniform in the limiting parameters are
not established.}

\begin{proposition}[Strict ADM mass bound]
\label{prop:adm-mass}
Let $n=5$ or $n=6$ and take smooth spherically symmetric initial
data satisfying the constraints on the GR branch. Assume that the
slice is complete, with a regular center and one asymptotically flat
end, and that $F\le1$ and the areal radius increases outward. If it
contains a marginal sphere of radius $R_h$, then
\begin{equation}
 2kG\mathcal M_\infty>R_h^{n-3}+\frac{\tilde\alpha}2R_h^{n-5}
 \label{form:mass-bound}
\end{equation}
Characteristic positivity is not needed for this mass inequality.
\end{proposition}

The positive mass-excess integral in Eq.~\eqref{form:mass-integral}
gives this inequality. For the radii admitted by Theorem~\ref{thm:formation}, its
construction approaches this strict lower bound. With Eq.~\eqref{form:radius-bound}, the infimum of $2kG\mathcal M_\infty$
over all five-dimensional data of this kind whose evolution forms a
marginal sphere with $C_T,C_S>0$ at formation is
$3\tilde\alpha/2$ and is not attained. In six dimensions it is zero:
there is no positive mass gap.

The data are specified by prescribing the slice geometry through its mass, $j$, and the
tilt $\varpi=-e_0R\ge0$; $\varpi/R$ is the angular
extrinsic curvature, so $\varpi$ measures how fast the spheres of the
slice are contracting. Eq.~\eqref{form:scalar-inversion} then reconstructs the scalar
derivatives algebraically. The sufficient reconstruction conditions are
listed in Appendix~\ref{app:realization}, in the paragraph
``Complete initial data.'' We use an exact FLRW or flat core and a
static exterior. On the slice, the mass is nondecreasing, $F\le1$,
and $\varpi>0$ wherever $F=1$.

For the scalar derivatives to be real, we impose strict positivity
of the discriminant in Eq.~\eqref{form:tilt-identity} in the
intervening matter region where $\mathcal J_m>0$. At each boundary
of the matter support, we use Eq.~\eqref{form:edge} to ensure
smoothness: the two scalar derivatives share a signed smooth factor
$\upsilon$. Replacing this factor by its absolute value
could destroy smoothness. These are conditions used to
construct the data, rather than extra assumptions on an arbitrary
collapse solution.
Increasing $\varpi$ makes the remaining discriminant positive where
needed, by Eq.~\eqref{form:tilt-identity}. It changes the slice's
orientation relative to radial null directions while leaving $F$,
$j$ and the characteristic conditions unchanged.

At $R_h$, the profile of $2kG\mathcal M$ defined by
Eq.~\eqref{form:tangent-profile}, using the tangent line in
Eq.~\eqref{form:d8}, is tangent to
$R^{n-3}+\tilde\alpha R^{n-5}/2$, giving $F$ a single
nondegenerate maximum. A sphere is marginal exactly when
$2kG\mathcal M$ equals this function of its radius, and trapped when
$2kG\mathcal M$ exceeds it. The prescribed profile lies on or below the tangent line
near contact. Since that line lies below the strictly convex curve
away from contact, the sphere at $R_h$ is marginal and every
neighboring sphere on the slice is untrapped. Outside the FLRW core the data use $j=2(K-1)$ for
$n=5$ and $j=z(5-z)/8$, $K=1+z$, for $n=6$, followed by an exterior
layer with dimensionless mass increase below $\eta_{\mathcal M}$
(Appendix~\ref{app:formation-construction}). The five-dimensional
profile saturates Eq.~\eqref{form:speed-optimize}; $K<2$ on the
tangent line is the only restriction requiring $\rho_h>1$. No restriction
on $\rho_h$ arises in six dimensions.

To see that the marginal sphere forms in the evolution, consider a
slice maximum $F=1$ of radius $R_h$. Each constructed data set satisfies
\begin{equation}
 F_{,\tau}(0,R_h)=\frac{2(n-3)R_h^2+(n-5)\tilde\alpha}
                      {R_h(R_h^2+\tilde\alpha)}>0
 \label{form:crossing}
\end{equation}
in the conformal coordinates of Eq.~\eqref{evol:metric} with $R=x$ on
the slice, independently of $j$ and of the tilt (Appendix~\ref{app:admissibility}).
For $n=5$ this is $4R_h/(R_h^2+\tilde\alpha)$.
The constrained evolution in Appendix~\ref{app:evolution} therefore
takes the maximum through $F=1$. Nearby earlier slices have $F<1$
everywhere and provide the required untrapped initial data. The
positive rate in Eq.~\eqref{form:crossing} comes in equal parts from
the contraction of the sphere and from the incoming scalar energy
flux through it.
At the formation event, in $u=\tau-x,\ v=\tau+x$, one has
$R_{,u}=-1$ and $R_{,v}=0$. The sign $R_{,u}<0$ persists, so the new
interval with $F>1$ has $R_{,v}<0$ and is future trapped.
The first marginal sphere is never a vacuum point: it carries
the incoming scalar flux
$\tilde\alpha kG(P+Q)^2=(n-3)\tilde\alpha/R_h^2+(n-5)\tilde\alpha^2/(2R_h^4)$
of Eq.~\eqref{form:marginal-flux}.
Mass monotonicity \cite{Maeda2008,Nozawa2008} then gives the strict
inequality in Eq.~\eqref{form:mass-bound}.

At fixed $R_h^2<(1+\sqrt2)\tilde\alpha$, five-dimensional masses
approaching the lower bound require $j>0$ on the exterior interval
$R_h<R<(2R_h^2\tilde\alpha+\tilde\alpha^2)^{1/4}$
if the characteristic conditions hold there. On every closed
subinterval, the slices then approach
a radial null direction except on a set of radii whose length tends to
zero by the integral estimate in Eq.~\eqref{form:near-minimizer}, extending
the requirement of nonzero scalar stress-energy in
Eq.~\eqref{form:vacuum-obstruction} from the marginal sphere to an
interval.

Each radius uses a different profile and a short evolution through
formation. Existence times and derivative bounds need not be uniform:
as $\rho_h\to1^+$ in five dimensions, $j<2$ and $K^2<j+2$ in
Eq.~\eqref{form:five-conditions} approach equality at $(K^2,j)=(4,2)$; as $\rho_h\to0$ in six dimensions, the angular curvature
$1/(\tilde\alpha \rho_h^2)$ diverges. Strict characteristic inequalities
persist briefly by continuity and approach their flat values at
infinity. Taking $\rho_h\to1^+$ and $\eta_{\mathcal M}\to0^+$ approaches
the five-dimensional radius and ADM-mass bounds in
Eqs.~\eqref{form:radius-bound} and \eqref{form:mass-bound}. The absence of a gap in six dimensions,
and its presence in five, is consistent with the numerical findings of
Ref.~\cite{Deppe2012}; in six dimensions the characteristic conditions
do not exclude arbitrarily small marginal spheres because the
admissible interval of $j$ remains nonempty as $K$ grows. For
$\rho_h$ near one, the five-dimensional data do not have uniformly
small curvature in Gauss--Bonnet units: the formation slice passes through the marginal sphere, where
$\tilde\alpha^2\mathcal R_{abcd}\mathcal R^{abcd}$ is close to $28$.
Paper~II gives a separate
construction, based on a narrow incoming pulse, in which the radius
bound is approached from complete data with arbitrarily weak curvature
at finite mass, and the characteristic conditions hold until
formation.
\section{Nonspherical marginal surfaces}
\label{sec:nonspherical-area}

The bound in Eq.~\eqref{form:radius-bound} depends on spherical
symmetry. A round sphere has one length scale, linking its area
directly to its intrinsic curvature. For a general surface these need
not be controlled by the same scale. We show that the radius bound
does not extend to a positive lower bound on the area of general
marginal surfaces. In spherical symmetry the
characteristic conditions restrict the curvature, but a surface that is
long in some directions and thin in others can have small area while
all its curvatures stay small. The construction below realizes this with
squashed three-spheres. For a closed three-surface $S$,
define its area radius by
\begin{equation}
 r_{\rm A}=\left(\frac{\operatorname{Area}(S)}{2\pi^2}\right)^{1/3}
 \label{area:radius}
\end{equation}
which agrees with $R$ on a round sphere.

Without symmetry the constraints can be solved as in
Section~\ref{sec:formation}: once the geometry is prescribed, the
scalar momentum is determined algebraically. Take the extrinsic curvature to be
proportional to the spatial metric, $\mathcal K_{ij}=c\gamma_{ij}$,
with constant $c>0$, and take $\Psi$ constant on the slice. With the
convention $\mathcal K=-\tfrac12\mathcal L_{e_0}\gamma$, a positive $c$
describes a contracting slice. The EGB momentum constraint then holds
identically: the Codazzi tensor of $\mathcal K$ vanishes because $c$ is
constant, and the scalar momentum density vanishes because $\Psi$ is
constant on the slice. The
Hamiltonian constraint determines $P$ from the geometry,
Eq.~\eqref{area:scalar}.

Every closed hypersurface $S$ of the slice has
$\operatorname{tr}_S\mathcal K=3c$, so its null expansions are
\begin{equation}
 \theta_\pm=\pm\mathcal H-3c
 \label{area:marginal}
\end{equation}
with $\mathcal H$ its mean curvature within the slice. Future marginal
surfaces ($\theta_+=0$) are the hypersurfaces of mean curvature $3c$,
so the desired marginal surface can be specified through the spatial
geometry before solving for the scalar data.

The sufficient condition on a prescribed smooth Riemannian metric
$q$ on $S^3$ is
\begin{equation}
 \lambda_q=\lambda_1\bigl(-\Delta_q+\tfrac12R_q\bigr)>0
 \label{area:condition}
\end{equation}
where $R_q$ is its intrinsic scalar curvature, $\Delta_q$ its
Laplacian, and $\lambda_1$ the lowest eigenvalue. Positive scalar
curvature is sufficient, although some metrics with negative scalar
curvature in part of the surface also satisfy this condition.

\begin{theorem}[Prescribed intrinsic metric]
\label{thm:prescribed-metric}
In five dimensions, fix $\alpha_{(2)}>0$ and a smooth metric
$q$ on $S^3$ satisfying Eq.~\eqref{area:condition}. Choose $\eta>0$
sufficiently small for this metric and the path used in
Appendix~\ref{app:area-prescribed}, and then choose $L$ so that
$\alpha_{(2)}/L^2\le a_q\eta^2$, with $a_q>0$ sufficiently small.
There are smooth EGB--scalar initial data on $\R^4$ whose slice
is complete and asymptotically flat, with a regular center. The slice
contains a future marginally outer trapped surface with induced
metric exactly $L^2q$. This surface is outermost and strictly stable. Near the surface,
$\mathcal K=(\eta/L)\gamma$, and the constructed scalar field is
constant on the slice. The surface has $\theta_+=0$ and
$\theta_-=-6\eta/L$. Outside a compact set the data agree with a
static GR-branch Boulware--Deser exterior. The full initial spacetime
curvature components, in an orthonormal frame with timelike vector
normal to the slice, are bounded by $C_q/L^2$. The data admit a short
strongly hyperbolic local evolution.
\end{theorem}

{The construction and stability proof are in
Appendices~\ref{app:area-prescribed} and \ref{app:area-stability};
Appendix~\ref{app:area-evolution} controls the full spacetime
curvature and local evolution. Strict stability means that the
operator varying $\theta_+$ under outward normal deformations within
the slice has positive principal eigenvalue; outermost means that
no marginally outer trapped surface encloses it. The order of
choices in the theorem is essential: the prescribed metric is
realized after sufficient rescaling, not at every absolute size. Appendix~\ref{app:area-prescribed} extends the construction to
families of prescribed metrics with common scale choices and
curvature bounds. This requires additional bounds on the metrics
and their eigenfunctions.}

At positive coupling, Eq.~\eqref{area:condition} is sufficient;
the converse proved in Eq.~\eqref{area:necessity} is weaker. A
strictly stable marginal surface isometric to $(S^3,L^2q)$, in data
with $\mathcal K=c\gamma$ near the surface and with spatial
curvature and $c^2$ bounded by $C_0/L^2$, obeys
$\lambda_q>-C\alpha_{(2)}/L^2$. Constancy of the matter scalar is
not a hypothesis of this converse. At zero coupling the conclusion
is $\lambda_q>0$. The spectral condition is the three-dimensional
analogue of that used by Mantoulidis and Schoen for apparent horizons
\cite{MantoulidisSchoen2015}.

The condition comes from the Hamiltonian constraint. For a field
constant on the slice, the scalar source of this constraint is
$3kGP^2$, twice the normal--normal component $\tfrac32kGP^2$ of the
right side of Eq.~\eqref{eq:field}. With $\mathcal K=c\gamma$ the
constraint requires this source to equal a combination of the
curvature of the slice and of the extrinsic curvature,
\begin{equation}
 3kGP^2={}^{(4)}\mathcal R+12c^2
 +\alpha_{(2)}\bigl[\mathcal G[\gamma]+4c^2\,{}^{(4)}\mathcal R+24c^4\bigr]
 \label{area:constraint}
\end{equation}
where ${}^{(4)}\mathcal R$ and $\mathcal G[\gamma]$ are the scalar
curvature and Gauss--Bonnet density of the spatial metric
(Appendix~\ref{app:area-completion}). For real $P$, the right side must be nonnegative; the construction
keeps it positive on and around the surface.

The operator in Eq.~\eqref{area:condition} controls the scalar
curvature of the slice when the surface is thickened with a radial
lapse $N_{\rm H}$, that is, when the surrounding metric is written as
$\gamma=N_{\rm H}^2\dd\varrho^2+\gamma_S(\varrho)$ as in
Appendix~\ref{app:area-prescribed}. Besides terms built from the
second fundamental form of the surfaces of constant $\varrho$ and the
radial derivative of their mean curvature, the scalar curvature of the
slice then contains the combination
$R_q-2N_{\rm H}^{-1}\Delta_qN_{\rm H}$, divided by $L^2$ on a surface
with induced metric $L^2q$ (Eq.~\eqref{area:collar}). Choose
$N_{\rm H}$ to be the positive eigenfunction for the lowest
eigenvalue, $(-\Delta_q+\tfrac12R_q)N_{\rm H}=\lambda_qN_{\rm H}$.
Dividing this equation by $N_{\rm H}/2$ gives
\begin{equation}
 R_q-2N_{\rm H}^{-1}\Delta_qN_{\rm H}=2\lambda_q
 \label{area:lapse-identity}
\end{equation}
Positivity of $\lambda_q$ therefore allows the constraint to be
solved for a real scalar momentum $P$. A smooth path of metrics and positive lapses joins this
geometry to a round exterior, and the condition $\alpha_{(2)}/L^2\le a_q\eta^2$ makes
the Gauss--Bonnet correction small enough to preserve positivity of
the right-hand side of Eq.~\eqref{area:constraint}.

\begin{corollary}[Berger surfaces with small area]
\label{cor:berger-area}
Fix $\alpha_{(2)}>0$, then a sufficiently small
$\eta_{\rm H}>0$ and a sufficiently large $L_{\rm H}$ such that
$\alpha_{(2)}/L_{\rm H}^2$ is small compared with $\eta_{\rm H}^2$.
For every sufficiently small $\varepsilon_{\rm H}>0$, there are
smooth initial data on a complete asymptotically flat slice with a
regular center. The slice contains a future marginally outer trapped
surface $S_{\varepsilon_{\rm H}}$, which is outermost and strictly stable. It has $\theta_+=0$,
$\theta_-=-6\eta_{\rm H}/L_{\rm H}<0$, and induced metric
$L_{\rm H}^2q_{\varepsilon_{\rm H}}$, where
\begin{equation}
 \begin{gathered}
 q_{\varepsilon_{\rm H}}=\sigma_1^2+\sigma_2^2+\varepsilon_{\rm H}^2\sigma_3^2,\qquad
 \lambda_{q_{\varepsilon_{\rm H}}}=\tfrac12R_{q_{\varepsilon_{\rm H}}}=4-\varepsilon_{\rm H}^2\\
 \operatorname{Area}(S_{\varepsilon_{\rm H}})
   =2\pi^2L_{\rm H}^3\varepsilon_{\rm H},\qquad
 r_{\rm A}=L_{\rm H}\varepsilon_{\rm H}^{1/3}\longrightarrow0
 \end{gathered}
 \label{area:small}
\end{equation}
The one-forms $\sigma_i$ are normalized by
$\sigma_1^2+\sigma_2^2+\sigma_3^2=\dd\Omega_3^2$.
All members have one common round exterior and one common ADM mass.
With $L_{\rm H}$ fixed as $\varepsilon_{\rm H}\to0^+$, the full
initial spacetime curvature components in the slice-normal
orthonormal frame are bounded by $C/L_{\rm H}^2$, independently of
$\varepsilon_{\rm H}$. Each fixed member admits a local strongly
hyperbolic evolution with positive physical kinetic coefficients
and a common hyperbolicity covector for gravity and matter. Neither
the injectivity radius nor the existence time is asserted to be
uniform as $\varepsilon_{\rm H}\to0^+$.
\end{corollary}

This result uses the explicit construction of
Appendix~\ref{app:area-geometry}, its common exterior in
Appendix~\ref{app:area-completion}, and the evolution estimates of
Appendix~\ref{app:area-evolution}. Although the Berger metrics
satisfy Eq.~\eqref{area:condition} uniformly, their shrinking fibers
violate the metric bounds required for common constants in the
general construction. We therefore use the explicit construction
to keep $L_{\rm H}$ fixed. The common mass is supplied by that
construction; it is not an arbitrary preassigned ADM mass.

The Hopf fibration writes $S^3$ as a family of circles, the Hopf
fibers, over a two-sphere. The Berger metric shortens these circles to
length of order $L_{\rm H}\varepsilon_{\rm H}$ while keeping the
two-sphere at scale $L_{\rm H}$. Its curvature stays of
order $1/L_{\rm H}^2$, while its volume, and hence the area of the
marginal surface, is proportional to $\varepsilon_{\rm H}$.
The topology remains $S^3$ for every $\varepsilon_{\rm H}>0$, but the
fiber lengths tend to zero. Bounded curvature therefore supplies no
positive lower bound for these lengths or for the area. The limit
degenerates along the fibers; no uniform evolution time as
$\varepsilon_{\rm H}\to0$ is asserted. Prescribed metrics of positive scalar curvature on
$S^3$ already occur as outermost minimal boundaries of
time-symmetric asymptotically flat general-relativistic data
\cite{CabreraMiao2018}. Here the data are not time symmetric, the
marginal surface lies inside complete data with a regular center, and
the constraints are the exact EGB--scalar ones.
Nonspherical vacuum EGB initial data
with common apparent horizons have also been constructed numerically
in the time-symmetric case \cite{Yoshino2011}.
The result constructs marginal surfaces on an initial slice.
Formation from initially untrapped data, the existence of an event
horizon, and trapping of every gravitational polarization remain open.
\section{Tidal curvature after characteristic loss}
\label{sec:tidal-prediction}

The continuation result of Section~\ref{sec:curvature-continuation}
follows the spherical solution smoothly through its first
characteristic loss, but it does not determine how tidal measurements
depend on the initial data after that event. Smooth spherical
continuation cannot answer this question, because it suppresses the
nonspherical degrees of freedom whose propagation has failed. We therefore
use a separate five-dimensional construction: a contracting homogeneous
scalar core and exact nonspherical solutions near it. Their symmetry
reduces the field equations sufficiently to follow these particular
solutions through and beyond characteristic loss, including their nonlinear effect on the core.
The initial data are smooth, complete, asymptotically flat and initially
untrapped. Their local curvature can be made arbitrarily weak, although
their total mass need not be small.

If the theory is to predict what an observer measures after a
characteristic loss, then small changes in the initial data should
produce small changes in the measurements, at least as long as the
curvature stays bounded. The
construction shows that this cannot be guaranteed together with
classical existence and local geometric uniqueness: a family of
initial data converges, together with all its derivatives, while the
tidal matrices measured by a fixed set of observers at a chosen later
time do not converge to the reference values. The mechanism is the
amplification of short-wavelength gravitational disturbances after
the spatial principal coefficient becomes negative, and the disturbances can be made
as small initially as one likes because the amplification grows
without bound with the wave number.

For the linearized equations about a fixed background, Reall,
Tanahashi and Way argued that where hyperbolicity fails the initial
value problem is ill posed, since the solutions do not depend
continuously on their initial data \cite{Reall2014}. For nonlinear
equations, hyperbolicity is necessary for a well-posed
noncharacteristic Cauchy problem \cite{Metivier2005}, and for
quasilinear first-order systems that are hyperbolic initially, its
loss at later times implies a Hadamard instability
\cite{LernerNguyenTexier2018}. These results do not by themselves
determine tidal measurements in exact developments of the EGB
equations. Comparing two spacetimes requires a geometric
identification of their points, gauge-fixed EGB equations can fail to
be strongly hyperbolic where the geometric theory is well posed
\cite{Papallo2017,Kovacs2020}, and in the full equations a growing
perturbation can modify its own background or end in a curvature
singularity. The construction therefore uses exact developments of the
full equations, whose curvature stays bounded, and compares
geometrically defined tidal tensors at a specified observation time.

A comparison of tidal measurements must specify corresponding observers
and frames without relying on a later coordinate identification. Mark
a compact ball of initial positions, launch the future unit-normal
geodesics, and parallel transport spatial orthonormal frames chosen by
one fixed prescription on the initial slice. Carry this initial marking
with the data under coordinate changes. Observer labels and elapsed
proper time then identify the measurements to be compared. Let $u^a$ be an
observer's unit tangent and $e_i^a$ the transported spatial frame.
The measured tidal tensor is
\begin{equation}
 \mathcal E_{ij}=\mathcal R_{abcd}e_i^a u^b e_j^c u^d
 \label{tidal:observable}
\end{equation}
These are the curvature components governing the relative acceleration
of neighboring freely falling observers.

{\begin{theorem}[Discontinuous dependence of tidal curvature]
\label{thm:tidal-discontinuity}
Fix the positive coupling in five dimensions and a smooth contracting
homogeneous scalar core of the type constructed in
Appendix~\ref{app:weak-reference}. Choose its initial slice $t=t_-$
with sufficiently small, fixed curvature, and set $t=0$ at its first
characteristic loss. The core extends to smooth, complete,
asymptotically flat, initially untrapped constrained data $D_*$.
For each sufficiently small fixed $T>0$, choose a ball of marked
observers that remains inside the common homogeneous region through $T$.
Specifically, let $\mathscr B$ be the ball $r\le r_{\rm b}$ of initial
positions in Cartesian coordinates, with $s(t_-)=1$, and choose
$r_{\rm b}>0$ so that $r_{\rm b}+\bar c_+d_{\rm c}(T)<R_g$.
Here $R_g,\bar c_+$ satisfy Eq.~\eqref{tidal:d39}, and the comoving
distance $d_{\rm c}$ is defined by Eq.~\eqref{tidal:d40}.
There are smooth, complete, asymptotically flat, initially untrapped
constrained data $D_{k_{\rm w}}$ and exact nonspherical developments
$\mathcal U_{k_{\rm w}}$, together with a development $\mathcal U_*$
of $D_*$, with the following properties.

The differences $D_{k_{\rm w}}-D_*$ have support in one fixed compact
set and tend to zero in $C^m$ for every fixed finite $m$ as
$k_{\rm w}\to\infty$ along the constructed sequence. The exterior data,
including all derivatives, agree exactly. The developments are smooth
on neighborhoods of $\mathcal O=\mathscr B\times[t_-,T]$, where
observer labels and elapsed proper time give the identification above.
Their metrics converge in $C^1$ on this region and all curvature
components in the prescribed orthonormal frames are uniformly bounded.
Nevertheless,
\begin{equation}
 \liminf_{k_{\rm w}\to\infty}\sup_{\mathcal O}\max_{i,j}
 \left|\tilde\alpha\bigl(\mathcal E_{ij}[\mathcal U_{k_{\rm w}}]
                         -\mathcal E_{ij}[\mathcal U_*]\bigr)\right|>0
 \label{tidal:separation}
\end{equation}
More precisely, the tidal matrix fails to converge to its reference
value at every fixed marked observer at the chosen time $T$.
For the same sequence, the fields converge smoothly on fixed compact subregions of $\mathcal O$ with $t<T$, including those with $0<t<T$.
\end{theorem}

Appendix~\ref{app:tidal-prediction} proves the theorem using the exact
reduced equations. For fixed reference data and $t_-$, each sequence is constructed for
one sufficiently small observation time $T>0$. Its positive amplitude
$\epsilon_{\rm a}$ is then chosen sufficiently small, depending on
these choices, and held fixed as $k_{\rm w}\to\infty$. The normalization
in Eq.~\eqref{tidal:seed} depends on $T$, so different observation
times require different sequences; the tidal lower bound need not be
uniform as $T\to0$. Smooth convergence here means convergence in
every fixed smooth seminorm. It gives neither a bound uniform over
all derivative orders nor an analytic-norm estimate.

Equation~\eqref{tidal:separation} alone is a uniform-norm statement.
The stronger assertion at each fixed observer follows from
Eq.~\eqref{tidal:d46}: at $T$ the perturbed tidal tensor has an eigenvalue
spread bounded away from zero, whereas the reference tensor is
isotropic. Thus a change of orthonormal measuring frame cannot remove
the discrepancy. The same sequence does not have nonconvergent tidal
measurements at every time after the transition.}

Metric convergence in $C^1$ controls the metric and connection, but
curvature contains second derivatives. In this construction the final
metric anisotropy is of order $k_{\rm w}^{-2}$, with a small fixed
coefficient. Its first derivatives vanish as the wave number increases,
while two derivatives can leave a finite curvature contribution.
Initially, however, the data converge with every fixed number of
derivatives. The distinction in Eq.~\eqref{tidal:separation} is generated
by the evolution, rather than inserted as an initial failure of
curvature convergence.

In this section, $s(t)$ denotes the reference core's scale factor,
and $H_c=\dot s/s<0$ its Hubble rate. The core is isotropic, so
Eq.~\eqref{sym:center} holds at every point, and all five gravitational
polarizations have squared phase speed
\begin{equation}
 c_g^2=\frac{2-K^2}{K^2},\qquad K=1+\tilde\alpha H_c^2
 \label{tidal:frw-speed}
\end{equation}
Contraction increases $K$ from a value close to one. The initially
weakly curved core reaches $K=\sqrt2$, where all gravitational
characteristic sectors degenerate at $(K^2,j)=(2,2)$ in
Fig.~\ref{fig:region}; we call this event the transition. The time coefficient $K$ stays positive and
$\tilde\alpha^2\mathcal R_{abcd}\mathcal R^{abcd}=88-48\sqrt2$
there, by Eq.~\eqref{cont:curvature-identity}. The loss is caused by
$Y=(2-K^2)/K$ in Eq.~\eqref{tidal:KY} changing sign on a regular background.

The exact nonspherical family uses a spatial basis that rotates with
wave number $k_{\rm w}$. Unequal scale factors in the rotating directions
produce Cartesian metric anisotropy with wave number $2k_{\rm w}$.
Translations combined with rotations preserve this geometry, so the
full equations reduce to ordinary differential equations for the scale
factors and scalar field. The matrix multiplying the logarithmic scale-factor accelerations in
Eq.~\eqref{tidal:acceleration-system} remains invertible near the transition. Although the spatial propagation coefficient
vanishes, the reduced equations still determine the time
accelerations, so the construction proceeds. The resulting exact solutions are a restricted family;
the reduction supplies no existence result for arbitrary spatial
perturbations after characteristic loss.

Before the transition, a linearized anisotropy mode oscillates. After
the spatial coefficient changes sign, it has a growing branch whose
rate increases with spatial frequency. Since $c_g^2$ crosses zero
linearly in time, the equation for the linearized anisotropy amplitude
$\phi$ near the transition has the form
$\ddot\phi-k_{\rm w}^2\,(\text{const})\,t\,\phi=0$ up to lower-order terms, which is Airy's equation
in the rescaled time $k_{\rm w}^{2/3}t$
(Appendix~\ref{app:tidal-prediction}); here $t$ is measured from the
transition in the units of that appendix. The transition
layer has $|t|\lesssim k_{\rm w}^{-2/3}$. This scale follows by balancing
two time derivatives against the $k_{\rm w}^2$ term, whose
coefficient is proportional to $t$. Beyond the layer, the leading
local growth rate is proportional to $k_{\rm w}\sqrt t$. Integrating
this rate gives a growth exponent proportional to
$k_{\rm w}t^{3/2}$ near the transition.

Before the transition the anisotropy is a standing gravitational wave
whose restoring force is proportional to the spatial coefficient. This
force vanishes at the transition, as for an oscillator whose spring
constant passes through zero and becomes negative, and afterwards the
mode grows instead of oscillating.
The initial anisotropy is chosen exponentially small in $k_{\rm w}$
so that this amplification produces a small fixed curvature difference
at the observation time. Any fixed number of initial spatial
derivatives contributes only powers of $k_{\rm w}$, which the
exponential smallness dominates. This gives the smooth convergence of
the initial data despite the later difference in tidal curvature. The initial
slice with weak curvature and the observation interval are fixed as
$k_{\rm w}$ increases;
the nonlinear estimates in Eqs.~\eqref{tidal:d33},
\eqref{tidal:mean-bounds} and \eqref{tidal:bootstrap} have constants
independent of $k_{\rm w}$. Without this uniformity, an
error that is small at each fixed wave number could overwhelm the tidal
difference along the sequence used in Eq.~\eqref{tidal:separation}.

Linear amplification alone would not establish
Eq.~\eqref{tidal:separation}. The anisotropy also changes the other scale
factors, which enter its own evolution equation. The appendix controls
these coupled nonlinear terms while retaining a uniformly invertible
acceleration matrix. The leading tidal difference survives the
nonlinear corrections, and all curvature components in the chosen orthonormal frames remain
bounded. Control of the metric alone would be insufficient here:
the principal coefficients depend on curvature, and the nonlinear
estimates must bound the combinations containing two derivatives.
The estimates also bound changes in the accelerations of the other
logarithmic scale factors in Eq.~\eqref{tidal:d10}. Their coupled
evolution is retained in the anisotropy equation.

The embedding in complete data is also part of the argument. The
initial perturbation is cut off inside a homogeneous core large enough
to contain the relevant pre-transition domains of dependence. The
constraints are solved exactly across the cutoff, and the exterior
data are unchanged. Before characteristic loss, local geometric uniqueness identifies
the interior development with the homogeneous solution. After that event,
that interior is supplied by the exact reduced equations. The complete
initial slice and $\mathcal O$ lie in a development with complete
spacelike Cauchy slices for the spacetime metric. Outside the core,
these slices remain in an early regular part of the exterior. This
arrangement does not require evolving the whole exterior to the same
late time reached in the homogeneous interior. Being a Cauchy surface
for the spacetime metric concerns its causal curves; it does not restore
the gravitational characteristic cones or assert well-posed EGB
evolution after characteristic loss.

{\begin{corollary}[Existence, uniqueness and continuous dependence]
\label{cor:tidal-continuation}
In a neighborhood of $D_*$ in the smooth-data topology specified
above, a continuation theory cannot satisfy all three requirements:
\begin{enumerate}
\item Classical existence on a common observation region containing
the marked central observers from $t_-$ to some positive time after
the transition.
\item Local geometric uniqueness among classical developments of the
same initial data on overlapping regions connecting the initial
slice to those observation points.
\item Continuous dependence of $\tilde\alpha\mathcal E_{ij}$ on the
initial data in the supremum norm on that common region.
\end{enumerate}
\end{corollary}

To apply the theorem, first suppose that the common region in the
corollary exists. Choose a sufficiently small positive $T$ in its
interior and construct the sequence for that $T$. The uniqueness
requirement would identify each proposed development with the
corresponding exact solution along the marked observers through $T$.
Equation~\eqref{tidal:separation} would then contradict continuous
dependence. Here geometric uniqueness after characteristic loss is
an assumption of the comparison. The conclusion concerns the three
requirements together; it does not decide which fails. The exact
comparison solutions extend smoothly through the transition, so the
result implies neither generic inextendibility nor nonexistence of
smooth extensions, and leaves possible selection rules unresolved.

Time averages over fixed windows within $[t_-,T]$ approach their
reference values despite the tidal discrepancy at $T$. That discrepancy becomes
concentrated in an interval of order $1/k_{\rm w}$ ending at $T$.
More precisely, Eq.~\eqref{tidal:averages} gives convergence in
$L^p([t_-,T];L^\infty(\mathscr B))$ for every fixed finite $p\ge1$.
This also covers averages with any fixed bounded time weight
supported in $[t_-,T]$. A window ending at $T$ is covered; a window
centered at $T$ and extending beyond it is not. Normalized windows
whose widths shrink as $1/k_{\rm w}$ can retain a nonzero average
magnitude of the tidal-tensor difference. For fixed initial
separation and relative velocity, the geodesic-deviation solutions
and their first time derivatives converge on the same interval.
These estimates concern those specified measurements along this
family; they do not assert convergence of instantaneous relative
acceleration, peak-sensitive measurements or arbitrary detector
protocols.

Averaging an observable of the exact equations is distinct from
restricting an effective field theory to its allowed wavelengths or
modifying its evolution equations. The estimates do not construct a
unique effective continuation or establish continuous dependence for
arbitrary coarse-grained observables. After restoring the fixed length
unit used in the appendix, the proper spatial wave number in the core
is proportional to $k_{\rm w}/(L_0a_3)$ and is unbounded in this sequence.
Exponential smallness of the initial amplitude ensures convergence in
each fixed derivative norm; it does not establish a derivative
expansion uniform in frequency. Assessing the omitted operators and
their effect near characteristic degeneration requires a specified
EFT and its coefficients. The present construction neither proves
that those operators cure the failure nor that they necessarily
dominate it.}
\newpage
\section{Discussion}

{The characteristic inequalities derived here relate gravitational propagation on a dynamical spherical scalar background to two local invariants, $K$ and $j$. In five dimensions they give a sharp lower bound on the radius of a regular marginal sphere and, for nonnegative mass, a sharp upper bound on the Kretschmann invariant. The latter is not an algebraic bound on every curvature component in an arbitrary frame. Smooth spherical continuation follows from the separate PDE estimates under the hypotheses of the finite-domain theorem. These distinctions matter when comparing a regular background with the initial-value problem for nonspherical perturbations.

The characteristic conditions provide a common description of the spherical results. In the $(K^2,j)$ plane of Fig.~\ref{fig:region}, the radius and curvature bounds are approached at $(4,2)$, the static vacuum characteristic boundary lies at $(2,0)$, and regular central values reach simultaneous degeneration at $(2,2)$. The angular coefficients determine whether an evolving local state reaches the boundary where the scalar or tensor gravitational sector degenerates.

The five-dimensional radius bound has a direct propagation meaning.
At a marginal sphere, $K_h-1=\tilde\alpha/R_h^2$, so decreasing the
radius increases the Gauss--Bonnet contribution to the characteristic
equations. Near the limiting radius, the scalar gradient is necessarily
timelike and the minimum gravitational phase speed in its rest frame
tends to zero, by Eqs.~\eqref{form:five-conditions} and
\eqref{form:speed-bound}. A marginal sphere is defined by the null expansions of the spacetime
metric, while the additional radius restriction comes from the
gravitational characteristic cones. In six dimensions the characteristic
conditions admit arbitrarily large $K$, allowing the small-radius
formation result. This does not keep curvature small: the angular
curvature at a marginal sphere grows as $1/R_h^2$.

The ADM mass describes the asymptotic end; it does not by itself
determine the local characteristic cones. Those also depend on the
mass concentration and on $j$, which distinguishes temporal from
spatial variation of the scalar field. A timelike scalar gradient
allows admissible marginal spheres below the static-vacuum radius
threshold. Matter therefore changes the propagation restrictions
themselves, as well as contributing to the mass. Under the exterior
assumptions preceding Eq.~\eqref{form:vacuum-obstruction}, mass
monotonicity connects the mass at the marginal sphere to the ADM
mass. For each fixed ADM mass in the interval specified there, the scalar
stress-energy tensor cannot approach zero at the sphere while the characteristic
conditions hold.

Complete smooth data can be chosen separately for every permitted formation radius and every positive mass excess. This establishes that the five-dimensional restrictions are sharp and that arbitrarily small radii and masses can occur in the six-dimensional construction. The existence time need not be uniform as the limiting parameters are approached. The first marginal sphere is defined relative to the selected spherical foliation; its existence alone does not establish an event horizon. With the continuation hypotheses and strict characteristic inequalities throughout the preceding evolution, a future marginal sphere forces a later characteristic degeneration somewhere in the corresponding finite domain, while the spherical background remains smooth there.

The nonspherical examples show why the spherical radius bound cannot
be interpreted as a universal minimum area. A round sphere ties its
area to a single curvature scale. The Berger surfaces instead shorten
the Hopf circles while keeping the transverse scale fixed. Their area
can therefore decrease while the spacetime curvature components in
the slice-normal frame remain weak. The common
exterior in Corollary~\ref{cor:berger-area} makes the distinction
particularly clear: the ADM mass stays fixed while the area tends to
zero. Small marginal-surface area need not indicate either small total
mass or large local curvature. These examples are initial data; whether
such surfaces form from initially untrapped data remains open.

The nonspherical construction gives outermost strictly stable marginal surfaces with intrinsic metrics $L^2q$ under the sufficient spectral condition~\eqref{area:condition} and the stated order of parameter choices. The converse at positive coupling is weaker and does not make strict positivity of this spectral quantity a necessary condition. The Berger family's shrinking fibers violate the uniform metric bounds used to obtain common construction constants for general metrics $q$. Its estimates are therefore proved separately. The mass is fixed along the family, rather than freely prescribed in advance. The local existence time need not be uniform as the fibers shrink.

The exact solutions in Section~\ref{sec:tidal-prediction} show that constrained initial data can converge with every fixed number of derivatives while the tidal matrices retain a nonzero discrepancy at a chosen time $T$ after the reference transition. Each sequence is chosen for one fixed $T$; its fields converge smoothly at every fixed earlier time $t<T$. Corollary~\ref{cor:tidal-continuation} concerns a continuation theory with classical developments on a common observation region. If geometric uniqueness identifies those developments with the exact solutions of the same data along the marked observers, continuous dependence of the tidal matrix fails. The exact solutions extend smoothly through the transition by symmetry reduction to ODEs; this construction supplies no existence result for arbitrary nonspherical perturbations.

The physical mechanism is the amplification of short-wavelength
gravitational disturbances when the spatial propagation coefficient
changes sign. Before the transition the anisotropy oscillates;
afterwards its growing branch amplifies more rapidly at larger wave
number. An exponentially small initial disturbance can consequently
produce a finite tidal difference at the selected observation time.
The metric perturbation and its first derivatives still tend to zero,
but curvature contains two derivatives, which retain this difference.
The comparison uses freely falling observers and transported frames,
so the discrepancy concerns the tidal tensor governing relative
acceleration. Its nonzero eigenvalue spread cannot be removed by
changing the orthonormal measuring frame. Together
with spherical continuation, this shows why a smooth collapse background
and bounded curvature are insufficient to settle the predictivity of
the full equations: exact spherical evolution omits the gravitational
degrees of freedom responsible for the amplification.

A discrepancy in instantaneous tidal curvature can thus coexist with convergence of time averages, geodesic separation and relative velocity. Here the averaging weights are fixed and bounded, the initial separation and relative velocity are fixed, and the measurements lie in $[t_-,T]$. Normalized windows shrinking as $k_{\rm w}^{-1}$ can still retain a nonzero average magnitude of the tidal-tensor difference. These conclusions concern the measurements specified in Section~\ref{sec:tidal-prediction}; they do not establish well-posedness for general coarse-grained observables. The distinction can also be read directly from geodesic deviation.
The tidal tensor acts on the instantaneous separation to give relative
acceleration. Relative velocity and separation depend on its action
over the preceding trajectory. Here the interval over which the tidal difference is
appreciable shrinks as the wave number grows. A finite instantaneous
difference can therefore leave a vanishing difference in the accumulated
response, for the fixed initial separation and velocity used in the
comparison.

Treating EGB as an exact classical theory makes it possible to follow
these effects without expanding in the coupling, including the change
of sign of a propagation coefficient and the nonlinear response to a
growing disturbance. This is useful when assessing a proposed evolution of the exact
equations: continuing a smooth spherical solution is not by
itself evidence that nearby nonspherical data have a well-posed evolution.
An EFT interpretation requires an additional assessment of the terms
and wavelengths retained in that evolution.

At the five-dimensional radius endpoint, $\tilde\alpha/R_h^2$ tends to one, so the retained Gauss--Bonnet contribution is not perturbatively small. Sending the coupling to zero while keeping $R_h$ proportional to $\sqrt{\tilde\alpha}$ does not improve this dimensionless expansion parameter. Near characteristic degeneration, an omitted contribution small compared with the separate Einstein and Gauss--Bonnet terms can still affect the coefficient left by their cancellation. Assessing EFT reliability therefore requires both the sizes of omitted operators and the sensitivity of the characteristic calculation to them. Neither a universal numerical cutoff $1/\sqrt{\tilde\alpha}$ nor the preservation or removal of the degeneration follows without assumptions about the EFT coefficients and scales.

The unbounded-wave-number sequence also fails to remain below any fixed cutoff on proper spatial wave number after physical units are restored. Its exponentially small initial amplitude gives convergence of every fixed derivative norm, but does not establish an expansion uniform in frequency for the evolution of those modes. Averaging an observable of the exact solutions, restricting the wavelengths of an EFT, and modifying the evolution equations are distinct operations. The first is proved here; the other two require further dynamical input. Equation-fixing studies illustrate why evolution beyond the original hyperbolic regime can depend on the additional regulating scales \cite{Corman2024}. Our averaged convergence neither selects a unique effective evolution nor proves that omitted operators necessarily dominate or cure the characteristic degeneration. Conversely, the failure of the tidal tensors to converge to the reference values along the constructed sequence of exact EGB solutions does not establish the same failure in a specified EFT or ultraviolet completion.

For nonspherical collapse, it remains to determine whether the marginal
surfaces constructed here form from initially untrapped data and whether
an event horizon develops. A separate question is how general
nonspherical data evolve after characteristic loss when the special
symmetry of the tidal construction is absent.}

Paper~II~\cite{PaperII} follows narrow incoming pulses through the admissible region. There, total mass determines, to leading order in the pulse width, the radii of scalar gravitational characteristic degeneration and trapping. For masses below the static vacuum value, the pulse shape determines which sector fails first. The companion paper also studies small-amplitude dispersal and gives a quantitative onset condition for characteristic degeneration and approaches the radius bound of Section~\ref{sec:radius} from data with arbitrarily weak curvature at finite mass.

\section*{Acknowledgments}
We thank Dr.~Robert Mann for useful discussions.
\appendix

\section{Characteristics and local well-posedness}
\label{app:symbol}

This appendix derives the characteristic coefficients quoted in
Section~\ref{sec:characteristics} and establishes local
well-posedness under the strict characteristic conditions. We
calculate the physical principal symbol after removing the
pure-gauge amplitudes. Rotations orthogonal to the angular part of
the characteristic covector separate the tensor, vector and scalar
sectors. The reduction first uses only spherical symmetry; the background field
equations then express the coefficients in terms of $K$ and $j$.
Under Eq.~\eqref{sym:conditions}, we construct a positive energy form
that remains regular when characteristic frequencies coincide. This
gives local well-posedness for nearby nonspherical data as well.

\subsection{The characteristic symbol on spherical backgrounds}
\label{app:symbol-derivation}

Let angular indices $i,k,l,m$ range from $2$ to $n-1$. The
independent curvature components in an orthonormal spherical frame are
\begin{equation}
 \begin{gathered}
 \mathcal R_{0101}=\mathcal A,\qquad
 \mathcal R_{0i0k}=\mathcal B\delta_{ik},\qquad
 \mathcal R_{1i1k}=\mathcal C\delta_{ik},\qquad
 \mathcal R_{0i1k}=\mathcal D\delta_{ik}\\
 \mathcal R_{iklm}=\mathcal E
       (\delta_{il}\delta_{km}-\delta_{im}\delta_{kl}),\qquad
 \mathcal E=F/R^2
 \end{gathered}
 \label{app:curvature}
\end{equation}

\paragraph{The weighted symbol.}
In generalized Kronecker deltas, the geometric left-hand side of
the metric equation in Eq.~\eqref{eq:field} can be written as
\begin{equation}
 G^a{}_b+\alpha_{(2)}H^{{\rm GB}\,a}{}_b
 =-\tfrac14\delta^{acd}_{bef}\mathcal R^{ef}{}_{cd}
  -\tfrac18\alpha_{(2)}\delta^{acdgh}_{befpq}
   \mathcal R^{ef}{}_{cd}\mathcal R^{pq}{}_{gh}
 \label{app:deltaform}
\end{equation}
For an amplitude $h_{ab}$ with covector $\xi_a$, the principal part of
the curvature variation is
\begin{equation}
 \delta\mathcal R_{abcd}=\tfrac12\bigl(\xi_b\xi_ch_{ad}+\xi_a\xi_dh_{bc}
 -\xi_a\xi_ch_{bd}-\xi_b\xi_dh_{ac}\bigr)
 \label{app:d1}
\end{equation}
Both curvature factors in the quadratic term of
Eq.~\eqref{app:deltaform} contribute equally, so the principal symbol
is
\begin{equation}
 \mathsf P(\xi)[h]^a{}_b=-\tfrac14\mathcal W^{acd}{}_{bef}\,
 \delta\mathcal R^{ef}{}_{cd}
 \label{app:d2}
\end{equation}
with $\mathcal W$ from Eq.~\eqref{sym:lovelock}; sans-serif capital
letters denote matrices. It agrees with the principal symbol of
Ref.~\cite{Reall2014}, Eqs.~(3.3)--(3.6), once their trace reversal is
undone.

To evaluate the contraction, first take a frame with $\mathcal D=0$,
in which the curvature is diagonal on the two-dimensional planes
spanned by pairs of frame vectors. Antisymmetry of the five-index
delta then requires $\{b,e,f\}=\{a,c,d\}$: the nonzero terms of
$\delta^{acdgh}_{acdpq}\mathcal R^{pq}{}_{gh}$ have $\{p,q\}=\{g,h\}$
disjoint from $\{a,c,d\}$, and the two orders of $(p,q)$ give a factor
two. Hence
\begin{equation}
 \mathcal W^{acd}{}_{bef}=\delta^{acd}_{bef}\,w(\{a,c,d\})
 \label{app:d3}
\end{equation}
with $w$ from Eq.~\eqref{sym:weights}: a given set of three frame
directions receives curvature contributions only from planes in the
remaining directions. Counting these planes gives the four weights
$w_{01i}=K$, $w_{0ik}=X$, $w_{1ik}=Y$ and $w_{ikl}=C_T$ of
Table~\ref{tab:weights}; with
$\tilde\alpha=2(n-3)(n-4)\alpha_{(2)}$,
\begin{equation}
 \begin{gathered}
 \begin{aligned}
 X&=1+\frac{\tilde\alpha[2\mathcal C+(n-5)\mathcal E]}{n-3},&
 Y&=1+\frac{\tilde\alpha[-2\mathcal B+(n-5)\mathcal E]}{n-3}\\
 H&=\frac{2\tilde\alpha\mathcal D}{n-3},&
 K&=1+\tilde\alpha\mathcal E,
 \end{aligned}\\
 C_T=1+\frac{\tilde\alpha[-2\mathcal A+2(n-5)(\mathcal C-\mathcal B)
                 +(n-5)(n-6)\mathcal E]}
                    {(n-3)(n-4)}
 \end{gathered}
 \label{app:curvcoeff}
\end{equation}
These agree with the tensor coefficients in Eqs.~(3.5)--(3.6) of
Ref.~\cite{Cao2021}, whose coupling is $\alpha_{(2)}$.

In a general orthonormal frame $\mathcal D$ need not vanish, and the weights of
the triples $\{A,i,k\}$, with $A$ a time or radial index, form the
mixed tensor $G_{\rm orb}{}^A{}_B$ with rows $(X,H)$ and $(-H,Y)$, the
common block of Eq.~\eqref{sym:orbit}. The off-diagonal entries $\pm H$
come from the nonzero terms of $\delta^{0ikgh}_{1ikpq}\mathcal R^{pq}{}_{gh}$,
which have $\{g,h\}=\{1,l\}$ and $\{p,q\}=\{0,l\}$ with an angular
direction $l\notin\{i,k\}$; each such $l$ contributes $4\mathcal D$.

\paragraph{Sector reduction.}
The symbol, a matrix acting on all $n(n+1)/2$ metric
components, reduces to three small blocks. The
reduction rests on two facts: the symbol is a sum of contributions,
one for each triple of frame directions, and a perturbation can be
decomposed according to how it transforms under the rotations of the
sphere that fix the angular direction of the wave covector.
Perturbations transforming
differently cannot mix in a rotation-invariant quadratic form; this
separates the tensor, vector and scalar sectors.

In a frame with $\mathcal D=0$, contracting the deltas gives the
quadratic form of the symbol as a weighted sum over orthonormal
three-frames,
\begin{equation}
 \mathfrak q(\xi,h):=h^b{}_a\,\mathsf P(\xi)[h]^a{}_b
 =\sum_{\mathcal T} w(\mathcal T)\;\xi_{\mathcal T}\cdot{\rm adj}(h|_{\mathcal T})\cdot\xi_{\mathcal T}
 \label{app:adjugate}
\end{equation}
Here $\mathcal T$ runs over unordered triples of frame directions,
$h|_{\mathcal T}$ is the mixed $3\times3$ block of $h^a{}_b$, and
$\xi_{\mathcal T}$ is the restricted covector. The adjugate is the
transpose of the cofactor matrix, so $\xi\cdot{\rm adj}(M)\cdot\xi$ is
quadratic in the entries of $M$; contractions use the orthonormal
metric.

Equation~\eqref{app:adjugate} is Eq.~\eqref{sym:lovelock} written out.
In the Einstein part, $\tfrac12h^b{}_a\delta^{acd}_{bef}\xi^e\xi_ch^f{}_d$, the
generalized Kronecker delta is a sum of $3\times3$ determinants over
triples of frame directions. For each unordered triple $\mathcal T$,
the terms with $\{a,c,d\}=\mathcal T$ combine into the adjugate form
of the block $h|_{\mathcal T}$, a bordered determinant with border
$\xi_{\mathcal T}$:
\begin{equation}
 \xi\cdot{\rm adj}(M)\cdot\xi
 =-\det\begin{pmatrix}M&\xi\\ \xi^{\rm T}&0\end{pmatrix}
 \label{app:d4}
\end{equation}
Adding the Gauss--Bonnet part multiplies the contribution of each triple by
the curvature weight $w(\mathcal T)$ of the complementary planes, as
in Eq.~\eqref{sym:weights}. The bordered determinant also makes gauge
invariance manifest: adding $\xi_a\zeta_b+\xi_b\zeta_a$ to $h_{ab}$
amounts to row and column operations that leave the determinant
unchanged.

Now take $e_2$ along the angular part of $\xi$ and let
$V_3=\operatorname{span}\{e_0,e_1,e_2\}$, so that $\xi$ lies in $V_3$.
In the rest of the sector reduction, $\hat A,\hat B$ label directions
of $V_3$ (not only the orbit directions $A,B$ used elsewhere), and
$i,k$ label the $N_\perp=n-3$ transverse directions $e_3,\ldots,e_{n-1}$.

A triple made of three transverse directions has
$\xi_{\mathcal T}=0$ and does not contribute to
Eq.~\eqref{app:adjugate}. Every other triple is either
$\{e_0,e_1,e_2\}$, or a triple $\{\hat A,\hat B,i\}$ with one
transverse direction, or a triple $\{\hat A,i,k\}$ with two. In a
frame with $\mathcal D=0$, the weights found below
Eq.~\eqref{app:d3}, with $e_2$ as one of the angular directions,
assign to the three kinds
\begin{equation}
 \begin{gathered}
 w_{012}=K\\
 w_{12i}=Y,\qquad w_{02i}=X,\qquad w_{01i}=K\\
 w_{0ik}=X,\qquad w_{1ik}=Y,\qquad w_{2ik}=C_T
 \end{gathered}
 \label{app:triple-weights}
\end{equation}
For a fixed transverse direction $i$, the second line contains one
triple for each plane $\{\hat A,\hat B\}$ of $V_3$; for a fixed
transverse pair $\{i,k\}$, the third line contains one triple for each
direction $\hat A$ of $V_3$.

The weights of the last two kinds therefore form two $3\times3$
matrices on $V_3$,
\begin{equation}
 \mathsf T_3=G_{\rm orb}\oplus C_T,\qquad
 \mathsf K_3={\rm adj}(G_{\rm orb})\oplus K
 \label{app:d5}
\end{equation}
In $\mathsf T_3$ the direction $\hat A$ receives the weight of the
triples $\{\hat A,i,k\}$. In $\mathsf K_3$ the plane $\{\hat A,\hat B\}$
is identified with its orthogonal direction in $V_3$ and receives the
weight of the triples $\{\hat A,\hat B,i\}$; for instance, $e_0$
receives the weight $Y$ of the triples $\{e_1,e_2,i\}$. A frame with
$\mathcal D=0$ has $H=0$ and $G_{\rm orb}=\operatorname{diag}(X,Y)$,
and there Eq.~\eqref{app:triple-weights} gives
$\mathsf T_3=\operatorname{diag}(X,Y,C_T)$ and
$\mathsf K_3=\operatorname{diag}(Y,X,K)$ in the basis $e_0,e_1,e_2$.
The orbit block of $\mathsf K_3$ thus has the two entries of
$G_{\rm orb}$ exchanged, which is
${\rm adj}(G_{\rm orb})=\operatorname{diag}(Y,X)$.

Finally, for $3\times3$ matrices $M,M'$ on $V_3$ write
\begin{equation}
 Q_\xi(M)=\xi_a{\rm adj}(M)^a{}_b\xi^b,\qquad
 Q_\xi(M,M')=\tfrac12\bigl[Q_\xi(M+M')-Q_\xi(M)-Q_\xi(M')\bigr]
 \label{app:d6}
\end{equation}
for the adjugate form and its symmetric bilinear form. For
$\mathcal T=\{e_0,e_1,e_2\}$ the restricted covector
$\xi_{\mathcal T}$ is $\xi$ itself, so the term of this triple in
Eq.~\eqref{app:adjugate} is $K\,Q_\xi(h|_{\mathcal T})$.

For the tensor amplitude $(h_T)_{ik}$, vector amplitude $h_{\hat Ai}=x_{\hat A}$,
and scalar amplitude $h=h_3+h_\perp\delta_\perp$, respectively,
collecting the triples in Eq.~\eqref{app:adjugate} gives
\begin{equation}
 \begin{gathered}
 \mathfrak q_T=\mathsf T_3(\xi,\xi)\,\Sigma_2(h_T),\qquad
 \mathfrak q_V=(\xi\times x)\cdot\mathsf K_3\cdot(\xi\times x)\\
 \mathfrak q_S=K\,Q_\xi(h_3)+2N_\perp\,h_\perp Q_\xi(h_3,\mathsf K_3)
   +\tbinom{N_\perp}{2} h_\perp^2\,\mathsf T_3(\xi,\xi)
 \end{gathered}
 \label{app:sectors}
\end{equation}
Here $h_T$ is transverse and trace-free, $h_3$ is supported on $V_3$,
$\delta_\perp$ is the transverse identity, the first index of
$\mathsf K_3$ is lowered in the contraction, and
\begin{equation}
 \Sigma_2(h_T)=\sum_{i<k}\bigl[(h_T)_{ii}(h_T)_{kk}-(h_T)_{ik}^2\bigr],
 \qquad
 (\xi\times x)^c=\epsilon^{c\hat A\hat B}\xi_{\hat A}x_{\hat B}
 \label{app:d7}
\end{equation}

Each form in Eq.~\eqref{app:sectors} collects the triples to which the
corresponding amplitude contributes. The tensor amplitude enters only
the triples $\{\hat A,i,k\}$ with two transverse directions, so only
$\mathsf T_3$ appears in $\mathfrak q_T$. The vector amplitude enters
the triples $\{\hat A,\hat B,i\}$ with one transverse direction, hence
$\mathsf K_3$ in $\mathfrak q_V$. The scalar amplitude enters both
kinds, and the three terms of $\mathfrak q_S$ collect the triples with
zero, one and two transverse directions: the single triple
$\{e_0,e_1,e_2\}$ has weight $K$, for each of the $N_\perp$ transverse
directions $i$ the triples $\{\hat A,\hat B,i\}$ contribute a cross
term between $h_3$ and $h_\perp$, and for each of the
$\binom{N_\perp}{2}$ transverse pairs $\{i,k\}$ the triples
$\{\hat A,i,k\}$ contribute $h_\perp^2\,\mathsf T_3(\xi,\xi)$. Transverse rotational symmetry forbids cross terms
between sectors.

The derivation used a frame with $\mathcal D=0$. Such a frame exists
on the nonempty open set
$(\mathcal B+\mathcal C)^2>4\mathcal D^2$ of spherical curvature
components: a boost removes the mixed component whenever
$|2\mathcal D/(\mathcal B+\mathcal C)|<1$. On scalar solutions this
includes every nonnull scalar gradient. Both sides of
Eq.~\eqref{app:sectors} are polynomials in the curvature components
and transform covariantly under orbit boosts. The identity therefore
extends to every spherical metric, including null scalar gradients
and orbit blocks with no diagonal orthonormal frame. For a trace-free amplitude,
$\Sigma_2(h_T)=-\tfrac12|h_T|^2$, where
$|h_T|^2=\sum_{i,k}(h_T)_{ik}^2$, so the tensor form $\mathfrak q_T$ is
degenerate exactly when $\mathsf T_3(\xi,\xi)=0$. Thus
$G_T=\mathsf T_3$.

The vector form is algebraically that of Maxwell theory in $2+1$
dimensions, with $x$ in the role of the vector potential. Let
$\mathsf C_\xi$ represent the map $x\mapsto\xi\times x$, so that
$\mathfrak q_V=x\cdot\mathsf C_\xi^{\rm T}\mathsf K_3\mathsf C_\xi\cdot x$.
The matrix $\mathsf C_\xi^{\rm T}\mathsf K_3\mathsf C_\xi$ annihilates
the gauge amplitude $x\propto\xi$, so its kernel is always at least
one-dimensional. A physical characteristic requires a second kernel direction,
that is, a rank of at most one, which is the vanishing of its
adjugate. Since the adjugate of a product is the product of the
adjugates in reverse order, and ${\rm adj}(\mathsf C_\xi)=\xi\otimes\xi$,
\begin{equation}
 {\rm adj}\bigl(\mathsf C_\xi^{\rm T}\mathsf K_3\mathsf C_\xi\bigr)
 =-Q_\xi(\mathsf K_3)\,\xi\otimes\xi
 \label{app:d8}
\end{equation}
Now ${\rm adj}(\mathsf K_3)=K G_{\rm orb}\oplus\Delta$ (for a
$2\times2$ matrix $M$, ${\rm adj}({\rm adj}\,M)=M$ and
$\det{\rm adj}\,M=\det M$, and $\det G_{\rm orb}=\Delta$), so
\begin{equation}
 Q_\xi(\mathsf K_3)=K\,G_V(\xi,\xi),\qquad
 G_V=G_{\rm orb}\oplus\frac{\Delta}{K}
 \label{app:d9}
\end{equation}
The physical characteristic equation of the vector sector is
therefore $G_V(\xi,\xi)=0$.

For the scalar sector, the transverse trace couples to the amplitudes
within $V_3$. The following shift removes that coupling:
\begin{equation}
 \mathfrak q_S=K\,Q_\xi\Bigl(h_3+\frac{N_\perp}{K} h_\perp\mathsf K_3\Bigr)
   -h_\perp^2\Bigl[N_\perp^2G_V(\xi,\xi)-\tbinom{N_\perp}{2}G_T(\xi,\xi)\Bigr]
 \label{app:completesquare}
\end{equation}
This is a completion of the square, using the bilinear form of
$Q_\xi$ and $Q_\xi(\mathsf K_3)=KG_V(\xi,\xi)$.

The first term is the Einstein quadratic form in three dimensions.
It has no physical kernel for $\xi\ne0$: in three dimensions the
principal symbol of the linearized Einstein tensor determines that of the
linearized Riemann tensor. If the latter vanishes, the amplitude is a
coordinate transformation.
A characteristic amplitude therefore has shifted $h_3$ equal to a
pure gauge amplitude $\xi_{\hat A}\zeta_{\hat B}+\xi_{\hat B}\zeta_{\hat A}$.
The shift $(N_\perp/K)h_\perp\mathsf K_3$ depends only on $h_\perp$
and the background, so it commutes with gauge transformations, which
act on $h_3$ only. The gauge part thus contributes nothing to the first term,
and the only remaining condition is the vanishing of the second.

The scalar sector is therefore characteristic exactly when
$G_S(\xi,\xi)=0$, where
\begin{equation}
 \tbinom{N_\perp+1}{2}G_S=N_\perp^2G_V-\tbinom{N_\perp}{2}G_T
 \label{app:d10}
\end{equation}
Since $N_\perp^2-\binom{N_\perp}{2}=\binom{N_\perp+1}2$, the orbit
blocks on the right combine into $\binom{N_\perp+1}2G_{\rm orb}$, so all
three factors share the block $G_{\rm orb}$, and
\begin{equation}
 C_V=\frac{\Delta}{K},\qquad
 C_S=\frac{2(n-3)\Delta-(n-4)KC_T}{(n-2)K}
 \label{app:physicalcoeff}
\end{equation}
This gives Eqs.~\eqref{sym:factors} and \eqref{sym:dependence} for
every covector, including $\omega=0$, and for every spherically
symmetric metric with $K\ne0$. The time components of the principal metric equations
$\mathsf P(\xi)[h]^a{}_b=0$, with $\mathsf P$ defined in Eq.~\eqref{app:d2},
follow from the spatial ones by the principal Bianchi identity; in
Eq.~\eqref{app:adjugate} this is the statement that $\mathfrak q$ is a form on
amplitudes modulo gauge.

\paragraph{Temporal gauge.}
The blocks obtained so far describe the characteristic equations. For
the well-posedness argument of Appendix~\ref{app:hyperbolic} we also
need the explicit symbol matrices in a gauge that is regular on the
physical characteristics; we use the temporal gauge $h_{0a}=0$.
For $\omega\ne0$, the transformation
$h_{ab}\mapsto h_{ab}+\xi_a\zeta_b+\xi_b\zeta_a$ fixes $h_{0a}=0$
without residual gauge freedom. Write
$\mathcal Q=X\omega^2-2H\omega \xi_R-Y\xi_R^2=-G_{\rm orb}^{AB}\xi_A\xi_B$,
the orbit form of Eq.~\eqref{sym:factors} with reversed sign, and
$\mathfrak l=H\omega+Y\xi_R$. In this gauge Eq.~\eqref{app:sectors} gives the
tensor factor $\mathcal Q-C_T\xi_\perp^2$ \cite{Cao2021}. For
$i=3,\ldots,n-1$, the vector pair $(h_{1i},h_{2i})$ has block
\begin{equation}
 2\mathsf P_V=
 \begin{pmatrix}K\omega^2-Y\xi_\perp^2&\xi_\perp\mathfrak l\\
 \xi_\perp\mathfrak l&\mathcal Q\end{pmatrix},\qquad
 \det(2\mathsf P_V)=K\omega^2
       \left(\mathcal Q-\frac{\Delta}{K}\xi_\perp^2\right)
 \label{app:vectorblock}
\end{equation}
For $K,\Delta>0$, its characteristic kernel is one-dimensional:
the lower-right entry is $\Delta \xi_\perp^2/K>0$ if $\xi_\perp\ne0$,
and the upper-left entry is $K\omega^2>0$ otherwise. The determinant
follows from $Y\mathcal Q+\mathfrak l^2=\Delta\omega^2$, the coordinate
form of $\det G_{\rm orb}{}^A{}_B=\Delta$.

Let $h_\perp$ be the common transverse diagonal amplitude
$h_{ii}$, with no sum on $i=3,\ldots,n-1$.
For $(h_{11},h_{22},h_{12},h_\perp)$, the scalar block is
\begin{equation}
 2\mathsf P_S=
 \begin{pmatrix}
 0&-K\omega^2&0&(n-3)(-K\omega^2+Y\xi_\perp^2)\\
 -K\omega^2&0&0&-(n-3)\mathcal Q\\
 0&0&K\omega^2&-(n-3) \xi_\perp\mathfrak l\\
 -K\omega^2+Y\xi_\perp^2&-\mathcal Q&-2\xi_\perp\mathfrak l&
                      -(n-4)(\mathcal Q-C_T\xi_\perp^2)
 \end{pmatrix}
 \label{app:scalarblock}
\end{equation}
After its rows are multiplied by $\operatorname{diag}(1,1,2,n-3)$, it is
the Hessian of $\mathfrak q_S$; the row weights account for the two
equal off-diagonal entries $h_{12}=h_{21}$ and the $n-3$ equal
transverse diagonal entries, and without them the block is not
symmetric. The first three rows determine
\begin{equation}
 \begin{aligned}
 h_{11}&=-\frac{(n-3)\mathcal Q}{K\omega^2}h_\perp,\qquad
 h_{22}=(n-3)\left(-1+\frac{Y\xi_\perp^2}{K\omega^2}\right)h_\perp\\
 h_{12}&=\frac{(n-3)\xi_\perp\mathfrak l}{K\omega^2}h_\perp
 \end{aligned}
 \label{app:scalar-amplitudes}
\end{equation}
This is the temporal-gauge version of completing the square in
Eq.~\eqref{app:completesquare}. Substitution in the fourth row
gives $(n-2)(\mathcal Q-C_S\xi_\perp^2)h_\perp=0$, with $C_S$ from
Eq.~\eqref{app:physicalcoeff}. Only one scalar gravitational amplitude
remains. No separate computation is needed at $\omega=0$, since
Eq.~\eqref{app:sectors} holds for every $\xi\ne0$. At a purely angular
covector normalized by $\xi_\perp=1$, the vector form has determinant
$-\Delta$ in $(x_0,x_1)$, and
Eq.~\eqref{app:completesquare} leaves $-\binom{n-2}2C_Sh_\perp^2$.

\paragraph{Using the background equations.}
Everything above holds for an arbitrary spherically symmetric metric.
The curvature components $\mathcal A,\ldots,\mathcal E$ are so far
independent quantities; on an actual solution of the field equations
with the minimally coupled scalar field they are determined by $K$ and
its stress-energy tensor, and $\Delta$ and the angular coefficients $C_I$ then depend on
the scalar field only through $j$.

The field equations, with the scalar stress-energy tensor of
Eq.~\eqref{eq:field}, fix the curvature components:
\begin{equation}
 \begin{gathered}
 \tilde\alpha\mathcal E=K-1,\qquad
 \tilde\alpha\mathcal B=U_R+\frac{\mathcal E_\Psi}{2K},\qquad
 \tilde\alpha\mathcal C=-U_R+\frac{\mathcal E_\Psi}{2K},\qquad
 \tilde\alpha\mathcal D=\frac{\tilde\alpha kGPQ}{K}\\
 \tilde\alpha\mathcal A=U_R-\frac{(n-1)b_0(K^2-1)}{4K^3}
   +\frac{(n-2)j}{2K}-\frac{j^2}{2K^3}
 \end{gathered}
 \label{app:background}
\end{equation}
where $\mathcal E_\Psi=\tilde\alpha kG(P^2+Q^2)$ and
\begin{equation}
 4KU_R=2(n-3)(K-1)+(n-5)(K-1)^2=(n-5)K^2+4K-(n-1)
 \label{app:d11}
\end{equation}
The relation for $\mathcal E$ is the definition of $K$, since
$\mathcal E=F/R^2$. The stress tensor has
$T^0{}_0=-T^1{}_1=-\tfrac12(P^2+Q^2)$, $T^0{}_1=-PQ$ and
$T^i{}_i=\tfrac12(P^2-Q^2)$, with no sum on the angular index $i$.

The relations for $\mathcal B$, $\mathcal C$ and $\mathcal D$ come from
the orbit components of the field equations. Evaluated on the
curvature of Eq.~\eqref{app:curvature}, Eq.~\eqref{app:deltaform}
gives
\begin{equation}
 \begin{aligned}
 \tilde\alpha\bigl(G^0{}_0+\alpha_{(2)}H^{{\rm GB}\,0}{}_0\bigr)
 &=-(n-2)K(\tilde\alpha\mathcal C+U_R)\\
 \tilde\alpha\bigl(G^1{}_1+\alpha_{(2)}H^{{\rm GB}\,1}{}_1\bigr)
 &=(n-2)K(\tilde\alpha\mathcal B-U_R)\\
 G^0{}_1+\alpha_{(2)}H^{{\rm GB}\,0}{}_1&=-(n-2)K\mathcal D
 \end{aligned}
 \label{app:d12}
\end{equation}
These components are linear in $\mathcal B,\mathcal C,\mathcal D$
with coefficients proportional to $K=w_{01i}$, because each of these
curvature components involves a single angular direction $e_i$ and the
Gauss--Bonnet term pairs it with the angular planes orthogonal to
$e_i$, the planes that determine $w_{01i}$. The terms in $\mathcal E$
alone are collected in $U_R$. Equating the components to $(n-2)kG$
times those of the stress tensor gives the three relations.

The relation for $\mathcal A$ comes from the angular component. It is
linear in $\mathcal A$, with coefficient $K$, and contains
$\mathcal B,\mathcal C,\mathcal D$ only through
$\mathcal B-\mathcal C$ and $\mathcal B\mathcal C-\mathcal D^2$:
\begin{equation}
 \begin{aligned}
 \tilde\alpha\bigl(G^i{}_i+\alpha_{(2)}H^{{\rm GB}\,i}{}_i\bigr)
 &=K\tilde\alpha\mathcal A+[(n-5)K+2]\,\tilde\alpha(\mathcal B-\mathcal C)
   +2\tilde\alpha^2(\mathcal B\mathcal C-\mathcal D^2)\\
 &\quad-\frac{(n-3)(n-4)}{2}(K-1)-\frac{(n-5)(n-6)}{4}(K-1)^2
 \end{aligned}
 \label{app:d13}
\end{equation}
The scalar-field source in the $i$--$i$ component of the metric equation
in Eq.~\eqref{eq:field}, multiplied by $\tilde\alpha$, is
$(n-2)kG\tilde\alpha T^i{}_i=(n-2)j/2$. The orbit relations give
$\tilde\alpha(\mathcal B-\mathcal C)=2U_R$ and
$\tilde\alpha^2(\mathcal B\mathcal C-\mathcal D^2)=j^2/(4K^2)-U_R^2$,
so the scalar field enters the angular equation only through $j$, and
solving it for $\mathcal A$ gives the last relation in
Eq.~\eqref{app:background}.

Substitution in Eq.~\eqref{app:curvcoeff} gives Eq.~\eqref{sym:base}
and the $C_T$ line of Eq.~\eqref{sym:general}; with
Eq.~\eqref{app:physicalcoeff} this gives the $C_S$ line, and hence
Eqs.~\eqref{sym:five} and \eqref{sym:six}. These relations hold in
every orthonormal spherical frame, including frames with
$\mathcal D\ne0$, and involve no division by $P^2-Q^2$, so they apply
also to null scalar gradients.

The double-null metric equations in Eqs.~\eqref{evol:radius},
\eqref{evol:angular} and \eqref{evol:constraints} also give the curvature
relations in Eq.~\eqref{app:background}. For example, kinematically
$\mathcal B-\mathcal C=\Box_2R/R$, where $\Box_2$ is the wave operator
of the two-dimensional orbit metric, because
$\mathcal R^{Ai}{}_{Bi}=-\nabla^A\nabla_BR/R$ (no sum on $i$) for a
spherical metric. In the conformal coordinates of
Eq.~\eqref{evol:metric}, $\Box_2R=-2e^fR_{,uv}$, and
Eq.~\eqref{evol:radius}, with $\mathcal V_{\mathrm{grav}}$ from
Eq.~\eqref{evol:potential}, gives $\Box_2R=2\mathcal V_{\mathrm{grav}}$.
This agrees with $\tilde\alpha(\mathcal B-\mathcal C)=2U_R$ because
Eq.~\eqref{evol:potential} with $\tilde\alpha F=(K-1)R^2$ gives
$U_R=\tilde\alpha\mathcal V_{\mathrm{grav}}/R$.

In terms of the orbit eigenvalues of Eq.~\eqref{sym:orbit},
\begin{equation}
 C_T=\frac{\lambda_-[(n-2)K-(n-3)\lambda_+]}{(n-4)K}
   +\frac{(n-5)(n-1-2K^2)}{(n-3)(n-4)K}
 \label{app:lambdaform}
\end{equation}
In five dimensions the second term vanishes and $b_0=2$, so $C_T$ is
proportional to $\lambda_-=(2-j)/(2K)$, as are
$\Delta=\lambda_+\lambda_-$, $C_V=\Delta/K$ and, by
Eq.~\eqref{app:physicalcoeff}, $C_S$. Each of $\Delta,C_T,C_V,C_S$
therefore carries the factor $2K\lambda_-=2-j$. For $n\ge6$ the second
term vanishes only at $K^2=(n-1)/2$, the value of $K^2$ at which a
regular center undergoes characteristic loss
(Eq.~\eqref{sym:center}).

Three limits check these formulas. In static vacuum, $P=Q=0$.
With $c_I=C_I/X$, five dimensions give
\begin{equation}
 X=Y=\frac1K,\qquad H=0,\qquad
 (c_T,c_V,c_S)=\left(3-\frac2{K^2},\frac1{K^2},\frac2{K^2}-1\right)
 \label{app:vacuum-five-check}
\end{equation}
In six dimensions,
\begin{equation}
 \begin{aligned}
 X=Y&=\frac{K^2+5}{6K},& H&=0\\
 c_T&=\frac{-K^4+50K^2-25}{4K^2(K^2+5)},&
 c_V&=\frac{K^2+5}{6K^2}\\
 c_S&=\frac{3(K^2-5)^2}{8K^2(K^2+5)}
 \end{aligned}
 \label{app:vacuum-six-check}
\end{equation}
These agree with Eqs.~(5.13) and (5.17) of Ref.~\cite{Reall2014}
for zero cosmological constant and $q=K$.

At a regular center, the scalar gradient has no radial component and
rotations about the center act transitively on the spatial
directions. Hence $Q=0$, $\mathcal A=\mathcal B$, $\mathcal C=\mathcal E$
and $\mathcal D=0$. With $\mathcal E_\Psi=j$, the relation
$\mathcal C=\mathcal E$ in Eq.~\eqref{app:background} reads
$j=2K(K-1+U_R)=(n-1)(K^2-1)/2$;
then $\mathcal A=\mathcal B$ holds identically. With this $j$,
$b_0+j=(n-3)K^2$ and $b_0-j=n-1-2K^2$, so $X=\lambda_+=K$,
$Y=\lambda_-$ and $H=0$, and Eqs.~\eqref{sym:general} and
\eqref{app:physicalcoeff} give $C_T=C_V=C_S=\lambda_-$. This gives
Eq.~\eqref{sym:center}. Flat FLRW with Hubble rate $H_c$ is, in the
comoving frame, isotropic at every point, with $F=H_c^2R^2$, and
satisfies Eq.~\eqref{sym:center} everywhere, in agreement with
Eq.~\eqref{form:flrw}.

For a null scalar gradient, $P=\pm Q$, the formulas remain regular:
\begin{equation}
 \begin{gathered}
 X=\frac{b_0+2\tilde\alpha kG P^2}{(n-3)K},\qquad
 Y=\frac{b_0-2\tilde\alpha kG P^2}{(n-3)K},\qquad
 H=\frac{\pm2\tilde\alpha kG P^2}{(n-3)K}\\
 j=0,\qquad \Delta=\frac{b_0^2}{(n-3)^2K^2}
 \end{gathered}
 \label{app:null-gradient-check}
\end{equation}
For every $n\ge5$ the unnormalized $C_I$ equal their vacuum values
at the same $K$, while $X,Y,H$ still depend on $P^2$ and $PQ$
through the scalar stress-energy tensor.

The common time and radial part of the inverse effective metrics is
\begin{equation}
 \begin{pmatrix}-X&H\\H&Y\end{pmatrix}
 =\frac{b_0}{(n-3)K}\begin{pmatrix}-1&0\\0&1\end{pmatrix}
 -\frac{\tilde\alpha kG}{(n-3)K}
 \begin{pmatrix}(P^2+Q^2)&-2PQ\\-2PQ&(P^2+Q^2)\end{pmatrix}
 \label{app:commoncone}
\end{equation}
The last matrix is positive semidefinite, with eigenvalues
$(P-Q)^2$ and $(P+Q)^2$; this is Eq.~\eqref{sym:orbit}
written in the chosen orthonormal frame. Since $b_0>0$, every time and radial covector
that is timelike for matter is timelike for the common block.
If $C_T,C_S>0$, Eq.~\eqref{app:physicalcoeff} gives $C_V,\Delta>0$, and the
covector dual to $e_0$ is timelike for every factor. If $C_I<0$ and
$\Delta>0$, the gravitational form has reversed signature, but every
covector timelike for matter still gives a negative value of
$G_I^{ab}\xi_a\xi_b$. It cannot be a hyperbolic time covector for that
factor. This gives Eq.~\eqref{sym:conditions}.

The two conditions express the positivity of one symmetric form.
Since $\mathfrak q$ is the quadratic form of a symmetric symbol, an
amplitude is characteristic exactly when it is a stationary point of
$\mathfrak q$ on the gauge-fixed amplitudes. Eliminating some of the
amplitudes at their stationary values therefore leaves a reduced form
whose degeneracies are the characteristic equations of the remaining
ones.

We apply this at the angular covector $\xi=e_2$, in the gauge
$h_{2a}=0$. Let $h^A{}_B$ be the mixed time and radial block of the
amplitude and $s_{ik}=h_{ik}$ its transverse block, with trace-free
part $s_0$. In the notation of Eq.~\eqref{app:sectors}, $h_3$
then reduces to the block $h^A{}_B$, while $h_T=s_0$ and
$h_\perp={\rm tr}\,s/(n-3)$. At this
covector $Q_\xi(M)$ is the determinant of the time and radial block of
$M$, and $\mathsf T_3(\xi,\xi)=C_T$. For $2\times2$ blocks the
bilinear form of the determinant is $\tfrac12{\rm tr}[{\rm adj}(M)M']$,
and the time and radial block of $\mathsf K_3$ is
${\rm adj}(G_{\rm orb})$; since
${\rm tr}[{\rm adj}(h)\,{\rm adj}(G_{\rm orb})]={\rm tr}(G_{\rm orb}h)$,
this gives $Q_\xi(h_3,\mathsf K_3)=\tfrac12{\rm tr}(G_{\rm orb}h)$. The
tensor and scalar forms of Eq.~\eqref{app:sectors} therefore add up to
\begin{equation}
 \mathfrak q_T+\mathfrak q_S
 =K\det(h^A{}_B)+({\rm tr}\,s)\,{\rm tr}(G_{\rm orb}h)+C_T\Sigma_2(s)
 \label{app:d14}
\end{equation}
so the time and radial amplitudes couple to $s$ only through
${\rm tr}\,s$.

At fixed $s$ this form is stationary in the $2\times2$ block when
$K\,{\rm adj}(h)+({\rm tr}\,s)\,G_{\rm orb}=0$, because the gradients of
$\det(h^A{}_B)$ and ${\rm tr}(G_{\rm orb}h)$ with respect to $h^A{}_B$
are ${\rm adj}(h)^B{}_A$ and $G_{\rm orb}{}^B{}_A$. Since
${\rm adj}({\rm adj}\,M)=M$ for $2\times2$ matrices, the stationary
point is
\begin{equation}
 h^A{}_B=-\frac{{\rm tr}\,s}{K}\,{\rm adj}(G_{\rm orb})^A{}_B
 \label{app:d15}
\end{equation}
which corresponds to a symmetric $h_{AB}$. There
$\det(h^A{}_B)=\Delta({\rm tr}\,s)^2/K^2$ and
${\rm tr}(G_{\rm orb}h)=-2\Delta\,{\rm tr}\,s/K$, because
$G_{\rm orb}\,{\rm adj}(G_{\rm orb})=\Delta\,\mathsf 1$, so the first
two terms add up to $-C_V({\rm tr}\,s)^2$. The reduced form is
therefore $-\mathfrak A(s)$, with
\begin{equation}
 \mathfrak A(s)=C_V({\rm tr}\,s)^2-C_T\Sigma_2(s)
 =\frac{C_T}2|s_0|^2+\frac{(n-2)C_S}{2(n-3)}({\rm tr}\,s)^2
 \label{app:angularform}
\end{equation}
The second form follows from
$\Sigma_2(s)=\tfrac12[({\rm tr}\,s)^2-|s|^2]$,
$|s|^2=|s_0|^2+({\rm tr}\,s)^2/(n-3)$ and Eq.~\eqref{sym:dependence}.

The tensor and scalar angular characteristic equations are therefore
the degeneracy conditions of $\mathfrak A$, namely $C_T=0$ for the
trace-free part and $C_S=0$ for the trace. The vector condition is
$\Delta=0$, as obtained above. Under Eq.~\eqref{sym:conditions},
$\Delta>0$ follows from Eq.~\eqref{sym:dependence}, so the vector
sector supplies no additional inequality. Equation~\eqref{sym:conditions}
is equivalent to positive definiteness of $\mathfrak A$;
nondegeneracy alone also allows negative $C_T$ or $C_S$. The matter scalar does not change this form. With minimal coupling,
the stress-tensor variation contains at most first derivatives of
the scalar amplitude, and the metric variation of $\Box\Psi$
contains at most first derivatives of the metric amplitude.
There are therefore no mixed metric--scalar terms in the principal
symbol.

\paragraph{A frame-independent form of the coefficients.}
The coefficients of Eq.~\eqref{app:curvcoeff} can also be expressed
without choosing an orthonormal frame; this form is not used
elsewhere in the paper. Differentiate the mass formula in Eq.~\eqref{def:mass} with respect to
$F$, treating $R$ and $F$ as independent variables. With
$\mathcal M_F=\partial_F(2kG\mathcal M)=R^{n-3}+\tilde\alpha R^{n-5}F$ and
$\mathfrak T_{AB}=\partial_R\mathcal M_F\,g_{AB}-2\partial_F\mathcal M_F\,\nabla_A\nabla_BR$,
\begin{equation}
 \begin{gathered}
 K=\frac{\mathcal M_F}{R^{n-3}},\qquad
 (n-3)R^{n-4}G_{\rm orb}^{AB}=({\rm tr}\,\mathfrak T)g^{AB}-\mathfrak T^{AB}\\
 (n-3)(n-4)R^{n-5}C_T=\partial_R^2\mathcal M_F-2\partial_R\partial_F\mathcal M_F\,\Box_2R
   +\partial_F\mathcal M_F\,{}^{(2)}\!R
 \end{gathered}
 \label{app:generating}
\end{equation}
for every spherical metric. Here the partial derivatives treat $R$ and
$F$ as independent arguments; $\Box_2$ and ${}^{(2)}\!R$ are the wave
operator and scalar curvature of $g_{AB}$, and the trace of
$\mathfrak T_{AB}$ uses this metric. These identities follow from
Eq.~\eqref{app:curvcoeff} once $\mathcal B$, $\mathcal C$ and
$\mathcal D$ are written in terms of $\nabla_A\nabla_BR$ and
$\mathcal A$ in terms of ${}^{(2)}\!R$; they require no field
equations. They relate the characteristic coefficients to the same
mass formula used to locate marginal spheres. In static vacuum,
$\mathcal M_F$ evaluated on the solution is the function $T$ of
Ref.~\cite{TakahashiSoda2010}, and these identities reduce to
expressions in $T$ and its first two radial derivatives.
\subsection{Local well-posedness near a spherical background}
\label{app:hyperbolic}

Let $(g,\Psi)$ be a smooth spherical solution of Eq.~\eqref{eq:field}
in $n\ge5$ dimensions, with $K,C_T,C_S>0$ at a reference point.
We restrict to a sufficiently small patch on which these inequalities
remain strict. The proof below chooses a foliation and a modified
harmonic formulation adapted to this solution; it does not assume
weak coupling or claim strong hyperbolicity in an arbitrary gauge.
We construct a definite energy form that remains regular when
characteristic frequencies coincide and extends to nearby
nonspherical fields and their derivatives. The construction uses
the variational form of the physical sectors and the modified
harmonic equations, whose auxiliary metrics separate the gauge
and constraint modes from the physical modes. Spectral projectors
then combine these groups into a positive symmetrizer.

First choose a time
direction with no zero physical frequency. Equation~\eqref{sym:base}
gives
\begin{equation}
 X-|H|=
 \frac{b_0+\tilde\alpha kG(|P|-|Q|)^2}{(n-3)K}>0,
 \qquad b_0=\frac{n-1+(n-5)K^2}{2}>0
 \label{hyp:time}
\end{equation}
where the second expression is Eq.~\eqref{def:jb0} rewritten.
The vector $e_0-(H/X)e_1$ has squared norm $-1+H^2/X^2<0$ for the
matter metric, and squared norm $-1/X<0$ for the effective metrics,
whose time--radial block is the inverse of
$\bigl(\begin{smallmatrix}-X&H\\H&Y\end{smallmatrix}\bigr)$. It is
therefore timelike for matter and for every gravitational sector, and
no physical frequency vanishes along it. Choose a local
spherical slice with normal parallel to this vector; such a slice
exists locally because the vector field is spherically symmetric and
tangent to the two-dimensional orbit space, where every smooth
nonvanishing vector field is orthogonal to a family of curves. Its normal
covector is also timelike for every factor by
Eq.~\eqref{app:commoncone}, so $X>0$ in the adapted orthonormal
frame; $Y>0$ there because the radial covector of this frame
annihilates the vector above, which is timelike for every factor. At
the center $H=0$ and a regular spherical slice suffices.
These choices persist on a small patch.

Use the time $t$ of Gaussian normal coordinates, with frequency
$\omega$, for the reference
metric and regular spatial coordinates at the center. Keep this foliation and these coordinates
fixed for nearby fields. The spherical equations supply the reference
coefficients; their extension uses only variational symmetry and the
principal Bianchi identities, without assuming a nonspherical
factorization into effective metrics.

Write $\boldsymbol\xi$ for the spatial part of the covector, with
$|\boldsymbol\xi|^2=\xi_R^2+\xi_\perp^2$. For
$\boldsymbol\xi\ne0$, the physical gravitational roots obey
\begin{equation}
 \omega_I^\pm=
 \frac{H\xi_R\pm\sqrt{\Delta\xi_R^2+XC_I\xi_\perp^2}}{X},
 \qquad
 X\omega_I^\pm-H\xi_R
 =\pm\sqrt{\Delta\xi_R^2+XC_I\xi_\perp^2}
 \label{hyp:roots}
\end{equation}
Since $Y,C_I>0$, the two roots have opposite signs. Thus the temporal
gauge used in Eqs.~\eqref{app:vectorblock} and
\eqref{app:scalarblock} is regular on every physical characteristic.

Let $\mathsf P_\star(\xi)$ be the principal symbol before fixing the
gauge, with the relative normalization of its equations obtained by
varying the action. For the metric equation in
Eq.~\eqref{eq:field}, the corresponding scalar equation is
$-2(n-2)kG\Box\Psi=0$. Raise the metric equation's indices and count
each off-diagonal metric component twice. With these weights,
$\mathsf P_\star$ is real symmetric. In the matrix argument, $h$ denotes
the combined metric and scalar amplitude, and a dagger denotes conjugate
transpose. The form
$h^\dagger\partial_\omega\mathsf P_\star h$ is unchanged by adding a
pure-gauge amplitude to a characteristic amplitude. To see this,
differentiate $\mathsf P_\star(\xi)(\xi_a\zeta_b+\xi_b\zeta_a)=0$ with
respect to $\omega$ and use
symmetry together with $\mathsf P_\star h=0$.

The gravitational blocks in Eqs.~\eqref{app:sectors},
\eqref{app:vectorblock} and \eqref{app:scalarblock}, together with
the matter equation in the variational normalization just specified, give
\begin{equation}
\begin{aligned}
 h_T^\dagger\partial_\omega\mathsf P_\star h_T
 &= (X\omega-H\xi_R)|h_T|^2\\
 h_V^\dagger\partial_\omega\mathsf P_\star h_V
 &=\frac{2K\omega^2}{K\omega^2-Y\xi_\perp^2+\mathcal Q}
 (X\omega-H\xi_R)(|h_{1i}|^2+|h_{2i}|^2)\\
 h_S^\dagger\partial_\omega\mathsf P_\star h_S
 &=(n-3)(n-2)(X\omega-H\xi_R)|h_\perp|^2\\
 \delta\Psi^*\partial_\omega\mathsf P_{\star,\Psi}\delta\Psi
 &=4(n-2)kG\omega|\delta\Psi|^2
\end{aligned}
\label{hyp:physical-form}
\end{equation}
Here $h_T,h_V,h_S$ are the gravitational amplitudes. The norm
$|h_T|^2$ counts both off-diagonal entries, and the vector expression
is summed over $i=3,\ldots,n-1$. On a vector characteristic,
$\mathcal Q=C_V\xi_\perp^2$ and $2\mathsf P_V$ has rank one and
positive trace: its lower-right entry is positive for
$\xi_\perp\ne0$, and its upper-left entry is $K\omega^2>0$
otherwise. Dividing the derivative of its determinant by the trace
gives the derivative of the zero eigenvalue. Counting the two
off-diagonal entries gives the vector line of
Eq.~\eqref{hyp:physical-form}, with positive prefactor.

For the scalar line, weight the rows of $2\mathsf P_S$ by
$\operatorname{diag}(1,1,2,n-3)$ and substitute
Eq.~\eqref{app:scalar-amplitudes}. The resulting symmetric quadratic
form is $(n-3)(n-2)(\mathcal Q-C_S\xi_\perp^2)|h_\perp|^2$.
Differentiate at fixed $h_\perp$ and restore the factor $1/2$:
derivatives of the eliminated amplitudes multiply their vanishing
equations. The matter line follows from
$\mathsf P_{\star,\Psi}=2(n-2)kG(\omega^2-|\boldsymbol\xi|^2)$.
The absence of principal cross terms means that these forms add at
coincident roots, with the same sign by Eq.~\eqref{hyp:roots}.

For radial covectors, the sector decomposition uses an arbitrary
angular direction. At $\xi_\perp=\mathcal Q=0$,
Eq.~\eqref{app:scalar-amplitudes} gives $h_{11}=h_{12}=0$ and
$h_{22}=-(n-3)h_\perp$, with squared spatial norm
$(n-3)(n-2)|h_\perp|^2$. The vector prefactor is two. Together with
the tensor amplitudes these give
$(X\omega-H\xi_R)h_{ab}^*h^{ab}$, $h_{0a}=0$, on all
$n(n-3)/2$ gravitational polarizations, independently of the angular
frame. Every spatial direction is radial at the center. Including
the matter amplitude, the form $h^\dagger\partial_\omega\mathsf P_\star h$
is positive definite on the space of physical characteristic amplitudes in temporal gauge at
positive frequency and negative definite there at negative frequency.

We now fix the gauge using the modified harmonic equations of
Ref.~\cite{Kovacs2020}. Gauge fixing is unavoidable: before it, every
pure-gauge amplitude $\xi_a\zeta_b+\xi_b\zeta_a$ lies in the kernel of
$\mathsf P_\star$ for every covector, so the symbol is never
invertible in the time direction and no energy estimate can be set up.
The modified harmonic condition adds to the equations a term that
makes the symbol invertible, at the cost of introducing two new kinds
of modes, coordinate perturbations and violations of the gauge
condition, whose propagation is governed by two auxiliary metrics.
We choose their characteristic frequencies to be separate from the
physical ones, so that gauge, constraint and physical modes can be
treated as independent groups.
Choose two distinct positive constants $\tilde a,\hat a$ and the auxiliary
inverse metrics
\begin{equation}
 \widetilde g^{ab}=g^{ab}-\tilde a\,e_0^ae_0^b,
 \qquad
 \widehat g^{ab}=g^{ab}-\hat a\,e_0^ae_0^b
 \label{hyp:auxiliary-metrics}
\end{equation}
where $e_0$ is the unit normal to the chosen foliation. For unit spatial
covectors the absolute physical frequencies have a positive lower
bound on a small compact patch. Choose both auxiliary speeds
$(1+\tilde a)^{-1/2}$ and $(1+\hat a)^{-1/2}$ below this bound and
below the matter speed. Their
four roots are then distinct from one another and from all physical
roots. (Only this separation of the roots, together with a common
time direction, is used below; Appendix~\ref{app:tidal-prediction}
instead takes both auxiliary speeds above every physical speed,
with $-1<\tilde a,\hat a<0$, which serves the same purpose.) The
auxiliary metrics depend algebraically on the evolving
metric and the fixed foliation. Prescribed gauge source functions can
accommodate the reference coordinates without changing the principal
symbol.

The principal gauge functional and gauge-fixing map are
\begin{equation}
\begin{aligned}
 (\mathsf G(\xi)h)^a
 &=\widetilde g^{bc}\left(\xi_b h^a{}_c
                  -\tfrac12\xi^a h_{bc}\right)\\
 (\mathsf Z(\xi)\zeta)^{ab}
 &=\widehat g^{c(a}\xi_c\zeta^{b)}
                  -\tfrac12\widehat g^{ab}\xi_c\zeta^c,
 &\mathsf P&=\mathsf P_\star-\mathsf Z\mathsf G
\end{aligned}
\label{hyp:gauge-symbol}
\end{equation}
They act only on the metric components and use the same weights as
$\mathsf P_\star$, counting each off-diagonal component twice. Let $\mathsf B_\xi$
undo these weights and contract the raised metric equation with
$\xi_b$. The principal Bianchi identity gives
\begin{equation}
 \mathsf B_\xi\mathsf P
 =-\tfrac12(\widehat g^{ab}\xi_a\xi_b)\mathsf G,
 \qquad
 \mathsf G(\xi)(\xi_a\zeta_b+\xi_b\zeta_a)
 =(\widetilde g^{bc}\xi_b\xi_c)\zeta^a
 \label{hyp:gauge-identities}
\end{equation}
These identities also hold for nearby nonspherical fields and
derivatives. At the spherical reference solution, their first consequence is that the
gauge-fixed symbol has no kernel away from the physical and auxiliary
cones. Suppose
$\mathsf Ph=0$ at a covector $\xi$ on none of these cones. Since
$\widehat g^{ab}\xi_a\xi_b\neq0$, the first identity gives $\mathsf Gh=0$, and hence $\mathsf P_\star h=0$. As
$\xi$ is not physical, $h$ must then be pure gauge,
$h_{ab}=\xi_a\zeta_b+\xi_b\zeta_a$, and the second identity gives
\begin{equation}
 0=\mathsf Gh=(\widetilde g^{bc}\xi_b\xi_c)\,\zeta
 \label{hyp:d1}
\end{equation}
so $\zeta=0$ because $\widetilde g^{bc}\xi_b\xi_c\neq0$. In particular,
$\mathsf P(\dd t)$ is invertible and remains so for nearby fields.

On the auxiliary cones the same identities produce the gauge and
constraint modes. Each $\widetilde g$ root has $n$ independent
pure-gauge modes. At a $\widehat g$ root one seeks, for each
prescribed gauge amplitude $\zeta$, an amplitude $h$ with
\begin{equation}
 \mathsf P_\star h=\mathsf Z\zeta,\qquad \mathsf Gh=\zeta
 \label{hyp:d2}
\end{equation}
so that $\mathsf Ph=0$. The source $\mathsf Z\zeta$ is compatible with
the principal Bianchi identity $\mathsf B_\xi\mathsf P_\star=0$ because
contracting Eq.~\eqref{hyp:gauge-symbol} with $\xi_b$ gives
$\mathsf B_\xi\mathsf Z\zeta=\tfrac12(\widehat g^{ab}\xi_a\xi_b)\zeta$,
which vanishes at such a root. Since this covector is not physical,
the field equations determine the amplitude modulo pure gauge, and the
second identity in Eq.~\eqref{hyp:gauge-identities} fixes the
pure-gauge part so that $\mathsf Gh=\zeta$. This gives $n$ constraint
modes.

Every physical amplitude has a unique representative satisfying
$\mathsf G h=0$. The $n(n-3)/2$ gravitational polarizations and the
matter scalar each have two frequencies, and each of the four
auxiliary roots carries $n$ modes. Together these give
\begin{equation}
 2\Bigl[\frac{n(n-3)}2+1\Bigr]+4n=n(n+1)+2
 \label{hyp:d3}
\end{equation}
independent eigenvectors, the full dimension of the evolution matrix
$\mathsf M$ defined below, which rewrites the second-order system as a
first-order one. The auxiliary cones remain separated for nearby fields, so the same
construction gives their gauge and constraint modes.

The remaining step is to control the physical modes near the
reference, including repeated roots. We use the energy form
$\mathsf E$ defined below. At the spherical reference solution it is positive
on each positive-frequency physical eigenspace, negative on each
negative-frequency one, and zero on pure-gauge modes. It is therefore
not a norm on the full state space. We will prove that the evolution
matrix is self-adjoint for this form on each physical group. Once
the form is definite on those groups, the spectral theorem gives
real frequencies and complete eigenspaces together. Write the two
symbols and the evolution matrix as
\begin{equation}
\begin{aligned}
 \mathsf P_\star(\omega)
 &=\omega^2\mathsf P_{\star,2}
       +\omega\mathsf P_{\star,1}+\mathsf P_{\star,0},
 &\mathsf E&=\begin{pmatrix}
 \mathsf P_{\star,1}&\mathsf P_{\star,2}\\
 \mathsf P_{\star,2}&0\end{pmatrix}\\
 \mathsf P(\omega)
 &=\omega^2\mathsf P_2+\omega\mathsf P_1+\mathsf P_0,
 &\mathsf M&=\begin{pmatrix}
 0&\mathsf 1\\-\mathsf P_2^{-1}\mathsf P_0&-\mathsf P_2^{-1}\mathsf P_1
 \end{pmatrix}
\end{aligned}
\label{hyp:evolution-matrix}
\end{equation}
Separate the positive and negative physical frequencies of the
reference solution and the four auxiliary groups by contours.
A spectral projector selects the invariant subspace containing all
generalized eigenvectors whose frequencies lie inside a contour;
its definition does not require diagonalizability. For nearby
coefficients, define the physical invariant spaces by
\begin{equation}
 \mathsf\Pi_\pm=\frac{1}{2\pi\mathrm i}
 \oint_{\Gamma_\pm}(\lambda\mathsf 1-\mathsf M)^{-1}\,\dd\lambda,
 \qquad V_\pm=\operatorname{ran}\mathsf\Pi_\pm
 \label{hyp:physical-riesz}
\end{equation}
The projectors are smooth and have constant rank while the contours
remain separated from the spectrum. The signs label the reference
groups without assuming that nearby frequencies are real.

The Bianchi identity controls the gauge functional on these entire
spaces. With the spatial covector fixed, write
\begin{equation}
 \mathsf G(\omega)=\mathsf G_0+\omega\mathsf G_1,\qquad
 \mathsf Z(\omega)=\mathsf Z_0+\omega\mathsf Z_1,\qquad
 \mathsf B_\xi(\omega)=\mathsf B_{\xi,0}+\omega\mathsf B_{\xi,1}
 \label{hyp:linear-gauge-maps}
\end{equation}
and put $q_{\widehat g}(\omega)=\widehat g^{ab}\xi_a\xi_b$.
The block row $\mathsf J=(\mathsf 1\;0)$ selects $h$ from a state
$z=(h,v)$, so $\mathsf J\mathsf M=(0\;\mathsf 1)$ and
\begin{equation}
 \mathsf P_0\mathsf J+\mathsf P_1\mathsf J\mathsf M
       +\mathsf P_2\mathsf J\mathsf M^2=0
 \label{hyp:companion-identity}
\end{equation}
Expand the first identity of Eq.~\eqref{hyp:gauge-identities} in
powers of $\omega$ and replace each term $\mathsf F_r\omega^r$ by
$\mathsf F_r\mathsf J\mathsf M^r$. Its left side becomes
\begin{equation}
 \sum_{i=0}^1\sum_{j=0}^2
       \mathsf B_{\xi,i}\mathsf P_j\mathsf J\mathsf M^{i+j}
 =\sum_{i=0}^1\mathsf B_{\xi,i}
       (\mathsf P_0\mathsf J+\mathsf P_1\mathsf J\mathsf M
              +\mathsf P_2\mathsf J\mathsf M^2)\mathsf M^i=0
\end{equation}
Since $q_{\widehat g}$ has scalar coefficients, its right side is
$-\tfrac12(\mathsf G_0\mathsf J+\mathsf G_1\mathsf J\mathsf M)
q_{\widehat g}(\mathsf M)$. Thus, with the block row
$\mathsf L=(\mathsf G_0\;\mathsf G_1)$,
\begin{equation}
 \mathsf Lq_{\widehat g}(\mathsf M)=0
 \label{hyp:companion-bianchi}
\end{equation}
This identity holds on the full state space and was obtained without
interchanging matrix coefficients.

The roots of $q_{\widehat g}$ belong to the separated auxiliary
groups, so $q_{\widehat g}(\mathsf M)$ is invertible on $V_\pm$.
This remains true in the presence of a Jordan block: the polynomial
on that block is a nonzero scalar multiple of the identity plus a
nilpotent matrix. Equation~\eqref{hyp:companion-bianchi} and invariance
of $V_\pm$ therefore give
\begin{equation}
 \mathsf L|_{V_\pm}=0,\qquad
 \mathsf L\mathsf M|_{V_\pm}=0
 \label{hyp:physical-gauge-states}
\end{equation}
For $z=(h,v)\in V_+\oplus V_-$, write $\mathsf Mz=(v,w)$.
The definition of $\mathsf M$ gives $\mathsf P_0h+\mathsf P_1v+\mathsf P_2w=0$.
Using $\mathsf P=\mathsf P_\star-\mathsf Z\mathsf G$ and
Eq.~\eqref{hyp:physical-gauge-states} reduces it to
\begin{equation}
 \mathsf P_{\star,0}h+\mathsf P_{\star,1}v+\mathsf P_{\star,2}w
 =\mathsf Z_0\mathsf Lz+\mathsf Z_1\mathsf L\mathsf Mz=0
 \label{hyp:physical-variational-states}
\end{equation}
Consequently, for any two states $z_i=(h_i,v_i)$ in these spaces,
\begin{equation}
 z_2^\dagger\mathsf E\mathsf Mz_1
 =-h_2^\dagger\mathsf P_{\star,0}h_1
       +v_2^\dagger\mathsf P_{\star,2}v_1
 \label{hyp:physical-self-adjoint}
\end{equation}
The coefficient matrices on the right are real symmetric. The
restricted form $\mathsf E\mathsf M$ is therefore Hermitian, proving
self-adjointness on the entire physical invariant subspace.

At the reference, a physical eigenvector $z=(h,\omega h)$ satisfies
\begin{equation}
 z^\dagger\mathsf Ez
 =h^\dagger(\mathsf P_{\star,1}+2\omega\mathsf P_{\star,2})h
 =h^\dagger\partial_\omega\mathsf P_\star h
\end{equation}
This form is definite on each reference eigenspace, with the sign of
its frequency, by Eq.~\eqref{hyp:physical-form}. Self-adjointness
makes distinct eigenspaces orthogonal. The complete set of modes counted in Eq.~\eqref{hyp:d3}
then shows that $\mathsf E$ is positive definite on $V_+$ and
negative definite on $V_-$, including coincident sector frequencies
and radial directions.

Definiteness persists on the ranges of the projectors for nearby coefficients: restricting a
nearby projector to its reference range identifies the two spaces
invertibly, and the pulled-back form varies continuously. Thus
$\mathsf M|_{V_+}$ is self-adjoint in the positive inner product
defined by $\mathsf E$, and $\mathsf M|_{V_-}$ is self-adjoint in that
defined by $-\mathsf E$. The finite-dimensional spectral theorem gives
real frequencies and a complete basis of eigenvectors on each group,
also at repeated roots. There are no Jordan chains. Since the groups
remain separated from zero, their frequency signs are unchanged.

A symmetrizer is a positive matrix $\mathsf S$ with
$\mathsf S\mathsf M=\mathsf M^\dagger\mathsf S$ for the first-order
evolution matrix $\mathsf M$. It makes $\langle z,\mathsf Sz\rangle$
a conserved energy of the frozen-coefficient system with state
vector $z$. Let $\mathsf\Pi_i$, $i=1,\ldots,4$, denote the projectors
onto the auxiliary groups. Together with the physical projectors
they give
\begin{equation}
 \mathsf S=
 \mathsf\Pi_+^\dagger\mathsf E\mathsf\Pi_+
 -\mathsf\Pi_-^\dagger\mathsf E\mathsf\Pi_-
 +\sum_{i=1}^4\mathsf\Pi_i^\dagger\mathsf\Pi_i,
 \qquad
 \mathsf S\mathsf M=\mathsf M^\dagger\mathsf S
 \label{hyp:symmetrizer}
\end{equation}
The second identity holds on each group separately. The projectors
commute with $\mathsf M$, so
\begin{equation}
 \begin{aligned}
 \mathsf S\mathsf M-\mathsf M^\dagger\mathsf S
 &=\mathsf\Pi_+^\dagger(\mathsf E\mathsf M-\mathsf M^\dagger\mathsf E)\mathsf\Pi_+
 -\mathsf\Pi_-^\dagger(\mathsf E\mathsf M-\mathsf M^\dagger\mathsf E)\mathsf\Pi_-\\
 &\quad+\sum_{i=1}^4\mathsf\Pi_i^\dagger(\mathsf M-\mathsf M^\dagger)\mathsf\Pi_i
 \end{aligned}
 \label{hyp:symmetrizer-split}
\end{equation}
and it suffices to show that each matrix in parentheses vanishes as a
form on the range of the corresponding projector.

The physical terms vanish by
Eq.~\eqref{hyp:physical-self-adjoint}.

On an auxiliary group $\mathsf M$ is a multiple of the identity, since
the group consists of a single real eigenvalue with a complete set of
eigenvectors, so any positive form on it, in particular
$\mathsf\Pi_i^\dagger\mathsf\Pi_i$, symmetrizes it. With
$\mathsf M\mathsf\Pi_i=\omega\mathsf\Pi_i$ and $\omega$ real,
$\mathsf\Pi_i^\dagger\mathsf M\mathsf\Pi_i=\omega\mathsf\Pi_i^\dagger\mathsf\Pi_i
=\mathsf\Pi_i^\dagger\mathsf M^\dagger\mathsf\Pi_i$, which makes the
auxiliary terms of Eq.~\eqref{hyp:symmetrizer-split} vanish.

The value of $\langle z,\mathsf Sz\rangle$ at a state $z$ is the sum
of $\langle\mathsf\Pi_+z,\mathsf E\mathsf\Pi_+z\rangle$,
$-\langle\mathsf\Pi_-z,\mathsf E\mathsf\Pi_-z\rangle$ and the squared
norms $|\mathsf\Pi_iz|^2$. These terms are nonnegative, because
$\mathsf E$ is positive on the positive physical group and negative on
the negative one. They all vanish only if every projection of $z$
vanishes, and then $z=0$, since the six projectors add up to the
identity. Thus $\mathsf S$ is a positive symmetrizer, smooth in the
field derivatives and unit spatial covector, with uniform upper and
lower bounds on a small compact patch,
\begin{equation}
 C^{-1}|z|^2\le\langle z,\mathsf Sz\rangle\le C|z|^2
 \label{hyp:symmetrizer-bounds}
\end{equation}
with a constant $C\ge1$. Choose a sufficiently small closed
neighborhood of the field and derivative values of the reference solution that preserves the
spectral gaps and definiteness. Its product with the sphere of unit
spatial covectors is compact, so continuity and positivity of
$\mathsf S$ give these bounds throughout the neighborhood. Group projectors
remain smooth when characteristic frequencies of different sectors
coincide and when the spatial covector is radial. This establishes strong hyperbolicity near the
spherical reference solution, including nonspherical fields and derivatives.

For frozen coefficients, a spatial Fourier transform with covector
$\boldsymbol\xi$ turns the principal part of the gauge-fixed
equations into
$\mathsf P_2\partial_t^2h+\mathrm i|\boldsymbol\xi|\mathsf P_1\partial_th
-|\boldsymbol\xi|^2\mathsf P_0h=0$, with $\mathsf P_1$ and
$\mathsf P_0$ evaluated at the unit covector
$\boldsymbol\xi/|\boldsymbol\xi|$. The state vector
$z=(h,\partial_th/(\mathrm i|\boldsymbol\xi|))$ then obeys
$\partial_tz=\mathrm i|\boldsymbol\xi|\mathsf Mz$, and hence
\begin{equation}
 \frac{\dd}{\dd t}\langle z,\mathsf Sz\rangle
 =\mathrm i|\boldsymbol\xi|\,\langle z,(\mathsf S\mathsf M-\mathsf M^\dagger\mathsf S)z\rangle=0
 \label{hyp:conservation}
\end{equation}
By Eq.~\eqref{hyp:symmetrizer-bounds}, the conserved quantity
$\langle z,\mathsf Sz\rangle$ is equivalent to the squared norm of the
state. For variable coefficients the standard energy
argument turns this into an estimate of the solution in terms of its
data on a short time interval.

The gauge-fixed EGB--scalar equations are linear in the second time
derivatives. Their coefficient matrix $\mathsf P_2$, defined in
Eq.~\eqref{hyp:evolution-matrix}, is invertible, so these time derivatives
are smooth functions of the remaining derivatives, which we substitute into the symbol
and symmetrizer. Appendix~A of Ref.~\cite{Kovacs2020} permits this
derivative dependence; its additional invertibility condition on
$\mathsf M$ holds because all frequencies are nonzero.

Take initial data for the chosen modified harmonic equations on
a relatively compact part of the initial slice. They consist of
$(\mathbf u,\partial_t\mathbf u)$, with $\mathbf u=(g_{ab},\Psi)$.
Their differences from the reference data must be sufficiently
small in
\begin{equation*}
 H^{N_{\rm s}+1}\times H^{N_{\rm s}}.
\end{equation*}
Here $N_{\rm s}$ is sufficiently large for the local existence
theorem in Appendix~A of Ref.~\cite{Kovacs2020}, and the Sobolev
norms act on $n-1$ spatial variables. The reduced equations then
admit a unique local solution on a smaller patch, with
\begin{equation*}
 \mathbf u\in C_tH^{N_{\rm s}+1}\cap C_t^1H^{N_{\rm s}}\cap C_t^2H^{N_{\rm s}-1}.
\end{equation*}
The solution depends continuously on the initial data in these norms.

To recover the original EGB--scalar equations, require the initial
data to satisfy the geometric constraints and impose compatible
initial conditions for the modified harmonic gauge.

The divergence of the reduced metric equation, together with the
scalar equation, gives a homogeneous wave equation for the gauge
condition with principal metric $\widehat g$. The initial gauge
choice and constraints set its value and normal derivative to zero,
so the solution satisfies the original equations. Comparing modified
harmonic coordinates gives uniqueness up to diffeomorphism. This is
a local result under strict characteristic inequalities; its
existence time need not remain uniform as a factor tends to zero.

\subsection{The speed bound near the minimum radius}
\label{app:speed}

For $j>0$ the scalar gradient is timelike, and we choose the
orthonormal scalar rest frame, in which $Q=0$ and, by
Eq.~\eqref{sym:base}, $H=0$, $X=\lambda_+$ and $Y=\lambda_-$. We use
the five-dimensional variables
$\varepsilon=1-j/2$ and $\kappa_I$ of Eqs.~\eqref{form:kappas} and
\eqref{form:five-metrics}, for which $K\lambda_-=\varepsilon$ and
$K\lambda_+=2-\varepsilon$ by Eq.~\eqref{form:orbit-eigenvalues}. In this frame
$\tilde\alpha kG(\Psi^a\xi_a)^2=j\omega^2=2(1-\varepsilon)\omega^2$,
so Eq.~\eqref{form:five-metrics} gives
\begin{equation}
 K\,G_I^{ab}\xi_a\xi_b=-(2-\varepsilon)\omega^2
 +\varepsilon\bigl(\xi_R^2+\kappa_I\xi_\perp^2\bigr)
 \label{hyp:d4}
\end{equation}
The characteristics of sector $I$ therefore obey
$\omega^2=v_R^2\xi_R^2+v_I^2\xi_\perp^2$, where $v_R$ is the common
radial phase speed and $v_I$ the angular phase speed in sector $I$:
\begin{equation}
 v_R^2=\frac{\lambda_-}{\lambda_+}=\frac{\varepsilon}{2-\varepsilon},\qquad
 v_I^2=\kappa_Iv_R^2,\qquad
 v_R^2=\frac{v_T^2+v_S^2}{2},\qquad
 v_V^2=\frac{v_R^2+v_S^2}{2}
 \label{form:speeds}
\end{equation}
The last two relations follow from $\kappa_T+\kappa_S=2$ and
$\kappa_V=(1+\kappa_S)/2$ in Eq.~\eqref{form:kappa-relations}.

For an oblique covector the squared phase speed
$\omega^2/(\xi_R^2+\xi_\perp^2)$ is the average of $v_R^2$ and $v_I^2$
with weights $\xi_R^2$ and $\xi_\perp^2$, so in each sector it lies
between $v_R^2$ and $v_I^2$. By the two averages in
Eq.~\eqref{form:speeds}, neither $v_R^2$ nor $v_V^2$ is smaller than
the lesser of $v_T^2$ and $v_S^2$. This value is also below one, the
squared speed of the matter scalar, because $j>0$ gives $\varepsilon<1$
and hence $v_R^2<1$. Therefore
\begin{equation}
 v_{\min}^2=\min\{v_T^2,v_S^2\}\le v_S^2
 \label{hyp:d5}
\end{equation}
In this rest frame, phase and group speeds coincide for purely radial
or purely angular propagation. For oblique directions the group
velocity, with components $(v_R^2\xi_R,v_I^2\xi_\perp)/\omega$, need not
be parallel to the covector, but its squared magnitude is again an
average of $v_R^2$ and $v_I^2$, with weights $v_R^2\xi_R^2$ and
$v_I^2\xi_\perp^2$; the group speeds therefore have the same minimum.

At fixed $K$, Eq.~\eqref{form:square} gives
$\kappa_S=2(2-\varepsilon)/K^2-1$, so that
\begin{equation}
 v_S^2=\kappa_Sv_R^2=\frac{2\varepsilon}{K^2}-\frac{\varepsilon}{2-\varepsilon}
 \label{hyp:d6}
\end{equation}
and
\begin{equation}
 \frac{(2-K)^2}{K^2}-v_S^2
 =\frac{2(2-\varepsilon-K)^2}{(2-\varepsilon)K^2}\ge0
 \label{form:speed-optimize}
\end{equation}
with equality exactly at $\varepsilon=2-K$, that is, $j=2(K-1)$.
Since $K<2$ in the admissible region, taking the nonnegative square
root gives $v_{\min}\le v_S\le(2-K)/K$.

The bound is attained. For $1<K<2$ the point $\varepsilon=2-K$ has
$0<\varepsilon<1$, hence $j>0$, and $0<\kappa_S=2/K-1<1$, so it
satisfies Eq.~\eqref{form:five-conditions}. There
$v_T^2-v_S^2=2(1-\kappa_S)v_R^2>0$, so $v_{\min}=v_S=(2-K)/K$.
Substituting $K_h=1+\tilde\alpha/R_h^2$, for which
$(2-K_h)/K_h=(R_h^2-\tilde\alpha)/(R_h^2+\tilde\alpha)$, gives
Eq.~\eqref{form:speed-bound}.
\section{Spherical equations and marginal-sphere formation}
\label{app:evolution}

We construct the complete constrained data of
Section~\ref{sec:formation} in conformal coordinates. After giving
the evolution equations and the local existence argument, we solve the
constraints algebraically in terms of the radial geometry, $j$ and
the slice orientation. The construction applies in both dimensions
and joins a regular FLRW center to a static vacuum end.

\subsection{Conformal equations}

In the metric
\begin{equation}
 \dd s^2=e^{2\sigma}(-\dd\tau^2+\dd x^2)+R^2\dd\Omega_{n-2}^2,\qquad
 u=\tau-x,\quad v=\tau+x,\quad f=\log 2-2\sigma
 \label{evol:metric}
\end{equation}
where we call $\tau$ the conformal time (the time coordinate of the
conformally flat orbit metric) and $u,v$ the conformal null
coordinates, the double-null equations of Refs.~\cite{Maeda2008,Nozawa2008}
can be written as follows after substituting the canonical scalar
stress-energy tensor and eliminating the second derivatives from the sources:
\begin{align}
 R_{,uv}&=-e^{-f}\mathcal V_{\mathrm{grav}}
 \label{evol:radius}\\
 \Psi_{uv}&=-\frac{(n-2)}{2R}(R_{,u}\Psi_v+R_{,v}\Psi_u)
 \label{evol:scalar}\\
 f_{uv}&=e^{-f}(\mathcal V_{\mathrm{grav},R}-2\mathcal V_{\mathrm{grav}}\mathcal V_{\mathrm{grav},F})
       +\frac{(n-2)kG }{K}\Psi_u\Psi_v
       -\frac{2\tilde\alpha (kG)^2e^f}{K^3}\Psi_u^2\Psi_v^2
 \label{evol:angular}
\end{align}
where
\begin{equation}
 F=1+2e^fR_{,u}R_{,v},\qquad K=1+\frac{\tilde\alpha F}{R^2},\qquad
 \mathcal V_{\mathrm{grav}}(R,F)=
 \frac{(n-3)F+(n-5)\tilde\alpha F^2/(2R^2)}{2R K}
 \label{evol:potential}
 \end{equation}
The derivatives $\mathcal V_{\mathrm{grav},R}$ and
$\mathcal V_{\mathrm{grav},F}$ treat $R$ and $F$ as independent
arguments. The angular component of the metric equation in Eq.~\eqref{eq:field},
after substituting Eqs.~\eqref{evol:radius} and \eqref{evol:constraints}, is
$K e^f$ times the residual of Eq.~\eqref{evol:angular}. Thus
its highest-derivative coefficient remains nonzero for $K>0$,
even when an angular characteristic coefficient vanishes. The
coefficients $C_T,C_S$ that vanish at a characteristic loss never
appear as divisors in these equations, so such a loss need not stop
the spherical evolution.

The remaining constraints are
\begin{equation}
 \mathscr C_u=R_{,uu}+f_uR_{,u}+\frac{kG R}{K}\Psi_u^2=0,\qquad
 \mathscr C_v=R_{,vv}+f_vR_{,v}+\frac{kG R}{K}\Psi_v^2=0
 \label{evol:constraints}
\end{equation}
The constraints propagate by
\begin{align}
 \partial_v \mathscr C_u
 &=-2\mathcal V_{\mathrm{grav},F}R_{,v}\mathscr C_u
   -\frac{2\tilde\alpha kG e^fR_{,u}\Psi_u^2}{R K^2}\mathscr C_v
 \nonumber\\
 \partial_u \mathscr C_v
 &=-2\mathcal V_{\mathrm{grav},F}R_{,u}\mathscr C_v
   -\frac{2\tilde\alpha kG e^fR_{,v}\Psi_v^2}{R K^2}\mathscr C_u
 \label{evol:propagation}
\end{align}
This is a homogeneous linear system for the constraint residuals.
Zero initial residuals therefore remain zero. Its coefficients contain
no inverse power of either null expansion, so the same conclusion
holds when a null expansion vanishes.

\subsection{Cauchy data}
\label{app:cauchy-data}

On the slice at which the marginal sphere forms, we set
\begin{equation}
 R(0,x)=x,\qquad e^{\sigma(0,x)}=a(x)^{-1},\qquad
 R_{,\tau}(0,x)=-V(x),\qquad
 V^2=1-\frac{1-F}{a^2}
 \label{evol:generaldata}
\end{equation}
Here $a$ fixes the radial proper distance on the slice, and $V$ fixes
its contraction: $e_1R=a$ and $e_0R=-aV$. For the construction in Section~\ref{sec:formation},
choose $a>0$ and the contracting sign $V\ge0$.
On this slice, $P=a\Psi_\tau$ and $Q=a\Psi_x$. The initial
constraints become
\begin{align}
 \mathcal M_{,x}&=\frac12 R^{n-2}(P^2+Q^2+2VPQ)\nonumber\\
f_\tau&=2V_x+Vf_x-\frac{2kG RPQ}{a^2K},\qquad
 f=\log 2+2\log a
 \label{evol:generalconstraints}
\end{align}
The first equation constrains the scalar derivatives; the second
determines $f_\tau$.
The scalar itself follows by integrating $\Psi_x=Q/a$.

\subsection{Local existence}
\label{app:local-development}

Away from the center, write
$\mathbf y=(R,f,\Psi)$ and
$\mathbf y_{uv}=\mathbf N(\mathbf y,\mathbf y_u,\mathbf y_v)$,
with the smooth sources $\mathbf N$ of Eqs.~\eqref{evol:radius}--\eqref{evol:angular}
on $R,K>0$. Setting
$\mathbf l_\pm=\mathbf y_\tau\pm\mathbf y_x$ and
$\mathbf z=(\mathbf y,\mathbf l_+,\mathbf l_-)$ gives
\begin{equation}
 \mathbf y_\tau=\tfrac12(\mathbf l_++\mathbf l_-),\qquad
 (\partial_\tau\mp\partial_x)\mathbf l_\pm=4\mathbf N(\mathbf z)
 \label{evol:characteristic-system}
\end{equation}
where $\mathbf N(\mathbf z)$ means
$\mathbf N(\mathbf y,\mathbf l_-/2,\mathbf l_+/2)$.
On the null triangle above a finite initial interval,
\begin{align}
 \mathbf y(\tau,x)&=\mathbf y_0(x)
  +\tfrac12\int_0^\tau(\mathbf l_++\mathbf l_-)(s,x)\,\dd s
  \nonumber\\
 \mathbf l_\pm(\tau,x)&=\mathbf l_{\pm,0}(x\pm\tau)
  +4\int_0^\tau\mathbf N\bigl(\mathbf z(s,x\pm(\tau-s))\bigr)\,\dd s
 \label{evol:characteristic-integrals}
\end{align}
The source $\mathbf N$ is locally Lipschitz on a compact neighborhood
of the initial state, so for short times the right-hand side is a
contraction, and its unique fixed point is the local solution.
In the system differentiated any number of times, the highest derivatives
of $\mathbf z$ enter linearly with bounded coefficients, so
Gr\"onwall's inequality bounds them on the same interval; the solution
is therefore as smooth as the data. The relation
$\mathbf y_x=(\mathbf l_+-\mathbf l_-)/2$ holds initially and is
preserved, since $\partial_\tau[\mathbf y_x-(\mathbf l_+-\mathbf l_-)/2]=0$.

For continuous dependence on a fixed reference triangle, localize
$\mathbf N$ to a compact subset of $R,K>0$ containing the reference
values. The resulting globally Lipschitz source gives
\begin{equation}
 \sup_{\tau\le T}|\mathbf z-\overline{\mathbf z}|
 \le e^{CT}\sup|\mathbf z_0-\overline{\mathbf z}_0|
 \label{evol:first-jet-dependence}
\end{equation}
The suprema are over the triangle and its initial base; $C$ depends
on the localized source. Sufficiently close data keep the solution
where that source agrees with the original one. This
estimate requires closeness only of the fields and their first
derivatives. Higher derivatives of the perturbed data may be large;
they remain finite for each smooth solution by the differentiated
equations. These also give higher-order continuous dependence when
the corresponding initial derivatives are close.

The same argument continues a solution through a finite endpoint
at positive radius if $\mathbf z$ stays in a compact subset of
$R,K>0$. Finitely many overlapping triangles cover the compact
domains used below. For formation, the data are exact FLRW (or flat)
near the center and exactly static outside a compact set, so the
solution on the triangles covering the intermediate annulus is matched
by uniqueness to the exact central and exterior solutions for a time
shorter than the overlap widths; Eq.~\eqref{evol:propagation}
preserves the constraints. This spherical argument applies in both dimensions without
positivity of the angular characteristic coefficients.
Nonspherical evolution requires Appendix~\ref{app:hyperbolic}.

\subsection{Solving the spherical constraints}
\label{app:realization}

We prescribe the geometry of the slice, through the mass function,
the invariant $j$ and the tilt, and determine the scalar data
algebraically: the mass constraint and the prescribed value of $j$
are two quadratic relations for the two scalar derivatives $P$ and
$Q$ at each radius, and the momentum constraint then fixes $f_\tau$.
Real solutions exist where a discriminant is nonnegative, and the
tilt is a free parameter that can be chosen to make it positive. This allows the construction of Section~\ref{sec:formation} to start
from a prescribed mass profile.
Use the Cauchy geometry and constraints in
Eqs.~\eqref{evol:generaldata}--\eqref{evol:generalconstraints}. Write
$\rho=R/\sqrt{\tilde\alpha}$ and use the dimensionless mass
\begin{equation}
 m=\frac{2kG\mathcal M}{\tilde\alpha^{(n-3)/2}}
 =\rho^{n-3}F+\frac12\rho^{n-5}F^2,\qquad
 K^2=1+\frac{2m}{\rho^{n-1}},\qquad
 m_{\rm marg}=\rho^{n-3}+\frac12\rho^{n-5}
 \label{form:mass-variables}
\end{equation}
Here $m_{\rm marg}(\rho)$ is the mass of a marginal sphere of radius
$\rho$. We use the smooth step
\begin{equation}
 \mathcal S(y)=\frac{e^{-1/y}}{e^{-1/y}+e^{-1/(1-y)}}\quad(0<y<1),
 \qquad \mathcal S(y)=0\quad(y\le0),\qquad \mathcal S(y)=1\quad(y\ge1)
 \label{form:step}
\end{equation}
It is smooth and nondecreasing, and
$\mathcal S$ and $1-\mathcal S$ are squares of smooth functions. With
\begin{equation}
 \mathcal J_m=\frac{2kG\tilde\alpha\mathcal M_{,R}}{R^{n-2}}
 =\rho^{2-n}m_\rho
 =\tilde\alpha kG(P^2+Q^2+2VPQ),\qquad j=\tilde\alpha kG(P^2-Q^2)
 \label{form:data-constraints}
\end{equation}
the scalar reconstruction is algebraic. To see this, write the energy
combination as
$2\mathcal J_m=\tilde\alpha kG[(1+V)(P+Q)^2+(1-V)(P-Q)^2]$
and use $j=\tilde\alpha kG(P+Q)(P-Q)$; this gives a quadratic equation for
$(P+Q)^2$. One root is
\begin{equation}
 \tilde\alpha kG(P+Q)^2=\frac{\mathcal J_m+
 \sqrt{\mathcal J_m^2-(1-V^2)j^2}}{1+V},\qquad
 P-Q=\frac{j}{\tilde\alpha kG(P+Q)}
 \label{form:scalar-inversion}
\end{equation}

We describe the slice by $m$, $j$ and the tilt
\begin{equation}
 \varpi=aV=-e_0R,\qquad
 a^2=1-F+\varpi^2,\qquad V=\frac{\varpi}{a},\qquad 1-V^2=\frac{1-F}{a^2}
 \label{form:tilt}
\end{equation}
where $e_0=a\partial_\tau$ and $e_1=a\partial_x$ on the slice. The
angular extrinsic curvature is $\varpi/R$. Increasing $\varpi$ changes the orientation of the slice while leaving
$F$, $K$ and $j$ fixed. The slice approaches a radial null orientation
as $V\to1$, or equivalently as $\varpi^2/(1-F)\to\infty$ where $F<1$.
The discriminant $\mathfrak d=\mathcal J_m^2-(1-V^2)j^2$ obeys the identity
\begin{equation}
 a^2\mathfrak d=\mathcal J_m^2\varpi^2-(1-F)(j^2-\mathcal J_m^2)
 \label{form:tilt-identity}
\end{equation}

\paragraph{Pointwise solvability.}
Set $\pi_\pm=\sqrt{\tilde\alpha kG}\,(P\pm Q)$, so that
$\mathcal J_m=\frac12[(1+V)\pi_+^2+(1-V)\pi_-^2]$ and
$j=\pi_+\pi_-$. For $|V|<1$,
\begin{equation}
 2\mathcal J_m-2\sqrt{1-V^2}\,|j|
 =\bigl(\sqrt{1+V}\,|\pi_+|-\sqrt{1-V}\,|\pi_-|\bigr)^2
 \label{form:amgm}
\end{equation}
Thus real data exist exactly when
$\mathcal J_m\ge\sqrt{1-V^2}|j|$. For $\pi_+\ne0$, take either
nonnegative root $\pi_+^2=(\mathcal J_m\pm\sqrt{\mathfrak d})/(1+V)$ and set
$\pi_-=j/\pi_+$. When $j=0<\mathcal J_m$, the zero root instead gives
$\pi_-^2=2\mathcal J_m/(1-V)$; when $\mathcal J_m=0$ and $|V|<1$,
both amplitudes vanish.

At $V=1$, $\mathcal J_m=\pi_+^2$: any $j$ is allowed for
$\mathcal J_m>0$, whereas $\mathcal J_m=0$ requires $j=0$ and
leaves $\pi_-$ unrestricted. Combining the two cases, every constrained slice with $F\le1$ (so $|V|\le1$) has $\mathcal J_m\ge0$ and
$\mathcal J_m>0$ wherever $j\ne0$: for $|V|<1$ this is Eq.~\eqref{form:amgm}, and at $V=\pm1$ it is $\mathcal J_m=\pi_\pm^2$ with $j=\pi_+\pi_-$. For $\mathcal J_m>0$,
Eq.~\eqref{form:tilt-identity} gives
$\mathfrak d>0$ exactly when
$\varpi^2>(1-F)(j^2/\mathcal J_m^2-1)$. A sufficiently large tilt
therefore suffices wherever $j/\mathcal J_m$ is finite, and any
$\varpi>0$ suffices if $|j|\le\mathcal J_m$. Smoothness at a matter
boundary requires the additional condition below.

\paragraph{Where the matter vanishes and at the center.}
Near a boundary point of the matter region $\{\mathcal J_m>0\}$, let
$\mathcal J_m=\upsilon^2\mu$ and $j=\upsilon^2\iota$, with $\upsilon,\mu,\iota$ smooth,
$\mu>0$ and $\varpi^2>(1-F)(\iota^2/\mu^2-1)$. Then $\mathfrak d=\upsilon^4\tilde{\mathfrak d}$ with
$\tilde{\mathfrak d}=\mu^2-(1-V^2)\iota^2>0$, and
\begin{equation}
 \pi_+=\upsilon\sqrt{\frac{\mu+\sqrt{\tilde{\mathfrak d}}}{1+V}},\qquad
 \pi_-=\upsilon\,\iota\sqrt{\frac{1+V}{\mu+\sqrt{\tilde{\mathfrak d}}}}
 \label{form:edge}
\end{equation}
are smooth solutions. The signed smooth factor $\upsilon$ must be retained:
replacing it by $|\upsilon|$ could introduce a cusp at a zero. Changing both
scalar derivatives by an overall sign leaves the constraints unchanged,
so their sign can be chosen to agree with this smooth continuation. A matter
density proportional to the square of a smooth cutoff is the case
$\iota=0$. At the center we use an exact contracting FLRW core: on
some $[0,\rho_c]$,
\begin{equation}
 m=c_{\rm F}\rho^{n-1},\qquad j=(n-1)c_{\rm F},\qquad \varpi=\rho\sqrt{z_c},\qquad
 z_c+\frac{z_c^2}2=c_{\rm F}>0
 \label{form:flrw-core}
\end{equation}
Then $K=1+z_c$, $a\equiv1$, $\mathcal J_m\equiv j$,
$\mathfrak d=V^2\mathcal J_m^2$, $\pi_+=\pi_-=\sqrt{\mathcal J_m}$ and $Q=0$.
These are the homogeneous data of Eq.~\eqref{form:flrw} in Appendix~\ref{app:exact} below. The choice
$m=j=\varpi=0$ gives flat data.

\paragraph{Complete initial data.}
For $n\in\{5,6\}$, choose smooth real functions $m,j,\varpi$ of $\rho\ge0$,
with $\varpi\ge0$. Take the FLRW data
of Eq.~\eqref{form:flrw-core}, or flat data, on $[0,\rho_c]$, and
$m=m_\infty$, $j=\varpi=0$ on $[\rho_e,\infty)$.
Require $m_\rho\ge0$, $m\le m_{\rm marg}$ and $\varpi>0$ where
$F=1$, ensuring $a>0$. In the intervening matter region
require $\mathcal J_m>0$ wherever $j\ne0$ and
$\varpi^2>(1-F)(j^2/\mathcal J_m^2-1)$ there. At each boundary of
$\{\mathcal J_m>0\}$ outside the core, require $\mathcal J_m=\upsilon^2\mu$ and $j=\upsilon^2\iota$, with
$\upsilon,\mu,\iota$ smooth, $\mu>0$, and
$\varpi^2>(1-F)(\iota^2/\mu^2-1)$, as assumed immediately before
Eq.~\eqref{form:edge}.

Define $a,V$ by Eq.~\eqref{form:tilt} and reconstruct $P,Q$ by
Eq.~\eqref{form:scalar-inversion}, with signs chosen to match
Eq.~\eqref{form:edge}; use the exact core values and set $P=Q=0$
in vacuum. Integrate $\Psi_x=Q/a$ with $\Psi\to0$ at infinity,
set $\Psi_\tau=P/a$, and obtain $f_\tau$ from the second constraint
in Eq.~\eqref{evol:generalconstraints}. The first is satisfied by
construction. Away from the center, the discriminant is positive where
$j\ne0$ and equals $\mathcal J_m^2$ where $j=0<\mathcal J_m$. At $F=1$, which is a
maximum on the slice, Eq.~\eqref{form:marginal-flux} gives
$\mathcal J_m>0$, so $V=1$ causes no singularity. Equation~\eqref{form:edge}
gives smooth scalar data where $\mathcal J_m$ vanishes, and the exact
core gives regular data at the center.

The exterior is a time-symmetric Boulware--Deser slice with metric
$\dd R^2/(1-F)+R^2\dd\Omega_{n-2}^2$ and constant scalar.
Since $1-F=1-m_\infty\rho^{3-n}+O(\rho^{4-2n})$, its proper radial
length is infinite and the end is asymptotically flat. The data are
therefore smooth and complete on $\mathbb R^{n-1}$, with mass $m$,
invariant $j$, angular extrinsic curvature $\varpi/R$, and
$2kG\mathcal M_\infty=\tilde\alpha^{(n-3)/2}m_\infty$.

Conversely, Eq.~\eqref{form:amgm} gives $m_\rho\ge0$ and $j=0$
where $\mathcal J_m=0$ on any smooth constrained slice with
$F\le1$, $e_1R>0$ and $\varpi\ge0$. The central, exterior and matter-boundary conditions above suffice
to construct smooth complete initial data; reversing time gives the expanding case.

The characteristic inequalities depend only on $K,j$, so they can be
imposed before reconstruction. On compact sets with positive lower
bounds for $\rho,\mathcal J_m,a^2\mathfrak d$, the data depend smoothly on
$(\rho,m,m_\rho,j,\varpi)$ and their derivatives. At matter boundaries
and the center the same statement holds in the smooth variables of
Eqs.~\eqref{form:edge} and \eqref{form:flrw-core}.

\subsection{Marginal spheres and admissibility}
\label{app:admissibility}

\paragraph{Marginal mass.}
Equation~\eqref{form:mass-variables} gives
\begin{equation}
 m-m_{\rm marg}=(F-1)\Bigl[\rho^{n-3}+\frac12\rho^{n-5}(F+1)\Bigr]
 \label{form:marginal-mass}
\end{equation}
Hence $F\le1$ if and only if $m\le m_{\rm marg}$, and a zero of $F-1$
is a nondegenerate maximum of $F$ if and only if $m-m_{\rm marg}$ has
a nondegenerate maximum there. The marginal mass is convex, with
$m_{\rm marg}''=2$ for $n=5$ and $6\rho$ for $n=6$; its tangent lines
lie strictly below it away from the point of contact. One can therefore arrange a first marginal sphere by making $m$ touch
$m_{\rm marg}$ from below at the chosen radius. A tangent-line mass
profile has exactly this property. Smooth interpolation away from the
contact point can then preserve both $m_\rho\ge0$ and
$m\le m_{\rm marg}$.

\paragraph{Formation rate.}
The double-null equations give
\begin{equation}
 \mathcal M_{,u}=-e^fR^{n-2}R_{,v}\Psi_u^2,\qquad
 \mathcal M_{,v}=-e^fR^{n-2}R_{,u}\Psi_v^2
 \label{form:mass-flux}
\end{equation}
To obtain the first, differentiate Eq.~\eqref{def:mass} along $u$,
using $\partial_F(2kG\mathcal M)=KR^{n-3}$ and $F_{,u}$ from
Eq.~\eqref{evol:potential}, and eliminate $R_{,uu}$ and $R_{,uv}$ with
Eqs.~\eqref{evol:constraints} and \eqref{evol:radius}; the terms
without $\Psi$ cancel because
$\partial_R(2kG\mathcal M)=2KR^{n-3}\mathcal V_{\mathrm{grav}}$ at fixed
$F$. The second follows in the same way.

On the slice of Eq.~\eqref{evol:generaldata}, where $e^f=2a^2$,
$P=a\Psi_\tau$ and $Q=a\Psi_x$, one has $R_{,u}=-\tfrac12(1+V)$,
$R_{,v}=\tfrac12(1-V)$, $2a\Psi_u=P-Q$ and $2a\Psi_v=P+Q$. With
$\partial_\tau=\partial_u+\partial_v$ and
$\partial_x=\partial_v-\partial_u$, Eq.~\eqref{form:mass-flux} gives
\begin{equation}
 \mathcal M_{,x}=\tfrac12R^{n-2}(P^2+Q^2+2VPQ),\qquad
 \mathcal M_{,\tau}=R^{n-2}\bigl[\tfrac12V(P^2+Q^2)+PQ\bigr]
 \label{form:d3}
\end{equation}
The expression for $\mathcal M_{,x}$ is the first constraint in
Eq.~\eqref{evol:generalconstraints}.

Now let a point on the slice satisfy $F=1$ and $F_{,x}=0$. Then $a=\varpi>0$
and $V=1$ by Eq.~\eqref{form:tilt}, so $R_{,\tau}=-1$ and
$\mathcal M_{,\tau}=\mathcal M_{,x}=\tfrac12R^{n-2}(P+Q)^2$. Write
$F=F(R,\mathcal M)$ using Eq.~\eqref{def:mass}. Since $R_{,x}=1$,
\begin{equation}
 F_{,\tau}=-F_{,R}+F_{,\mathcal M}\mathcal M_{,\tau},\qquad
 0=F_{,x}=F_{,R}+F_{,\mathcal M}\mathcal M_{,x}
 \label{form:d4}
\end{equation}
and subtracting the second relation from the first gives
$F_{,\tau}=-2F_{,R}$, with $F_{,R}$ taken at fixed $\mathcal M$.
Differentiating Eq.~\eqref{def:mass} at fixed $\mathcal M$ gives, at
$F=1$,
\begin{equation}
 F_{,R}=-\frac{2(n-3)R^2+(n-5)\tilde\alpha}{2R(R^2+\tilde\alpha)}
 \label{form:d5}
\end{equation}
This gives Eq.~\eqref{form:crossing}, independently of $j$, the tilt
and the rest of the data; for $n=6$ it equals
$(6+\tilde\alpha/R_h^2)/[R_h(1+\tilde\alpha/R_h^2)]$.

At such a point the slope of the mass is that of the marginal mass,
$m_\rho=m_{\rm marg}'(\rho_h)=(n-3)\rho_h^{n-4}+\tfrac12(n-5)\rho_h^{n-6}$,
because $F=1$ and $F_{,\rho}=0$ in Eq.~\eqref{form:mass-variables}.
With $V=1$, Eq.~\eqref{form:data-constraints} then gives
\begin{equation}
 \mathcal J_m=\tilde\alpha kG(P+Q)^2=\frac{n-3}{\rho_h^2}+\frac{n-5}{2\rho_h^4}>0
 \label{form:marginal-flux}
\end{equation}
Since $P+Q=2a\Psi_v$, this is the incoming scalar flux, the source of
$\mathcal M_{,v}$ in Eq.~\eqref{form:mass-flux}.

Suppose now that $F\le1$ on the slice, with equality exactly at
finitely many points, and that the data agree with the prescribed
exact core and static exterior near the two ends. At a contact
$x=R_h$, Eq.~\eqref{form:crossing} gives $F_{,\tau}(0,R_h)>0$.
By continuity, $F_{,\tau}\ge c>0$ on a spatial neighborhood of
the contact and a time interval about zero. For negative $\tau$,
\begin{equation}
 F(\tau,x)=F(0,x)-\int_\tau^0F_{,\tau}(s,x)\,\dd s
 \le1+c\tau<1
 \label{form:d6}
\end{equation}
At the contact point, the forward integral gives
\begin{equation}
 F(\tau,R_h)=1+\int_0^\tau F_{,\tau}(s,R_h)\,\dd s
 \ge1+c\tau>1\qquad(\tau>0)
 \label{form:d7}
\end{equation}
Continuity in $x$ gives an interval with $F>1$ on each slice with sufficiently small positive $\tau$. This argument does not require a
nondegenerate spatial maximum.

Away from neighborhoods of the points satisfying $F(0,x)=1$, $F$ is bounded away from one
on the compact part of the initial slice. In the exact regular FLRW or flat core, $F<1$ persists for a
sufficiently short common time interval.
For the outer tail, write the static vacuum function as
$F_{\rm ext}(R)$ and set $A(R)=1-F_{\rm ext}(R)>0$.
The exterior is time-symmetric initially. Its optical coordinates
in Eq.~\eqref{form:vacuum-optical}, with $r_*'(R)=A(R)^{-1}$, give
\begin{equation}
 r_*(R(\tau,x))
 =\frac12\bigl[r_*(x-\tau)+r_*(x+\tau)\bigr]
\end{equation}
Since $r_*$ is increasing, $R(\tau,x)\ge x-|\tau|$.
If the static end starts at $R_e$ with $F_{\rm ext}(R_e)<1$,
then for $x\ge R_e+\delta$ and $|\tau|<\delta$ both points $x-\tau$ and $x+\tau$ on the initial slice lie in that end and $R(\tau,x)>R_e$.
The static function is nonincreasing in $R$, so
$F(\tau,x)\le F_{\rm ext}(R_e)<1$ throughout this tail.
The bounded remaining region is covered by the compact part of
the slice.

Taking a common smaller time interval for these regions and the
finitely many neighborhoods of the marginal spheres gives $\tau_1>0$ such that
$F<1$ on the whole leaf for $-\tau_1\le\tau<0$, while every contact
has an interval with $F>1$ on each leaf $0<\tau\le\tau_1$.
At contact $R_{,u}=-1$ and $R_{,v}=0$. The first sign persists
nearby, so $F>1$ implies $R_{,v}<0$: these intervals are future
trapped. The foliation first forms marginal spheres at $\tau=0$,
exactly at the prescribed radii.

\paragraph{Five dimensions.}
For $K\ge1$, Eq.~\eqref{form:five-conditions} reads
\begin{equation}
 j<2,\qquad j>K^2-2,\qquad j<3K^2-2
 \label{form:five-polygon}
\end{equation}
In the $(K^2,j)$ plane its closure is the quadrilateral with vertices
$(1,-1)$, $(1,1)$, $(4/3,2)$ and $(4,2)$. Since $K^2$ is affine in $m$
at fixed $\rho$, the admissible set is an intersection of three open
half-planes in $(m,j)$. It is not convex in $(K,j)$: the boundary segment $j=3K^2-2$, on which $C_T=0$, is the graph
of a convex function of $K$, so the region below it is not convex. A useful choice
\begin{equation}
 j_{\rm sel}=2(K-1)
 \label{form:five-selection}
\end{equation}
is admissible for $1\le K<2$, since $(2K-2)-(K^2-2)=K(2-K)$,
$(3K^2-2)-(2K-2)=K(3K-2)$ and $2-(2K-2)=2(2-K)$. It is the equality
case of Eq.~\eqref{form:speed-optimize}. The FLRW value
$j=2(K^2-1)$ is admissible if and only if $K^2<2$.

\paragraph{Six dimensions.}
For every $K^2\ge1$ the admissible values of $j$ form an open
interval. To see this, write $S_6$ and $T_6$ of Eq.~\eqref{sym:six} as
\begin{equation}
 \begin{aligned}
 S_6&=-12\Bigl(j-\frac{2K^2}3\Bigr)^2+\frac{25K^4-90K^2+225}3\\
 T_6&=4(j-2K^2)^2-(17K^4-50K^2+25)
 \end{aligned}
 \label{form:six-squares}
\end{equation}
The last term of $S_6$ is positive, since
$25K^4-90K^2+225=25(K^2-9/5)^2+144$, so $S_6>0$ on an interval
$(j^S_-,j^S_+)$ about $2K^2/3$, with
$j^S_\pm=2K^2/3\pm\sqrt{25K^4-90K^2+225}/6$. The quadratic $T_6$ has
real roots in $j$ if and only if $17K^4-50K^2+25\ge0$. This polynomial
vanishes at $K^2=(25\pm10\sqrt2)/17$, and $(25-10\sqrt2)/17<1$, so for
$K^2\ge1$ the roots of $T_6$ are real exactly when
\begin{equation}
 K^2\ge\frac{25+10\sqrt2}{17}\approx2.30
 \label{form:six-real-roots}
\end{equation}
For $1\le K^2<(25+10\sqrt2)/17$, $T_6$ has no real root in $j$ and is
positive throughout that range.

When the roots of $T_6$ are real, they are
$j^T_\pm=2K^2\pm\frac12\sqrt{17K^4-50K^2+25}$.
Their midpoint is $2K^2$, whereas the polynomial $S_6(j)$ has its maximum at
$j=2K^2/3$. The threshold in Eq.~\eqref{form:six-real-roots} exceeds
two, and at this midpoint of the two roots of $T_6$
\begin{equation}
 \left.S_6\right|_{j=2K^2}=-13K^4-30K^2+75\le-37<0\qquad(K^2\ge2)
\end{equation}
Since $2K^2>2K^2/3$, this gives
$j^S_+<2K^2\le j^T_+$. Thus the upper component of $T_6>0$ cannot
meet the interval $S_6>0$, including when the two roots of $T_6$ coincide.
The admissible values of $j$ are therefore
\begin{equation}
 j^S_-<j<j^S_+\quad\Bigl(K^2<\frac{25+10\sqrt2}{17}\Bigr),\qquad
 j^S_-<j<\min(j^S_+,j^T_-)\quad\Bigl(K^2\ge\frac{25+10\sqrt2}{17}\Bigr)
 \label{form:six-interval}
\end{equation}

A useful choice is
\begin{equation}
 j_{\rm sel}=\frac{z(5-z)}8,\qquad K=1+z
 \label{form:six-j}
\end{equation}
Substituting it in Eq.~\eqref{sym:six} gives
\begin{align}
 T_6&=24+86z+z^2+\frac{17}{16}z^2(z-5)^2\nonumber\\
 S_6&=16(z-1)^2(z+2)+16+10z
       +\frac{21}{16}z^2+\frac{31}{8}z^3
       +\frac{13}{16}z^4
 \label{form:six-positive-polynomials}
\end{align}
Both expressions are positive for $z\ge0$, since each of their terms is
nonnegative there and the terms $24$ and $16$ are positive. Their sum
is $8\Delta_6$ by Eq.~\eqref{sym:dependence} with $n=6$, so
$\Delta_6>0$ as well. The interval in Eq.~\eqref{form:six-interval}
therefore contains $j_{\rm sel}$ and is nonempty for every $K\ge1$.

For the FLRW value $j=\frac52(K^2-1)$ and the vacuum value $j=0$,
Eq.~\eqref{sym:six} gives
\begin{equation}
 \begin{aligned}
 T_6&=8K^2(5-2K^2),& S_6&=16K^2(5-2K^2)&&\bigl(j=\tfrac52(K^2-1)\bigr)\\
 T_6&=-K^4+50K^2-25,& S_6&=3(K^2-5)^2&&(j=0)
 \end{aligned}
 \label{form:six-flrw-vacuum}
\end{equation}
Hence the FLRW value is admissible if and only if $K^2<5/2$. The roots
of $K^4-50K^2+25$ are $K^2=25\pm10\sqrt6$, with $25-10\sqrt6<1$, so
the vacuum value is admissible if and only if $K^2<25+10\sqrt6$ and
$K^2\ne5$.

Unlike the five-dimensional set, the six-dimensional admissible set is
not convex in $(K^2,j)$. The points $(1,-1)$ and $(9,0)$ are
admissible, with $(T_6,S_6)=(44,20)$ and $(344,48)$, but their
midpoint $(5,-1/2)$ has $S_6=-43<0$.

\subsection{Construction of the initial data and the mass bound}
\label{app:formation-construction}

\paragraph{Choice of profiles.}
The construction has three parts. The mass profile of an FLRW core with small curvature is joined to the
tangent line of $m_{\rm marg}$ at the intended formation radius. Beyond
that radius the mass increases by an arbitrarily small amount before
becoming constant. The profile $j$ follows an admissible value until
the geometry permits a smooth transition to vacuum; the tilt is chosen so that the constraint equations have real
solutions for the scalar derivatives throughout.

Let $n=5$ with $\rho_h>1$, or $n=6$ with $\rho_h>0$, and let
$\eta_{\mathcal M}$ be the permitted excess in the dimensionless
mass $m$. A smaller $\eta_{\mathcal M}$ gives a
stronger statement, so we may assume $0<\eta_{\mathcal M}\le1$ and,
if $n=5$, $\eta_{\mathcal M}<\frac12(3\rho_h^2+1)(\rho_h^2-1)$. This
condition ensures
$m_{\rm marg}(\rho_h)+\eta_{\mathcal M}<\frac32\rho_h^4$, which keeps
$K<2$ on the whole exterior.

Just inside the chosen marginal radius $\rho_h$, the mass follows the tangent line
of the marginal mass,
\begin{equation}
 m_{\rm tan}=m_{\rm marg}(\rho_h)+s_h(\rho-\rho_h),\qquad
 s_h=m_{\rm marg}'(\rho_h)
 \label{form:d8}
\end{equation}
that is, $s_h=2\rho_h$ for $n=5$ and $s_h=3\rho_h^2+\frac12$ for
$n=6$. Since $m_{\rm marg}$ is strictly convex, this line touches
$m_{\rm marg}$ at $\rho_h$ and lies below it elsewhere
(Appendix~\ref{app:admissibility}).

Choose $\rho_a<\rho_b<\rho_h$ with $m_{\rm tan}>0$ on
$[\rho_a,\rho_h]$ and, for $n=5$, $m_{\rm tan}<\frac32\rho^4$ there.
On the tangent line these conditions mean $K>1$ and, for $n=5$, $K<2$,
since $K^2=1+2m/\rho^{n-1}$; in this range the values $j_{\rm sel}$ of
Eqs.~\eqref{form:five-selection} and \eqref{form:six-j} are admissible.
Both conditions hold at $\rho=\rho_h$, and hence on $[\rho_a,\rho_h]$
for $\rho_a$ close enough to $\rho_h$. For $n=5$ the second one at
$\rho=\rho_h$ is
\begin{equation}
 m_{\rm tan}(\rho_h)=\rho_h^2+\frac12<\frac32\rho_h^4
 \label{form:tangent-radius}
\end{equation}
that is, $(3\rho_h^2+1)(\rho_h^2-1)>0$, or $\rho_h>1$, which is the
radius bound $K_h=1+1/\rho_h^2<2$ of Eq.~\eqref{form:radius-bound}.
This is the only place where the radius bound enters.

Inside $\rho_a$ the mass is that of the FLRW core, $c_{\rm F}\rho^{n-1}$,
for which $K^2=1+2c_{\rm F}$ and $j=(n-1)c_{\rm F}$ by
Eq.~\eqref{form:flrw-core}. Choose $c_{\rm F}$ such that
\begin{equation}
 c_{\rm F}\rho^{n-1}<m_{\rm tan}\quad(\rho_a\le\rho\le\rho_b),\qquad
 c_{\rm F}\rho_b^4<\frac12,\qquad
 0<c_{\rm F}<c_{\max}
 \label{form:core-conditions}
\end{equation}
with $c_{\max}=\frac12$ for $n=5$ and $\frac34$ for $n=6$. The first
condition keeps the interpolated mass below the tangent line, since the
mass defined in Eq.~\eqref{form:tangent-profile} is a convex combination
of $c_{\rm F}\rho^{n-1}$ and $m_{\rm tan}$ on $[\rho_a,\rho_b]$. The
second keeps the core mass $c_{\rm F}\rho^{n-1}$ below $m_{\rm marg}$ for
$0<\rho\le\rho_b$, where
$c_{\rm F}\rho^{n-1}<\frac12\rho^{n-5}<m_{\rm marg}$. The third makes
the FLRW values admissible. These are $j=2(K^2-1)$ for $n=5$ and
$j=\frac52(K^2-1)$ for $n=6$, admissible exactly when $K^2<2$ and
$K^2<\frac52$, respectively (Appendix~\ref{app:admissibility}), that
is, when $c_{\rm F}<c_{\max}$. Since $m_{\rm tan}$ has a positive minimum
on $[\rho_a,\rho_b]$, every sufficiently small $c_{\rm F}>0$ satisfies
Eq.~\eqref{form:core-conditions}.

With these choices, define the mass profile by
\begin{equation}
 \begin{gathered}
 m=(1-\mathcal S_0)c_{\rm F}\rho^{n-1}+\mathcal S_0 m_{\rm tan}\quad(\rho\le \rho_h),\qquad
 \mathcal S_0=\mathcal S\Bigl(\frac{\rho-\rho_a}{\rho_b-\rho_a}\Bigr)\\
 m_\rho=s_h\Bigl[1-\mathcal S\Bigl(\frac{\rho-\rho_h}{d_h}\Bigr)\Bigr]
 +\epsilon_{\rm m}\mathcal S\Bigl(\frac{\rho-\rho_h}{d_h}\Bigr)[1-\mathcal S(\rho-\rho_2)]^2
 \quad(\rho\ge \rho_h)
 \end{gathered}
 \label{form:tangent-profile}
\end{equation}
where
\begin{equation}
 d_h=\frac{\eta_{\mathcal M}}{4s_h},\qquad
 \epsilon_{\rm m}=\min\Bigl\{s_h,\ \frac{\eta_{\mathcal M}}{4(\rho_2+1-\rho_h)}\Bigr\}
 \label{form:d9}
\end{equation}
and $\rho_2\ge3+\max\{\rho_h,\rho_{\rm vac}\}$ is arbitrary, with
$\rho_{\rm vac}=(2m_{\rm marg}(\rho_h)+2)^{1/4}$ for $n=5$ or
$((m_{\rm marg}(\rho_h)+1)/2)^{1/5}$ for $n=6$. Inside $\rho_h$ the
mass interpolates between the FLRW core and the tangent line on
$[\rho_a,\rho_b]$. Beyond $\rho_h$ its slope decreases from $s_h$ to
$\epsilon_{\rm m}\le s_h$ over the width $d_h$, and the growth is
switched off between $\rho_2$ and $\rho_2+1$. The radius
$\rho_{\rm vac}$ is chosen so that, beyond it, the mass is small
enough compared with $\rho^{n-1}$ for $j$ to be brought to zero
within the admissible region (see below).

Next choose the invariant and tilt as
\begin{equation}
 \begin{aligned}
 j&=\bigl[(1-\mathcal S_1)(n-1)c_{\rm F}+\mathcal S_1j_{\rm sel}(K)\bigr]\,
   [1-\mathcal S(\rho-\rho_2+2)]\\
 \varpi&=\bigl[(1-\mathcal S_2)\rho\sqrt{z_c}+\mathcal S_2\bar\varpi\bigr]\,[1-\mathcal S(\rho-\rho_2+1)]
 \end{aligned}
 \label{form:j-w-profile}
\end{equation}
with $\mathcal S_1=\mathcal S(4\rho/\rho_a-1)$,
$\mathcal S_2=\mathcal S(4\rho/\rho_a-2)$, $j_{\rm sel}$ from
Eq.~\eqref{form:five-selection} or \eqref{form:six-j}, and
\begin{equation}
 \bar\varpi^2=1+\sup_{[\rho_a/2,\rho_2-1]}(1-F)\Bigl(\frac{j^2}{\mathcal J_m^2}-1\Bigr)_+
 \label{form:d10}
\end{equation}
The supremum is finite because $m_\rho>0$ on $(0,\rho_2+1)$. The
invariant $j$ changes from its FLRW value $(n-1)c_{\rm F}$ to
$j_{\rm sel}$ on $[\rho_a/4,\rho_a/2]$ and is switched off between
$\rho_2-2$ and $\rho_2-1$. The tilt changes from its core value
$\rho\sqrt{z_c}$ to the constant $\bar\varpi$ on $[\rho_a/2,3\rho_a/4]$
and is switched off between $\rho_2-1$ and $\rho_2$.

The core and exterior conditions hold on $[0,\rho_a/4]$ and
$[\rho_2+1,\infty)$. In the inner transition,
$m_\rho=(1-\mathcal S_0)(n-1)c_{\rm F}\rho^{n-2}+\mathcal S_0s_h
+\mathcal S_0'(m_{\rm tan}-c_{\rm F}\rho^{n-1})>0$.
Both terms in the convex combination lie below $m_{\rm marg}$;
beyond $\rho_h$, $m_\rho\le s_h$ keeps $m\le m_{\rm tan}$.
Thus $F\le1$, with equality only at $\rho_h$. Since $m''(\rho_h)=0$, this
maximum is nondegenerate, with
$F_{,RR}(R_h)=-2/(R_h^2+\tilde\alpha)$ for $n=5$ and
$-6/(R_h^2+\tilde\alpha)$ for $n=6$. The mass excess satisfies
$0<m_\infty-m_{\rm marg}(\rho_h)\le s_hd_h+
\epsilon_{\rm m}(\rho_2+1-\rho_h)\le\eta_{\mathcal M}/2$.

For scalar reconstruction, consider first the interpolation interval
$[\rho_a/4,3\rho_a/4]$, where the mass is still FLRW, so that
$\mathcal J_m=(n-1)c_{\rm F}$ and $K=1+z_c$ are constant. There $j$
interpolates between $(n-1)c_{\rm F}$ and $j_{\rm sel}(1+z_c)$, and
both values lie in $(0,(n-1)c_{\rm F}]$ (for $n=5$,
$j_{\rm sel}=2z_c\le4(z_c+z_c^2/2)$; for $n=6$, $j_{\rm sel}\le5z_c/8$),
so $0<j\le\mathcal J_m$ and Eq.~\eqref{form:tilt-identity} gives
$a^2\mathfrak d\ge\mathcal J_m^2\varpi^2>0$. The choice of $\bar\varpi$ in Eq.~\eqref{form:d10} ensures
$\varpi^2>(1-F)(j^2/\mathcal J_m^2-1)$ on $[3\rho_a/4,\rho_2-1]$,
with $j=0$ beyond and $\varpi(\rho_h)>0$. The only outer boundary of the matter support is at
$\rho_2+1$, where $\mathcal J_m=\epsilon_{\rm m}\rho^{2-n}
[1-\mathcal S(\rho-\rho_2)]^2$ and $j=0$, as required by
Eq.~\eqref{form:edge}.

The characteristic inequalities hold during the inner interpolation
because the FLRW and selected values lie in the same allowed
$j$ interval. On $[\rho_a/2,\rho_2-2]$, $j=j_{\rm sel}$; in five
dimensions its condition $m<\frac32\rho^4$ follows from the choices
of $\rho_a,c_{\rm F}$ up to $\rho_h$ and from
$m\le m_\infty<\frac32\rho_h^4$ beyond. On $[\rho_2-2,\infty)$ the
choice of $\rho_2$ gives $K^2<2$ for $n=5$ and $K^2<5$ for $n=6$,
so the entire interpolation from $j_{\rm sel}$ to zero is admissible.
Compactness and the static end give positive lower bounds for $C_T$
and $C_S$ for each fixed choice of the data.

These profiles satisfy the conditions of the constraint construction in
Appendix~\ref{app:realization}. By Eq.~\eqref{form:crossing} and
Appendix~\ref{app:local-development}, the evolution first forms a
marginal sphere at $\tau=0$, exactly at $R_h=\rho_h\sqrt{\tilde\alpha}$,
and the characteristic inequalities are strict on a time interval
containing $\tau=0$. For $n=5$, $j_h=2(K_h-1)$ because
$\rho_a/2<\rho_h<\rho_2-2$, so the minimum propagation speed of the scalar gravitational mode
attains the bound in Eq.~\eqref{form:speed-bound}. For $n=6$, $F/R^2=1/(\tilde\alpha \rho_h^2)$
at $R_h$. This establishes Eqs.~\eqref{form:five-existence} and
\eqref{form:six-existence}.
As $\rho_h\to1^+$ in five dimensions, the allowed interval at the marginal
sphere, $(K_h^2-2,2)$, shrinks to the corner $(K^2,j)=(4,2)$, so $C_T$ and $C_S$ at the marginal sphere tend to zero for every
such sequence of data.

\paragraph{Mass bound.}
Consider smooth initial data in spherical symmetry that satisfy the
constraints. Assume that the initial slice is complete, has a regular
center and one asymptotically flat end, and obeys $F\le1$ and
$e_1R>0$. Suppose also that $F=1$ at $\rho_h$. Then
\begin{equation}
 m_\infty-m_{\rm marg}(\rho_h)=\int_{\rho_h}^\infty\rho^{n-2}\mathcal J_m\,\dd\rho>0
 \label{form:mass-integral}
\end{equation}
The integrand is nonnegative by Eq.~\eqref{form:amgm} and strictly positive near $\rho_h$ by
Eq.~\eqref{form:marginal-flux}.
That formula is written for the contracting orientation $V=1$.
At an expanding marginal sphere, reversing time gives the same
positive expression for $\mathcal J_m$ with $(P-Q)^2$ in place of
$(P+Q)^2$, so the mass inequality has the same conclusion.
The identity $m(\rho_h)=m_{\rm marg}(\rho_h)$ gives
Eq.~\eqref{form:mass-bound}. ADM-mass conservation gives the same
bound at first formation on a spherical foliation with outward
increasing $R$, since that leaf has $F\le1$. The preceding data
approach but never attain the bound for every $\rho_h>1$ in five
dimensions and $\rho_h>0$ in six dimensions.

\paragraph{Data with mass close to the bound.}
Suppose in addition that
$m_\infty\le m_{\rm marg}(\rho_h)+\eta_{\mathcal M}$. By mass
monotonicity the mass on $[\rho_h,\infty)$ then lies between
$m_{\rm marg}(\rho_h)$ and $m_{\rm marg}(\rho_h)+\eta_{\mathcal M}$. Let
$F_{\rm BD}$ be the static function of mass $m_{\rm marg}(\rho_h)$.
Equation~\eqref{form:mass-variables} gives
$\partial m/\partial F=\rho^{n-3}K\ge\rho^{n-3}$ wherever $m\ge0$, so
$0<\partial F/\partial m\le\rho^{3-n}$ and
\begin{equation}
 \rho^{n-3}|F-F_{\rm BD}|\le\eta_{\mathcal M}\qquad(\rho\ge\rho_h)
 \label{form:d11}
\end{equation}
Thus the mass function and $F$ approach those of the static exterior.
This does not by itself imply that the scalar stress-energy tensor or the slice's
extrinsic curvature is small.

For $n=5$ and $1<\rho_h^2<1+\sqrt2$, the data are not close to vacuum
if Eq.~\eqref{form:five-conditions} holds on
$[\rho_h,(2\rho_h^2+1)^{1/4})$, an interval that is nonempty because
$\rho_h^4<2\rho_h^2+1$. There
$m\ge m_{\rm marg}(\rho_h)=\rho_h^2+\tfrac12$, so
$K^2=1+2m/\rho^4\ge1+(2\rho_h^2+1)/\rho^4$, and admissibility forces
\begin{equation}
 j>K^2-2\ge\frac{2\rho_h^2+1}{\rho^4}-1>0
 \label{form:d12}
\end{equation}
By Eqs.~\eqref{form:amgm} and \eqref{form:tilt},
$\mathcal J_m\ge\sqrt{1-V^2}\,j$ with
$1-V^2=(1-F)/(1-F+\varpi^2)$. Since $m_\rho=\rho^3\mathcal J_m$ and the
mass increases by at most $\eta_{\mathcal M}$ beyond $\rho_h$,
integrating over this interval gives
\begin{equation}
 \int_{\rho_h}^{(2\rho_h^2+1)^{1/4}}\rho^3\sqrt{\frac{1-F}{1-F+\varpi^2}}
 \Bigl(\frac{2\rho_h^2+1}{\rho^4}-1\Bigr)\dd\rho\le\eta_{\mathcal M}
 \label{form:near-minimizer}
\end{equation}

Let $I$ be a closed subinterval strictly inside these limits.
The weight $\rho^3[(2\rho_h^2+1)/\rho^4-1]$ has a positive
minimum $c_I$ on $I$. Since $(1-F)/(1-F+\varpi^2)=1-V^2$,
the integral estimate gives
\begin{equation}
 \int_I\sqrt{1-V^2}\,\dd\rho
 \le\frac{\eta_{\mathcal M}}{c_I}
\end{equation}
For each fixed $0<\epsilon<1$, the integrand is at least
$\sqrt{2\epsilon-\epsilon^2}$ where $|V|\le1-\epsilon$.
The length of that set therefore satisfies
\begin{equation}
 \left|\{\rho\in I:|V(\rho)|\le1-\epsilon\}\right|
 \le\frac{\eta_{\mathcal M}}
 {c_I\sqrt{2\epsilon-\epsilon^2}}
\end{equation}
Thus $|V|\to1$ in measure on $I$. For the contracting slices in
the construction, $V\ge0$, so $V\to1$: the slice approaches the
contracting radial null direction.

The bounded-tilt estimate follows from Eq.~\eqref{form:near-minimizer}.
Equation~\eqref{form:d11} and $F_{\rm BD}<1$ for $\rho>\rho_h$
give a positive lower bound for $1-F$ on $I$ when
$\eta_{\mathcal M}$ is small. For each fixed $L\ge0$, the factor
$\sqrt{(1-F)/(1-F+\varpi^2)}$ therefore has a positive lower bound
where $|\varpi|\le L$, independent of the particular datum and of
sufficiently small mass excess. The length of this set is at most
a constant times $\eta_{\mathcal M}$.
No pointwise rate at either endpoint is asserted. This extends
Eq.~\eqref{form:vacuum-obstruction} from the marginal sphere to an
exterior interval.

For $\rho_h^2>1+\sqrt2$ and small $\eta_{\mathcal M}$, the exterior
matter can instead be confined to $[\rho_h,\rho_h+d_h]$, with exact
static vacuum beyond. Keep the inner construction and omit the
$\epsilon_{\rm m}$ term in Eq.~\eqref{form:tangent-profile}.
Since $2m_{\rm marg}(\rho_h)=2\rho_h^2+1<\rho_h^4$, choose
$\eta_{\mathcal M}$ small enough that
$2[m_{\rm marg}(\rho_h)+\eta_{\mathcal M}]<\rho_h^4$.
Then $K^2=1+2m/\rho^4<2$ throughout the exterior, so the admissible
interval contains both $j_{\rm sel}$ and zero.

To complete the reconstruction, $j$ must vanish before the density
does. Write $y=(\rho-\rho_h)/d_h$. In Eq.~\eqref{form:j-w-profile},
replace the outer cutoff factors multiplying $j$ and $\varpi$ by
\begin{equation}
 1-\mathcal S(4y-1),\qquad 1-\mathcal S(4y-2)
 \label{form:thin-layer-cutoffs}
\end{equation}
respectively, and replace Eq.~\eqref{form:d10} by
\begin{equation}
 \bar\varpi^2=1+
 \sup_{[\rho_a/2,\,\rho_h+d_h/2]}
 (1-F)\left(\frac{j^2}{\mathcal J_m^2}-1\right)_+
 \label{form:thin-layer-tilt}
\end{equation}
The invariant is zero by $y=1/2$; only then does the tilt decrease,
reaching zero at $y=3/4$. The supremum is finite because
$\mathcal J_m>0$ throughout its closed interval. It ensures a
positive discriminant until $j$ vanishes, after which
$\mathfrak d=\mathcal J_m^2>0$ wherever matter remains.
At $y=1$, the smooth factor $\upsilon=\sqrt{1-\mathcal S(y)}$,
with $\mu=s_h\rho^{2-n}$ and $\iota=0$ in Eq.~\eqref{form:edge},
extends the scalar derivatives smoothly by zero. Thus the static end
begins at $\rho_h+d_h$.
The data near $\rho_h$ are unchanged, so $F$ retains its nondegenerate
maximum there, the formation rate is still Eq.~\eqref{form:crossing},
and the speed bound in Eq.~\eqref{form:speed-bound} remains attained. Since $0\le m_\rho\le s_h$
for $\rho>\rho_h$, comparison with $m_{\rm tan}$ still gives $F<1$ there,
and the positive mass excess is at most $s_hd_h=\eta_{\mathcal M}/4$.

For $n=6$ and $K_h^2>5$, that is, $\rho_h^2<1/(\sqrt5-1)$, the exterior
contains points with $K^2=5$, since $K\to1$ at infinity, and every
admissible exterior has $j>0$ there, because $S_6(5,j)=4j(20-3j)$ is
positive only for $0<j<20/3$.

\subsection{Exact central and exterior solutions}
\label{app:exact}

The constructed data agree with exact solutions at both ends: a
contracting homogeneous (FLRW) core at the center, which supplies a
regular center with nonzero scalar energy density, and the static
vacuum solution outside the matter, which supplies the asymptotically
flat end.
The central solution is contracting flat FLRW, with
\begin{equation}
 H_c^2+\frac{\tilde\alpha}{2}H_c^4=\frac{kG P_c^2}{n-1},\qquad
 \dot P_c=-(n-1)H_cP_c,\qquad
 \dot H_c=-\frac{kG P_c^2}{1+\tilde\alpha H_c^2}
 \label{form:flrw}
\end{equation}
Here $H_c$ is the Hubble rate and $P_c$ is the scalar proper-time
derivative in the homogeneous core. Dots denote central proper-time
derivatives. With the initial scale factor normalized
to one, the choices $H_c=-\sqrt{z_c/\tilde\alpha}$ and
$kG P_c^2=(n-1)(z_c+z_c^2/2)/\tilde\alpha$ reproduce the core data
of Eq.~\eqref{form:flrw-core}.

For the vacuum end we use the static GR-branch solution of
Ref.~\cite{Boulware1985}.
To embed its prescribed radial interval in conformal coordinates, write
$A=1-F>0$, define the tortoise coordinate $r_*$ by $\partial_Rr_*=A^{-1}$, and set its static time on the
initial slice to $t_{\rm s}(R)$, with $\partial_Rt_{\rm s}=-V/A$. Then
\begin{equation}
 \partial_R(t_{\rm s}-r_*)=-\frac{1+V}{A}<0,\qquad
 \partial_R(t_{\rm s}+r_*)=\frac{1-V}{A}>0
 \label{form:vacuum-optical}
\end{equation}
so $t_{\rm s}\mp r_*$ are monotone along the slice, and increasing
functions of them serve as the conformal null coordinates $u,v$, with
$u=-R$ and $v=R$ on the slice.
These coordinates give Eq.~\eqref{evol:generaldata}. The local existence argument in Appendix~\ref{app:local-development}
then gives a smooth solution about the complete constructed
slice. Its time interval may shrink as the limiting radius or mass is
approached.
\section{Curvature and spherical continuation}
\label{app:curvature-continuation}

This appendix has two parts. The first derives the curvature bound of
Eq.~\eqref{cont:curvature-bounds} from the characteristic conditions.
The second shows that a bound on $K$ alone is enough to continue a
smooth spherical solution: null integration
controls the solution away from the axis, while a maximum principle
for the normalized scalar equations controls the scalar derivatives
at the center. An auxiliary regular system then supplies the
higher-derivative bounds needed to pass through a central endpoint.

\subsection{The curvature identity and bound}
\label{cont:algebra}

In five dimensions, Eq.~\eqref{app:curvature} gives
\begin{equation}
 \mathcal R_{abcd}\mathcal R^{abcd}
 =4\mathcal A^2
  +12(\mathcal B^2+\mathcal C^2-2\mathcal D^2)
  +12\mathcal E^2
 \label{cont:contraction}
\end{equation}
For $n=5$, Eq.~\eqref{app:curvcoeff} reads
$C_T=1-\tilde\alpha\mathcal A$, $X=1+\tilde\alpha\mathcal C$,
$Y=1-\tilde\alpha\mathcal B$, $H=\tilde\alpha\mathcal D$ and
$K=1+\tilde\alpha\mathcal E$, which gives
Eq.~\eqref{cont:deviation}. Equation~\eqref{sym:base} gives
\begin{equation}
 \tilde\alpha^2(\mathcal B^2+\mathcal C^2-2\mathcal D^2)
 =(X-1)^2+(Y-1)^2-2H^2
 =\frac{2(K-1)^2+j^2/2}{K^2}
 \label{cont:mixed-curvature}
\end{equation}
since $(\tilde\alpha kG)^2[(P^2+Q^2)^2-4P^2Q^2]=j^2$. Substitution gives
Eq.~\eqref{cont:curvature-identity}.
Every term on the right-hand side of that identity is nonnegative,
but this does not bound the individual orthonormal components
$\mathcal B,\mathcal C,\mathcal D$: their Lorentzian contraction in
Eq.~\eqref{cont:mixed-curvature} allows cancellations. The smooth
continuation result below uses the spherical evolution equations and
separate estimates on their solutions.
Thus an upper bound
$\tilde\alpha^2\mathcal R_{abcd}\mathcal R^{abcd}\le C$ gives
\begin{equation}
 1\le K\le1+\sqrt{C/12},\qquad
 |j|\le K\sqrt{C/6}
 \label{cont:invariant-controls}
\end{equation}
provided $K\ge1$.

For $K\ge1$, Eq.~\eqref{form:five-conditions} is equivalent to
\begin{equation}
    1\le K<2,\qquad K^2-2<j<\min(2,3K^2-2)
 \label{cont:d6}
\end{equation}
With $K(\lambda_++\lambda_-)=2$, $4\Delta=K(C_T+3C_S)$ and
$C_T+C_S=2\lambda_-$, Eq.~\eqref{cont:curvature-identity} becomes
\begin{equation}
 28-\tilde\alpha^2\mathcal R_{abcd}\mathcal R^{abcd}
 =\frac{12(K-1)(2-K)(K^2+K+2)}{K^2}+24\Delta+4C_T(2-C_T)
 \label{cont:twentyeight}
\end{equation}
On the admissible set the first term is nonnegative. Since
$4\Delta=K(C_T+3C_S)>KC_T\ge C_T$, the second exceeds $6C_T$. Finally,
$C_T<2\lambda_-=(2-j)/K<(4-K^2)/K\le3$, because $C_S>0$ and $j>K^2-2$.
Therefore
\begin{equation}
 28-\tilde\alpha^2\mathcal R_{abcd}\mathcal R^{abcd}
 >\frac{12(K-1)(2-K)(K^2+K+2)}{K^2}+2C_T(7-2C_T)>0
 \label{cont:d7}
\end{equation}
which gives Eq.~\eqref{cont:curvature-bounds} and an explicit
positive lower bound for $28-\tilde\alpha^2\mathcal R_{abcd}\mathcal R^{abcd}$.
At the corner $(K^2,j)=(4,2)$ all three terms on the right-hand side of Eq.~\eqref{cont:twentyeight} vanish.

At any admissible marginal sphere,
$K_h=1+\tilde\alpha/R_h^2$ and $K_h^2-2<j_h<2$.
Consequently $R_h^2\to\tilde\alpha^+$ forces
$(K_h^2,j_h)\to(4,2)$. Equation~\eqref{cont:curvature-identity}
then gives $\tilde\alpha^2\mathcal R_{abcd}\mathcal R^{abcd}\to28$.
The data constructed in Appendix~\ref{app:formation-construction}
approach this limit as $\rho_h\to1^+$, so the constant is sharp;
Paper~II realizes it from complete initial data
with arbitrarily weak curvature.

\subsection{Conformal equations and first-derivative bounds}
\label{cont:first-derivatives}

The continuation argument is by contradiction. Suppose the
smooth spherical solution could not be extended beyond some time.
We show that, as long as $K$ stays bounded, the fields and their
first derivatives stay bounded up to that time, and that these bounds
suffice to extend the solution, a contradiction. Away from the
center this follows by integrating the equations along null rays. Near
the center the $1/R$ terms of the scalar equation require a different
argument. The normalized scalar equations give a maximum principle:
the maximum of the two scalar derivative magnitudes on a compact
central triangle is attained on its regular axis. The axis values
therefore control the whole triangle. Higher derivatives near the
center are then obtained from an auxiliary regular system
(Appendix~\ref{cont:polar} onward).

Consider smooth initial data in spherical symmetry for the
five-dimensional EGB--scalar equations with $\tilde\alpha>0$.
Work on the GR branch and require the data to satisfy the constraints.
Assume that the initial slice has a regular center, is untrapped,
and has areal radius increasing outward. Choose the conformal coordinates of
Eq.~\eqref{evol:metric} with $R(0,x)=x$, and reflect across the center.
The signed radius is odd in $x$; the scalar and conformal factor are
even. For a fixed finite $v_b>0$, define
\begin{equation}
 \mathscr D_{v_b}
 =\{\,0\le\tau\le v_b,\ |x|\le v_b-\tau\,\},
 \qquad u=\tau-x,\quad v=\tau+x
 \label{cont:domain}
\end{equation}
This is the domain of dependence of $[-v_b,v_b]$ for the spherical
equations. Assume that the development is smooth near this closed
initial interval. If $K$ remains bounded on the smooth part of
$\mathscr D_{v_b}$, the spherical solution extends smoothly throughout
$\mathscr D_{v_b}$, including its axis and terminal vertex. By
Eq.~\eqref{cont:invariant-controls} a bound on the Kretschmann scalar
suffices. No bound on $j$ is assumed, and characteristic positivity is
not required.
This is a statement on the specified finite domain with a smooth
neighborhood of its closed initial base. It requires neither complete
initial data nor an asymptotically flat end. Its conclusion concerns
the exactly spherical equations and does not imply well-posed
nonspherical evolution when a gravitational characteristic condition fails.
It is therefore enough to establish continuation under
\begin{equation}
 1\le K\le K_*,\qquad K_*\ge2
 \label{cont:bounded-K}
\end{equation}
Suppose smooth continuation first fails in Eq.~\eqref{cont:domain}.
All earlier closed slices are regular.
On the initial slice, $F<1$ and $R(0,x)=x$ imply $|V|<1$ in
Eq.~\eqref{evol:generaldata}. The mass constraint in
Eq.~\eqref{evol:generalconstraints} then gives $\mathcal M_x\ge0$;
regularity gives $\mathcal M=0$ at the center.
The ingoing constraint and the mass equation give, on the physical
half of the reflected domain,
\begin{equation}
 \partial_u(e^fR_{,u})
 =-\frac{kG R}{K}e^f\Psi_u^2\le0,\qquad
 \partial_v(2kG\mathcal M)
 =-2kG R^3e^fR_{,u}\Psi_v^2\ge0
 \label{cont:mass-monotonicity}
\end{equation}
Initially $R_{,u}<0$, and the first equation preserves this sign.
Each outgoing ray starts either on the initial slice or on the regular
axis. Its initial mass is nonnegative, so the second equation preserves
$\mathcal M\ge0$ and hence $K\ge1$. Reflection handles the other half.

In five dimensions, Eqs.~\eqref{evol:radius}--\eqref{evol:angular}
reduce to
\begin{align}
 R_{,uv}&=-\frac{e^{-f}R(K-1)}{\tilde\alpha K},
 &F&=1+2e^fR_{,u}R_{,v}\nonumber\\
 f_{uv}&=\frac{e^{-f}}{\tilde\alpha}(1-C_T),
 &j&=2\tilde\alpha kG e^f\Psi_u\Psi_v\nonumber\\
 \Psi_{uv}&=-\frac{3}{2R}
 (R_{,u}\Psi_v+R_{,v}\Psi_u)
 \label{cont:conformal-equations}
\end{align}
Set $\varkappa=e^{-f}/(-R_{,u})$. In the Bondi-type chart $(v,R)$
the metric is
$\dd s^2=-A\varkappa^2\dd v^2+2\varkappa\,\dd v\,\dd R+R^2\dd\Omega_3^2$,
so $\varkappa$ is the lapse of that chart. The ingoing constraint, the radius equation
and the expression for $F$ give
\begin{equation}
 \partial_u\log\varkappa=\frac{kG R\Psi_u^2}{K R_{,u}}\le0,\qquad
 0\le\partial_v\log(-R_{,u})=\frac{\varkappa R(K-1)}{\tilde\alpha K}
 <\frac{\varkappa R}{\tilde\alpha},\qquad
 R_{,v}=\frac{A\varkappa}{2}
 \label{cont:lapse-identities}
\end{equation}
The middle bound uses only $K\ge1$: the Gauss--Bonnet factor
$(K-1)/K$ is less than one whatever the mass.

\paragraph{Rectangles away from the axis.}
Let $\mathscr Q=[u_0,u_1]\times[v_0,v_1]$ lie in the physical half.
Suppose the solution is smooth on $\mathscr Q$ except possibly at the
corner $(u_1,v_1)$, and near the past edges $\Sigma_{\rm out}=\{u_0\}\times[v_0,v_1]$ and
$\Sigma_{\rm in}=[u_0,u_1]\times\{v_0\}$, and that $R\ge R_0>0$ on
$\mathscr Q$ away from the corner. Every integration below runs from an edge
to a point other than the corner. Since $R_{,u}<0$,
$R\le R_{\max}:=\max_{\Sigma_{\rm out}}R$.
Equation~\eqref{cont:lapse-identities} bounds
\begin{equation}
 \varkappa\le\varkappa_0:=\max_{\Sigma_{\rm out}}\varkappa,\qquad
 -R_{,u}\le\bar\nu:=
 \max_{\Sigma_{\rm in}}(-R_{,u})
 e^{\varkappa_0R_{\max}(v_1-v_0)/\tilde\alpha}
 \label{cont:d1}
\end{equation}
Hence $e^{-f}\le\varkappa_0\bar\nu$, and the radius equation bounds
$|R_{,uv}|\le\varkappa_0\bar\nu R_{\max}/\tilde\alpha$.
Integration bounds $R_{,v}$. With $R\ge R_0$, the scalar equation
is a pair of linear null transport equations with bounded
coefficients; Gr\"onwall's inequality bounds $\Psi_u$ and $\Psi_v$
on the finite rectangle.

To control $f$, substitute $C_T$ from Eq.~\eqref{sym:five} into $f_{uv}=e^{-f}(1-C_T)/\tilde\alpha$:
\begin{equation}
 \tilde\alpha f_{uv}=e^{-f}\Bigl(1-\frac3K+\frac2{K^3}\Bigr)
 +\frac{3\tilde\alpha kG\Psi_u\Psi_v}{K}
 -\frac{2\tilde\alpha^2(kG)^2e^f\Psi_u^2\Psi_v^2}{K^3}
 \label{cont:f-source}
\end{equation}
For $K\ge1$ the bracket lies in $[1-\sqrt2,1)$ and the last
term is nonpositive. Thus $f_{uv}$ has an upper bound without
assuming an upper bound on $e^f$. Double integration bounds $f$
above; the bound on $e^{-f}$ bounds it below. The source is then
bounded in absolute value, so integration bounds $f_u,f_v$.
Finally, $F=1+2e^fR_{,u}R_{,v}$ and
$K=1+\tilde\alpha F/R^2$ bound $K$.

The fields and first derivatives now lie in a bounded region with
$R,K>0$, where the conformal equations have smooth coefficients.
Appendix~\ref{app:local-development} gives continuation past the corner.
No prior upper bound on $K$ or $j$ is needed away from the axis.

\paragraph{At the center.}
Near the axis, the $1/R$ scalar coefficient requires a different
estimate. Let $(T,T)$ be a first central endpoint and, for some
$0\le u_{\rm c}<T$, set $\mathscr N=\{u_{\rm c}\le u\le v<T\}$, with
$8(T-u_{\rm c})^2(K_*-1)\le\tilde\alpha$.
Since $R_{,uv}\le0$, $R_{,v}$ is at most its initial value
$(1-V)/2<1$, so $R<v-u\le T-u_{\rm c}$. Consequently
$F\le1/8$, $A\ge1/2$ and $R_{,v}=A\varkappa/2>0$ there. Define
\begin{equation}
 \Upsilon_+=\frac{\Psi_v}{R_{,v}},\qquad \Upsilon_-=-\frac{\Psi_u}{R_{,u}},\qquad
 \Gamma_R=\frac{R_{,uv}}{R_{,u}R_{,v}}
 =\frac{2R(K-1)}{\tilde\alpha KA}\ge0
 \label{cont:normalized-scalar}
\end{equation}
The scalar and radius equations give
\begin{equation}
 \begin{aligned}
 \partial_u(R^{3/2}\Upsilon_+)
 &=R_{,u}\Bigl[\tfrac32R^{1/2}\Upsilon_--\Gamma_R R^{3/2}\Upsilon_+\Bigr]\\
 \partial_v(R^{3/2}\Upsilon_-)
 &=R_{,v}\Bigl[\tfrac32R^{1/2}\Upsilon_+-\Gamma_R R^{3/2}\Upsilon_-\Bigr]
 \end{aligned}
 \label{cont:averaging}
\end{equation}

For a point $\mathsf p\in\mathscr N$, both the outgoing segment from the
axis point $(u_{\mathsf p},u_{\mathsf p})$ to $\mathsf p$ and the ingoing segment from $\mathsf p$ to
$(v_{\mathsf p},v_{\mathsf p})$ lie in $\mathscr N$ and end at regular axis
points before $(T,T)$. Parametrize each by the areal radius $r=R\in(0,R_{\mathsf p}]$ and
integrate Eq.~\eqref{cont:averaging} outward from the axis. With
$w_R(r)=\exp[-\int_r^{R_{\mathsf p}}\Gamma_R(y)\,\dd y]\in(0,1]$,
\begin{equation}
 \Upsilon_-(\mathsf p)=\int_0^{R_{\mathsf p}}w_R\,\Upsilon_+\,\frac{3r^{1/2}\,\dd r}{2R_{\mathsf p}^{3/2}},\qquad
 \Upsilon_+(\mathsf p)=\int_0^{R_{\mathsf p}}w_R\,\Upsilon_-\,\frac{3r^{1/2}\,\dd r}{2R_{\mathsf p}^{3/2}}
 \label{cont:d8}
\end{equation}
where each weight and integrand is evaluated along its own ray:
outgoing in the first formula and ingoing in the second.
The weight $3r^{1/2}/(2R_{\mathsf p}^{3/2})$ integrates to one over
$[0,R_{\mathsf p}]$, and $0<w_R\le1$, so each
expression is an average with possible damping.
Along each segment the mass equation reads
$\dd(2kG\mathcal M)/\dd r=kGr^3A\Upsilon_\pm^2$, with $\Upsilon_+$ on the
outgoing and $\Upsilon_-$ on the ingoing segment, and
$2kG\mathcal M=(K^2-1)r^4/(2\tilde\alpha)$.

The maximum bound follows directly from the differential equations.
Expanding Eq.~\eqref{cont:averaging} gives
\begin{align*}
 \partial_u\Upsilon_+
 &=R_{,u}\left[\frac{3}{2R}(\Upsilon_--\Upsilon_+)
                  -\Gamma_R\Upsilon_+\right]\\
 \partial_v\Upsilon_-
 &=R_{,v}\left[\frac{3}{2R}(\Upsilon_+-\Upsilon_-)
                  -\Gamma_R\Upsilon_-\right]
\end{align*}
For $\mathsf p\in\mathscr N$, consider the compact triangle
$\{u_{\mathsf p}\le u\le v\le v_{\mathsf p}\}$.
The solution is smooth there, including its axis, because
$v_{\mathsf p}<T$. Let $M$ be the maximum of
$\max(|\Upsilon_+|,|\Upsilon_-|)$ on this triangle. If $M=0$, or if
the triangle consists of a single axis point, the axis bound is
immediate. Suppose instead that $M>0$ is attained at an off-axis
point $\mathsf q$.

If $\sigma\Upsilon_+(\mathsf q)=M$, with $\sigma=\pm1$, increasing
$u$ at fixed $v$ stays inside the triangle for a sufficiently small
step. The one-sided maximum condition gives
$\partial_u(\sigma\Upsilon_+)(\mathsf q)\le0$, whereas the equation gives
\begin{equation*}
 \partial_u(\sigma\Upsilon_+)(\mathsf q)
 =(-R_{,u})\left[\frac{3}{2R}(M-\sigma\Upsilon_-)
                       +\Gamma_R M\right]\ge0
\end{equation*}
All terms on the right are evaluated at $\mathsf q$. Both terms in
brackets are nonnegative, so equality requires
$\Gamma_R(\mathsf q)=0$ and $\sigma\Upsilon_-(\mathsf q)=M$.

If $\sigma\Upsilon_-(\mathsf q)=M$, decrease $v$ at fixed $u$ instead.
The one-sided maximum condition is now
$\partial_v(\sigma\Upsilon_-)(\mathsf q)\ge0$, while
\begin{equation*}
 \partial_v(\sigma\Upsilon_-)(\mathsf q)
 =-R_{,v}\left[\frac{3}{2R}(M-\sigma\Upsilon_+)
                       +\Gamma_R M\right]\le0
\end{equation*}
Thus again $\Gamma_R(\mathsf q)=0$ and both signed scalar derivatives
equal $M$. The two inward variations remain available on either null
edge and at their corner, so these cases cover all off-axis maxima.

Since $R>0$ and $A>0$ off the axis, $\Gamma_R(\mathsf q)=0$ implies
$K(\mathsf q)=1$ and hence $\mathcal M(\mathsf q)=0$.
The mass is nonnegative at the nearby points with the same $u$ and
smaller $v$. Its one-sided derivative at $\mathsf q$ must therefore
satisfy $\partial_v(2kG\mathcal M)(\mathsf q)\le0$. But the mass
equation gives
\begin{equation*}
 \partial_v(2kG\mathcal M)(\mathsf q)
 =kG R^3 A R_{,v}\Upsilon_+^2(\mathsf q)>0
\end{equation*}
This contradiction excludes a positive off-axis maximum. The
maximum is attained on the regular axis, using only points strictly
before $T$.

On the axis
$\Upsilon_+=\Upsilon_-$, $A=1$, $j=\tilde\alpha kG\Upsilon_+^2$, and the regular-center identity in Eq.~\eqref{sym:center} gives $K^2-1=j/2$. With
\begin{equation}
 J_{\rm c}=\sup_{u_{\rm c}\le s<T}j(s,s)
 \label{cont:d2}
\end{equation}
the maximum principle above, the relation
$j=\tilde\alpha kG\,A\,\Upsilon_+\Upsilon_-$ with $0<A\le1$, and integration
of the mass equation along outgoing segments give
\begin{equation}
 \tilde\alpha kG\max(\Upsilon_+^2,\Upsilon_-^2)\le J_{\rm c},\qquad
 |j|\le J_{\rm c},\qquad K^2-1\le\tfrac12J_{\rm c}
 \qquad\text{on }\mathscr N
 \label{cont:central-first}
\end{equation}
and at axis points the left sides equal $j$, $j$ and $j/2$. Under
Eq.~\eqref{cont:bounded-K}, $J_{\rm c}\le2(K_*^2-1)$.

The factor $\varkappa$ is now controlled from the edge
$\Sigma_{\rm c}=\{u=u_{\rm c}\}$. By Eq.~\eqref{cont:lapse-identities},
$\varkappa\le\max_{\Sigma_{\rm c}}\varkappa$, and along ingoing rays
\begin{equation}
 \frac{\dd\log\varkappa}{\dd R}=\frac{kGR\Upsilon_-^2}{K}
 \le\frac{2(K_*^2-1)R}{\tilde\alpha}
 \label{cont:d3}
\end{equation}
which bounds $\varkappa$ below. On the axis, $F=0$ and
$R_{,u}=-R_{,v}$ give $R_{,v}=\varkappa/2$ and $e^{-f}=\varkappa^2/2$.
Along outgoing rays, Eq.~\eqref{cont:lapse-identities} increases
$-R_{,u}$ by at most the factor
$\exp[(T-u_{\rm c})^2\max_{\Sigma_{\rm c}}\varkappa/\tilde\alpha]$.
Thus $-R_{,u}$, $R_{,v}=A\varkappa/2$ and $e^{\pm f}$ have finite
positive upper and lower bounds on $\mathscr N$.

The scalar derivatives $\Psi_u=-\Upsilon_-R_{,u}$ and
$\Psi_v=\Upsilon_+R_{,v}$ are then bounded; no lower bound on their
magnitudes is needed.
Equation~\eqref{cont:f-source} with Eq.~\eqref{cont:central-first}
bounds $f_{uv}$. Integration from $\Sigma_{\rm c}$ bounds $f_v$, and
integration from the axis, where $f_u=f_v$, bounds $f_u$. The
constraints
\begin{equation}
 R_{,uu}=-f_uR_{,u}-\frac{kG R}{K}\Psi_u^2,\qquad
 R_{,vv}=-f_vR_{,v}-\frac{kG R}{K}\Psi_v^2
 \label{cont:central-constraints}
\end{equation}
then bound the second radius derivatives. These bounds on the fields and their first derivatives depend only
on $K_*$, $T-u_{\rm c}$ and the data on $\Sigma_{\rm c}$.

Write $\tau=(u+v)/2$ and $x=(v-u)/2$. These estimates give
$R_{,x}\ge c>0$ and $R_{,\tau}=O(x)$. We can therefore use the areal radius as a spatial coordinate near
the axis. To remove the mixed time--radius metric component, choose
a polar time satisfying
\begin{equation}
 t_x=\frac{R_{,\tau}}{R_{,x}}t_\tau
 \label{cont:polar-time}
\end{equation}
with proper-time data on the axis. Its characteristics are the level
curves of $t$, along which $\dd\tau/\dd x=-R_{,\tau}/R_{,x}=O(x)$; the
characteristic leaving the axis at time $\tau_0$ therefore reaches a
given $x$ at a time differing from $\tau_0$ by $O(x^2)$, so points of
the past cone close enough to the axis lie on characteristics that
leave the axis before the endpoint. The derivative
$\partial_\tau(R_{,\tau}/R_{,x})$ is bounded using
Eq.~\eqref{cont:central-constraints}, the mixed derivative
$R_{,uv}$ from Eq.~\eqref{evol:radius}, and $R_{,x}\ge c>0$. Differentiating the
characteristic equation with respect to its axis starting time
therefore gives a linear equation with bounded coefficient. Its
Jacobian remains finite and nonzero, so this change of coordinates
is regular. The bounded conformal factor also makes the endpoint
proper time finite.

\subsection{Polar chart near the center}
\label{cont:polar}

In the rest of the central argument we use the dimensionless variables
of Eq.~\eqref{form:mass-variables} with $n=5$,
$\rho=R/\sqrt{\tilde\alpha}$ and $m=2kG\mathcal M/\tilde\alpha$, together with
$\varphi=\sqrt{kG}\,\Psi$ and the polar time $t$ measured in units of
$\sqrt{\tilde\alpha}$ and normalized by $B(t,0)=1$. In these variables
the metric near the center takes the polar form
\begin{equation}
 \tilde\alpha^{-1}\dd s^2
 =-\frac{B^2}{A}\dd t^2+A^{-1}\dd\rho^2+\rho^2\dd\Omega_3^2,
 \qquad A=1-F,\qquad K=1+\frac{F}{\rho^2}
 \label{cont:polar-metric}
\end{equation}
so that $B$ is the coordinate speed of radial null rays. The local time normalization at the center replaces the usual normalization at infinity.
Use Cartesian spatial coordinates $\boldsymbol x\in\mathbb R^4$
with $\rho=|\boldsymbol x|$ in this polar chart; $\nabla$ and
$\operatorname{div}$ are the Euclidean gradient and divergence. Set
$\mathbf U=(\Pi,\boldsymbol\Phi)$, where
$\boldsymbol\Phi=\nabla\varphi$, $\Phi=\partial_\rho\varphi$ and
$\Pi=B^{-1}\varphi_t$. Then $\Pi\pm\Phi=\sqrt{\tilde\alpha kG}\,\Upsilon_\pm$,
so $|\mathbf U|^2=\tilde\alpha kG(\Upsilon_+^2+\Upsilon_-^2)/2$, and
Eq.~\eqref{cont:central-first} gives $|\mathbf U|\le U_{\max}:=\sqrt{2(K_*^2-1)}$
on a past null cone sufficiently close to the central endpoint. Denote its radius by $\rho_{\rm cone}(t)$ and its endpoint
time by $T_{\rm c}$:
\begin{equation}
 \rho_{\rm cone}'(t)=-B_{\rm phys}(t,\rho_{\rm cone}(t)),\qquad
 \rho_{\rm cone}(T_{\rm c})=0
 \label{cont:central-cone}
\end{equation}
The subscript $\mathrm{phys}$ refers to the original solution; the
auxiliary solution used below is denoted by $\mathrm{aux}$.

The continuation argument below will imply that failure to extend
through the central endpoint forces $j(s,s)$ to be unbounded as
$s\to T$. The reason is that a bound $J$ on $j$ along the axis
already gives the required bound on $K$. If the solution fails to
extend and $j<2$ at an earlier axis point, the center must reach
$j=2$ and $K^2=2$ before $T$. At that crossing,
Eq.~\eqref{sym:five} gives $\Delta=C_T=C_V=C_S=0$.

To obtain the $K$ bound from the axis bound, choose $K_*$ with
$K_*^2>1+J/2$ and $u_{\rm c}\in[0,T)$ satisfying
$8(T-u_{\rm c})^2(K_*-1)\le\tilde\alpha$.
Let $s_b$ be the supremum of the $s<T$ for which $K\le K_*$ on
$\mathscr N\cap\{v\le s\}$. For each such $s$, this set contains the
compact triangle $\{u_{\mathsf p}\le u\le v\le v_{\mathsf p}\}$ for every
point $\mathsf p$ in it. Equation~\eqref{cont:central-first} therefore
holds throughout the set and gives $K^2-1\le J/2<K_*^2-1$.
By continuity, $s_b=T$.

The scalar derivatives are now bounded in the cone of
Eq.~\eqref{cont:central-cone}, expressed in polar coordinates. To control
higher derivatives at the tip, we next construct the auxiliary solution.

\subsection{Auxiliary equations}

Near the center the scalar equation is
quasilinear: its coefficient $B$ is determined by the mass, and the
mass by the scalar derivatives themselves. Bounds on the first
derivatives do not immediately give bounds on higher derivatives
through the tip of the cone, because the estimates for the original
system could degenerate there. We therefore compare the solution with an
auxiliary system in which the mass source $|\mathbf U|^2$ is replaced
by a bounded smooth function of it, cut off at a small radius. The
auxiliary system has smooth solutions for all time. Inside the cone,
where $|\mathbf U|$ is already known to be bounded, the cutoff is
inactive, so the auxiliary and physical solutions coincide there. The smoothness of the auxiliary solution through
the tip is then inherited by the physical one. The cutoff is fixed once
and for all, independently of the derivative order.

Choose a smooth nonnegative function $\Theta_{\rm a}(z)$ equal to $z$
for $0\le z\le4U_{\max}^2$ and constant for sufficiently large $z$.
Let $\chi_{\rm a}(\rho)$ be a smooth radial function between zero and one,
equal to one for $\rho\le\rho_\chi/2$ and zero for $\rho\ge\rho_\chi$.
Define the bounded auxiliary source
\begin{equation}
 \mathfrak e=\chi_{\rm a}\Theta_{\rm a}(|\mathbf U|^2),\qquad
 0\le\mathfrak e\le\mathfrak e_*:=\max\Theta_{\rm a}
 \label{cont:auxiliary-source}
\end{equation}
Only the radial mass source is changed. The auxiliary equations are
\begin{align}
 m_\rho&=\rho^3A\mathfrak e,\qquad
 \bar m=\frac{m}{\rho^4},\qquad K=\sqrt{1+2\bar m},
 \qquad F=\rho^2(K-1),\qquad A=1-F\nonumber\\
 \frac{B_\rho}{B}&=\frac{2\rho(K-1)}{KA},\qquad B(t,0)=1\nonumber\\
 \Pi_t&=\operatorname{div}(B\boldsymbol\Phi),\qquad
 \boldsymbol\Phi_t=\nabla(B\Pi)
 \label{cont:auxiliary-equations}
\end{align}
The first line is the radial mass constraint with the cutoff source,
the second is the lapse constraint, and the third is the scalar wave
equation written as a first-order system. Take smooth compactly supported radial data, with $\boldsymbol\Phi$
a gradient. Choose $\rho_\chi$ once, so that
$\rho_\chi^2\sqrt{\mathfrak e_*/2}<1/2$. While $A>0$, the mass equation gives
$0\le m\le(\mathfrak e_*/4)\min(\rho^4,\rho_\chi^4)$. A first zero of $A$
would require $m=\rho^2+1/2\ge1/2$, contradicting this bound.
At every radius,
\begin{align}
 1\le K&\le\sqrt{1+\mathfrak e_*/2},\qquad
 0\le F\le\rho_\chi^2\sqrt{\mathfrak e_*/2}\nonumber\\
 A&\ge A_{\min}:=1-\rho_\chi^2\sqrt{\mathfrak e_*/2}>\frac12,\qquad
 1\le B\le
 \exp\!\left(\frac{\mathfrak e_*\rho_\chi^2}{2A_{\min}}\right)
 \label{cont:auxiliary-metric}
\end{align}
The bound on $B$ follows from $K-1\le m/\rho^4\le(\mathfrak e_*/4)\min(1,\rho_\chi^4/\rho^4)$
by integrating the lapse equation from the center.

The rest of this subsection establishes that the auxiliary system
has smooth solutions for all time. The argument is the usual one for
a wave equation with coefficients determined by the solution: bounds
on the coefficient $B$ and its derivatives in terms of the scalar
state, an energy estimate for the scalar system with those
coefficients, an iteration giving local existence, and a linear
growth inequality that prevents the energies from becoming infinite
in finite time.

Differentiation of the lapse constraint gives
\begin{equation}
 \operatorname{div}(\nabla B)=\widehat s,\qquad
 \widehat s=\frac{2B}{K^3A}
 \left[(1+\rho^2(K-1)^2)\mathfrak e+
 2(K-1)^2(2K+1)\right]
 \label{cont:lapse-source}
\end{equation}
This source is bounded. Outside the cutoff the mass is spatially
constant, $K-1=O(\rho^{-4})$ and $\widehat s=O(\rho^{-8})$.
Radial integration yields
\begin{equation}
 \frac{B_\rho}{\rho}
 =\rho^{-4}\int_0^\rho y^3\widehat s(y)\,\dd y,\qquad
 B_{\rho\rho}=\widehat s-3\frac{B_\rho}{\rho}
 \label{cont:lapse-low}
\end{equation}
Thus $B,\partial B,\partial^2B$ have uniform pointwise bounds.
Here and below, $\partial$ and $\partial^2$ denote all first and all
second Cartesian spatial derivatives.

The mass equation and its first derivative are
\begin{align}
 (\boldsymbol x\cdot\nabla+4)\bar m&=A\mathfrak e\nonumber\\
 \left(\boldsymbol x\cdot\nabla+5+
 \frac{\rho^2\mathfrak e}{K}\right)\bar m_i
 &=A\mathfrak e_i-2x_i(K-1)\mathfrak e
 \label{cont:mass-derivative}
\end{align}
The extra zeroth-order coefficient is nonnegative.

The remaining coefficient estimates use radial averages. For a radial
function $g$ define
\begin{equation}
 (\operatorname{Av}g)(\rho)
 =\rho^{-4}\int_0^\rho y^3g(y)\,\dd y
 =\int_0^1z^3g(z\rho)\,\dd z
 \label{cont:d4}
\end{equation}
All norms in the central argument use $\dd^4\boldsymbol x$,
including the angular area $2\pi^2$. Radial integration and
H\"older's inequality give
\begin{equation}
 \begin{gathered}
 \|\operatorname{Av}g\|_{L^p}
 \le\frac{p}{4(p-1)}\|g\|_{L^p},\qquad 1<p\le\infty\\
 g-4\operatorname{Av}g
 =\rho^{-4}\!\int_0^\rho y^4g'(y)\,\dd y,\qquad
 \rho\,\operatorname{Av}|g|\le c_0\|g\|_{L^4}
 \end{gathered}
 \label{cont:averages}
\end{equation}
where $c_0=4^{-3/4}(2\pi^2)^{-1/4}$.
Integrating the mass equation and Eq.~\eqref{cont:mass-derivative}
from the center gives
$0\le\bar m\le\operatorname{Av}\mathfrak e$ and
$|\partial\bar m|\le
\operatorname{Av}(|\partial\mathfrak e|+2\rho(K-1)\mathfrak e)$.
The nonnegative coefficient in Eq.~\eqref{cont:matching-mass} below
similarly gives $|\delta\bar m|\le\operatorname{Av}|\delta\mathfrak e|$,
where $\delta$ denotes the difference of two solutions.
The bounded smooth source satisfies
$|\delta\mathfrak e|\le C|\delta \mathbf U|$ and
$|\partial\mathfrak e|\le C(1+|\partial \mathbf U|)$, with constants fixed
by the cutoffs.

Differentiate $\partial_\rho\log B=2\rho(K-1)/(KA)$ once and
eliminate $\rho\bar m_\rho$ using $A\mathfrak e-4\bar m$.
The ratio $(K-1)/(KA)\le2\bar m$ is smooth in $\bar m$
because $\rho^2(K-1)\le1/2$, and is $O(\rho^{-4})$
outside the cutoff. Together with Eq.~\eqref{cont:averages},
this gives
\begin{align}
 \|\nabla B\|_{W^{1,\infty}}&\le C,\qquad
 \|\nabla^3B\|_{L^4}\le C(1+\|\nabla \mathbf U\|_{L^4})
 \label{cont:aux-coeff}\\
 \|\delta B\|_\infty+\|\nabla\delta B\|_\infty+
 \|\nabla^2\delta B\|_{L^4}&\le C\|\delta \mathbf U\|_{L^4}
 \label{cont:aux-diff}
\end{align}
The constants depend only on the cutoffs. For differences,
integrate
$\delta\log B=\int_0^\rho2y\,\delta[(K-1)/(KA)]\,\dd y$,
using $|\delta[(K-1)/(KA)]|\le C|\delta\bar m|$
and the last bound in
Eq.~\eqref{cont:averages} inside the cutoff and decay outside.
The Hessian difference is bounded by
$C(|\delta\bar m|+|\delta\mathfrak e|+
\min(1,\rho^{-4})|\delta B|)$, so the $L^4$ bound follows from
the first radial-average estimate. The comparison must therefore
use $H^1\subset L^4$, rather than $L^2$, in four spatial dimensions.

Let
\begin{equation}
 E_0=\|\mathbf U\|_{L^2}^2,\qquad
 E_1=E_0+\|\partial\mathbf U\|_{L^2}^2,\qquad
 E_2=E_1+\|\partial^2\mathbf U\|_{L^2}^2
 \label{cont:d5}
\end{equation}
and consider the linear system $\Pi_t=\operatorname{div}(B\boldsymbol\Phi)$,
$\boldsymbol\Phi_t=\nabla(B\Pi)$ with a given lapse. Then
$\frac{\dd}{\dd t}E_0=2\int\Pi\,\boldsymbol\Phi\cdot\nabla B$, and
commuting with one and two Cartesian derivatives gives
\begin{equation}
 \frac{\dd E_1}{\dd t}\le C\|\nabla B\|_{W^{1,\infty}}E_1,\qquad
 \frac{\dd E_2^{1/2}}{\dd t}\le C\|\nabla B\|_{W^{1,\infty}}E_2^{1/2}
 +C\|\nabla^3B\|_{L^4}E_1^{1/2}
 \label{cont:aux-energy}
\end{equation}
using $H^1(\mathbb R^4)\subset L^4$. No time derivative of $B$ enters.

Local existence for the coupled system follows by iteration.
For smooth compactly supported radial data $\mathbf U_0$ with gradient
$\boldsymbol\Phi_0$, iterate the constraints and the linear scalar
system, starting from $\mathbf U^{[0]}=\mathbf U_0$: solve the constraints for the
lapse of the current iterate, then evolve the next iterate with that
lapse. Equations
\eqref{cont:aux-coeff} and \eqref{cont:aux-energy} give a common
interval $[0,T_0]$, depending only on $E_2(\mathbf U_0)$ and the fixed
cutoffs, such that
\begin{equation}
 \sup_{0\le t\le T_0}E_2(\mathbf U^{[\ell]}(t))
 \le4E_2(\mathbf U_0)
 \label{cont:aux-iteration-bound}
\end{equation}
The source in the equation for the difference of consecutive iterates
is $(\operatorname{div}(\delta B\,\boldsymbol\Phi^{[\ell]}),
\nabla(\delta B\,\Pi^{[\ell]}))$, with
$\delta B=B[\mathbf U^{[\ell]}]-B[\mathbf U^{[\ell-1]}]$.
By Eq.~\eqref{cont:aux-diff} and $H^1(\mathbb R^4)\subset L^4$,
its $H^1$ norm is at most
$CE_2(\mathbf U_0)^{1/2}\,\|\mathbf U^{[\ell]}-\mathbf U^{[\ell-1]}\|_{H^1}$, where the
constant depends only on the cutoffs.
Since the initial differences vanish,
\begin{equation}
 \|\mathbf U^{[\ell+1]}-\mathbf U^{[\ell]}\|_{C_tH^1}
 \le CT_0e^{CT_0}E_2(\mathbf U_0)^{1/2}
      \|\mathbf U^{[\ell]}-\mathbf U^{[\ell-1]}\|_{C_tH^1}
 \label{cont:aux-contraction}
\end{equation}
For smaller $T_0$ this is a contraction. Strong $H^1$ convergence,
the uniform $H^2$ bound and Eq.~\eqref{cont:aux-diff} give a unique
solution in $C_tH^1\cap L^\infty_tH^2$.

Global existence follows because the energy inequalities are
linear at the top order.
Substituting the coefficient estimates of Eq.~\eqref{cont:aux-coeff} into
Eq.~\eqref{cont:aux-energy} yields
\begin{equation}
 E_1'\le CE_1,\qquad E_2'\le C(1+E_1)(1+E_2)
 \label{cont:aux-growth}
\end{equation}
Gr\"onwall's inequality gives $E_1(t)\le E_1(0)e^{Ct}$ and
\begin{equation}
 1+E_2(t)
 \le[1+E_2(0)]
 \exp\!\left[C\int_0^t(1+E_1(s))\,\dd s\right]
 \label{cont:aux-gronwall}
\end{equation}
Thus $E_2$ is bounded on every finite interval, and the local
construction can be restarted indefinitely with the same cutoffs.

For higher derivatives, the Poisson equation for the lapse in
Eq.~\eqref{cont:lapse-source} again gains two spatial derivatives
relative to its source. Differentiating the
bounded smooth map $\mathbf U\mapsto\Theta_{\rm a}(|\mathbf U|^2)$ makes the highest
field derivative appear linearly; the other terms contain lower
orders. The differentiated energy equations therefore give a
linear inequality at each new order, with coefficients already
bounded at lower orders. Induction preserves smoothness on every
finite interval. The evolution equations then control time
derivatives, and $\operatorname{curl}\boldsymbol\Phi=0$ persists.

\subsection{Agreement with the original solution}

Fix the source function and spatial cutoff first. Then choose a late
$t_0<T_{\rm c}$ such that $\rho_{\rm cone}(t_0)<\rho_\chi/4$.
The auxiliary initial data will agree with the original solution on a
ball slightly larger than the base of this cone. The slice at $t_0$
is regular, since it precedes the first possible failure of continuation.
The inequalities $|\mathbf U|\le U_{\max}<2U_{\max}$ and
$\rho_{\rm cone}(t_0)<\rho_\chi/4$ allow a small enlargement of the
closed cone base on this regular slice.

Extend $\varphi$ and $\Pi$ from that ball to smooth functions of compact
support. Define the auxiliary vector datum by
$\boldsymbol\Phi=\nabla\varphi$ using the extended scalar, so it remains
a gradient and agrees with the original datum on the ball.

With $B(t,0)=1$, both constraints are integrated from the center.
The mass and lapse inside the matched ball therefore depend only on
the scalar field there. Compare the scalar equations on the original
ingoing null cone, whose boundary speed $-B_{\rm phys}$ gives a
nonpositive boundary flux in the difference energy estimate.

For each $t_1<T_{\rm c}$, compare the solutions on the original cone
truncated at $t_1$. The radii of its spatial slices are at least
$\rho_{\rm cone}(t_1)$, a positive lower bound because
$t_1<T_{\rm c}$; the balls still include the regular center. On this
cone $\rho<\rho_\chi/4$, where $\chi_{\rm a}=1$, and, for the original solution,
$|\mathbf U|\le U_{\max}$ by Eqs.~\eqref{cont:central-first} and \eqref{cont:bounded-K}. Thus
$\mathfrak e=\chi_{\rm a}\Theta_{\rm a}(|\mathbf U|^2)=|\mathbf U|^2$
there. The original solution therefore obeys the auxiliary system in
Eq.~\eqref{cont:auxiliary-equations} on this cone. Write
$\delta \mathbf U=\mathbf U_{\rm aux}-\mathbf U_{\rm phys}$, and
similarly $\delta\bar m$, $\delta B$ and $\delta\mathfrak e$.

The mass difference satisfies
\begin{equation}
 \left(\rho\partial_\rho+4+
 \frac{2\rho^2\mathfrak e_{\rm aux}}
 {K_{\rm aux}+K_{\rm phys}}\right)\delta\bar m
 =A_{\rm phys}\delta\mathfrak e
 \label{cont:matching-mass}
\end{equation}
This follows from the first line of Eq.~\eqref{cont:mass-derivative},
$(\rho\partial_\rho+4)\bar m=A\mathfrak e$, written for both solutions,
since $\delta(A\mathfrak e)=A_{\rm phys}\delta\mathfrak e
+\mathfrak e_{\rm aux}\delta A$ with $\delta A=-\rho^2\delta K$ and
$\delta K=2\delta\bar m/(K_{\rm aux}+K_{\rm phys})$. The source is
Lipschitz in $\mathbf U$, and the extra zeroth-order coefficient is
nonnegative. With $0<A_{\rm phys}\le1$, integration from the center
therefore gives
$|\delta\bar m|\le\operatorname{Av}|\delta\mathfrak e|
\le C\operatorname{Av}|\delta\mathbf U|$, as noted after
Eq.~\eqref{cont:averages}.

Both lapses are normalized at the center and both constraints are
integrated outward from it, so Eq.~\eqref{cont:aux-diff} applies on
central balls:
\begin{equation}
 \|\delta B\|_{L^\infty(\mathbb B)}
 +\|\nabla\delta B\|_{L^\infty(\mathbb B)}
 +\|\nabla^2\delta B\|_{L^4(\mathbb B)}
 \le C\|\delta \mathbf U\|_{L^4(\mathbb B)}
 \le C(t_1)\|\delta \mathbf U\|_{H^1(\mathbb B)}
 \label{cont:matching-lapse}
\end{equation}
where $\mathbb B=\mathbb B_{\rho_{\rm cone}(t)}$ is the Euclidean ball of radius $\rho_{\rm cone}(t)$
in $\mathbb R^4$, the slice of the cone at time $t$. The second
inequality is the Sobolev inequality $H^1\subset L^4$ on $\mathbb B$,
whose constant depends only on a positive lower bound for the radius,
here $\rho_{\rm cone}(t_1)$.

Subtract the scalar equations with $B_{\rm phys}$ as the principal
coefficient:
\begin{equation}
 \delta\Pi_t=\operatorname{div}(B_{\rm phys}\,\delta\boldsymbol\Phi)
 +\operatorname{div}(\delta B\,\boldsymbol\Phi_{\rm aux}),\qquad
 \delta\boldsymbol\Phi_t=\nabla(B_{\rm phys}\,\delta\Pi)
 +\nabla(\delta B\,\Pi_{\rm aux})
 \label{cont:matching-difference}
\end{equation}
Smoothness of the auxiliary solution and
Eq.~\eqref{cont:matching-lapse} bound the difference source, the terms
containing $\delta B$, in $H^1$ by $C(t_1)\|\delta \mathbf U\|_{H^1}$.
Since $\rho_{\rm cone}'=-B_{\rm phys}$, the moving-boundary energy flux
is nonpositive. In the time derivative of
$\int_{\mathbb B}|\delta\mathbf U|^2$, the boundary term from the
principal part of Eq.~\eqref{cont:matching-difference} and the one from
the motion of $\partial\mathbb B$ add up to
$-\int_{\partial\mathbb B}B_{\rm phys}(\delta\Pi-\delta\Phi)^2$, where
$\delta\Phi=\Phi_{\rm aux}-\Phi_{\rm phys}$ is the radial component of
$\delta\boldsymbol\Phi$. The first derivatives of $\delta\mathbf U$ obey
equations with the same principal part, and their boundary terms are
nonpositive by the Cauchy--Schwarz inequality. Hence
\begin{equation}
 \frac{\dd}{\dd t}\|\delta \mathbf U\|_{H^1(\mathbb B)}^2
 \le C(t_1)\|\delta \mathbf U\|_{H^1(\mathbb B)}^2
 \label{cont:matching-energy}
\end{equation}

The difference vanishes at $t=t_0$, so Gr\"onwall's inequality gives
\begin{equation}
 \|\delta\mathbf U(t)\|_{H^1(\mathbb B_{\rho_{\rm cone}(t)})}^2
 \le e^{C(t_1)(t-t_0)}\,
 \|\delta\mathbf U(t_0)\|_{H^1(\mathbb B_{\rho_{\rm cone}(t_0)})}^2=0,
 \qquad t_0\le t\le t_1
 \label{cont:matching-gronwall}
\end{equation}
Thus $\delta \mathbf U\equiv0$ on every truncated cone with
$t_1<T_{\rm c}$, and Eqs.~\eqref{cont:matching-mass} and
\eqref{cont:matching-lapse} give $\delta\bar m=\delta B=0$ there. The
constant $C(t_1)$ may grow without bound as $t_1\to T_{\rm c}$, but the
truncated cones exhaust the open cone, so the original solution
coincides with the auxiliary one on the whole open cone. Since the
auxiliary solution is smooth across $T_{\rm c}$, every derivative of
the original solution has a limit at the tip, namely the value of the
corresponding derivative of the auxiliary solution.

By continuity, $|\mathbf U|\le U_{\max}$ on the closed cone. The inequalities $|\mathbf U|<2U_{\max}$ and
$\rho<\rho_\chi/2$ therefore hold on a neighborhood of the closed cone,
where $\mathfrak e=|\mathbf U|^2$.
Choose a slightly larger auxiliary past null cone by moving its vertex
a sufficiently small amount beyond $T_{\rm c}$, so that the cone remains
in this neighborhood and its base lies inside the larger initial ball
where the data agree. Its slices are central balls, so every radial segment used to
integrate the constraints also lies where $\mathfrak e=|\mathbf U|^2$.
The remaining field equations follow. The auxiliary equations contain only
the radial mass equation; its temporal counterpart,
$m_t=2\rho^3AB\Pi\Phi$, holds because the defect satisfies
\begin{equation}
 \partial_\rho\bigl(m_t-2\rho^3A B\Pi\Phi\bigr)
 =-\frac{\rho|\mathbf U|^2}{K}
       \bigl(m_t-2\rho^3A B\Pi\Phi\bigr)
 \label{cont:temporal-defect}
\end{equation}
a homogeneous linear equation whose solution vanishes at the center
and hence everywhere. The angular field equation then follows from
the contracted Bianchi identity for $\rho>0$, where $A>0$, and extends
through the center by smoothness. No gravitational characteristic
coefficient is divided out at any stage.

The Cartesian metric
$g_{ij}=\delta_{ij}+(K-1)x_ix_j/A$ of the auxiliary solution is smooth
on the enlarged cone, including its tip. The auxiliary solution is
therefore a smooth spherical solution of the field equations on a
neighborhood of the closed original cone.

Agreement on the open original cone was proved above. To establish
agreement on the rest of the overlap, consider a point in the enlarged
cone that lies outside the original cone and away from the axis.
It has a past rectangle whose earlier edges lie in the open original
cone or close to the matched initial ball. Both solutions are defined
and agree on those earlier edges. Uniqueness for the double-null system
in Appendix~\ref{app:local-development}, together with the bounds of
Appendix~\ref{cont:first-derivatives}, gives agreement on the overlap
of their domains in the enlarged cone.

The auxiliary solution thus extends the original development without
assuming that the old polar chart already extends beyond its cone.

Finally the null coordinates are carried across the tip. Extend the
original $u$ and $v$ into the enlarged region by solving the
transport equations
\begin{equation}
 (\partial_t+B\partial_\rho)u=0,\qquad
 (\partial_t-B\partial_\rho)v=0
 \label{cont:optical-extension}
\end{equation}
which say that $u$ is constant along outgoing and $v$ along ingoing
radial null rays. Differentiating in $\rho$ shows that $u_\rho$ and
$v_\rho$ obey linear transport equations along the same rays, with
coefficients $-B_\rho$ and $B_\rho$; a solution of such an equation
cannot change sign, so $u_\rho<0$ and $v_\rho>0$ persist and
$(u,v)$ remain coordinates. The reflection $x\mapsto-x$ across the
center exchanges the two families of rays and preserves the
axis, where $u=v$, so the original conformal coordinates extend as
well.

\subsection{Continuation throughout the domain}

Continue the initial development to compact truncated triangles in
Eq.~\eqref{cont:domain}, using uniqueness on overlaps. Suppose the
first failure occurs at $\tau=T_*\le v_b$.

At a noncentral point $(u_*,v_*)$ of the limiting slice,
$R_{,u}<0$ gives $R(u,v)\le R(-v,v)=v$. Choose a past rectangle
$[u_0,u_*]\times[v_0,v_*]$ with $|u_*|<v_0<v_*$ and $-v_0<u_0<u_*$. Its earlier edge
$v=v_0$ has positive minimum radius. Whenever
$R<\sqrt{\tilde\alpha/(K_*-1)}$, the bound on $K$ gives $F<1$
and hence $R_{,v}>0$. The radius cannot decrease through this
threshold along an outgoing ray, so it has a positive lower bound
on the rectangle. Appendix~\ref{cont:first-derivatives} applies.

A first central endpoint is covered by the central construction;
every point of its small past triangle except the tip is already
smooth. On a sloping side of the domain, use a past rectangle with
regular neighborhoods of its earlier edges. A finite cover and
uniqueness extend the compact limiting slice, or the terminal
vertex when $T_*=v_b$, contradicting first failure. This gives
continuation under Eq.~\eqref{cont:bounded-K};
Eq.~\eqref{cont:invariant-controls} gives the curvature criterion.

\subsection{Characteristic loss after a marginal sphere}

For $K\ge1$, the strict characteristic conditions imply $K<2$ and $|j|<2$.
The continuation result therefore keeps the spherical background
smooth up to and through their first failure in any finite domain
of Eq.~\eqref{cont:domain}. This separates characteristic loss from
a curvature singularity.
Suppose these conditions hold throughout the evolution up to and
including a slice containing a future-trapped sphere at
$(u_{\rm tr},v_{\rm tr})$, with the development smooth in a neighborhood
of the closed initial base of $\mathscr D_{v_{\rm tr}}$.
If they persisted throughout the rest of $\mathscr D_{v_{\rm tr}}$,
continuation would give a regular axis point $(v_{\rm tr},v_{\rm tr})$.
On the future ingoing ray $v=v_{\rm tr}$, however,
\begin{equation}
 R_{,uv}=-\frac{e^{-f}R(K-1)}{\tilde\alpha K}\le0,\qquad
 R_{,v}(u,v_{\rm tr})\le R_{,v}(u_{\rm tr},v_{\rm tr})<0
 \label{cont:trapped-ray}
\end{equation}
This contradicts $R_{,v}>0$ at a regular axis. At a future
marginal sphere, $R_{,v}=0$, $F=1$ and $K>1$, so
$R_{,uv}<0$. A nearby point on the future ingoing ray is trapped
while the strict conditions still hold. The same conclusion follows:
a characteristic condition must fail at a later finite conformal time
somewhere in this compact domain. This argument does not specify
the location of the first failure of a characteristic inequality.
\section{Nonspherical marginal surfaces}
\label{app:area}

We construct the data of Section~\ref{sec:nonspherical-area} for
$n=5$ and $\tilde\alpha=4\alpha_{(2)}$. The scalar momentum solves
the constraints on a prescribed spatial geometry, which is then
joined to a round vacuum exterior. Shortening the Hopf circles of
three-spheres gives arbitrarily small marginal surfaces without
large curvature. We first give this explicit construction, then
extend it to prescribed intrinsic metrics. The radial distance
$\varrho$, round areal radius $R$ and fixed scale $L_{\rm H}$ have
dimensions of length; $\varepsilon_{\rm H}$ is dimensionless.

The construction gives exact constrained initial data and a short
local evolution for each fixed metric. It does not show that these
marginal surfaces form from untrapped data, that they lie on event
horizons, or that they trap every gravitational polarization. The
uniform estimates for the degenerating Berger family are proved
directly below. The construction for families of prescribed metrics in
Appendix~\ref{app:area-prescribed} gives common curvature bounds and
choices of scale under uniform bounds on the metrics and positive
lower bounds on the lowest eigenvalues and eigenfunctions. The
shrinking Berger metrics do not satisfy those assumptions.

\subsection{Geometry and the marginal surface}
\label{app:area-geometry}

The slice is built from the inside out as a sequence of radial
intervals. Starting from a regular center, the three-spheres of
constant radial distance are first round, then squashed by
shrinking their Hopf circles, which re-expand further out (the
marginal surface lies where the logarithmic derivative of the Hopf radius
reaches $3c_{\rm H}$, with the constant $c_{\rm H}$ fixed below), and
finally returned to round spheres that join a vacuum exterior. Every
interval is described by two warp factors, the length scale $a_\perp$
of the base directions and the radius $a_\parallel$ of the Hopf
circles.
Let $\sigma_i$ be the usual left-invariant one-forms on $S^3$, normalized by
$\dd\sigma_1=-2\sigma_2\wedge\sigma_3$ cyclically, so
$\sum_i\sigma_i^2$ is the unit round metric. With proper radial
distance $\varrho$, set
\begin{equation}
 \gamma=\dd\varrho^2+a_\perp(\varrho)^2(\sigma_1^2+\sigma_2^2)
                  +a_\parallel(\varrho)^2\sigma_3^2,
 \qquad
 \lambda_\perp=\frac{a_\perp'}{a_\perp},\quad
 \lambda_\parallel=\frac{a_\parallel'}{a_\parallel}
 \label{area:warps}
\end{equation}
A surface of constant $\varrho$ is a three-sphere. Its two transverse
length scales are $a_\perp$, and its Hopf circle has length scale
$a_\parallel$. The quantities $\lambda_\perp,\lambda_\parallel$
measure their fractional change with proper distance. Primes on
these quantities and the warp factors denote $\varrho$ derivatives;
derivatives of one-variable profiles refer to their displayed
argument.

The sectional curvatures of a constant-$\varrho$ three-sphere are $4/a_\perp^2-3a_\parallel^2/a_\perp^4$ and
$a_\parallel^2/a_\perp^4$ twice. The radial sectional curvatures of
$\gamma$ are $-(\lambda_\perp'+\lambda_\perp^2)$ twice and
$-(\lambda_\parallel'+\lambda_\parallel^2)$; the tangential ones subtract
$\lambda_\perp^2$ and $\lambda_\perp\lambda_\parallel$, respectively.
The remaining curvature components, those with one radial and
three tangential indices, are, up to sign and a factor at most
two, $(a_\parallel/a_\perp^2)(\lambda_\perp-\lambda_\parallel)$.
In particular, these expressions contain no negative powers of
$a_\parallel$ when $a_\perp$ and the logarithmic derivatives are
controlled. The Hopf circles can therefore shrink without forcing
large spatial curvature. The scalar curvature is
\begin{equation}
 \begin{split}
 {}^{(4)}\mathcal R={}&\frac8{a_\perp^2}
 -\frac{2a_\parallel^2}{a_\perp^4}
 -4\lambda_\perp'-2\lambda_\parallel'
 -6\lambda_\perp^2-2\lambda_\parallel^2
 -4\lambda_\perp\lambda_\parallel
 \end{split}
 \label{area:scalarcurvature}
\end{equation}

All interpolations use the smooth decreasing cutoff
\begin{equation}
 \chi_{\rm H}(y)=1-\mathcal S(y)
 \label{area:cutoff}
\end{equation}
with $\mathcal S$ the smooth step of Eq.~\eqref{form:step}.
It equals one for $y\le0$, zero for $y\ge1$, has all derivatives vanishing at the endpoints,
and satisfies $\int_0^1\chi_{\rm H}\,\dd y=1/2$ by the symmetry
$\mathcal S(1-y)=1-\mathcal S(y)$.
First fix a sufficiently small dimensionless number $\eta_{\rm H}>0$.
At the prescribed coupling, choose $L_{\rm H}$ sufficiently large that
$\alpha_{(2)}/L_{\rm H}^2$ is small compared with $\eta_{\rm H}^2$,
and set $c_{\rm H}=\eta_{\rm H}/L_{\rm H}$.
We then vary $\varepsilon_{\rm H}$ while keeping these choices fixed.
All constants below are independent of $\varepsilon_{\rm H}$.
On each interpolation interval, $x$ denotes the dimensionless radial
coordinate specified for that interval.

To start from a regular center, define $f_{\rm H}(x)=x$ for $x\le1/2$,
\begin{equation}
 f_{\rm H}(x)=\frac12+\int_0^{x-1/2}\chi_{\rm H}(y)\,\dd y
 \quad(1/2\le x\le3/2),\qquad f_{\rm H}(x)=1\quad(x\ge3/2)
 \label{area:d29}
\end{equation}
and take $a_\perp=a_\parallel=L_{\rm H}f_{\rm H}(\varrho/L_{\rm H})$
through $\varrho_{\rm C}=3L_{\rm H}/2$. This metric is Euclidean near the
center and joins the round cylinder smoothly. Its scalar curvature is
\begin{equation}
 {}^{(4)}\mathcal R=\frac6{L_{\rm H}^2}
 \left(\frac{1-(f_{\rm H}')^2}{f_{\rm H}^2}
                   -\frac{f_{\rm H}''}{f_{\rm H}}\right)\ge0
 \label{area:d30}
\end{equation}
where these derivatives are with respect to $x$. Its curvature is
$O(L_{\rm H}^{-2})$.

We next shorten the Hopf circles. Keep $a_\perp=L_{\rm H}$ until
the interpolation back to round spheres below, so that
$\lambda_\perp=0$ on the next four intervals and only
$\lambda_\parallel$ is prescribed there. The first interval starts at
$\varrho_{\rm C}$ and switches on the contraction of the Hopf circles:
\begin{equation}
 \lambda_\parallel=-4c_{\rm H}\bigl[1-\chi_{\rm H}(x)\bigr],\qquad
 x=c_{\rm H}(\varrho-\varrho_{\rm C})\in[0,1]
 \label{area:d1}
\end{equation}
Since $\dd\varrho=\dd x/c_{\rm H}$ and
$\int_0^1(1-\chi_{\rm H})\,\dd x=1/2$, $\log a_\parallel$ decreases by
$2$ across this interval. Starting from $a_\parallel=L_{\rm H}$, the
terminal values are therefore $a_\parallel=L_{\rm H}e^{-2}$ and
$\lambda_\parallel=-4c_{\rm H}$.
The second interval continues with this constant logarithmic
derivative for a length $D_{\rm H}$ to be chosen below.

The third interval begins at $\varrho_{\rm B}$ and changes
$\lambda_\parallel$ from $-4c_{\rm H}$ to $4c_{\rm H}$:
\begin{equation}
 \begin{gathered}
 \lambda_\parallel=c_{\rm H}W_{\rm H}(x),\qquad
 W_{\rm H}(x)=4-8\chi_{\rm H}(x),\qquad
 x=c_{\rm H}(\varrho-\varrho_{\rm B})\in[0,1]\\
 W_{\rm H}(x_{\rm h})=3,\qquad
 q_{\rm H}=\exp\left(\int_0^{x_{\rm h}}W_{\rm H}(x)\,\dd x\right)
 \end{gathered}
 \label{area:d31}
\end{equation}
Since $\chi_{\rm H}'<0$ in the interior, $W_{\rm H}$
increases strictly from $-4$ to $4$. There is therefore a unique
$x_{\rm h}\in(0,1)$ with $W_{\rm H}(x_{\rm h})=3$, and
$W_{\rm H}'(x_{\rm h})>0$. The orbit at $x_{\rm h}$, where
$\lambda_\parallel=3c_{\rm H}$, will be the marginal surface. On this
interval $\dd\log a_\parallel=W_{\rm H}\,\dd x$, so $q_{\rm H}$ is the
ratio of the Hopf radius at $x_{\rm h}$ to that at $\varrho_{\rm B}$.
It lies between zero and one because the integral defining it is
negative: since $\int_0^1\chi_{\rm H}\,\dd x=1/2$ and $W_{\rm H}\ge3$
on $[x_{\rm h},1]$,
\begin{equation}
 \int_0^1W_{\rm H}\,\dd x=0,\qquad\text{so}\qquad
 \int_0^{x_{\rm h}}W_{\rm H}\,\dd x=-\int_{x_{\rm h}}^1W_{\rm H}\,\dd x<0
 \label{area:d2}
\end{equation}

We choose $D_{\rm H}$ so that the Hopf radius at $x_{\rm h}$ is exactly
$\varepsilon_{\rm H}L_{\rm H}$, that is,
$a_\parallel=\varepsilon_{\rm H}L_{\rm H}/q_{\rm H}$ at
$\varrho_{\rm B}$. The second interval multiplies $a_\parallel$ by
$e^{-4c_{\rm H}D_{\rm H}}$, starting from $L_{\rm H}e^{-2}$, so its
length must be
\begin{equation}
 D_{\rm H}=\frac1{4c_{\rm H}}
 \log\frac{q_{\rm H}e^{-2}}{\varepsilon_{\rm H}},
 \qquad 0<\varepsilon_{\rm H}<q_{\rm H}e^{-2}
 \label{area:d32}
\end{equation}
At $x_{\rm h}$ the orbit then has
\begin{equation}
 a_\parallel=\varepsilon_{\rm H}L_{\rm H},\qquad
 \lambda_\parallel=3c_{\rm H},\qquad
 \lambda_\parallel'=c_{\rm H}^2W_{\rm H}'(x_{\rm h})>0
 \label{area:d3}
\end{equation}
Because $\int_0^1W_{\rm H}\,\dd x=0$, the Hopf radius at the end of
the third interval is again $\varepsilon_{\rm H}L_{\rm H}/q_{\rm H}$.
The fourth interval has $\lambda_\parallel=4c_{\rm H}$ and length
$(4c_{\rm H})^{-1}\log[q_{\rm H}/(e\varepsilon_{\rm H})]$, which is
positive since $\varepsilon_{\rm H}<q_{\rm H}e^{-2}$, and ends at
$a_\parallel=L_{\rm H}/e$.

To make the orbit at $x_{\rm h}$ future marginal, we prescribe the
contraction of the slice in time as well as its spatial geometry.
Throughout the interior and all subsequent intervals up to the vacuum
change of slicing below, set $\mathcal K=c_{\rm H}\gamma$, using
$\mathcal K=-\tfrac12\mathcal L_{e_0}\gamma$ as elsewhere in this
paper. Consider an orbit with outward unit normal
$e_1=\partial_\varrho$ within the slice. Its second fundamental form,
half the $\varrho$ derivative of its induced metric, is
$a_\perp^2\lambda_\perp(\sigma_1^2+\sigma_2^2)
+a_\parallel^2\lambda_\parallel\sigma_3^2$, so its mean curvature
within the slice is
\begin{equation}
 \mathcal H=2\lambda_\perp+\lambda_\parallel
 \label{area:d4}
\end{equation}
The same symbol is used for the mean curvature of every three-surface
below.

The expansions of the future null normals $e_0\pm e_1$ are
$\pm\mathcal H$ minus the trace of $\mathcal K$ over the orbit, which is
$3c_{\rm H}$ because $\mathcal K=c_{\rm H}\gamma$ and the orbit is
three-dimensional:
\begin{equation}
 \theta_\pm=\pm\mathcal H-3c_{\rm H}
 \label{area:expansions}
\end{equation}
An orbit is therefore future marginal, $\theta_+=0$, exactly when
$\mathcal H=3c_{\rm H}$. On the four intervals above,
$\mathcal H=\lambda_\parallel$, so the orbit at $x_{\rm h}$ is
marginal. Its induced metric is
$L_{\rm H}^2(\sigma_1^2+\sigma_2^2)
+\varepsilon_{\rm H}^2L_{\rm H}^2\sigma_3^2
=L_{\rm H}^2q_{\varepsilon_{\rm H}}$, so it is the future marginal
surface in Eq.~\eqref{area:small}, with $\theta_-=-6c_{\rm H}<0$.
Beyond $x_{\rm h}$ the logarithmic derivative $\lambda_\parallel$
exceeds $3c_{\rm H}$ on the rest of the third interval and equals
$4c_{\rm H}$ on the fourth, so all three-spheres up to the next
interpolation have $\theta_+>0$.

We now restore round three-spheres while keeping their outgoing
expansion positive. Starting at $\varrho_{\rm I}$, take an
interval of length $1/(2c_{\rm H})$, parameterized by
$x=2c_{\rm H}(\varrho-\varrho_{\rm I})\in[0,1]$, and prescribe
\begin{equation}
 \lambda_\perp=2c_{\rm H}(1-\chi_{\rm H}(x)),\qquad
 \partial_\varrho\log(a_\parallel/a_\perp)=4c_{\rm H}\chi_{\rm H}(x)
 \label{area:d33}
\end{equation}
The initial ratio is $e^{-1}$; integration gives
$a_\perp=a_\parallel=R_{\rm b}=e^{1/2}L_{\rm H}$ at the endpoint, with
$\lambda_\perp=\lambda_\parallel=2c_{\rm H}$ and all higher derivatives
matching the round metric with $R'=2c_{\rm H}R$.
Here $\theta_+=3c_{\rm H}-2c_{\rm H}\chi_{\rm H}\ge c_{\rm H}$.
Outside the cap and through this interpolation,
$L_{\rm H}\le a_\perp\le e^{1/2}L_{\rm H}$,
$0<a_\parallel\le a_\perp$, and
\begin{equation}
 |\lambda_\perp|+|\lambda_\parallel|\le Cc_{\rm H},\quad
 |\lambda_\perp'|+|\lambda_\parallel'|\le Cc_{\rm H}^2,\qquad
 {}^{(4)}\mathcal R\ge\frac6{eL_{\rm H}^2}-Cc_{\rm H}^2
                         \ge\frac3{eL_{\rm H}^2}
 \label{area:corebounds}
\end{equation}
The curvature formulas above therefore give a uniform
$C/L_{\rm H}^2$ bound. Decreasing $\varepsilon_{\rm H}$ lengthens
the intervals in which the Hopf circles contract and expand; it does
not increase this bound. These intervals have finite length for
every fixed $\varepsilon_{\rm H}>0$.

\subsection{Exact constraints and asymptotically flat completion}
\label{app:area-completion}

Having fixed the spatial geometry, we now solve the exact
Einstein--Gauss--Bonnet constraints on it. With the extrinsic
curvature proportional to the metric and the scalar field constant on
the slice, the momentum constraints hold identically and the
Hamiltonian constraint becomes an algebraic equation for the scalar
momentum $P$, which has a real solution wherever its geometric side
is positive. We then switch the matter off and change the slicing so
that the data become the static vacuum solution outside a compact
set.

In the $4+1$ decomposition of the EGB equations \cite{Torii2008}, the
constraints involve the spatial projection of the spacetime Riemann
tensor,
\begin{equation}
 \widehat R_{ijkl}={}^{(4)}\mathcal R_{ijkl}
   +\mathcal K_{ik}\mathcal K_{jl}-\mathcal K_{il}\mathcal K_{jk}
 \label{area:projected}
\end{equation}
by the Gauss equation, with $\widehat R_{ij}$ and $\widehat R$ its
contractions with $\gamma^{ij}$. The intrinsic curvature of the spatial
metric is written ${}^{(4)}\mathcal R_{ijkl}$. Twice the
normal--normal component of Eq.~\eqref{eq:field} is the Hamiltonian
constraint. When both free indices in Eq.~\eqref{app:deltaform} are
normal, all summed indices are spatial, so this constraint contains
only $\widehat R_{ijkl}$. With the scalar source
$2(n-2)kG\,T_{00}[\Psi]$ for $n=5$, it reads
\begin{equation}
 \widehat R+\alpha_{(2)}\bigl(\widehat R^2-4|\widehat R_{ij}|^2
 +|\widehat R_{ijkl}|^2\bigr)=3kG\bigl(P^2+|D\Psi|^2\bigr)
 \label{area:d5}
\end{equation}
where $D$ is the covariant derivative of $\gamma$. Write
$\mathcal G[\gamma]=({}^{(4)}\mathcal R)^2
-4|{}^{(4)}\mathcal R_{ij}|^2+|{}^{(4)}\mathcal R_{ijkl}|^2$ for the
same combination built from the spatial metric.

Let $\mathcal K=c\gamma$ with a constant $c$ on a region of an arbitrary
slice $(M^4,\gamma)$. Then
\begin{equation}
 \widehat R_{ijkl}={}^{(4)}\mathcal R_{ijkl}
 +c^2(\gamma_{ik}\gamma_{jl}-\gamma_{il}\gamma_{jk}),\qquad
 \widehat R_{ij}={}^{(4)}\mathcal R_{ij}+3c^2\gamma_{ij}
 \label{area:d6}
\end{equation}
In four spatial dimensions the tensor
$\gamma_{ik}\gamma_{jl}-\gamma_{il}\gamma_{jk}$ has trace $3\gamma_{jl}$
over $i,k$, its contraction with ${}^{(4)}\mathcal R^{ijkl}$ is
$2\,{}^{(4)}\mathcal R$, and its squared norm is $24$. Hence
\begin{equation}
 \begin{gathered}
 \widehat R^2=({}^{(4)}\mathcal R)^2+24c^2\,{}^{(4)}\mathcal R+144c^4,\qquad
 |\widehat R_{ij}|^2=|{}^{(4)}\mathcal R_{ij}|^2
 +6c^2\,{}^{(4)}\mathcal R+36c^4\\
 |\widehat R_{ijkl}|^2=|{}^{(4)}\mathcal R_{ijkl}|^2
 +4c^2\,{}^{(4)}\mathcal R+24c^4
 \end{gathered}
 \label{area:d7}
\end{equation}
and in the Gauss--Bonnet combination the coefficients of
$c^2\,{}^{(4)}\mathcal R$ and $c^4$ add up to $24-24+4=4$ and
$144-144+24=24$:
\begin{equation}
 \begin{split}
 \widehat R&={}^{(4)}\mathcal R+12c^2\\
 \widehat R^2-4|\widehat R_{ij}|^2+|\widehat R_{ijkl}|^2
 &=\mathcal G[\gamma]+4c^2\,{}^{(4)}\mathcal R+24c^4
 \end{split}
 \label{area:gbshift}
\end{equation}

The Codazzi tensor
$D_i\mathcal K_{jk}-D_j\mathcal K_{ik}=c(D_i\gamma_{jk}-D_j\gamma_{ik})$
vanishes, and hence so does the geometric part of the EGB momentum
constraint \cite{Torii2008}. The Codazzi equation expresses the mixed
components $\mathcal R_{0ijk}$ of the spacetime curvature through this
tensor, and in the form of Eq.~\eqref{app:deltaform} every term of the
normal--spatial component of the field equations contains exactly one
such component. Take $\Psi$ constant on the region. The scalar momentum
density $T_{0i}[\Psi]=P\,D_i\Psi$ then vanishes, and the source of the
Hamiltonian constraint reduces to $3kGP^2$. Substituting
Eq.~\eqref{area:gbshift}, the exact Hamiltonian constraint is solved by
\begin{equation}
 \begin{gathered}
 P=\sqrt{\frac{\mathcal C_{\rm H}[\gamma,c]}{3kG}}\\
 \mathcal C_{\rm H}[\gamma,c]={}^{(4)}\mathcal R+12c^2
 +\alpha_{(2)}\bigl[\mathcal G[\gamma]
       +4c^2\,{}^{(4)}\mathcal R+24c^4\bigr]
 \end{gathered}
 \label{area:scalar}
\end{equation}
wherever $\mathcal C_{\rm H}>0$. At a boundary of the matter support we arrange
$\mathcal C_{\rm H}=\chi_{\rm m}^2$ times a smooth positive function, with
$\chi_{\rm m}\ge0$ a smooth cutoff, so that $P$ vanishes smoothly. Set the
constant scalar to zero.

For the preceding geometry, $\mathcal C_{\rm H}\ge6c_{\rm H}^2$. The
curvature of this geometry is bounded by $C/L_{\rm H}^2$ and
$c_{\rm H}^2=\eta_{\rm H}^2/L_{\rm H}^2$, so the Gauss--Bonnet
contribution to $\mathcal C_{\rm H}$ is bounded by
$C\alpha_{(2)}/L_{\rm H}^4$. Wherever ${}^{(4)}\mathcal R\ge0$ this
gives
\begin{equation}
 \mathcal C_{\rm H}\ge12c_{\rm H}^2-\frac{C\alpha_{(2)}}{L_{\rm H}^4}
 =\Bigl(12-\frac{C\alpha_{(2)}}{\eta_{\rm H}^2L_{\rm H}^2}\Bigr)c_{\rm H}^2
 \ge6c_{\rm H}^2
 \label{area:d8}
\end{equation}
where the last step uses our choice $\alpha_{(2)}/L_{\rm H}^2\ll\eta_{\rm H}^2$.
The scalar curvature is nonnegative in the cap and, by
Eq.~\eqref{area:corebounds}, on the intervals that follow it. It is also
nonnegative for the round comparison metric satisfying
$\partial_\varrho R=2c_{\rm H}R$ through $R=3R_{\rm b}/2$, which is
used to define the matter cutoff below: there
\begin{equation}
 {}^{(4)}\mathcal R=\frac6{R^2}-48c_{\rm H}^2
 \ge\frac{8/(3e)-48\eta_{\rm H}^2}{L_{\rm H}^2}>0
 \label{area:d9}
\end{equation}
since $R_{\rm b}^2=eL_{\rm H}^2$ and $\eta_{\rm H}$ is small. As
$\mathcal C_{\rm H}$ is smooth and
bounded below by $6c_{\rm H}^2$, $P$ is real and smooth, including at
the center, where the metric is flat and
$\mathcal C_{\rm H}=12c_{\rm H}^2+24\alpha_{(2)}c_{\rm H}^4$.

\paragraph{Round constraint identity.}
For round data, the constraints can be solved directly in terms of
mass and tilt. Let $F$ be determined by Eq.~\eqref{def:mass}, put
$A=1-F$, and choose a smooth $\varpi$ with $A+\varpi^2>0$. Set
\begin{equation}
 a=\sqrt{A+\varpi^2},\qquad
 \gamma=\frac{\dd R^2}{a^2}+R^2\dd\Omega_3^2,\qquad
 k_r=\partial_R\varpi,\qquad k_t=\frac{\varpi}{R}
 \label{area:round-ansatz}
\end{equation}
Here $k_r,k_t$ are the radial and angular eigenvalues of
$\mathcal K$, and $\partial_\varrho R=a$. The intrinsic radial and
angular sectional curvatures are $-aa_R/R$ and $(1-a^2)/R^2$.
In the Gauss equation, adding $k_rk_t$ and $k_t^2$, respectively,
cancels the tilt because $aa_R=A_R/2+\varpi\varpi_R$. Thus, with
radial index $1$ and angular indices $i,j,k,l$,
\begin{equation}
 \widehat R_{1i1j}=\frac{F_R}{2R}\delta_{ij},\qquad
 \widehat R_{ijkl}=\frac{F}{R^2}
 (\delta_{ik}\delta_{jl}-\delta_{il}\delta_{jk})
 \label{area:round-projection}
\end{equation}
The only possible Codazzi coefficient also cancels:
\begin{equation}
 \partial_\varrho k_t+\frac aR(k_t-k_r)
 =a\left(\frac{\varpi_R}{R}-\frac{\varpi}{R^2}\right)
 +\frac aR\left(\frac{\varpi}{R}-\varpi_R\right)=0
 \label{area:round-codazzi}
\end{equation}
Consequently $\mathcal R_{0ijk}=0$ and the momentum constraint holds
for $\Psi=0$. There are three projected sectional curvatures of
each kind in Eq.~\eqref{area:round-projection}. Their scalar
contraction is $6(F_R/(2R)+F/R^2)$ and their Gauss--Bonnet
combination is $24(F_R/(2R))(F/R^2)$. The Hamiltonian constraint is
therefore
\begin{equation}
 \begin{split}
 3kGP^2
 &=\frac{3F_R}{R}+\frac{6F}{R^2}
   +\frac{3\tilde\alpha FF_R}{R^3}\\
 &=\frac3{R^3}\partial_R
   \left(R^2F+\frac{\tilde\alpha F^2}{2}\right)
 =\frac{6kG\mathcal M_R}{R^3}
 \end{split}
 \label{area:round-constraint}
\end{equation}
In particular, $P^2=2\mathcal M_R/R^3$. When $\varpi=cR$, the
extrinsic curvature is $c\gamma$ and
$F=1-(\partial_\varrho R)^2+c^2R^2$. In the notation of
Appendix~\ref{app:realization}, this gives
\begin{equation}
 \mathcal C_{\rm H}=\frac{6kG\mathcal M_R}{R^3},\qquad
 \mathcal J_m=\frac{\tilde\alpha}3\mathcal C_{\rm H}=j,\qquad
 \mathfrak d=V^2\mathcal J_m^2
 \label{area:roundidentity}
\end{equation}
Thus the direct solution agrees with the reconstruction from
$(m,\mathcal J_m,cR)$ in Appendix~\ref{app:realization}.
The round spheres have mean curvature $3\partial_\varrho R/R$,
which exceeds $3c$ if and only if $F<1$.

\paragraph{Completion.}
Consider a round region with $\mathcal K=c\gamma$ and $\partial_\varrho R\ge(1+\mu_1)cR$ for some
$\mu_1>0$, on an annulus $R_1\le R\le R_2$ where $\mathcal C_{\rm H}>0$, and
let $\mathcal M_{\rm col}$ be its mass there. We first turn off the
matter while keeping $\mathcal K=c\gamma$, and then change the
slicing in vacuum so that $\mathcal K$ vanishes at infinity. Choose
radii $R_2<R_3<R_4$ and set
\begin{equation}
 \begin{gathered}
 \chi_{\rm m}=\chi_{\rm H}\Bigl(\frac{R-R_1}{R_2-R_1}\Bigr),\qquad
 \mathcal M(R)=\mathcal M_{\rm col}(R_1)
 +\int_{R_1}^{R}\partial_{\bar R}\mathcal M_{\rm col}(\bar R)\,\chi_{\rm m}(\bar R)^2\,\dd\bar R\\
 \varpi=cR\,\chi_{\rm s},\qquad
 \chi_{\rm s}=\chi_{\rm H}\Bigl(\frac{R-R_3}{R_4-R_3}\Bigr)
 \end{gathered}
 \label{area:completion}
\end{equation}
with $\mathcal M$ constant for $R\ge R_2$. The cutoff $\chi_{\rm m}$
decreases from one at $R_1$ to zero at $R_2$ and switches off the
growth of the mass; $\chi_{\rm s}$ decreases from one at $R_3$ to zero
at $R_4$ and switches off the tilt, and with it the extrinsic
curvature. On the GR branch,
\begin{equation}
 F=\frac{4kG\mathcal M}{R^2+\sqrt{R^4+4kG\tilde\alpha\mathcal M}},
 \qquad A=1-F
 \label{area:round}
\end{equation}
For $R\ge R_1$, Eqs.~\eqref{area:round-ansatz} and
\eqref{area:round-constraint} give the exact data
\begin{equation}
 \begin{gathered}
 \gamma=\frac{\dd R^2}{A+\varpi^2}+R^2\dd\Omega_3^2,\qquad
 \mathcal K^R{}_R=\partial_R\varpi,\qquad
 \mathcal K^i{}_k=\frac{\varpi}R\,\delta^i{}_k\\
 P=\chi_{\rm m}\sqrt{\frac{2\partial_R\mathcal M_{\rm col}}{R^3}},\qquad \Psi=0
 \end{gathered}
 \label{area:bending}
\end{equation}
with $P=0$ for $R\ge R_2$ and angular indices $i,k$. The data have
$\mathcal K=c\gamma$ through $R_3$, change slicing in vacuum on
$[R_3,R_4]$, and are static Boulware--Deser beyond.

These data are smooth, their round spheres are untrapped, and they
have one asymptotically flat end. By construction $\mathcal M_R=\chi_{\rm m}^2\mathcal M_{{\rm col},R}\ge0$,
and Eq.~\eqref{area:round-constraint} gives the value of $P$ in
Eq.~\eqref{area:bending}. On the annulus $[R_1,R_2]$,
$\mathcal C_{\rm H}=\chi_{\rm m}^2\mathcal C_{{\rm H},{\rm col}}$;
the positive square root of $\mathcal C_{{\rm H},{\rm col}}$ is smooth, so
multiplication by $\chi_{\rm m}$ switches off $P$ smoothly.
Since $\mathcal M\le\mathcal M_{\rm col}$ on
$[R_1,R_2]$ and is constant beyond, $F<1$ throughout the completion.
For comparison with the general scalar reconstruction,
$\Psi=0$ gives $j=\mathcal J_m$, so
$a^2\mathfrak d=\mathcal J_m^2\varpi^2>0$ where $\chi_{\rm m}>0$. At the outer boundary of the matter support,
$R=R_2$, the function $\mathcal J_m$ has the form
$\mathcal J_m=\chi_{\rm m}^2\mathcal J_{\rm col}$ required by
Eq.~\eqref{form:edge}, with $\upsilon=\chi_{\rm m}$ and
$\mu=\iota=\mathcal J_{\rm col}$, where
\begin{equation}
 \mathcal J_{\rm col}=\frac{2kG\tilde\alpha\,\partial_R\mathcal M_{\rm col}}{R^3}
 \label{area:d10}
\end{equation}
is the value of $\mathcal J_m$ computed from $\mathcal M_{\rm col}$.
The cutoff is constant to all orders at both endpoints, so the data
match the inner solution smoothly at $R_1$ and vacuum at $R_2$.

Finally, since $A>0$, every round sphere has
\begin{equation}
 \theta_+=\frac{3\bigl[\sqrt{A+\varpi^2}-\varpi\bigr]}{R}>0
 \label{area:d11}
\end{equation}
so no round sphere of the completion is marginal or trapped. At large
radius, $A=1-2kG\mathcal M_\infty/R^2+O(R^{-6})$ gives one
asymptotically flat end with ADM mass $\mathcal M_\infty=\mathcal M(R_2)$.
A common round annulus therefore gives a common exterior and ADM mass.

Suppose further that the curvature of the original spatial metric is at most
$C_0/L^2$, with $c=\eta/L$, $\eta\le1$ and $L\ge\sqrt{\tilde\alpha}$,
and that $L/C_0\le R\le C_0L$ on the annulus. Then the curvature of the
completed slice, $|\mathcal K|^2$ and $kGP^2$ are at most $C/L^2$, with
$C$ depending only on $C_0$, $\mu_1$ and the ratios $R_3/R_2$, $R_4/R_2$. On the
matter region $0\le\partial_R\mathcal M\le\partial_R\mathcal M_{\rm col}$
and $F\le F_{\rm col}$, so the upper bounds on $\mathcal M$ and
$\mathcal M_R$ inherited from the original metric are preserved. The branch factor also
stays bounded away from zero. For nonnegative mass, $K\ge1$.
If the matching mass is negative, $\mathcal C_{{\rm H},{\rm col}}>0$
and Eq.~\eqref{area:round-constraint} give
\begin{equation}
 \frac{F_{\rm col}}{R^2}
 +K\frac{F_{{\rm col},R}}{2R}>0,\qquad
 1-K<\frac{C_1\tilde\alpha}{L^2}K
 \label{area:completion-branch}
\end{equation}
at $R_1$, where the projected sectional curvatures bound $C_1$
in terms of $C_0$. Thus $K(R_1)\ge(1+C_1)^{-1}$.
On the completed profile,
$K^2=1+4kG\tilde\alpha\mathcal M/R^4$ increases while
$\mathcal M<0$, since $\mathcal M_R\ge0$; after the mass becomes
nonnegative, $K\ge1$. Differentiating Eq.~\eqref{def:mass} now gives
\begin{equation}
 \frac{F_R}{R}
 =\frac{2kG\mathcal M_R/R^3-2F/R^2}{K}
 \label{area:completion-derivative}
\end{equation}
and bounds both projected sectional curvatures by $C/L^2$.
The intrinsic curvatures subtract $\varpi\varpi_R/R$ and
$\varpi^2/R^2$, respectively; these and $|\mathcal K|^2$ obey
the same bound for the prescribed slicing cutoff. No derivative
of $\chi_{\rm m}$ enters this estimate, so no lower bound on the
width $R_2-R_1$ is needed.

The projected curvature in Eq.~\eqref{area:round-projection} and
the vanishing mixed curvature in Eq.~\eqref{area:round-codazzi}
are independent of $\varpi$. Appendix~\ref{app:area-evolution}
uses these identities to bound the remaining spacetime curvature
on the whole slice, including the change of slicing in vacuum.

The tilt itself also obeys useful geometric bounds. In the vacuum,
$A$ increases with $R$ if
$\mathcal M_\infty>0$, as in both applications below. At $R_2$,
Eq.~\eqref{area:round} makes $F$ an increasing function of the mass,
and $\mathcal M(R_2)\le\mathcal M_{\rm col}(R_2)$; for the original
round data, $1-F_{\rm col}=(\partial_\varrho R)^2-c^2R^2$. Hence
\begin{equation}
 A(R)\ge A(R_2)\ge1-F_{\rm col}(R_2)
 \ge\bigl[(1+\mu_1)^2-1\bigr]c^2R_2^2\qquad(R\ge R_2)
 \label{area:d12}
\end{equation}
With $\mathcal M=\mathcal M_\infty$, Eq.~\eqref{def:mass} reads
$2kG\mathcal M_\infty=R^2F+\tilde\alpha F^2/2$. The right side is at
least $R^2F$, which gives the first bound below at every radius. It
increases with $F$ on the GR branch, where its $F$ derivative
$R^2+\tilde\alpha F=R^2K$ is positive, so evaluating it at $R_2$,
where $F<1$, gives the second:
\begin{equation}
 F\le\frac{2kG\mathcal M_\infty}{R^2},\qquad
 2kG\mathcal M_\infty<R_2^2+\frac{\tilde\alpha}2
 \label{area:d13}
\end{equation}
With $\varpi\le cR$, the boost is therefore bounded:
\begin{equation}
 \frac{\varpi^2}A\le\frac{(R/R_2)^2}{(1+\mu_1)^2-1}\quad
 (R_2\le R\le R_4),
 \qquad
 A\ge1-\frac{R_2^2}{R_3^2}\Bigl(1+\frac{\tilde\alpha}{2R_2^2}\Bigr)\quad
 (R\ge R_3)
 \label{area:boost}
\end{equation}
These bounds describe the slicing; the curvature estimate above
does not require a boost factor.

For Appendix~\ref{app:area-geometry}, the round annulus begins at
$R_{\rm b}=e^{1/2}L_{\rm H}$ with
$\partial_\varrho R=2c_{\rm H}R$, so $\mu_1=1$ and
$F_{\rm col}=1-3c_{\rm H}^2R^2$. Choose
$R_1=5R_{\rm b}/4$, $R_2=3R_{\rm b}/2$,
$R_3=2R_{\rm b}$ and $R_4=3R_{\rm b}$ in Eq.~\eqref{area:completion}.
The data become static beyond $3R_{\rm b}$. Their exterior and
ADM mass are independent of $\varepsilon_{\rm H}$, as are the
boost bound~\eqref{area:boost} and the $O(L_{\rm H}^{-2})$
bounds on the scalar stress-energy tensor and curvature in the slice-normal frame.
Thus the construction supplies one common ADM mass as
$\varepsilon_{\rm H}$ varies. It does not prescribe an arbitrary
ADM mass independently of $L_{\rm H}$, $\eta_{\rm H}$ and the
completion.

\subsection{Stability and outermostness}
\label{app:area-stability}

A marginal surface is strictly stable if an outward deformation
within the slice makes its outgoing expansion positive at first
order. The deformation can be chosen with the profile of the positive
principal eigenfunction of the stability operator; strict stability
means that its principal eigenvalue is positive. Deforming inward
then gives outer trapped surfaces just inside it. A marginal surface
is outermost if no other marginally outer trapped surface encloses it. Both properties are expected
of an apparent horizon, and both follow from the construction: the
mean curvature of the nearby three-spheres increases through the
marginal value at the surface, and the three-spheres outside it have
positive outgoing expansion.
Let $\mathcal K=c\gamma$ near a closed hypersurface $S$, with
outward unit normal $\nu$, second fundamental form $\mathrm{II}$
and trace-free part $\mathring{\mathrm{II}}$. Since
$\operatorname{tr}_S\mathcal K=3c$, a future marginal surface has
$\mathcal H=3c$ and $\theta_-=-6c$. Its stability operator is
the mean-curvature variation operator,
\begin{equation}
 \mathcal L_Sf=-\Delta_Sf
 -\bigl(|\mathrm{II}|^2+\operatorname{Ric}_\gamma(\nu,\nu)\bigr)f
 \label{area:stability}
\end{equation}
It is self-adjoint because $\mathcal K(\nu,\cdot)$ has no tangential
part \cite{AnderssonMarsSimon2008}. The Gauss equation gives
\begin{equation}
 \mathcal L_S=-\Delta_S+\frac12\Bigl(R_S-|\mathring{\mathrm{II}}|^2
 -\bigl({}^{(4)}\mathcal R+12c^2\bigr)\Bigr)
 \label{area:gauss-stability}
\end{equation}

\paragraph{Surrounding surfaces.}
For $\gamma=N_{\rm H}^2\dd\varrho^2+\gamma_S(\varrho)$,
$N_{\rm H}>0$, one has
$\nu=N_{\rm H}^{-1}\partial_\varrho$,
$\mathrm{II}=(2N_{\rm H})^{-1}\partial_\varrho\gamma_S$ and
$\partial_\varrho\mathcal H=\mathcal L_SN_{\rm H}$.
Integration by parts yields
\begin{equation}
 \int_S f\mathcal L_S f
 =\int_S N_{\rm H}^2\left|\dd(f/N_{\rm H})\right|^2
  +\int_S\frac{\partial_\varrho\mathcal H}{N_{\rm H}}f^2
 \label{area:d14}
\end{equation}
so $\lambda_1(\mathcal L_S)\ge
\inf_S(\partial_\varrho\mathcal H/N_{\rm H})$.
If $\partial_\varrho\mathcal H\ge0$ and is positive somewhere,
the principal eigenfunction gives strict positivity of $\lambda_1$.
For the homogeneous Berger orbits, $N_{\rm H}=1$ and
\begin{equation}
 \mathcal L_S=-\Delta_S+c_{\rm H}^2W_{\rm H}'(x_{\rm h}),\qquad
 \lambda_1(\mathcal L_S)=c_{\rm H}^2W_{\rm H}'(x_{\rm h})>0
 \label{area:berger-eigenvalue}
\end{equation}
This controls arbitrary angular variations.

The exterior foliation by closed surfaces with $\theta_+>0$
also establishes outermostness. Any enclosing closed surface
with $\theta_+\le0$ touches the furthest foliation leaf it meets from inside.
At contact its mean curvature is at least that of the leaf, and
the two tangential traces of $\mathcal K$ agree. Its outgoing
expansion must therefore be positive, a contradiction.
The foliation extends through the round completion to infinity,
so $S_{\varepsilon_{\rm H}}$ is outermost relative to the unique
end, even among enclosing weakly outer trapped surfaces.
Together with Appendix~\ref{app:area-evolution}, this establishes
Eq.~\eqref{area:small} and the statements accompanying it in
Section~\ref{sec:nonspherical-area}.

\subsection{Prescribing the intrinsic metric of a marginal surface}
\label{app:area-prescribed}

We now explain why Eq.~\eqref{area:condition} is sufficient to
prescribe the intrinsic metric of a marginal surface. The difficulty
is to join that metric to a regular center and a round exterior while
keeping $\mathcal C_{\rm H}>0$, so that the scalar momentum
in Eq.~\eqref{area:scalar} is real. A radial lapse chosen from the positive lowest eigenfunction $u_q$ in Eq.~\eqref{area:d16} supplies the
required positive contribution to the Hamiltonian constraint.

\paragraph{Positivity of $\mathcal C_{\rm H}$.}
In a region covered by the surrounding surfaces above, the Gauss
equation and the mean-curvature variation formula give,
for every constant $c$,
\begin{equation}
 \begin{gathered}
 {}^{(4)}\mathcal R+12c^2=\mathcal S_{\rm H}
 -\frac{2}{N_{\rm H}}\partial_\varrho\mathcal H
 -\frac43(\mathcal H^2-9c^2)-|\mathring{\mathrm{II}}|^2\\
 \mathcal S_{\rm H}=R_{\gamma_S}
       -\frac{2}{N_{\rm H}}\Delta_{\gamma_S}N_{\rm H}
 \end{gathered}
 \label{area:collar}
\end{equation}
For $c=\eta/L$, $\eta\le1$, and curvature bounded by $C_0/L^2$,
the Gauss--Bonnet correction in Eq.~\eqref{area:scalar} is
$O(\alpha_{(2)}/L^4)$. Thus
$\mathcal C_{\rm H}\ge\varsigma_0/(2L^2)$ if
$\mathcal S_{\rm H}\ge\varsigma_0/L^2$ and
\begin{equation}
 2N_{\rm H}^{-1}|\partial_\varrho\mathcal H|
 +\tfrac43|\mathcal H^2-9c^2|+|\mathring{\mathrm{II}}|^2
 \le\frac{\varsigma_0}{4L^2}
 \label{area:d15}
\end{equation}
provided $\alpha_{(2)}/L^2$ is small depending on
$\varsigma_0,C_0$. Where ${}^{(4)}\mathcal R\ge0$, instead
$\mathcal C_{\rm H}\ge6c^2$ for
$\alpha_{(2)}/L^2\le a_{\rm R}\eta^2$, with $a_{\rm R}$ depending on $C_0$.
A spatially constant rescaling by $e^{2b}$ multiplies
$\mathcal S_{\rm H}$ by $e^{-2b}$; the construction below controls
this accumulated scale change and the $C^2$ size of spatially
varying factors.

\paragraph{Connecting to positive scalar curvature.}
Let $u_q>0$ be the lowest eigenfunction in
Eq.~\eqref{area:condition}, normalized by $\max u_q=1$:
\begin{equation}
 \left(-\Delta_q+\frac12R_q\right)u_q=\lambda_q u_q
 \label{area:d16}
\end{equation}
The parameter $p\in[0,1]$ below labels a path of spatial metrics;
it is not a time coordinate. Set $q_p=u_q^pq$. The three-dimensional
conformal transformation formulas give
\begin{equation}
 \left(-\Delta_{q_p}+\frac12R_{q_p}\right)u_q^{1-p}
 =u_q^{1-2p}\left[\lambda_q+\frac34p(2-p)
                      |\dd\log u_q|_q^2\right]
 \label{area:conformal-path}
\end{equation}
Since $0<u_q\le1$, the bracket is at least $\lambda_q$ and
$u_q^{-p}\ge1$, so the positive function $u_q^{1-p}$ satisfies
$(-\Delta_{q_p}+R_{q_p}/2)u_q^{1-p}\ge\lambda_qu_q^{1-p}$.
Pairing this inequality with the positive lowest eigenfunction of the
self-adjoint operator on the left gives
$\lambda_1(-\Delta_{q_p}+R_{q_p}/2)\ge\lambda_q$.
At $p=1$,
$R_{u_qq}=u_q^{-1}(2\lambda_q+\tfrac32|\dd\log u_q|_q^2)>0$.
Connect $u_qq$ to the round metric through positive-scalar-curvature
metrics on $S^3$ \cite[Corollary 1.1]{Marques2012}, choosing the path
smooth and constant near its endpoints.

The spectral condition allows negative scalar curvature locally.
For $q=e^{2b_{\rm e}}q_{\rm round}$, for instance,
\begin{equation}
 \Bigl(-\Delta_q+\frac12R_q\Bigr)e^{-2b_{\rm e}}
 =3e^{-4b_{\rm e}}\bigl(1-|\dd b_{\rm e}|^2\bigr),\qquad
 R_q=e^{-2b_{\rm e}}\bigl(6-4\Delta_{S^3}b_{\rm e}-2|\dd b_{\rm e}|^2\bigr)
 \label{area:mplus-example}
\end{equation}
with derivatives taken in the round metric. Choose $b_{\rm e}$ as a
smooth function of the distance $r$ to a fixed point of $S^3$ in the
round metric, equal to $r^2/2$ near $r=0$, constant outside $r=1/2$,
and with $|\dd b_{\rm e}|\le1/2$. Then $\lambda_q>0$ by the same
pairing argument, applied to the positive function $e^{-2b_{\rm e}}$,
while $R_q=-6$ at $r=0$.

\paragraph{Constructing the surrounding spatial geometry.}
We use one path of metrics and lapses, first to reach the prescribed
metric from a regular cap and then to return to round spheres.
Concatenate the pairs $(q_p,u_q^{1-p})$ of
Eq.~\eqref{area:conformal-path} with the positive-scalar-curvature
path from $u_qq$ to the round metric, using lapse one on the latter
part. Reparametrize both parts smoothly, with all derivatives
vanishing at the join and the endpoints. Denote the resulting pair
by $(q_s,N_s)$, $0\le s\le1$. Then
\begin{equation}
 \begin{gathered}
 (q_0,N_0)=(q,u_q),\qquad
 (q_1,N_1)=(q_{\rm round},1),\qquad
 \min u_q\le N_s\le1\\
 R_{q_s}-2N_s^{-1}\Delta_{q_s}N_s\ge\varsigma_{\rm P}>0
 \end{gathered}
 \label{area:paired-path}
\end{equation}
The last inequality follows from Eq.~\eqref{area:conformal-path}
on the first part and positive scalar curvature on the second.
For fixed $q$, the path, its lapse and all their derivatives are
bounded. Set $c=\eta/L$ and $\mathcal K=c\gamma$ throughout the
interior.

The joins near the marginal surface will require a small
angle-dependent conformal factor. For a function $v$ on $S^3$,
independent of $s$, define
\begin{equation}
 q_s^{(v)}=e^{2\chi_{\rm H}(s)v}q_s
 \label{area:path-factor}
\end{equation}
This has initial metric $e^{2v}q$ and final metric $q_{\rm round}$.
For any three-metric $h$, positive lapse $N$ and function $f$, the
conformal formulas give
\begin{equation}
 \begin{split}
 R_{e^{2f}h}-2N^{-1}\Delta_{e^{2f}h}N
 =e^{-2f}\bigl[&R_h-2N^{-1}\Delta_hN-4\Delta_hf\\
 &-2|\dd f|_h^2-2\langle\dd f,\dd\log N\rangle_h\bigr]
 \end{split}
 \label{area:path-source-transform}
\end{equation}
With $f=\chi_{\rm H}(s)v$, this shows that the lower bound in
Eq.~\eqref{area:paired-path} changes by at most $C\|v\|_{C^2}$.
All the functions $v$ used below are $O(\eta)$ in every fixed
angular derivative norm. Choosing $\eta$ small therefore leaves
at least $\varsigma_{\rm P}/2$ along the entire path.

We record the radial estimates once for both traversals. Write
\begin{equation}
 \gamma=N_s^2\dd\varrho^2+L^2e^{2b_{\rm r}(\varrho)}q_s^{(v)},\qquad
 s=s(\varrho),\qquad
 \tau_s=\operatorname{tr}_{q_s^{(v)}}\partial_s q_s^{(v)}
 \label{area:path-metric}
\end{equation}
Here $b_{\rm r}$ depends only on $\varrho$, while $v$ depends only on
the angles. Direct differentiation gives
\begin{equation}
 \begin{split}
 \mathrm{II}^i{}_j
 &=\frac1{N_s}\left[b_{\rm r}'\delta^i{}_j
 +\frac{s'}2(q_s^{(v)})^{ik}\partial_s(q_s^{(v)})_{kj}\right]\\
 \mathcal H&=\frac{3b_{\rm r}'+s'\tau_s/2}{N_s}
 \end{split}
 \label{area:path-shape}
\end{equation}
Traverse the path over length $L/\eta=1/c$, with a fixed smooth
profile flat at its endpoints. Thus $s'=O(c)$ and $s''=O(c^2)$.
If $b_{\rm r}'=O(c)$ and $b_{\rm r}''=O(c^2)$, the fixed path bounds and
Eq.~\eqref{area:path-shape} imply
$\mathcal H=O(c)$, $\partial_\varrho\mathcal H=O(c^2)$ and
$|\mathring{\mathrm{II}}|=O(c)$. In particular,
\begin{equation}
 \frac2{N_s}|\partial_\varrho\mathcal H|
 +\frac43|\mathcal H^2-9c^2|+|\mathring{\mathrm{II}}|^2
 \le\frac{C\eta^2}{L^2}
 \label{area:path-bounds}
\end{equation}
Tangential spatial curvature is bounded by the curvature of
$L^2e^{2b_{\rm r}}q_s^{(v)}$ and the quadratic terms in $\mathrm{II}$.
The components with one radial index involve angular derivatives
of $\mathrm{II}$; those with two radial indices also involve
$N_s^{-1}\partial_\varrho\mathrm{II}$ and
$N_s^{-1}\operatorname{Hess}_{\gamma_S}N_s$.
For bounded $b_{\rm r}$ these are all $O(L^{-2})$. The lapse Hessian
need not be small in $\eta$; its fixed bound suffices for curvature,
while Eq.~\eqref{area:collar} gives the lower bound needed for
$\mathcal C_{\rm H}>0$.

Fix $\mu_{\rm P}>0$ and first specify the interval containing the
marginal surface. Let
$W_{\rm P}=(1+\mu_{\rm P})(1-\chi_{\rm H})$ and choose
$x_{\rm P}\in(0,1)$ with $W_{\rm P}(x_{\rm P})=1$.
On an interval of length $L$, set
\begin{equation}
 \begin{gathered}
 \gamma_S=e^{2b_{\rm H}}L^2q,\qquad N_{\rm H}=u_q,\qquad
 \partial_\varrho b_{\rm H}
 =cW_{\rm P}\bigl((\varrho-\varrho_{\rm A})/L\bigr)u_q\\
 b_{\rm H}(\varrho_{\rm h})=0,\qquad
 \varrho_{\rm h}=\varrho_{\rm A}+Lx_{\rm P}
 \end{gathered}
 \label{area:d18}
\end{equation}
Its initial and final angular factors are $O(\eta)$; write the
initial one as $b_{\rm A}$. The second fundamental form is pure
trace and $\mathcal H=3cW_{\rm P}$. At
$S=\{\varrho=\varrho_{\rm h}\}$ the induced metric is exactly
$L^2q$, with $\theta_+=0$ and $\theta_-=-6c$.
Moreover $\mathcal L_Su_q=3\eta W_{\rm P}'(x_{\rm P})/L^2$,
so Eq.~\eqref{area:d14} and $\max u_q=1$ give
\begin{equation}
 \lambda_1(\mathcal L_S)\ge\frac{3\eta W_{\rm P}'(x_{\rm P})}{L^2}>0,
 \qquad
 \mathcal S_{\rm H}|_S
 =\frac{R_q-2\Delta_qu_q/u_q}{L^2}
 =\frac{2\lambda_q}{L^2}
 \label{area:mots-zone}
\end{equation}
Throughout this interval,
$\mathcal S_{\rm H}=[2\lambda_q+O(\eta)]/L^2$ by
Eq.~\eqref{area:path-source-transform}. The largest radial term
in Eq.~\eqref{area:collar} is $O(\eta)/L^2$, from
$\partial_\varrho\mathcal H$; the others are $O(\eta^2)/L^2$.
After $S$, $W_{\rm P}>1$ and hence $\mathcal H>3c$.

To join the regular center to this interval, start with the cap
of Appendix~\ref{app:area-geometry} at scale $L$. Its cylindrical
end has metric $L^2q_{\rm round}$. Follow
Eq.~\eqref{area:path-metric} backwards from $s=1$ to $s=0$, using
$v=b_{\rm A}$ and $b_{\rm r}=0$, over length $L/\eta$.
At the first end the lapse is one and the metric is round; at the
second they are $u_q$ and $e^{2b_{\rm A}}L^2q$.
All derivatives of the path vanish at its endpoints, so the metric
matches the cap and the start of Eq.~\eqref{area:d18} to every order.
Equations~\eqref{area:paired-path}--\eqref{area:path-bounds} give
$\mathcal S_{\rm H}\ge\varsigma_{\rm P}/(2L^2)$, while the
remaining terms in Eq.~\eqref{area:collar} are $O(\eta^2)/L^2$
on this whole traversal.

Only one short interpolation is needed before the outward
traversal. At the end of Eq.~\eqref{area:d18}, keep $q$ and
$N_{\rm H}=u_q$ fixed over length $L$ and set
\begin{equation}
 \partial_\varrho b_{\rm H}
 =c(1+\mu_{\rm P})\bigl[(1-\beta)u_q+\beta\bigr],\qquad
 \beta=1-\chi_{\rm H}\bigl((\varrho-\varrho_+)/L\bigr)
 \label{area:outward-interpolation}
\end{equation}
Here $\varrho_+$ is the start of this interval. The derivative
changes smoothly from $c(1+\mu_{\rm P})u_q$ to the
angle-independent value $c(1+\mu_{\rm P})$, and
\begin{equation}
 \mathcal H=3c(1+\mu_{\rm P})\bigl[(1-\beta)+\beta/u_q\bigr]
 \ge3c(1+\mu_{\rm P})>3c
 \label{area:outward-expansion}
\end{equation}
The final angular factor, denoted $b_{\rm C}$, is still
$O(\eta)$. The quantity $\mathcal S_{\rm H}$ remains
$[2\lambda_q+O(\eta)]/L^2$, and the largest of the remaining
terms in Eq.~\eqref{area:collar} is $O(\eta)/L^2$ because
$\beta'=O(L^{-1})$.

Now follow Eq.~\eqref{area:path-metric} from $s=0$ to $s=1$,
with $s'\ge0$, $v=b_{\rm C}$ and $b_{\rm r}$ initially zero. Choose
one constant $T_{\rm H}$ bounding $|\tau_s|$ for all sufficiently
small $\eta$, and prescribe
\begin{equation}
 b_{\rm r}'=c(1+\mu_{\rm P})+\frac{T_{\rm H}}6s',\qquad
 0\le b_{\rm r}\le b_{\rm f}:=1+\mu_{\rm P}+\frac{T_{\rm H}}6
 \label{area:return-dilation}
\end{equation}
The interval has length $1/c$, so integration gives exactly
$b_{\rm r}=b_{\rm f}$ at its final endpoint. Equation~\eqref{area:path-shape}
now gives
\begin{equation}
 \mathcal H
 =\frac{3c(1+\mu_{\rm P})+(T_{\rm H}+\tau_s)s'/2}{N_s}
 \ge3c(1+\mu_{\rm P})>3c
 \label{area:return-mean-curvature}
\end{equation}
The angle-independent dilation does not introduce angular
curvature terms. Along the return path,
\begin{equation}
 \mathcal S_{\rm H}\ge
 \frac{e^{-2b_{\rm f}}\varsigma_{\rm P}}{2L^2}
 \label{area:return-source}
\end{equation}
and the other terms in Eq.~\eqref{area:collar} obey
Eq.~\eqref{area:path-bounds}.
At the initial join, the metric, lapse and $b_{\rm r}'$ match
Eq.~\eqref{area:outward-interpolation}. At the final join, the
small angular factor has disappeared, $N_s=1$, and
\begin{equation}
 R=Le^{b_{\rm f}},\qquad
 \partial_\varrho R=(1+\mu_{\rm P})cR
 \label{area:round-endpoint}
\end{equation}
All higher derivatives agree with the round metric satisfying
$\partial_\varrho R=(1+\mu_{\rm P})cR$, because all derivatives
of the path vanish at its endpoint.

Fix the path, $\mu_{\rm P}$ and $T_{\rm H}$ first. The latter
can be chosen before the final value of $\eta$, since
$b_{\rm C}=O(\eta)$ with coefficients already fixed.
Choose $\eta>0$ small enough that the $O(\eta)$ and $O(\eta^2)$
terms on the right of Eq.~\eqref{area:collar} are smaller in magnitude
than the positive lower bounds for $\mathcal S_{\rm H}$, and that
the round annulus described below has $\mathcal C_{\rm H}>0$ and $0<F<1$.
Its radius ratios and the bound on $R/L$ are already fixed. Finally take
$\alpha_{(2)}/L^2\le a_q\eta^2$ with $a_q>0$ sufficiently small.
The spatial curvature bound makes the Gauss--Bonnet correction
$O(\alpha_{(2)}/L^4)$, so $\mathcal C_{\rm H}>0$ on every path
and interpolation. The cap has nonnegative scalar curvature and
retains $\mathcal C_{\rm H}\ge6c^2$ under the same scale choice.
Equation~\eqref{area:scalar} therefore supplies a smooth real
scalar momentum on the whole interior.

Continue the round metric with
$\partial_\varrho R=(1+\mu_{\rm P})cR$ over an annulus of fixed
radius ratios. There
${}^{(4)}\mathcal R=6/R^2-12(1+\mu_{\rm P})^2c^2$ and
$F=1-[(1+\mu_{\rm P})^2-1]c^2R^2<1$.
Since $R/L$ is bounded independently of $\eta$, choosing $\eta$
sufficiently small as above keeps $\mathcal C_{{\rm H},{\rm col}}>0$.
The completion of Appendix~\ref{app:area-completion} then applies
with $\mu_1=\mu_{\rm P}$. The resulting data are complete, smooth
and asymptotically flat with a regular center. The positive
outgoing expansion from $S$ through the return path and completion
makes $S$ outermost by the contact argument above. Their curvature
is bounded by $C_q/L^2$, and Appendix~\ref{app:area-evolution}
supplies local evolution.

A family of prescribed metrics can use the same constants in the
estimates for the metric paths and curvature, and the same choices of
$L$ and $\eta$, under the following assumptions. Both $\lambda_q$
and $\min u_q$ have positive lower bounds independent of $q$.
The metrics are uniformly elliptic and obey common bounds, as do
their derivatives.
To obtain a uniform positive lower bound in
Eq.~\eqref{area:paired-path}, set
$\lambda_* =\inf_q\lambda_q>0$. The endpoint metrics $u_qq$ have
scalar curvature at least $2\lambda_*$ by
Eq.~\eqref{area:conformal-path}. The uniform bounds and common
ellipticity make this set precompact in $C^2$. Scalar curvature is
uniformly continuous in that topology on the bounded set.
Consequently finitely many smooth representatives suffice, with
each endpoint joined to a nearby representative by a straight
metric segment of scalar curvature at least $\lambda_*$.
Join each representative once to the round metric through
positive scalar curvature. The finitely many fixed paths have a
common positive minimum; the short segments and these paths have
uniform derivative bounds. Smooth flat reparametrizations give
the paths used in Eq.~\eqref{area:paired-path} with uniform constants.
No continuous choice of representative across the family is needed.
At fixed $L,\eta,\mu_{\rm P}$ and common $T_{\rm H}$,
Eq.~\eqref{area:round-endpoint} gives the same round radius and
the same derivatives at the endpoint for every member. Applying the same
round continuation and cutoffs gives one exterior and ADM mass.

For the Berger family, the direct path has trace variation
$O(|\log\varepsilon_{\rm H}|)$, so the restriction $\alpha_{(2)}/L^2\le a_q\eta^2$ is
not uniform as the Hopf circles shrink. The explicit construction
in Appendix~\ref{app:area-geometry} instead varies only
$a_\parallel$ while
$R_{\gamma_S}=8/a_\perp^2-2a_\parallel^2/a_\perp^4\ge6/a_\perp^2$.
Its longer radial intervals preserve uniform curvature bounds.
Here $L_{\rm H}$ and $\eta_{\rm H}$ are fixed before
$\varepsilon_{\rm H}$ tends to zero. In particular, neither a
uniform positive injectivity radius nor a uniform local existence
time is part of this small-area conclusion.

\paragraph{A necessary condition.}
For a strictly stable marginal surface isometric to $(S^3,L^2q)$
in data with $\mathcal K=c\gamma$, Eqs.~\eqref{area:d5},
\eqref{area:gbshift} and \eqref{area:gauss-stability} give, as quadratic forms,
\begin{equation}
 \begin{split}
 -\Delta_S+\frac{R_S}{2}
 ={}&\mathcal L_S+\frac{|\mathring{\mathrm{II}}|^2}{2}
       +\frac{3kG}{2}(P^2+|D\Psi|^2)\\
 &-\frac{\alpha_{(2)}}2
    \bigl[\mathcal G+4c^2\,{}^{(4)}\mathcal R+24c^4\bigr]
 \end{split}
 \label{area:converse-identity}
\end{equation}
The scalar terms are nonnegative. Hence
\begin{equation}
 \lambda_q>-\frac{\alpha_{(2)}L^2}2\sup_S
 \bigl|\mathcal G+4c^2\,{}^{(4)}\mathcal R+24c^4\bigr|
 \label{area:necessity}
\end{equation}
If curvature and $c^2$ are bounded by $C_0/L^2$, this is
$\lambda_q>-C\alpha_{(2)}/L^2$.
At positive coupling this is weaker than the sufficient condition
$\lambda_q>0$ used in the construction. The converse does not require
$\Psi$ to be constant: the term $|D\Psi|^2$ in
Eq.~\eqref{area:converse-identity} is nonnegative.
At zero coupling, the scalar
dominant energy condition and the stability inequality of
Refs.~\cite{GallowaySchoen2006,AnderssonMarsSimon2008} give
$\lambda_1(-\Delta_S+R_S/2)\ge\lambda_1(\mathcal L_S)>0$
even without $\mathcal K=c\gamma$.

\subsection{Physical kinetic terms and local evolution}
\label{app:area-evolution}

This subsection shows that the data evolve, at least for a short
time, and that their full initial spacetime curvature is bounded
uniformly as the Hopf circles shrink. The first part bounds the curvature
components with two normal legs, which the constraints do not
contain, from the evolution equations; the second checks that the
kinetic terms of all five gravitational polarizations are positive,
which, together with the weak-coupling result of
Ref.~\cite{Kovacs2020} invoked below, gives local well-posedness.
To control full spacetime curvature, write
$|\mathcal R_{abcd}|_{e_0}$ for the positive norm of orthonormal
components with timelike leg $e_0$. In this subsection the index $0$
refers to the unit normal $e_0$ of the initial slice, Latin indices
$i,j,\ldots$ are spatial, hats denote the projected curvature of
Eq.~\eqref{area:projected}, and $\delta^{iabc}_{jdef}$ is the
generalized Kronecker delta of Section~\ref{sec:characteristics}.
This is a bound in frames whose timelike vector is the slice
normal, and hence also in frames differing by a spatial rotation.
It is not a bound on components in arbitrarily boosted frames.

In an orthonormal frame adapted to the slice, the spacetime curvature
has spatial components, mixed components $\mathcal R_{0ijk}$ with one
normal index, and the electric part $E_{ij}=\mathcal R_{0i0j}$. The
spatial components are the projected curvature of
Eq.~\eqref{area:projected}. In the interior,
$\mathcal K=c\gamma$ with constant $c$, so the Codazzi tensor
$D_i\mathcal K_{jk}-D_j\mathcal K_{ik}$ vanishes. In the round
completion, Eq.~\eqref{area:round-codazzi} gives the same result,
including through the vacuum change of slicing. Thus the mixed
components vanish on the entire initial slice. The spatial field
equations determine its electric part:
\begin{equation}
 -E^i{}_j+\delta^i{}_j\operatorname{tr}E
 +\alpha_{(2)}\delta^{iabc}_{jdef}E_a{}^d\widehat R_{bc}{}^{ef}
 =\frac{3kG P^2+\widehat R}{2}\delta^i{}_j-\widehat R^i{}_j
 \label{area:electric}
\end{equation}
This is the spatial component of Eq.~\eqref{eq:field} written with
Eq.~\eqref{app:deltaform}, with the stress
$T^i{}_j[\Psi]=\tfrac12P^2\delta^i{}_j$ of a scalar constant on the
slice. The terms with one normal index up and one down in the same
curvature factor are linear in $E$ and form the left side; the purely
spatial Einstein terms give
$\widehat R^i{}_j-\tfrac12\delta^i{}_j\widehat R$. Antisymmetry
excludes quadratic electric terms, since two normal indices cannot
both be up. The Gauss--Bonnet terms with the two normal indices in
different factors are products of mixed components, which vanish, and
the purely spatial Gauss--Bonnet tensor vanishes in four dimensions.

To solve Eq.~\eqref{area:electric} for $E$, note that its Einstein
part $E^i{}_j\mapsto-E^i{}_j+\delta^i{}_j\operatorname{tr}E$ has
eigenvalues $-1$ on trace-free tensors and $\operatorname{tr}\delta-1=3$
on their trace part, so its inverse has norm one. Its
correction, the generalized Kronecker delta contracted with
$\widehat R$, is bounded by $24\alpha_{(2)}|\widehat R_{ijkl}|$, since
for fixed $i,j,a,d$ the summed pairs $b,c$ and $e,f$ must complete
$\{i,a\}$ and $\{j,d\}$ to all four spatial directions, so that each
matrix entry is either zero or $\pm4\alpha_{(2)}$ times a single
component of $\widehat R$. This bound is small,
because $|\widehat R_{ijkl}|\le C/L_{\rm H}^2$ and
$\alpha_{(2)}/L_{\rm H}^2$ is small, and the exact inverse is
therefore uniformly bounded. The source and therefore $E$ are
$O(L_{\rm H}^{-2})$. This single inversion bounds all initial
spacetime curvature uniformly in $\varepsilon_{\rm H}$. For
prescribed geometry the same argument uses $C_q/L^2$ throughout.

For the physical kinetic terms, split the gravitational
variational symbol of Appendix~\ref{app:hyperbolic} into Einstein
and Gauss--Bonnet parts, $\mathsf P_\star=
\mathsf P_\star^{\rm E}+\mathsf P_\star^{\rm GB}$. The Gauss--Bonnet
part of $\mathcal W$ in Eq.~\eqref{sym:lovelock} is linear in the
curvature, and the curvature variation is quadratic in $\xi$, so
\begin{equation}
 \|\mathsf P_\star^{\rm GB}(\xi)\|
 \le C\alpha_{(2)}|\mathcal R_{abcd}|_{e_0}|\xi|_{e_0}^2
 \label{area:symbolbound}
\end{equation}
Freeze coefficients and rotate a unit spatial covector to $e^1$, so
that $\xi_a=(\omega,1,0,0,0)$ with temporal frequency $\omega$. We
impose the gauge conditions
\begin{equation}
 h_{1a}=0\qquad(a=0,\ldots,4)
 \label{area:d23}
\end{equation}
They require no division by $\omega$: the gauge transformation
$h_{ab}\mapsto h_{ab}+\xi_a\zeta_b+\xi_b\zeta_a$ changes $h_{1a}$ by
$\zeta_a+\xi_a\zeta_1$, so the choice $\zeta_1=-h_{11}/2$,
$\zeta_i=-h_{1i}$ for $i=2,3,4$, and $\zeta_0=-h_{10}+\omega h_{11}/2$
imposes them. The remaining amplitudes are
the lapse $h_{00}$, the transverse shifts $h_{0i}$ and the spatial
block, which we split into its trace-free part and trace,
\begin{equation}
 h_{ij}=(h_{\rm tf})_{ij}+\frac{h_{\rm tr}}{\sqrt3}\,\delta_{ij}
 \qquad(i,j=2,3,4)
 \label{area:d24}
\end{equation}

The lapse equation, the row of $\mathsf P_\star$ belonging to
$h_{00}$, is the principal part of the Hamiltonian constraint. By
Eq.~\eqref{app:deltaform} it involves only the variation of
$\widehat R_{ijkl}$, so it contains neither temporal frequency nor
lapse or transverse shift. Its Einstein part is one half of the
principal symbol of the variation of $\widehat R$; since $h_{11}=0$,
\begin{equation}
 \tfrac12\bigl(\xi_i\xi_jh_{ij}-|\boldsymbol\xi|^2\operatorname{tr}h\bigr)
 =-\tfrac12(h_{22}+h_{33}+h_{44})=-\frac{\sqrt3}2\,h_{\rm tr}
 \label{area:d25}
\end{equation}
Its Gauss--Bonnet part is bounded by Eq.~\eqref{area:symbolbound}.
The lapse equation therefore eliminates the trace as
$h_{\rm tr}=\mathsf D_{\rm tr}h_{\rm tf}$, where
$\|\mathsf D_{\rm tr}\|=O(\alpha_{(2)}|\mathcal R_{abcd}|_{e_0})$.

Let $\mathsf R_{\rm H}$ insert this spatial perturbation and
$\mathsf Z_{\rm H}$ insert the normalized transverse shifts
$\sqrt2h_{0i}$, $i=2,3,4$, into the metric amplitude.
Using Eq.~\eqref{hyp:evolution-matrix}, define
\begin{equation}
 \mathsf W_{\rm sh}=\mathsf Z_{\rm H}^T\mathsf P_{\star,0}\mathsf Z_{\rm H},
 \qquad
 \mathsf V_{\rm sh}=\mathsf Z_{\rm H}^T\mathsf P_{\star,1}\mathsf R_{\rm H}
 \label{area:d26}
\end{equation}
The lapse drops out of the quadratic action, because its row
vanishes on these amplitudes and has no temporal entries. The shift
block is independent of temporal frequency and $\mathsf P_{\star,2}$
annihilates temporal metric components. Hence $\omega^2$ multiplies
only $\mathsf R_{\rm H}^T\mathsf P_{\star,2}\mathsf R_{\rm H}$, and the
shifts couple to $h_{\rm tf}$ through $2\omega\mathsf V_{\rm sh}$ and
through terms without $\omega$. Their stationary values are
$-\omega\mathsf W_{\rm sh}^{-1}\mathsf V_{\rm sh}h_{\rm tf}$ up to
terms without $\omega$. Eliminating the shifts in the quadratic
action therefore gives the coefficient of $\omega^2$,
\begin{equation}
 \mathsf K_{\rm phys}
 =\mathsf R_{\rm H}^{T}\mathsf P_{\star,2}\mathsf R_{\rm H}
 -\mathsf V_{\rm sh}^{T}\mathsf W_{\rm sh}^{-1}\mathsf V_{\rm sh}
 \label{area:kinetic}
\end{equation}

At zero coupling $\mathsf W_{\rm sh}=\mathsf 1/2$ and
$\mathsf V_{\rm sh}=0$. The Einstein part of the $\omega^2$
coefficient on spatial amplitudes is
$\tfrac12[|h|^2-(\operatorname{tr}h)^2]$, where
$|h|^2=\sum_{i,j}h_{ij}^2=|h_{\rm tf}|^2+h_{\rm tr}^2$ and
$(\operatorname{tr}h)^2=3h_{\rm tr}^2$, so
\begin{equation}
 \mathsf R_{\rm H}^{T}\mathsf P_{\star,2}^{\rm E}\mathsf R_{\rm H}
 =\frac{\mathsf 1}2-\mathsf D_{\rm tr}^{T}\mathsf D_{\rm tr}
 \label{area:d27}
\end{equation}
By Eq.~\eqref{area:symbolbound} the Gauss--Bonnet parts of
$\mathsf W_{\rm sh}$, $\mathsf V_{\rm sh}$ and
$\mathsf R_{\rm H}^T\mathsf P_{\star,2}\mathsf R_{\rm H}$ are
$O(\alpha_{(2)}|\mathcal R_{abcd}|_{e_0})$, as is
$\mathsf D_{\rm tr}$. Consequently
\begin{equation}
 \mathsf K_{\rm phys}=\frac{\mathsf 1}2
 +O\bigl(\alpha_{(2)}|\mathcal R_{abcd}|_{e_0}\bigr)>0
 \label{area:d28}
\end{equation}
on all five gravitational polarizations. The principal equation obtained by varying the trace amplitude
$h_{\rm tr}$, defined in Eq.~\eqref{area:d24}, recovers $h_{00}$;
the principal Bianchi identity $\mathsf B_\xi\mathsf P_\star=0$
supplies the equations corresponding to the gauge-fixed components
$h_{1a}$ in Eq.~\eqref{area:d23}. This bound is uniform over unit spatial covectors and
independent of the Hopf radius. For a prescribed intrinsic metric of the marginal surface, it
holds with $C_q/L^2$ in place of $C/L_{\rm H}^2$.

The curvature bound places the data in the weak-coupling regime of
the modified harmonic formulation of Ref.~\cite{Kovacs2020}: by
Eq.~\eqref{area:symbolbound}, the Gauss--Bonnet part of the symbol is
smaller than the Einstein part by a factor
$O(\alpha_{(2)}/L_{\rm H}^2)$.
Choose fixed, separated auxiliary cones and the initial time
vector proportional to $e_0$. The normal is a common hyperbolicity
covector, and unit spatial covectors have frequencies bounded away
from zero, as required by Appendix~A of that reference.
The minimally coupled scalar satisfies a wave equation with positive
kinetic coefficient and no mixed metric--scalar second derivatives.
The direct-sum symmetrizer remains smooth when characteristic
frequencies coincide. Conservation of the scalar stress-energy tensor
and the Lovelock Bianchi identity propagate the initially vanishing
gauge constraints.

For each fixed $\varepsilon_{\rm H}>0$, use the modified harmonic
gauge with a prescribed source,
$\mathscr H^a=-\widetilde g^{bc}\Gamma[g]^a_{bc}-f_{\rm g}^a=0$,
where $\widetilde g$ is the auxiliary metric of
Eq.~\eqref{hyp:auxiliary-metrics} and $f_{\rm g}^a$ is the value of
$-\widetilde g^{bc}\Gamma[g]^a_{bc}$ for a smooth reference metric
equal to static Boulware--Deser in the far exterior. Match the initial
lapse and shift there and choose their time derivatives to impose the
gauge. The gauge source does
not change the principal symbol. The perturbations and the field equations evaluated on the reference fields vanish outside a compact
set. Coefficient derivatives are bounded in the exterior, so local
Sobolev existence gives
$T_{\varepsilon_{\rm H}}>0$. Shorten this interval so that the
curvature remains in the weak-coupling regime, the physical kinetic
terms remain positive, and the time covector remains a common hyperbolicity covector. Neither the
injectivity radius nor this existence time is asserted uniform in
$\varepsilon_{\rm H}$.

Finally, replace $(\alpha_{(2)},\tilde\alpha)$ by
$(\lambda_\alpha\alpha_{(2)},\lambda_\alpha\tilde\alpha)$,
$0\le\lambda_\alpha\le1$, in
Eqs.~\eqref{area:scalar}--\eqref{area:bending}, fixing the core
geometry and cutoffs. The positivity of $\mathcal C_{\rm H}$ in
the matter region, smooth vanishing of $P$ at its boundary, and the
weak-coupling bounds remain valid; Eq.~\eqref{area:round} is
smooth at $\lambda_\alpha=0$. This connects the exact constrained
data continuously to Einstein--scalar data.
With $L_{\rm H},\eta_{\rm H}$ and coupling fixed, shrinking the
Hopf circles gives $r_{\rm A}/\sqrt{\tilde\alpha}\to0$ by
Eq.~\eqref{area:small}; every member remains smooth with positive area.
\section{Exact solutions with different tidal curvatures}
\label{app:tidal-prediction}

We derive Eq.~\eqref{tidal:separation} by placing a contracting
homogeneous scalar core inside complete data with weak curvature, constructing
exact anisotropic solutions near it, and comparing tidal measurements
in a common observation region.

Write $\alpha=\alpha_{(2)}$, so $\tilde\alpha=2(n-3)(n-4)\alpha=4\alpha$
in five dimensions. We use the rescaled scalar $\Psi$ of
Section~\ref{sec:characteristics} throughout. In five dimensions
$\kappa_5^2=3A_3kG$ and $\psi=\Psi/\sqrt{A_3}$, so the scalar kinetic
term $\kappa_5^2(\nabla\psi)^2$ of the action reads
$3kG(\nabla\Psi)^2$. The wave-number parameter $k_{\rm w}$ is distinct
from the constant $k$ of Eq.~\eqref{def:mass}.

The reference data and their initial time $t_-$ are fixed first.
We then fix a sufficiently small observation time $T>0$ after the
reference transition, choose a sufficiently small positive
$\epsilon_{\rm a}$, and take an unbounded sequence of $k_{\rm w}$ with
$\epsilon_{\rm a}$ held fixed. Constants may depend on the fixed reference data and $T$, but are
independent of $k_{\rm w}$ and uniform for $0<\epsilon_{\rm a}\le\epsilon_0$,
with a sufficiently small fixed $\epsilon_0$.
Several estimates below involve fractional powers of $k_{\rm w}$ and
make sense only for dimensionless quantities. We therefore fix one
length $L_0$ (a fixed multiple of $\sqrt{\tilde\alpha}$ will do) and
measure all lengths, times and frequencies in units of it, without
introducing new symbols for the dimensionless quantities. The combination
$\tilde\alpha\mathcal E_{ij}$ appearing in Eq.~\eqref{tidal:separation}
is dimensionless and is unaffected by this choice.

\subsection{Complete reference data with weak curvature}
\label{app:weak-reference}

The reference solution is a contracting homogeneous scalar core. It
is the simplest background on which the transition can be followed
exactly: its characteristic coefficients depend on time only, all
sectors degenerate together at $K=\sqrt2$, and the transition is
reached from arbitrarily weak curvature by starting the contraction
early enough. The core is then embedded in asymptotically flat data
by switching off the scalar time derivative and the contraction
outside a large radius.

Let $s(t)$ be the scale factor of a spatially flat, contracting
homogeneous solution, with $t$ its proper time and $H_c=\dot s/s$.
The scalar equation and the two independent gravitational equations
give
\begin{equation}
 12H_c^2+24\alpha H_c^4=3kG\dot\Psi^2,\qquad
 s^4\dot\Psi=\mathsf q_*,\qquad
 \dot H_c=-4H_c^2\frac{1+2\alpha H_c^2}{1+4\alpha H_c^2}
 \label{tidal:frw}
\end{equation}
Here $\mathsf q_*$ is the conserved homogeneous scalar charge; the radial
derivative $Q$ of the spherical scalar vanishes in this solution.
The three relations follow from Eq.~\eqref{form:flrw} with $n=5$ and
$P_c=\dot\Psi$: the first is the Hamiltonian constraint, the second
integrates the scalar equation $\partial_t(s^4\dot\Psi)=0$, and the
third follows from the spatial gravitational equation, which with the
constraint reads $\dot H_c(1+\tilde\alpha H_c^2)=-kG\dot\Psi^2$, by
eliminating $\dot\Psi^2$ with the first.

On the initial slice $t=t_-$, set $s(t_-)=1$ and
$H_c(t_-)=-H_-$, with $H_-\sqrt{\tilde\alpha}$ small and fixed. Direct
linearization of the full physical symbol gives five trace-free
transverse polarizations with polynomial proportional to
$-K\omega^2+Y|\boldsymbol\xi|^2$, where
\begin{equation}
 K=1+4\alpha H_c^2,\qquad
 Y=1+4\alpha(\dot H_c+H_c^2)=\frac{2-K^2}{K}
 \label{tidal:KY}
\end{equation}
The last form of $Y$ follows from the third relation of
Eq.~\eqref{tidal:frw}, which gives
$\tilde\alpha(\dot H_c+H_c^2)=-(K-1)(K+2)/K$. These factors agree with
the values $X=K$ and $Y=\lambda_-$ of Eq.~\eqref{sym:center} for $n=5$,
which flat FLRW satisfies at every point (Appendix~\ref{app:symbol}).
By Eq.~\eqref{sym:weights-five}, $K-1$ is $\tilde\alpha$ times the sectional
curvature $H_c^2$ of a spatial plane, and $Y-1$ is $\tilde\alpha$ times
that of a time--space plane, $\dot H_c+H_c^2$.

Set $t=0$ at the transition, where $K=\sqrt2$ and $Y=0$. Since $H_c<0$ and, by
Eq.~\eqref{tidal:frw}, $\dot H_c<0$,
\begin{equation}
 \dot K=2\tilde\alpha H_c\dot H_c>0,\qquad
 \dot Y=-\Bigl(1+\frac2{K^2}\Bigr)\dot K<0
 \label{tidal:d1}
\end{equation}
Thus the spatial coefficient $Y$ changes sign at $t=0$, while the time
coefficient $K$ increases and stays positive. The background stays regular on
a fixed short interval after $t=0$: the third relation of
Eq.~\eqref{tidal:frw} is a smooth first-order equation for $H_c$, and the value
of $H_c$ at $t=0$, fixed by $\tilde\alpha H_c^2=\sqrt2-1$, does not depend on $H_-$.

The Kretschmann scalar is four times the sum of the squared sectional
curvatures of the ten planes spanned by pairs of comoving frame vectors
(Section~\ref{sec:curvature-continuation}), four time--space planes
and six spatial ones, so
\begin{equation}
 \tilde\alpha^2\mathcal R_{abcd}\mathcal R^{abcd}
 =16\tilde\alpha^2(\dot H_c+H_c^2)^2+24\tilde\alpha^2H_c^4
 =16(Y-1)^2+24(K-1)^2
 \label{tidal:d2}
\end{equation}
At $t=0$ this is $16+24(\sqrt2-1)^2=88-48\sqrt2$, the value quoted in
Section~\ref{sec:tidal-prediction}.

To embed the core in asymptotically flat data, choose smooth
$\chi_{\rm c},\chi_{\rm V}\in[0,1]$, equal to one for arguments
at most $1/2$ and zero for arguments at least $3/5$.
In the notation of Appendix~\ref{app:cauchy-data}, prescribe
\begin{align}
 &Q=0,\quad P=P_0\chi_{\rm c}(H_-R),\quad
 P_0^2=\frac{4H_-^2+2\tilde\alpha H_-^4}{kG},\qquad
 \mathcal M(R)=\frac12\int_0^R r^3P(r)^2\,\dd r
 \label{tidal:core-data}\\
 &R^2F+\frac{\tilde\alpha}{2}F^2=2kG\mathcal M,\qquad
 V=H_-R\,\chi_{\rm V}(H_-R),\qquad a^2=\frac{1-F}{1-V^2}
 \label{tidal:core-metric}\\
 &f_{,\tau}=2V_{,R}+Vf_{,R},\qquad
 f=\log2+2\log a,\qquad \Psi_{,\tau}=P/a
 \label{tidal:core-time}
\end{align}
Here $P_0^2$ is the value of $\dot\Psi^2$ at $t_-$ given by the first
relation of Eq.~\eqref{tidal:frw}, and $\mathcal M$ integrates the
first constraint of Eq.~\eqref{evol:generalconstraints} with $Q=0$
and $n=5$. The first relation of Eq.~\eqref{tidal:core-metric} is the
mass formula of Eq.~\eqref{def:mass} for $n=5$, and the last one is
Eq.~\eqref{evol:generaldata} solved for $a^2$.
Equation~\eqref{tidal:core-time} is the second constraint of
Eq.~\eqref{evol:generalconstraints} with $Q=0$, together with
$P=a\Psi_{,\tau}$.

The mass formula is quadratic in $F$, and we take its GR-branch root,
the one with $K>0$. For $H_-R\le1/2$ both cutoffs equal one, so
$\mathcal M=P_0^2R^4/8$ and
\begin{equation}
 R^2F+\frac{\tilde\alpha}{2}F^2=2kG\mathcal M
 =H_-^2R^4+\frac{\tilde\alpha}{2}H_-^4R^4
 \label{tidal:d3}
\end{equation}
The GR-branch root is therefore $F=H_-^2R^2$, and $V=H_-R$ gives $a=1$
for $R\le1/(2H_-)$. These are exactly the homogeneous data at $t_-$: on
a flat FLRW slice $F=H_c^2R^2$ (Appendix~\ref{app:symbol}),
$e_0R=H_cR=-aV$ and $P=\dot\Psi$.

For $H_-R<3/5$ the bound $\chi_{\rm c}\le1$ gives
$\mathcal M\le P_0^2R^4/8$, hence $F\le H_-^2R^2<9/25$, because the
left side of the mass formula increases with $F$ on the GR branch
(its $F$-derivative is $R^2+\tilde\alpha F=R^2K>0$). For
$H_-R\ge3/5$, $\mathcal M$ is constant, and differentiating the mass
formula gives
\begin{equation}
 F_{,R}=-\frac{2RF}{R^2+\tilde\alpha F}<0
 \label{tidal:d4}
\end{equation}
Hence $0\le F\le9/25$ everywhere. Since $\chi_{\rm V}$ vanishes for
$H_-R\ge3/5$, also $0\le V<3/5$. Both $1-F$ and $1-V^2$ are therefore
at least $16/25$, and $a^2$ stays positive. The slice has a regular
center and a static asymptotically flat exterior: near the center the
data are homogeneous, and for $H_-R\ge3/5$ they have $P=V=0$ and
constant $\mathcal M$, which is the time-symmetric Boulware--Deser
slice of Appendix~\ref{app:realization}.

The coordinate spheres have $\theta_+=3a(1-V)/R>0$, from
$\theta_\pm=3(e_0\pm e_1)R/R$ with $e_1R=a$ and $e_0R=-aV$. Any closed
embedded surface bounding a compact region is tangent from inside to a
coordinate sphere at its maximum radius. There its outward mean
curvature is no smaller, and its extrinsic-curvature contribution to
$\theta_+$ is the same, since the two surfaces share their tangent
space. Thus it cannot be weakly outer trapped.

Apart from the fixed small parameter $\tilde\alpha H_-^2$, the data
depend on $R$ only through $H_-R$. This holds for $V$ by definition,
and $P$ is $H_-$ times such a function, since
$kGP_0^2=H_-^2(4+2\tilde\alpha H_-^2)$. Hence $H_-^2\mathcal M$ is a
fixed function of $H_-R$, and so, by the mass formula multiplied by
$H_-^2$, are $F$ and $a$. Each radial derivative therefore brings a
factor $H_-$, and the initial curvature components in the slice-normal orthonormal frame are $O(H_-^2)$. The
finite ADM mass, in contrast, grows as $H_-^{-2}$ as $H_-\to0$; only
the local curvature is small.

The core must contain the past of the observation region through the
transition. Before $t=0$, $1<K<\sqrt2$, so Eq.~\eqref{tidal:frw-speed}
gives $0<c_g^2<1$: all physical gravitational speeds are at most the
matter speed. It therefore suffices to bound the comoving
distance $\int_{t_-}^0\dd t/s$ traveled at unit speed between $t_-$
and the transition; since $s(t_-)=1$, comoving and initial Cartesian
distances agree.

To evaluate this distance, set $z=\tilde\alpha H_c^2$ and
$z_-=\tilde\alpha H_-^2$. Eliminating $\dot\Psi$ between the first two
relations of Eq.~\eqref{tidal:frw} shows that
$H_c^2(1+2\alpha H_c^2)s^8$ is constant, and the third relation gives
the rate of change of $z$:
\begin{equation}
 z\Bigl(1+\frac z2\Bigr)s^8=z_-\Bigl(1+\frac{z_-}2\Bigr),\qquad
 \dot z=\frac{8z^{3/2}(1+z/2)}{\sqrt{\tilde\alpha}\,(1+z)}
 \label{tidal:d5}
\end{equation}
The first of these gives $s(z)$, and the second turns the integral over
$t$ into an integral over $z$, from $z_-$ to the transition value
$\sqrt2-1$:
\begin{equation}
 s(z)=\left[\frac{z_-(1+z_-/2)}{z(1+z/2)}\right]^{1/8},
 \qquad
 H_-\int_{t_-}^0\frac{\dd t}{s(t)}<\frac{\sqrt2}{3}<\frac12
 \label{tidal:buffer}
\end{equation}
The integral is evaluated as
\begin{align}
 H_-\int_{t_-}^0\frac{\dd t}{s(t)}
 &=\frac{z_-^{3/8}}{8(1+z_-/2)^{1/8}}
 \int_{z_-}^{\sqrt2-1}(1+z)z^{-11/8}(1+z/2)^{-7/8}\,\dd z\nonumber\\
 &<\frac{\sqrt2}{3(1+z_-/2)^{1/8}}
 \label{tidal:buffer-integral}
\end{align}
The bound uses $1+z\le\sqrt2$, $(1+z/2)^{-7/8}\le1$, and
$\int_{z_-}^\infty z^{-11/8}\dd z=\tfrac83z_-^{-3/8}$.
This distance is therefore less than $\sqrt2/(3H_-)$, which leaves a
fixed margin inside the core radius $1/(2H_-)$.

\subsection{The exact nonspherical system}

The nonspherical solutions are chosen so that the full field
equations reduce to ordinary differential equations. This is
achieved by a spatial frame that rotates in a plane as one moves
along an axis orthogonal to it: the metric is homogeneous in the sense
that a translation along the axis combined with a rotation in the plane
is a symmetry, but
in Cartesian coordinates its components oscillate with wave number
$2k_{\rm w}$. The wave number $k_{\rm w}$ is a free parameter, and the
tidal discrepancy arises in the limit $k_{\rm w}\to\infty$.

Let $x^1,\ldots,x^4$ be Cartesian coordinates on the initial core.
We use the rotating spatial coframe
\begin{equation}
 \begin{gathered}
 \begin{aligned}
 \vartheta^1&=\cos(k_{\rm w} x^3)\dd x^1+\sin(k_{\rm w} x^3)\dd x^2,&
 \vartheta^2&=-\sin(k_{\rm w} x^3)\dd x^1+\cos(k_{\rm w} x^3)\dd x^2\\
 \vartheta^3&=\dd x^3,&\vartheta^4&=\dd x^4,
 \end{aligned}\\
 g=-N^2\dd t^2+\sum_{i=1}^4 a_i^2(\vartheta^i)^2
 \end{gathered}
 \label{tidal:metric}
\end{equation}
with $\Psi=\Psi(t)$. Translating $x^3$ by any amount while rotating
$(x^1,x^2)$ through $k_{\rm w}$ times that amount leaves every
$\vartheta^i$ unchanged, and so do translations in $x^1$, $x^2$ and
$x^4$. These isometries act transitively on each slice of constant
$t$, so the geometry is homogeneous, and the full field equations
reduce exactly to ordinary differential equations for $a_i,N,\Psi$.
In Cartesian components, the $(x^1,x^2)$ part of the spatial metric is
\begin{equation}
 \frac{a_1^2+a_2^2}{2}\bigl[(\dd x^1)^2+(\dd x^2)^2\bigr]
 +\frac{a_1^2-a_2^2}{2}\bigl[\cos(2k_{\rm w}x^3)\bigl((\dd x^1)^2-(\dd x^2)^2\bigr)
 +2\sin(2k_{\rm w}x^3)\,\dd x^1\dd x^2\bigr]
 \label{tidal:d6}
\end{equation}
For $a_1\ne a_2$ the Cartesian metric therefore oscillates with wave
number $2k_{\rm w}$; for $a_1=a_2$ the rotation cancels.

We write
\begin{equation}
 V_{\rm hom}=\prod_{i=1}^4a_i,\qquad
 H_i=N^{-1}\partial_t\log a_i,\qquad
 k_1=\frac{k_{\rm w}a_1}{a_2a_3},\qquad
 k_2=\frac{k_{\rm w}a_2}{a_1a_3}
 \label{tidal:d7}
\end{equation}
Since $\vartheta^1\wedge\vartheta^2=\dd x^1\wedge\dd x^2$,
$V_{\rm hom}$ is the spatial volume element per unit coordinate volume.
The $H_i$ are the expansion rates of the four directions in proper time
(not to be confused with the coefficient $H$ of Eq.~\eqref{sym:base}).
On each slice,
\begin{equation}
 \dd\vartheta^1=-k_{\rm w}\,\vartheta^2\wedge\vartheta^3,\qquad
 \dd\vartheta^2=k_{\rm w}\,\vartheta^1\wedge\vartheta^3,\qquad
 \dd\vartheta^3=\dd\vartheta^4=0
 \label{tidal:d8}
\end{equation}
so the nonzero structure constants of the orthonormal coframe
$a_i\vartheta^i$ (no sum) are $k_1$ and $k_2$:
$\dd(a_1\vartheta^1)=-k_1\,(a_2\vartheta^2)\wedge(a_3\vartheta^3)$ and
$\dd(a_2\vartheta^2)=k_2\,(a_1\vartheta^1)\wedge(a_3\vartheta^3)$.

The spatial curvature follows from these structure constants. In the
orthonormal frame it is diagonal on the planes spanned by pairs of
frame vectors, with nonzero sectional curvatures
\begin{equation}
 r_{12}=\frac{(k_1-k_2)^2}{4},\quad
 r_{13}=\frac{(k_1-k_2)(k_1+3k_2)}4,\quad
 r_{23}=-\frac{(k_1-k_2)(3k_1+k_2)}4
 \label{tidal:sectionals}
\end{equation}
The spatial scalar curvature is twice their sum,
$R_{\rm sp}=-(k_1-k_2)^2/2$. All of them vanish when $k_1=k_2$, that
is, when $a_1=a_2$. The fourth spatial direction is a flat product
factor, so the intrinsic four-dimensional Gauss--Bonnet density
vanishes: its antisymmetrized contraction requires four distinct
spatial directions, while every curvature component with an index
along the fourth direction vanishes.

Up to a time boundary term, a
constant spatial volume and the overall factor $1/(2\kappa_5^2)$, the full action is
\begin{equation}
 \begin{split}
 L_{\rm red}=NV_{\rm hom}\bigg[&R_{\rm sp}-2\sum_{i<j}H_iH_j
 -8\alpha H_1H_2H_3H_4\\
 &-8\alpha H_4(r_{12}H_3+r_{13}H_2+r_{23}H_1)
 +3kG(\dot\Psi/N)^2\bigg]
 \end{split}
 \label{tidal:action}
\end{equation}
This action is obtained by substituting the structure equations,
including the mixed space--time curvature, into the EGB action and
discarding total time derivatives. With $N=1$ the extrinsic curvature
is diagonal in the frame, $\mathcal K_{ij}=-H_i\delta_{ij}$, so the
Einstein part reduces to
\begin{equation}
 R_{\rm sp}+\sum_iH_i^2-\Bigl(\sum_iH_i\Bigr)^2
 =R_{\rm sp}-2\sum_{i<j}H_iH_j
 \label{tidal:d9}
\end{equation}
The Gauss--Bonnet part has three pieces. The intrinsic
four-dimensional density vanishes, as noted above. The term quartic
in the extrinsic curvature is proportional to the product
$H_1H_2H_3H_4$. The term quadratic in the extrinsic curvature
multiplies each nonzero spatial sectional curvature $r_{ij}$ by the
product $H_kH_l$ of the rates in the two complementary directions;
since $r_{ij}\ne0$ only for $i,j\le3$, every such product contains
$H_4$.

The ansatz is invariant under reversing the signs of any two of
$\vartheta^1,\vartheta^2,\vartheta^3$, and independently under
reversing the sign of $\vartheta^4$. Every omitted off-diagonal or
momentum equation is odd under one of these symmetries, so variation
in $a_i,N,\Psi$ supplies the full system.
Varying the lapse before setting $N=1$ gives
\begin{equation}
 R_{\rm sp}+2\sum_{i<j}H_iH_j
 +8\alpha\left[3H_1H_2H_3H_4
 +H_4(r_{12}H_3+r_{13}H_2+r_{23}H_1)\right]
 =3kG(\dot\Psi/N)^2
 \label{tidal:lapse}
\end{equation}
The Bianchi identity propagates this constraint under the scale-factor
equations, and $V_{\rm hom}\dot\Psi=\mathsf q_{k_{\rm w}}$ is conserved in proper time; on the
reference solution $V_{\rm hom}=s^4$ and $\mathsf q_{k_{\rm w}}=\mathsf q_*$.

Within this homogeneous ansatz, we separate the anisotropy between
directions 1 and 2 from the remaining logarithmic scale-factor variables by writing
\begin{equation}
 \log a_1=\ell_{\rm m}+\beta,\qquad
 \log a_2=\ell_{\rm m}-\beta,\qquad
 \log a_3=\ell_3,\qquad
 \log a_4=\ell_4,\qquad
 \mathbf X=(\ell_{\rm m},\ell_3,\ell_4)
 \label{tidal:d10}
\end{equation}
The oscillating part of the Cartesian metric in Eq.~\eqref{tidal:d6} has amplitude
$(a_1^2-a_2^2)/2=e^{2\ell_{\rm m}}\sinh(2\beta)$, and on the reference
solution the three components of $\mathbf X$ all equal $\log s$, which we write as
$\mathbf X_*=(\log s,\log s,\log s)$.
Only $\ell_{\rm m}$ is an average, namely
$(\log a_1+\log a_2)/2$; $\ell_3$ and $\ell_4$ are individual logarithmic
scale factors. 
From now on $N=1$, so that
$H_1=\dot\ell_{\rm m}+\dot\beta$, $H_2=\dot\ell_{\rm m}-\dot\beta$,
$H_3=\dot\ell_3$ and $H_4=\dot\ell_4$. The volume element
$V_{\rm hom}=e^{2\ell_{\rm m}+\ell_3+\ell_4}$ does not depend on
$\beta$, and $k_{1,2}=k_{\rm w}e^{\pm2\beta-\ell_3}$, so
Eq.~\eqref{tidal:sectionals} gives
$R_{\rm sp}=-2r_{12}=-2k_{\rm w}^2e^{-2\ell_3}\sinh^2(2\beta)$ and
\begin{equation}
 r_{12}H_3+r_{13}H_2+r_{23}H_1
 =k_{\rm w}^2e^{-2\ell_3}\bigl[(\dot\ell_3-2\dot\ell_{\rm m})\sinh^2(2\beta)
 -2\dot\beta\sinh(4\beta)\bigr]
 \label{tidal:d11}
\end{equation}

The exact additional part of the action relative to $\beta=0$ is
\begin{equation}
 \begin{split}
 \frac{L_\beta}{V_{\rm hom}}={}&2(1+4\alpha\dot\ell_3\dot\ell_4)\dot\beta^2
 -2k_{\rm w}^2e^{-2\ell_3}[1+4\alpha\dot\ell_4(\dot\ell_3-2\dot\ell_{\rm m})]
       \sinh^2(2\beta)\\
 &+16\alpha k_{\rm w}^2e^{-2\ell_3}\dot\ell_4\dot\beta\sinh(4\beta)
 \end{split}
 \label{tidal:beta-action}
\end{equation}
Its first term collects the $\dot\beta^2$ parts of $-2H_1H_2$ and
$-8\alpha H_1H_2H_3H_4$, since $H_1H_2=\dot\ell_{\rm m}^2-\dot\beta^2$.
The other two terms come from $R_{\rm sp}$ and from the display
preceding Eq.~\eqref{tidal:beta-action}, multiplied by $-8\alpha H_4$. The rest of $L_{\rm red}$ does not depend
on $\beta$ or $\dot\beta$, so the equation for $\beta$ follows from
$L_\beta$ alone.

The last term of Eq.~\eqref{tidal:beta-action}, which is linear in
$\dot\beta$, contributes
$64\alpha k_{\rm w}^2V_{\rm hom}e^{-2\ell_3}\dot\ell_4\dot\beta\cosh(4\beta)$
both to $\partial_t(\partial L_\beta/\partial\dot\beta)$ and to
$\partial L_\beta/\partial\beta$, and these contributions cancel in the
Euler--Lagrange equation. Divided by 4, what remains is
\begin{equation}
 \partial_t\{V_{\rm hom}(1+4\alpha\dot\ell_3\dot\ell_4)\dot\beta\}
 +k_{\rm w}^2\sinh(4\beta)\bigl\{4\alpha\,\partial_t(V_{\rm hom}e^{-2\ell_3}\dot\ell_4)
 +V_{\rm hom}e^{-2\ell_3}[1+4\alpha\dot\ell_4(\dot\ell_3-2\dot\ell_{\rm m})]\bigr\}=0
 \label{tidal:d12}
\end{equation}
With
$\partial_t\log(V_{\rm hom} e^{-2\ell_3})=2\dot\ell_{\rm m}-\dot\ell_3+\dot\ell_4$,
the products $\dot\ell_3\dot\ell_4$ and $\dot\ell_{\rm m}\dot\ell_4$
cancel inside the braces, and the equation becomes
\begin{equation}
 \partial_t\{V_{\rm hom}(1+4\alpha\dot\ell_3\dot\ell_4)\dot\beta\}
 +V_{\rm hom} k_{\rm w}^2e^{-2\ell_3}
       [1+4\alpha(\ddot\ell_4+\dot\ell_4^2)]\sinh(4\beta)=0
 \label{tidal:beta}
\end{equation}

Like the time and spatial coefficients $K$ and $Y$, the factors
$1+4\alpha\dot\ell_3\dot\ell_4$ and $1+4\alpha(\ddot\ell_4+\dot\ell_4^2)$
in this equation each differ from one by $\tilde\alpha$ times a
spacetime sectional curvature: $\dot\ell_3\dot\ell_4$ is that of the
plane of the third and fourth directions, and
$\ddot\ell_4+\dot\ell_4^2=\ddot a_4/a_4$ that of the plane of the time
and fourth directions. Since the second factor involves a time--space
curvature, it contains the acceleration $\ddot\ell_4$, which must be
determined together with the other components of $\ddot{\mathbf X}$ from
Eq.~\eqref{tidal:action}. Replacing it by its reference value would
discard a nonlinear contribution needed below.

The reduced equations determine the accelerations as long as the
matrix multiplying them is invertible. Since $V_{\rm hom}$ contains no
velocities, this matrix, after division of the reduced equations by
$V_{\rm hom}$, is the velocity Hessian of $L_{\rm red}/V_{\rm hom}$.
In the four logarithmic scale factors, whose velocities are the $H_i$,
its diagonal entries vanish, because $L_{\rm red}/V_{\rm hom}$ is
linear in each $H_i$ separately. At isotropy, where $r_{ij}=0$ and
every $H_i$ equals $H_c$, each off-diagonal entry is
\begin{equation}
 \frac{\partial^2(L_{\rm red}/V_{\rm hom})}{\partial H_i\,\partial H_j}
 =-2-8\alpha H_c^2=-2K\qquad(i\ne j)
 \label{tidal:d13}
\end{equation}
The velocity Hessian therefore has the eigenvalue $-6K$ on
$(1,1,1,1)$ and the eigenvalue $2K$, with multiplicity three, on the
vectors whose components sum to zero. The anisotropy direction
$(1,-1,0,0)$ is such a vector and is orthogonal to the directions
$(1,1,0,0)$, $(0,0,1,0)$ and $(0,0,0,1)$ of $\ell_{\rm m}$, $\ell_3$
and $\ell_4$, so the Hessian is block diagonal in $(\mathbf X,\beta)$.
The $\beta$ block is $4K$, twice the eigenvalue because $\beta$ enters
two of the $\log a_i$, and the block for the three components of $\mathbf X$ is
\begin{equation}
 M_{\mathbf X,*}=K\begin{pmatrix}-4&-4&-4\\-4&0&-2\\-4&-2&0\end{pmatrix},
 \qquad
 M_{\mathbf X,*}^{-1}=K^{-1}\begin{pmatrix}
 1/12&-1/6&-1/6\\-1/6&1/3&-1/6\\-1/6&-1/6&1/3
 \end{pmatrix}
 \label{tidal:mean-matrix}
\end{equation}

At isotropy the velocity Hessian involves only the time coefficient
$K$; the spatial coefficient $Y$, which vanishes at the transition,
does not enter.
Since $K>0$, the Hessian is nonsingular at the reference transition and
remains so for sufficiently small curvature perturbations. The exact reduced
solutions can therefore continue through $Y=0$; as
stressed in Section~\ref{sec:tidal-prediction}, this is a statement
about the reduced ordinary differential equations only.

\subsection{Initial anisotropy and its linear evolution}

The anisotropy must be prescribed on the initial slice with weak curvature, long before
the transition, with a size chosen so that its amplification through
the transition produces a prescribed small effect at the observation
time. This requires control of the linearized anisotropy on the whole
interval: it oscillates before the transition, with its energy at any
two times related by a bound polynomial in $k_{\rm w}$, and grows
afterwards at a rate that we bound from above and below.

We first linearize Eq.~\eqref{tidal:beta} about the reference. The
anisotropy enters that equation only through the factors $\dot\beta$
and $\sinh(4\beta)$, so at linear order the other factors take their
reference values: $V_{\rm hom}=s^4$, $e^{-2\ell_3}=s^{-2}$, and the
factors $1+4\alpha\dot\ell_3\dot\ell_4$ and
$1+4\alpha(\ddot\ell_4+\dot\ell_4^2)$ become $K$ and $Y$ of
Eq.~\eqref{tidal:KY}, since all components of $\dot{\mathbf X}$ equal $H_c$. With
$\sinh(4\beta)\approx4\beta$, the linearized amplitude $\phi$ obeys
\begin{equation}
 (p_*\phi')'+k_{\rm w}^2w_*\phi=0,\qquad p_*=s^4K,\qquad w_*=4s^2Y
 \label{tidal:d14}
\end{equation}
and division by $p_*$ gives
\begin{equation}
 \phi''+b_*\phi'-k_{\rm w}^2 a_*\phi=0,\qquad
 b_*=4H_c+\dot K/K,\qquad a_*=-4Y/(s^2K)
 \label{tidal:linear}
\end{equation}
Dots and primes both denote derivatives with respect to the proper
time $t$.

The coefficient $a_*$ has the sign of $-Y$. Before $t=0$, $K<\sqrt2$, so
$Y>0$ and the last term of Eq.~\eqref{tidal:linear} is a restoring
force. Since $Y(0)=0$ and $\dot Y(0)<0$
by Eqs.~\eqref{tidal:KY} and \eqref{tidal:d1}, $a_*(0)=0$ and $a_*'(0)>0$, so
after $t=0$ this term drives growth. Near this simple zero of $a_*$, the
substitution $\hat t=k_{\rm w}^{2/3}t$ turns
Eq.~\eqref{tidal:linear} into
\begin{equation}
 \partial_{\hat t}^2\phi+k_{\rm w}^{-2/3}b_*(k_{\rm w}^{-2/3}\hat t)\,\partial_{\hat t}\phi
 -k_{\rm w}^{2/3}a_*(k_{\rm w}^{-2/3}\hat t)\,\phi=0
 \label{tidal:d15}
\end{equation}
Since
$k_{\rm w}^{2/3}a_*(k_{\rm w}^{-2/3}\hat t)=a_*'(0)\,\hat t
+O(k_{\rm w}^{-2/3}\hat t^2)$, for bounded $\hat t$ this is, to
leading order, Airy's equation
\begin{equation}
 \partial_{\hat t}^2\phi=a_*'(0)\,\hat t\,\phi
 \label{tidal:d16}
\end{equation}
The turning point therefore has the Airy scale $k_{\rm w}^{-2/3}$ and,
beyond it, the growing solution has the exponential factor
$\exp[k_{\rm w}\int_0^t\sqrt{a_*}\,\dd t']$
\cite[Chapter~11]{Olver1974}. We use the following exact bounds rather than
the asymptotic expansion.

It is convenient to work with wave numbers for which the linearized
mode has zero velocity both on the initial slice and at the transition. The mode
then enters the transition with the same data, $\phi(0)=1$ and
$\phi'(0)=0$, at every such wave number, which removes a phase ambiguity
from the later estimates; such wave numbers form an unbounded sequence.
To select wave numbers with $\phi'(t_-)=\phi'(0)=0$, we use the form
$(p_*\phi')'+k_{\rm w}^2w_*\phi=0$ of the linearized equation, in which
$p_*>0$, and $w_*>0$ for $t<0$. Start with $\phi(t_-)=1$,
$\phi'(t_-)=0$, and set
\begin{equation}
 \pi_\phi=\frac{p_*\phi'}{k_{\rm w}},\qquad\text{so that}\qquad
 \phi'=\frac{k_{\rm w}\pi_\phi}{p_*},\qquad
 \pi_\phi'=-k_{\rm w}w_*\phi
 \label{tidal:d17}
\end{equation}
Let $\Theta$ be the continuous polar angle of the point $(\phi,-\pi_\phi)$
in the plane, with $\Theta(t_-)=0$. The point never reaches the origin,
because $\phi=\pi_\phi=0$ at one time would force $\phi\equiv0$.
Differentiating $\tan\Theta=-\pi_\phi/\phi$ and using the system gives
\begin{equation}
 \Theta'=k_{\rm w}(w_*\cos^2\Theta+p_*^{-1}\sin^2\Theta)
 \label{tidal:angle}
\end{equation}

The factor multiplying $k_{\rm w}$ in Eq.~\eqref{tidal:angle} lies
between the smaller and the larger of $w_*$ and $p_*^{-1}$. Both are
bounded on $[t_-,0]$, and $w_*\ge0$ there, so $\Theta$ never
decreases; on any fixed closed interval contained in $[t_-,0)$ both have a positive lower bound, so
$\Theta'\ge ck_{\rm w}$ there. The endpoint angle $\Theta(0;k_{\rm w})$
is continuous in $k_{\rm w}$, and these bounds give
\begin{equation}
 ck_{\rm w}\le\Theta(0;k_{\rm w})\le Ck_{\rm w}
 \label{tidal:d18}
\end{equation}
Hence it tends to zero as $k_{\rm w}\to0^+$ and, by continuity,
takes every positive integer multiple of $\pi$; since it is at most
$Ck_{\rm w}$, the wave numbers at which it does so form an unbounded
sequence. At these wave numbers $\sin\Theta(0;k_{\rm w})=0$, that is,
$\pi_\phi(0)=0$. Thus $\phi'(0)=0$, and $\phi(0)\ne0$ because the point
$(\phi,-\pi_\phi)$ does not reach the origin; normalize to $\phi(0)=1$.
No periodic boundary condition restricts the wave numbers on the
Cartesian core.

Before the transition a positive energy controls both the mode and
its derivative, including at $w_*=0$:
\begin{equation}
 \mathcal E_{\rm reg}[\phi]
 =p_*^2\phi'^2+(1+k_{\rm w}^2p_*w_*)\phi^2
 \label{tidal:regularized-energy}
\end{equation}
The equation $(p_*\phi')'=-k_{\rm w}^2w_*\phi$ cancels the two mixed
terms when this energy is differentiated, leaving
\begin{equation}
 \mathcal E_{\rm reg}'=k_{\rm w}^2(p_*w_*)'\phi^2+2\phi\phi'
 \label{tidal:d19}
\end{equation}
By Eq.~\eqref{tidal:KY}, $p_*w_*=4s^6(2-K^2)$ decreases to zero on
$[t_-,0]$: both factors are nonnegative and decreasing there. Moreover,
$p_*$ is bounded away from zero, so
$2|\phi\phi'|\le C\mathcal E_{\rm reg}$. Thus
\begin{equation}
 -\left(C+\frac{k_{\rm w}^2|(p_*w_*)'|}{1+k_{\rm w}^2p_*w_*}\right)
 \mathcal E_{\rm reg}
 \le\mathcal E_{\rm reg}'\le C\mathcal E_{\rm reg}
 \label{tidal:transfer}
\end{equation}
The only frequency-dependent coefficient integrates exactly. For
$t_-\le t\le t'\le0$,
\begin{equation}
 \int_t^{t'}\frac{k_{\rm w}^2|(p_*w_*)'|}{1+k_{\rm w}^2p_*w_*}\,\dd v
 =\log\frac{1+k_{\rm w}^2(p_*w_*)(t)}{1+k_{\rm w}^2(p_*w_*)(t')}
 \label{tidal:d21}
\end{equation}
Consequently the energy at a later time is bounded by a
frequency-independent multiple of its earlier value, while the reverse
bound grows at most polynomially in $k_{\rm w}$:
\begin{equation}
 \mathcal E_{\rm reg}(t')\le C\mathcal E_{\rm reg}(t),\qquad
 \mathcal E_{\rm reg}(t)\le
 C\frac{1+k_{\rm w}^2(p_*w_*)(t)}{1+k_{\rm w}^2(p_*w_*)(t')}
 \mathcal E_{\rm reg}(t')
 \label{tidal:d20}
\end{equation}
These bounds hold for every solution of Eq.~\eqref{tidal:linear}. In particular, the energy at either time is at most $C(1+k_{\rm w}^2)$
times its value at the other, a factor subexponential in $k_{\rm w}$. For the normalized mode,
$\mathcal E_{\rm reg}(0)=1$. The weight in the backward bound then
cancels the weight of $\phi^2$ in the energy, giving
\begin{equation}
 |\phi(t)|\le C,\qquad |\phi'(t)|\le Ck_{\rm w}
 \qquad(t_-\le t\le0,\quad k_{\rm w}\ge1)
 \label{tidal:preloss-linear}
\end{equation}
No sign condition on the oscillating mode was used.

Fix a sufficiently short observation time $T>0$. Smoothness and the
simple zero of $a_*$ give constants $c_a,C_a,B>0$ such that
$c_at\le a_*(t)\le C_at$ and $|b_*(t)|\le B$ on $[0,T]$.
The divergence form $(p_*\phi')'=k_{\rm w}^2p_*a_*\phi$, together with
$\phi(0)=1$ and $\phi'(0)=0$, gives $\phi\ge1$ and $\phi'>0$ for
$t>0$. Its logarithmic derivative $r_\phi=\phi'/\phi$ therefore obeys
\begin{equation}
 r_\phi'+b_*r_\phi+r_\phi^2=k_{\rm w}^2a_*,\qquad r_\phi(0)=0
 \label{tidal:d24}
\end{equation}
A comparison function that covers the transition as well as the later
growth is
\begin{equation}
 f_\phi(t)=\frac{k_{\rm w}^2t^2}{1+k_{\rm w}t^{3/2}}
 \label{tidal:comparison-function}
\end{equation}
Indeed, putting $x=k_{\rm w}t^{3/2}$ in its derivative gives
\begin{equation}
 f_\phi'+f_\phi^2
 =k_{\rm w}^2t\,\frac{x^2+x/2+2}{(1+x)^2},\qquad
 \frac12k_{\rm w}^2t\le f_\phi'+f_\phi^2\le2k_{\rm w}^2t
 \label{tidal:d25}
\end{equation}
Choose $0<c\le\min(1,c_a/2)$ and $C\ge\max(1,2C_a)$.
The functions $ce^{-Bt}f_\phi$ and $Ce^{Bt}f_\phi$ are respectively
sub- and supersolutions of Eq.~\eqref{tidal:d24}: the added linear
terms have signs $b_*-B\le0$ and $b_*+B\ge0$, and the quadratic terms
are bounded using $c^2\le c$ and $C^2\ge C$. All three functions
vanish at $t=0$. Subtracting the differential inequalities and using
an integrating factor therefore gives
\begin{equation}
 c\,\frac{k_{\rm w}^2t^2}{1+k_{\rm w}t^{3/2}}
 \le\frac{\phi'}\phi\le
 C\,\frac{k_{\rm w}^2t^2}{1+k_{\rm w}t^{3/2}}
 \qquad(0\le t\le T)
 \label{tidal:global-growth}
\end{equation}
Here the fixed factors $e^{\pm BT}$ have been absorbed into the
constants. The two regimes meet at $t=k_{\rm w}^{-2/3}$, without a
matching argument. With
\begin{equation}
 \mu_{\rm g}(t)=\sqrt{t+k_{\rm w}^{-2/3}}
 \label{tidal:d22}
\end{equation}
the upper bound is at most $Ck_{\rm w}\mu_{\rm g}$ for $0\le t\le T$,
and the lower bound is comparable to $k_{\rm w}\mu_{\rm g}$ once
$t\ge k_{\rm w}^{-2/3}$. In particular
\begin{equation}
 ck_{\rm w}\mu_{\rm g}\le\phi'/\phi\le Ck_{\rm w}\mu_{\rm g}
 \quad(k_{\rm w}^{-2/3}\le t\le T),\qquad \phi(T)\ge e^{ck_{\rm w}}
 \label{tidal:growth}
\end{equation}
The last estimate follows by integrating the lower bound over
$[T/2,T]$ for sufficiently large $k_{\rm w}$.

We will also use an integrated version valid from $t=0$. The integral
of $k_{\rm w}\mu_{\rm g}$ over $[0,k_{\rm w}^{-2/3}]$ is the fixed
number $\tfrac23(2^{3/2}-1)$. Losing this bounded contribution and
integrating the lower bound in Eq.~\eqref{tidal:growth} gives
\begin{equation}
 \frac{\phi(t)}{\phi(t')}
 \ge C^{-1}\exp\left(ck_{\rm w}\int_{t'}^t\mu_{\rm g}(v)\,\dd v\right)
 \qquad(0\le t'\le t\le T)
 \label{tidal:integrated-growth}
\end{equation}
The constant prefactor allows $\phi'(0)=0$; no positive pointwise lower
bound for $\phi'/\phi$ at $t=0$ is required.

The Airy limit can be read separately from Eq.~\eqref{tidal:d15}.
Write $\Phi(\hat t)=\phi(k_{\rm w}^{-2/3}\hat t)$ and let
$\phi_{\rm Ai}$ solve Eq.~\eqref{tidal:d16} with value 1 and
derivative 0 at $\hat t=0$. On every fixed interval
$0\le\hat t\le\hat t_0$, the two first-order systems for
$(\Phi,\partial_{\hat t}\Phi)$ and
$(\phi_{\rm Ai},\partial_{\hat t}\phi_{\rm Ai})$ have bounded
coefficient matrices differing by $O(k_{\rm w}^{-2/3})$.
To see the resulting error explicitly, set $e_{\rm Ai}=\Phi-\phi_{\rm Ai}$.
Subtracting their second-order equations gives
\begin{equation}
 \begin{split}
 \partial_{\hat t}^2e_{\rm Ai}={}&k_{\rm w}^{2/3}a_*(k_{\rm w}^{-2/3}\hat t)e_{\rm Ai}
 -k_{\rm w}^{-2/3}b_*(k_{\rm w}^{-2/3}\hat t)\partial_{\hat t}e_{\rm Ai}\\
 &+\bigl[k_{\rm w}^{2/3}a_*(k_{\rm w}^{-2/3}\hat t)-a_*'(0)\hat t\bigr]
 \phi_{\rm Ai}\\
 &-k_{\rm w}^{-2/3}b_*(k_{\rm w}^{-2/3}\hat t)\partial_{\hat t}\phi_{\rm Ai}
 \end{split}
 \label{tidal:airy-difference}
\end{equation}
The last two terms are $O(k_{\rm w}^{-2/3})$ and the initial errors vanish.
Integrating the two first-order error equations and applying
Gr\"onwall's inequality therefore gives
\begin{equation}
 |\Phi-\phi_{\rm Ai}|+
 |\partial_{\hat t}\Phi-\partial_{\hat t}\phi_{\rm Ai}|
 \le C_{\hat t_0}k_{\rm w}^{-2/3}
 \qquad(0\le\hat t\le\hat t_0)
 \label{tidal:airy-convergence}
\end{equation}
The Wronskian $\mathrm{Ai}\,\mathrm{Bi}'-\mathrm{Ai}'\,\mathrm{Bi}=1/\pi$
gives the explicit limiting solution
\begin{equation}
 \phi_{\rm Ai}(\hat t)
 =\pi\bigl[\mathrm{Bi}'(0)\,\mathrm{Ai}(a_*'(0)^{1/3}\hat t)
 -\mathrm{Ai}'(0)\,\mathrm{Bi}(a_*'(0)^{1/3}\hat t)\bigr]
 \label{tidal:d23}
\end{equation}
Primes on the Airy functions denote derivatives of their arguments.
Both coefficients are positive, since $\mathrm{Ai}'(0)<0<\mathrm{Bi}'(0)$.
Eq.~\eqref{tidal:d16}, with the stated initial data, makes $\phi_{\rm Ai}$ positive, convex and strictly
increasing for $\hat t>0$. Its logarithmic derivative is positive
there and approaches $\sqrt{a_*'(0)\hat t}$ as $\hat t\to\infty$,
by the positive-argument Airy asymptotics
\cite[Chapter~11]{Olver1974}. At a fixed $\hat t_0>0$, the physical
logarithmic derivative is therefore
$k_{\rm w}^{2/3}\phi_{\rm Ai}'(\hat t_0)/\phi_{\rm Ai}(\hat t_0)$
up to a relative error tending to zero. These asymptotics are not
needed to start the growth comparison.

The leading exponential growth of $\phi$ follows from its logarithmic derivative $r_\phi=\phi'/\phi$, which satisfies Eq.~\eqref{tidal:d24}. On $[k_{\rm w}^{-2/3},T]$, put
$e=r_\phi-k_{\rm w}\sqrt{a_*}$. Then
\begin{equation}
 e'+(r_\phi+k_{\rm w}\sqrt{a_*}+b_*)e
 =-k_{\rm w}\bigl[(\sqrt{a_*})'+b_*\sqrt{a_*}\bigr]
 \label{tidal:action-error}
\end{equation}
Here $\sqrt{a_*}\asymp\sqrt t$ and
$|(\sqrt{a_*})'+b_*\sqrt{a_*}|\le C/\sqrt t$. The coefficient
of $e$ is at least $ck_{\rm w}\sqrt t$ for large $k_{\rm w}$.
Also $|e(k_{\rm w}^{-2/3})|\le Ck_{\rm w}^{2/3}$ by
Eq.~\eqref{tidal:global-growth}. Variation of constants followed by
integration in $t$ gives
\begin{equation}
 \int_{k_{\rm w}^{-2/3}}^T|e(t)|\,\dd t
 \le C+Ck_{\rm w}\int_{k_{\rm w}^{-2/3}}^T
 \frac{1}{\sqrt v}\frac{1}{k_{\rm w}\sqrt v}\,\dd v
 \le C(1+\log k_{\rm w})
 \label{tidal:action-integral}
\end{equation}
Indeed, a source at $v$ is damped by
$\exp[-ck_{\rm w}\int_v^t\sqrt z\,\dd z]$, whose integral over
$t\ge v$ is at most $C/(k_{\rm w}\sqrt v)$; the initial term has
bounded integral by the same estimate. On
$[0,k_{\rm w}^{-2/3}]$, both $\int r_\phi\,\dd t$ and
$k_{\rm w}\int\sqrt{a_*}\,\dd t$ are bounded. Thus, uniformly on
$[0,T]$,
\begin{equation}
 \log\phi(t)=k_{\rm w}\int_0^t\sqrt{a_*(v)}\,\dd v+O(1+\log k_{\rm w})
 \label{tidal:exponential-action}
\end{equation}
This states the leading exponential rate; it does not suppress the
polynomial prefactor present in the Airy asymptotics.

The size of the initial anisotropy is fixed backward from the
observation time: the linearized amplitude at time $T$ is to
be $\epsilon_{\rm a}/k_{\rm w}^2$. This gives a metric anisotropy of
order $k_{\rm w}^{-2}$ and, since curvature contains two derivatives
and each brings a factor of order $k_{\rm w}$, a curvature anisotropy
of order one. Choose a sufficiently small $\epsilon_{\rm a}>0$ after fixing $T$,
independently of $k_{\rm w}$. On the initial slice prescribe
\begin{equation}
 \beta_{k_{\rm w}}(t_-)=\frac{\epsilon_{\rm a}\phi(t_-)}{k_{\rm w}^2\phi(T)},\qquad
 \dot\beta_{k_{\rm w}}(t_-)=0,\qquad
 \mathbf X_{k_{\rm w}}(t_-)=\mathbf X_*(t_-),\quad \dot{\mathbf X}_{k_{\rm w}}(t_-)=\dot{\mathbf X}_*(t_-)
 \label{tidal:seed}
\end{equation}
Since $\dot\beta_{k_{\rm w}}(t_-)=0=\phi'(t_-)$, the linearized
evolution of these data is
$\epsilon_{\rm a}\phi(t)/(k_{\rm w}^2\phi(T))$, which equals
$\epsilon_{\rm a}/k_{\rm w}^2$ at $t=T$. The initial anisotropy is exponentially
small: Eq.~\eqref{tidal:preloss-linear} bounds $|\phi(t_-)|$ uniformly,
and Eq.~\eqref{tidal:growth} bounds $\phi(T)$ from below. Hence
\begin{equation}
 |\beta_{k_{\rm w}}(t_-)|
 \le C\epsilon_{\rm a}k_{\rm w}^{-2}e^{-ck_{\rm w}}
 \label{tidal:d26}
\end{equation}

Choose $\dot\Psi_{k_{\rm w}}(t_-)$ with the reference sign from the exact
constraint
\begin{equation}
 3kG\dot\Psi_{k_{\rm w}}(t_-)^2
 =12H_-^2+24\alpha H_-^4
 -2K_-k_{\rm w}^2\sinh^2(2\beta_{k_{\rm w}}(t_-))
 \label{tidal:initial-scalar}
\end{equation}
Here $K_-=1+\tilde\alpha H_-^2$ is the reference value of $K$ at $t_-$.
This is Eq.~\eqref{tidal:lapse} with $N=1$, evaluated on the data of
Eq.~\eqref{tidal:seed}: there $s(t_-)=1$, all four rates equal $-H_-$, and
$k_1-k_2=2k_{\rm w}\sinh(2\beta_{k_{\rm w}}(t_-))$. The right side is
positive for large $k_{\rm w}$, because its first two terms equal
$3kG\dot\Psi_*(t_-)^2>0$ by Eq.~\eqref{tidal:frw}, while the last is at
most $e^{-2ck_{\rm w}+o(k_{\rm w})}$.

Every fixed initial spatial derivative of the perturbation tends to
zero, since its size is a polynomial in $k_{\rm w}$ times
$e^{-ck_{\rm w}+o(k_{\rm w})}$. In Cartesian coordinates these initial data
change the metric by terms
proportional to $\cosh(2\beta_{k_{\rm w}})-1$ and to
$\sinh(2\beta_{k_{\rm w}})$ times $\cos(2k_{\rm w}x^3)$ or
$\sin(2k_{\rm w}x^3)$, so each spatial derivative brings at most a
factor $2k_{\rm w}$; the extrinsic curvature is $H_-$ times the
metric, and the scalar velocity changes by
$O(k_{\rm w}^2\beta_{k_{\rm w}}(t_-)^2)$.

\subsection{Nonlinear evolution at fixed frequency}

The linearized analysis is not enough, because the anisotropy feeds
back on $\mathbf X$, and $\mathbf X$ and its derivatives enter
Eq.~\eqref{tidal:beta}. This subsection shows that the feedback is
small throughout, with constants independent of $k_{\rm w}$, so that
the exact solution follows the linearized one up to the observation
time. After writing the reduced equations with a uniformly invertible
acceleration matrix, we use the linear energy to control the
nonlinear errors before the transition. A single Green-function
estimate then controls the interval from $t=0$ to the fixed observation
time, including the transition layer.

To control the exact nonlinear solution, we compare it with the
reference and with the linearized solution. Set
\begin{equation}
 \widehat\beta=k_{\rm w}^2\beta_{k_{\rm w}},\qquad
 E_{\rm a}(t)=\frac{\epsilon_{\rm a}\phi(t)}{\phi(T)},\qquad E_{\rm a,0}=E_{\rm a}(0)
 \label{tidal:d27}
\end{equation}
so that $E_{\rm a}$ is $k_{\rm w}^2$ times the linearized evolution of
the data of Eq.~\eqref{tidal:seed} and has the same data as
$\widehat\beta$ at $t_-$. On $[0,T]$ it increases from
$E_{\rm a}(0)=\epsilon_{\rm a}/\phi(T)$, which is exponentially small by
Eq.~\eqref{tidal:growth}, to $E_{\rm a}(T)=\epsilon_{\rm a}$. Write
$\delta\mathbf X=\mathbf X-\mathbf X_*$, with the same convention for
derivatives, and measure the errors in $\mathbf X$, its first derivative and the scalar charge by
\begin{equation}
 Z=|\delta\mathbf X|+|\delta\dot{\mathbf X}|
 +|\mathsf q_{k_{\rm w}}-\mathsf q_*|
 \label{tidal:d28}
\end{equation}
The scalar field enters the scale-factor equations only through
$\dot\Psi=\mathsf q_{k_{\rm w}}/V_{\rm hom}$, so the only error it adds
to those of the scale factors is the difference of the conserved
charges, which is constant in time.

The reduced equations are second order in $(\mathbf X,\beta)$. After
division by $V_{\rm hom}$ we write them as a linear system for the
accelerations $(\ddot{\mathbf X},\ddot\beta)$,
\begin{equation}
 \begin{pmatrix}M_{\mathbf X\mathbf X}&M_{\mathbf X\beta}\\
 M_{\mathbf X\beta}^{\rm T}&M_{\beta\beta}\end{pmatrix}
 \begin{pmatrix}\ddot{\mathbf X}\\ \ddot\beta\end{pmatrix}
 =\begin{pmatrix}\mathbf S\\ S_\beta\end{pmatrix}
 \label{tidal:acceleration-system}
\end{equation}
whose matrix and source depend on the variables and their velocities.
The matrix is the velocity Hessian of $L_{\rm red}/V_{\rm hom}$ in these
variables, and the last row of the system is Eq.~\eqref{tidal:beta}
multiplied by $4/V_{\rm hom}$. At $\beta=\dot\beta=0$ the matrix is
block diagonal, $M_{\mathbf X\beta}=0$, and on the reference its blocks
are $M_{\mathbf X\mathbf X}=M_{\mathbf X,*}$ of
Eq.~\eqref{tidal:mean-matrix} and $M_{\beta\beta}=4K$.

The anisotropy changes the matrix only through $L_\beta$. The first
term of Eq.~\eqref{tidal:beta-action} is quadratic in $\dot\beta$ and
has no factor $k_{\rm w}^2$; the other two contain $\beta$, are at least
quadratic in $(\beta,\dot\beta)$ and carry the factor $k_{\rm w}^2$.
The last term is linear in the components of $\dot{\mathbf X}$, so only the first two
change the $M_{\mathbf X\mathbf X}$ block. In them the components of $\dot{\mathbf X}$ enter through
the products $\dot\ell_3\dot\ell_4$ and $\dot\ell_{\rm m}\dot\ell_4$,
and the $M_{\mathbf X\mathbf X}$ block changes by $O(\dot\beta^2+k_{\rm w}^2\beta^2)$
in the entries
\begin{equation}
 \frac{\partial^2(L_\beta/V_{\rm hom})}{\partial\dot\ell_3\,\partial\dot\ell_4}
 =8\alpha\bigl[\dot\beta^2-k_{\rm w}^2e^{-2\ell_3}\sinh^2(2\beta)\bigr],\quad
 \frac{\partial^2(L_\beta/V_{\rm hom})}{\partial\dot\ell_{\rm m}\,\partial\dot\ell_4}
 =16\alpha k_{\rm w}^2e^{-2\ell_3}\sinh^2(2\beta)
 \label{tidal:mean-block-change}
\end{equation}
The other diagonal block comes from the first term alone; it is
$M_{\beta\beta}=4(1+4\alpha\dot\ell_3\dot\ell_4)$ and does not depend on
$(\beta,\dot\beta)$.

The off-diagonal block $M_{\mathbf X\beta}$ receives no contribution
from the second term of Eq.~\eqref{tidal:beta-action}, which contains
no $\dot\beta$. It has a zero entry for $\dot\ell_{\rm m}$ and the
entries
\begin{equation}
 \frac{\partial^2(L_\beta/V_{\rm hom})}{\partial\dot\beta\,\partial\dot\ell_3}
 =16\alpha\dot\ell_4\dot\beta,\qquad
 \frac{\partial^2(L_\beta/V_{\rm hom})}{\partial\dot\beta\,\partial\dot\ell_4}
 =16\alpha\dot\ell_3\dot\beta
 +16\alpha k_{\rm w}^2e^{-2\ell_3}\sinh(4\beta)
 \label{tidal:d29}
\end{equation}
so it has size $O(|\dot\beta|+k_{\rm w}^2|\beta|)$.

After the terms with accelerations are retained on the left, the
contribution of the anisotropy to the source $\mathbf S$ in the $\mathbf X$ equations is
bounded by
\begin{equation}
 C\{(1+k_{\rm w}^2)\dot\beta^2+k_{\rm w}^2\beta^2
                           +k_{\rm w}^2|\beta\dot\beta|\}
 \label{tidal:mean-source}
\end{equation}
It comes from the terms of
$\partial_t(\partial L_\beta/\partial\dot{\mathbf X})-\partial L_\beta/\partial\mathbf X$
without accelerations, divided by $V_{\rm hom}$. In these, the three
terms of Eq.~\eqref{tidal:beta-action} appear with their sizes $\dot\beta^2$,
$k_{\rm w}^2\beta^2$ and $k_{\rm w}^2|\beta\dot\beta|$, multiplied by
bounded functions of the components of $\mathbf X$ and $\dot{\mathbf X}$, and the time
derivatives of $\sinh^2(2\beta)$ and $\sinh(4\beta)$ add terms of sizes
$k_{\rm w}^2|\beta\dot\beta|$ and $k_{\rm w}^2\dot\beta^2$, respectively.
In particular, both $k_{\rm w}^2\dot\beta^2$ and the coupling term
$k_{\rm w}^2\beta\ddot\beta$ are present. Both come from the time
derivative of $\partial L_\beta/\partial\dot\ell_4$, which contains
$16\alpha k_{\rm w}^2V_{\rm hom}e^{-2\ell_3}\dot\beta\sinh(4\beta)$.
With
$\partial_t\log(V_{\rm hom}e^{-2\ell_3})=2\dot\ell_{\rm m}-\dot\ell_3+\dot\ell_4$,
\begin{equation}
 \begin{split}
 \partial_t\bigl[16\alpha k_{\rm w}^2V_{\rm hom}e^{-2\ell_3}\dot\beta\sinh(4\beta)\bigr]
 ={}&16\alpha k_{\rm w}^2V_{\rm hom}e^{-2\ell_3}\bigl[\sinh(4\beta)\,\ddot\beta
 +4\cosh(4\beta)\,\dot\beta^2\\
 &+(2\dot\ell_{\rm m}-\dot\ell_3+\dot\ell_4)\,\dot\beta\sinh(4\beta)\bigr]
 \end{split}
 \label{tidal:mixed-acceleration}
\end{equation}
The first term on the right couples $\ddot\beta$ to the equations for
$\mathbf X$. After division by $V_{\rm hom}$, its coefficient is the part
$16\alpha k_{\rm w}^2e^{-2\ell_3}\sinh(4\beta)$ of the last entry of
Eq.~\eqref{tidal:d29}, and the second is the source term of size
$k_{\rm w}^2\dot\beta^2$ in Eq.~\eqref{tidal:mean-source}.

The acceleration coefficients are smooth functions of $\mathbf X$,
$\dot{\mathbf X}$, $\widehat\beta$, $k_{\rm w}\dot\beta$ and
$k_{\rm w}^{-1}$; for instance
$k_{\rm w}^2\sinh(4\beta)=k_{\rm w}^2\sinh(4k_{\rm w}^{-2}\widehat\beta)$
and $\dot\beta=k_{\rm w}^{-1}(k_{\rm w}\dot\beta)$. They extend
smoothly to $k_{\rm w}^{-1}=0$ on a fixed small ball around the
reference values of $(\mathbf X,\dot{\mathbf X})$ and
$\widehat\beta=k_{\rm w}\dot\beta=0$, at whose center the matrix has the
nonsingular blocks $M_{\mathbf X,*}$ and $4K$.
The reference inverse in Eq.~\eqref{tidal:mean-matrix}, the reference $\beta$ block $4K$, and the smooth dependence of the full matrix in Eq.~\eqref{tidal:acceleration-system} therefore give an inverse uniformly bounded in $k_{\rm w}$ on this ball.

Interchanging directions 1 and 2 sends $\beta\mapsto-\beta$ and
leaves $\mathbf X$ unchanged: the map
$(x^1,x^2,x^3,x^4)\mapsto(x^2,x^1,-x^3,x^4)$ exchanges $\vartheta^1$
and $\vartheta^2$ in Eq.~\eqref{tidal:metric} and reverses
$\vartheta^3$, so it takes the solution with $(a_1,a_2)$ to the one
with $(a_2,a_1)$. Correspondingly, Eq.~\eqref{tidal:beta-action} is
even in $(\beta,\dot\beta)$; the equations for $\mathbf X$ are even in the
anisotropy, and the anisotropy equation is odd. Thus the anisotropy
first contributes to the $\mathbf X$ equations at quadratic order. Dividing Eq.~\eqref{tidal:beta} by
$V_{\rm hom}(1+4\alpha\dot\ell_3\dot\ell_4)$ and multiplying by
$k_{\rm w}^2$, we write the anisotropy equation as
Eq.~\eqref{tidal:linear} for $\widehat\beta$ with a remainder,
\begin{equation}
 \widehat\beta''+b_*\widehat\beta'-k_{\rm w}^2a_*\widehat\beta
 =R_{\widehat\beta}
 \label{tidal:d30}
\end{equation}
Since the equation is odd, its terms linear in $(\beta,\dot\beta)$
differ from those of Eq.~\eqref{tidal:linear} only through $\mathbf X$
and its derivatives, and its other terms are at least
cubic. The anisotropy remainder $R_{\widehat\beta}$ is therefore cubic
or contains a factor from $\delta\mathbf X$ or its derivatives.

Before the transition we need only the linear energy estimate already proved.
Let $G_{\rm ret}(t,t')$ solve Eq.~\eqref{tidal:linear} in $t$, with
$G_{\rm ret}(t',t')=0$ and $\partial_tG_{\rm ret}(t',t')=1$.
Its energy at $t=t'$ is $p_*(t')^2$, so the forward bound in
Eq.~\eqref{tidal:d20} gives
\begin{equation}
 |G_{\rm ret}(t,t')|+|\partial_tG_{\rm ret}(t,t')|\le C
 \qquad(t_-\le t'\le t\le0)
 \label{tidal:d31}
\end{equation}
This uniform forward bound will control the nonlinear forcing. It
does not give a frequency-independent bound for evolution backward in time.

We first keep the acceleration feedback explicit. Subtracting the reference equations from the first three rows of Eq.~\eqref{tidal:acceleration-system} and applying $M_{\mathbf X\mathbf X}^{-1}$ together with Eqs.~\eqref{tidal:mean-block-change}--\eqref{tidal:mean-source} gives
\begin{equation}
 \begin{split}
 |\delta\ddot{\mathbf X}|\le C\bigl\{&Z+(1+k_{\rm w}^2)\dot\beta^2
 +k_{\rm w}^2\beta^2+k_{\rm w}^2|\beta\dot\beta|\\
 &+(|\dot\beta|+k_{\rm w}^2|\beta|)|\ddot\beta|\bigr\}
 \end{split}
 \label{tidal:preloss-nonlinear}
\end{equation}
The last term comes from the coupling to $\ddot\beta$ through the
off-diagonal block in Eq.~\eqref{tidal:d29}. Eq.~\eqref{tidal:beta} gives the coupled
bound
\begin{equation}
 |\ddot\beta|\le C(|\dot\beta|+k_{\rm w}^2|\beta|)
 (1+|\delta\ddot{\mathbf X}|)
 \label{tidal:preloss-coupling}
\end{equation}
Its denominator $1+4\alpha\dot\ell_3\dot\ell_4$ is bounded away from
zero in the fixed ball on which the acceleration matrix is uniformly invertible. Both coefficient errors in
the anisotropy equation are $O(Z+|\delta\ddot{\mathbf X}|)$, since
Eq.~\eqref{tidal:beta} includes the components of $\ddot{\mathbf X}$. Subtracting
its reference linearization and using
$\sinh(4\beta)-4\beta=O(\beta^3)$ therefore gives
\begin{equation}
 |R_{\widehat\beta}|\le
 C(Z+|\delta\ddot{\mathbf X}|)
 (|\widehat\beta'|+k_{\rm w}^2|\widehat\beta|)
 +Ck_{\rm w}^{-2}|\widehat\beta|^3
 \label{tidal:remainder-general}
\end{equation}
All these bounds use the exact nonlinear equations.

On a maximal initial interval in this neighborhood, assume
$|\widehat\beta|\le C_0 E_{\rm a,0}$ and
$|\widehat\beta'|\le C_0 k_{\rm w}E_{\rm a,0}$, with $C_0$ a fixed constant
larger than twice the constants in Eq.~\eqref{tidal:preloss-linear}.
These inequalities hold for the initial data in Eq.~\eqref{tidal:seed}. In the following
estimates $C$ may depend on this fixed $C_0$. Since
$\beta=\widehat\beta/k_{\rm w}^2$, they give
\begin{equation}
 \begin{gathered}
 (1+k_{\rm w}^2)\dot\beta^2\le CE_{\rm a,0}^2,\qquad
 k_{\rm w}^2\beta^2\le CE_{\rm a,0}^2/k_{\rm w}^2,\qquad
 k_{\rm w}^2|\beta\dot\beta|\le CE_{\rm a,0}^2/k_{\rm w}\\
 |\dot\beta|+k_{\rm w}^2|\beta|\le CE_{\rm a,0}
 \end{gathered}
 \label{tidal:preloss-forcing}
\end{equation}
Substituting Eq.~\eqref{tidal:preloss-coupling} into
Eq.~\eqref{tidal:preloss-nonlinear} produces
$CE_{\rm a,0}^2|\delta\ddot{\mathbf X}|$ on the right. Since $E_{\rm a,0}$ is
exponentially small, this term is absorbed for large $k_{\rm w}$,
yielding
\begin{equation}
 |\delta\ddot{\mathbf X}|\le C(Z+E_{\rm a,0}^2),\qquad
 \dot Z\le C(Z+E_{\rm a,0}^2)
 \label{tidal:preloss-acceleration}
\end{equation}
The second inequality holds almost everywhere, because $Z$ is a sum
of norms and its charge term is constant. Initially $\mathbf X$ and $\dot{\mathbf X}$ agree exactly with the reference, and
Eq.~\eqref{tidal:initial-scalar} gives
\begin{equation}
 Z(t_-)=|\mathsf q_{k_{\rm w}}-\mathsf q_*|
 \le Ck_{\rm w}^2\beta(t_-)^2\le Ck_{\rm w}^{-2}E_{\rm a,0}^2
 \label{tidal:d32}
\end{equation}
Gr\"onwall's inequality on the fixed interval therefore gives
\begin{equation}
 Z+|\delta\ddot{\mathbf X}|\le CE_{\rm a,0}^2
 \label{tidal:d33}
\end{equation}
The constants do not depend on the endpoint of the current bootstrap
interval. Equation~\eqref{tidal:remainder-general} now yields
\begin{equation}
 |R_{\widehat\beta}|\le Ck_{\rm w}^2E_{\rm a,0}^3
 \label{tidal:preloss-remainder}
\end{equation}
The anisotropy error and its derivative vanish at $t_-$. Variation of
constants gives
\begin{equation}
 \widehat\beta(t)-E_{\rm a}(t)
 =\int_{t_-}^tG_{\rm ret}(t,t')R_{\widehat\beta}(t')\,\dd t'
 \label{tidal:preloss-variation}
\end{equation}
Differentiation replaces the kernel by $\partial_tG_{\rm ret}$;
there is no endpoint term because $G_{\rm ret}(t,t)=0$.
Using Eq.~\eqref{tidal:d31}, both errors are at most
$Ck_{\rm w}^2E_{\rm a,0}^3$. Since $k_{\rm w}^2E_{\rm a,0}^2\to0$, this strictly
improves the assumed bounds, in view of
$|E_{\rm a}|\le CE_{\rm a,0}$ and $|E_{\rm a}'|\le Ck_{\rm w}E_{\rm a,0}$ from
Eq.~\eqref{tidal:preloss-linear}. Equation~\eqref{tidal:d33} also keeps
$\mathbf X$ and $\dot{\mathbf X}$ in the ball where the acceleration
matrix is uniformly invertible. Ordinary differential
equation continuation thus closes the argument on all of $[t_-,0]$.

In particular the errors entering the transition satisfy
\begin{equation}
 \begin{gathered}
 Z(0)\le\delta_Z=CE_{\rm a,0}^2\\
 |\widehat\beta(0)-E_{\rm a,0}|+k_{\rm w}^{-2/3}|\widehat\beta'(0)|
 \le\delta_\beta=Ck_{\rm w}^2E_{\rm a,0}^3
 \end{gathered}
 \label{tidal:import}
\end{equation}
The anisotropy estimate holds without the derivative weight on the whole
pre-transition interval. At later times, an error in the initial
derivative is propagated by $G_{\rm ret}(t,0)$. After division by the
linear growth factor $\phi(t)$, Eq.~\eqref{tidal:green-bounds} gives the
weight $k_{\rm w}^{-2/3}=1/(k_{\rm w}\mu_{\rm g}(0))$ for this contribution.
This is the weight used in the initial-error bound of
Eq.~\eqref{tidal:homogeneous-error}.

We now start one bootstrap at $t=0$ and continue it to $T$:
\begin{equation}
 |\widehat\beta|\le2E_{\rm a},\qquad
 |\widehat\beta'|\le C_1k_{\rm w}\mu_{\rm g}E_{\rm a},\qquad
 Z<\rho
 \label{tidal:d34}
\end{equation}
Here $\rho>0$ is a fixed sufficiently small radius for the ball where the acceleration
matrix is uniformly invertible, and $C_1$ exceeds the upper constant in
$|E_{\rm a}'|\le Ck_{\rm w}\mu_{\rm g}E_{\rm a}$.
By Eq.~\eqref{tidal:import}, these inequalities hold strictly at $t=0$
for large $k_{\rm w}$. In the original variables they read
$|\beta|\le2E_{\rm a}/k_{\rm w}^2$ and
$|\dot\beta|\le C_1\mu_{\rm g}E_{\rm a}/k_{\rm w}$. Thus every term
of Eq.~\eqref{tidal:mean-source} is at most
$C\mu_{\rm g}^2E_{\rm a}^2$, using
$k_{\rm w}\mu_{\rm g}\ge k_{\rm w}^{2/3}\ge1$, and the mixed
acceleration block is $O(E_{\rm a})$.

At $\beta=\dot\beta=0$ the rows for $\ddot{\mathbf X}$ in the acceleration system differ
from those of the reference by $O(Z)$. With the uniform inverse,
Eqs.~\eqref{tidal:mean-source} and \eqref{tidal:beta} therefore imply
\begin{align}
 |\delta\ddot{\mathbf X}|&\le C(Z+\mu_{\rm g}^2E_{\rm a}^2+E_{\rm a}|\ddot\beta|)\nonumber\\
 \ddot\beta+b_*\dot\beta-k_{\rm w}^2a_*\beta&=R_\beta,\qquad
 |R_\beta|\le C(E_{\rm a}Z+E_{\rm a}|\delta\ddot{\mathbf X}|+k_{\rm w}^2|\beta|^3)
 \label{tidal:coupled-errors}
\end{align}
The second line follows from Eq.~\eqref{tidal:d30}, divided by
$k_{\rm w}^2$, and the estimate in Eq.~\eqref{tidal:remainder-general}, so
$R_\beta=R_{\widehat\beta}/k_{\rm w}^2$. In it, the coefficient of
$\dot\beta$ differs from $b_*$ by $O(Z+|\delta\ddot{\mathbf X}|)$, and
that of $\sinh(4\beta)/4$ differs from $-k_{\rm w}^2a_*$ by
$O(k_{\rm w}^2(Z+|\delta\ddot{\mathbf X}|))$, because
Eq.~\eqref{tidal:beta} contains $\ddot\ell_3$ and $\ddot\ell_4$. With
$|\dot\beta|+k_{\rm w}^2|\beta|\le CE_{\rm a}$ and
$\sinh(4\beta)-4\beta=O(\beta^3)$, this gives the bound on $R_\beta$.

The bound on $|\delta\ddot{\mathbf X}|$ follows by substituting the equation for $\ddot\beta$ in the second line of Eq.~\eqref{tidal:coupled-errors} into its first-line inequality. Since $a_*(0)=0$, we have $|a_*|\le Ct\le C\mu_{\rm g}^2$ on
$[0,T]$, and so
\begin{equation}
 |\ddot\beta|\le|b_*\dot\beta|+|a_*\widehat\beta|+|R_\beta|
 \le C\mu_{\rm g}^2E_{\rm a}+CE_{\rm a}(Z+|\delta\ddot{\mathbf X}|)
 \label{tidal:d35}
\end{equation}
Substitution therefore produces a term
$CE_{\rm a}^2|\delta\ddot{\mathbf X}|$. Since
$E_{\rm a}\le\epsilon_{\rm a}$, it is absorbed on the left for fixed
small $\epsilon_{\rm a}$, without dividing by $Y$, because $a_*$, the
coefficient that vanishes with $Y$, enters the bound on $\ddot\beta$
only as a factor. This gives the first bound in
Eq.~\eqref{tidal:mean-bounds} below.

The errors in $\dot{\mathbf X}$ and $\mathbf X$ follow by integrating twice. Both
integrations use Eq.~\eqref{tidal:integrated-growth}, which gives
\begin{equation}
 E_{\rm a}(t')^2\le CE_{\rm a}(t)^2
 \exp\left(-ck_{\rm w}\int_{t'}^t\mu_{\rm g}(v)\,\dd v\right)
 \qquad(0\le t'\le t\le T)
 \label{tidal:amplitude-decay}
\end{equation}
The derivative of the exponential with respect to $t'$ is
$ck_{\rm w}\mu_{\rm g}(t')$ times that exponential. Integrating it,
and then using that $\mu_{\rm g}$ increases, gives respectively
\begin{equation}
 \int_0^t\mu_{\rm g}E_{\rm a}^2\,\dd t'
 \le\frac{CE_{\rm a}(t)^2}{k_{\rm w}},\qquad
 \int_0^t\mu_{\rm g}^2E_{\rm a}^2\,\dd t'
 \le\frac{C\mu_{\rm g}(t)E_{\rm a}(t)^2}{k_{\rm w}}
 \label{tidal:d36}
\end{equation}
These estimates hold from $t=0$, including the transition layer.
Since $|\delta\dot{\mathbf X}|\le Z$, the bound $|\delta\ddot{\mathbf X}|\le C(Z+\mu_{\rm g}^2E_{\rm a}^2)$ just obtained from Eqs.~\eqref{tidal:coupled-errors} and \eqref{tidal:d35} yields
$\dot Z\le CZ+C\mu_{\rm g}^2E_{\rm a}^2$ almost everywhere.
Gr\"onwall's inequality, started from $Z(0)\le\delta_Z$, and the
second integral give
$Z\le C(\delta_Z+\mu_{\rm g}E_{\rm a}^2/k_{\rm w})$.
This bounds $|\delta\dot{\mathbf X}|$; integrating once more and using the
first integral bounds $|\delta\mathbf X|$.
Inserting these bounds and
$k_{\rm w}^2|\beta|^3\le8E_{\rm a}^3/k_{\rm w}^4$ into
Eq.~\eqref{tidal:coupled-errors} bounds the remainder. Altogether,
\begin{gather}
 |\delta\ddot{\mathbf X}|\le C(Z+\mu_{\rm g}^2E_{\rm a}^2),\qquad
 |\delta\dot{\mathbf X}|\le C(\delta_Z+\mu_{\rm g} E_{\rm a}^2/k_{\rm w}),\qquad
 |\delta \mathbf X|\le C(\delta_Z+E_{\rm a}^2/k_{\rm w}^2)
 \label{tidal:mean-bounds}\\
 |R_{\widehat\beta}|=k_{\rm w}^2|R_\beta|
 \le C(k_{\rm w}^2E_{\rm a}\delta_Z+k_{\rm w}^2\mu_{\rm g}^2E_{\rm a}^3)
 \label{tidal:nonlinear-remainder}
\end{gather}
The second term of Eq.~\eqref{tidal:nonlinear-remainder} comes mainly
from $E_{\rm a}|\delta\ddot{\mathbf X}|$, that is, from the change of
the components of $\ddot{\mathbf X}$ in Eq.~\eqref{tidal:beta}.

The anisotropy error is again controlled with the retarded Green
function of Eq.~\eqref{tidal:linear}. With
$I_b(t)=\int_0^t b_*(t')\,\dd t'$, it is
\begin{equation}
 G_{\rm ret}(t,t')=\phi(t)\phi(t')e^{I_b(t')}
              \int_{t'}^t e^{-I_b(t'')}\phi(t'')^{-2}\,\dd t''
 \label{tidal:green}
\end{equation}
It is built from $\phi$ and the second solution
$\phi(t)\int_0^te^{-I_b(t'')}\phi(t'')^{-2}\,\dd t''$ obtained by
reduction of order; the Wronskian of the two solutions is $e^{-I_b}$,
which fixes the normalization $\partial_tG_{\rm ret}(t',t')=1$.
Equation~\eqref{tidal:integrated-growth} and the monotonicity of
$\mu_{\rm g}$ bound the integral in the kernel for $0\le t'\le t\le T$:
\begin{equation}
 \begin{split}
 \int_{t'}^t e^{I_b(t')-I_b(t'')}
 \left(\frac{\phi(t')}{\phi(t'')}\right)^2\dd t''
 &\le C\int_{t'}^t
 \exp\left(-ck_{\rm w}\int_{t'}^{t''}\mu_{\rm g}(v)\,\dd v\right)\dd t''\\
 &\le C\int_{t'}^\infty
 e^{-ck_{\rm w}\mu_{\rm g}(t')(t''-t')}\,\dd t''
 \le\frac{C}{k_{\rm w}\mu_{\rm g}(t')}
 \end{split}
 \label{tidal:d37}
\end{equation}
Only the exponential majorant is extended to infinity; the reference
solution is used on $[0,T]$. Multiplication by
$\phi(t)/\phi(t')$ gives the bound for $G_{\rm ret}$ below.
Differentiating its exact formula gives
\begin{equation}
 \partial_tG_{\rm ret}(t,t')=\frac{\phi'(t)}{\phi(t)}G_{\rm ret}(t,t')
 +\frac{\phi(t')}{\phi(t)}e^{I_b(t')-I_b(t)}
 \label{tidal:d38}
\end{equation}
The global upper bound $\phi'/\phi\le Ck_{\rm w}\mu_{\rm g}$
controls the first term. The second satisfies the same estimate,
because $\phi$ and $\mu_{\rm g}$ increase. Hence
\begin{equation}
 |G_{\rm ret}(t,t')|\le\frac{C\phi(t)}{k_{\rm w}\mu_{\rm g}(t')\phi(t')},\qquad
 |\partial_tG_{\rm ret}(t,t')|
 \le\frac{C\mu_{\rm g}(t)\phi(t)}{\mu_{\rm g}(t')\phi(t')}
 \label{tidal:green-bounds}
\end{equation}
Both bounds hold for $0\le t'\le t\le T$, so the nonlinear comparison
starts at $t=0$. Variation of constants gives
\begin{equation}
 \begin{split}
 \widehat\beta(t)-E_{\rm a}(t)={}&[\widehat\beta(0)-E_{\rm a,0}]\phi(t)
 +\widehat\beta'(0)G_{\rm ret}(t,0)\\
 &+\int_0^tG_{\rm ret}(t,t')R_{\widehat\beta}(t')\,\dd t'
 \end{split}
 \label{tidal:bootstrap-sum}
\end{equation}
This expression has the required initial data because $\phi(0)=1$,
$\phi'(0)=0$, and the kernel vanishes at $t=0$ with unit derivative.
For the two homogeneous terms, Eq.~\eqref{tidal:green-bounds} and
$1/(k_{\rm w}\mu_{\rm g}(0))=k_{\rm w}^{-2/3}$ give the bound
\begin{equation}
 C\frac{|\widehat\beta(0)-E_{\rm a,0}|+k_{\rm w}^{-2/3}|\widehat\beta'(0)|}{E_{\rm a,0}}
 \le C\frac{\delta_\beta}{E_{\rm a,0}}
 \label{tidal:homogeneous-error}
\end{equation}
after division by $E_{\rm a}(t)$. Their derivatives obey the same
bound after division by $k_{\rm w}\mu_{\rm g}(t)E_{\rm a}(t)$.
For the integral term, use Eq.~\eqref{tidal:nonlinear-remainder}
and $\phi(t)/\phi(t')=E_{\rm a}(t)/E_{\rm a}(t')$. After division by
$E_{\rm a}(t)$, its absolute value is bounded by
\begin{equation}
 C\int_0^t\left[\frac{k_{\rm w}\delta_Z}{\mu_{\rm g}(t')}
 +k_{\rm w}\mu_{\rm g}(t')E_{\rm a}(t')^2\right]\dd t'
 \label{tidal:bootstrap-integral}
\end{equation}
The derivative of the integral has no endpoint term because
$G_{\rm ret}(t,t)=0$. After division by
$k_{\rm w}\mu_{\rm g}(t)E_{\rm a}(t)$, the derivative bound in
Eq.~\eqref{tidal:green-bounds} gives the same integral.
Finally $\int_0^T\mu_{\rm g}^{-1}\dd t\le2\sqrt T$, and
Eq.~\eqref{tidal:d36} bounds the integral containing $k_{\rm w}\mu_{\rm g}E_{\rm a}^2$. We obtain
\begin{equation}
 \begin{split}
 \frac{|\widehat\beta-E_{\rm a}|}{E_{\rm a}}
 +\frac{|\widehat\beta'-E_{\rm a}'|}{k_{\rm w}\mu_{\rm g}E_{\rm a}}
 &\le CE_{\rm a}(t)^2+Ck_{\rm w}\delta_Z+C\delta_\beta/E_{\rm a,0}\\
 &\le CE_{\rm a}(t)^2+C(k_{\rm w}^2+k_{\rm w})E_{\rm a,0}^2
 \end{split}
 \label{tidal:bootstrap}
\end{equation}
The growing amplitude must remain inside the integral:
replacing $E_{\rm a}(t')$ by $\epsilon_{\rm a}$ before integration
would give an uncontrolled term of order
$k_{\rm w}\epsilon_{\rm a}^2$. The remaining error tends to zero,
since
\begin{equation}
 C(k_{\rm w}^2+k_{\rm w})E_{\rm a,0}^2
 \le C(k_{\rm w}^2+k_{\rm w})\epsilon_{\rm a}^2e^{-2ck_{\rm w}}=o(1)
 \label{tidal:bootstrap-small}
\end{equation}

For small fixed $\epsilon_{\rm a}$ and large $k_{\rm w}$,
Eq.~\eqref{tidal:bootstrap} improves the assumed anisotropy bounds.
Since $E_{\rm a}\le\epsilon_{\rm a}$, it gives
$|\widehat\beta|\le(1+C\epsilon_{\rm a}^2+o(1))E_{\rm a}<2E_{\rm a}$
and
$|\widehat\beta'|\le|E_{\rm a}'|+(C\epsilon_{\rm a}^2+o(1))k_{\rm w}\mu_{\rm g}E_{\rm a}
<C_1k_{\rm w}\mu_{\rm g}E_{\rm a}$, once $C_1$ exceeds the constant $C$
in the upper bound derived from Eqs.~\eqref{tidal:global-growth} and \eqref{tidal:d22} on $[0,T]$, while
Eq.~\eqref{tidal:mean-bounds} keeps
$Z\le C(\delta_Z+\mu_{\rm g}E_{\rm a}^2/k_{\rm w})$ small. The solution
thus stays in the ball on which the acceleration matrix is uniformly
invertible, since $\widehat\beta$ and
$k_{\rm w}\dot\beta=\widehat\beta'/k_{\rm w}$ are $O(\epsilon_{\rm a})$.
This invertibility and positivity of the scale factors then extend each
exact ODE solution through $T$.

At $t=T$, where $E_{\rm a}=\epsilon_{\rm a}$,
Eq.~\eqref{tidal:bootstrap} gives $\widehat\beta(T)$ and
$|\widehat\beta'(T)|\le Ck_{\rm w}\epsilon_{\rm a}$. The second line of
Eq.~\eqref{tidal:coupled-errors}, with
$k_{\rm w}^2a_*\beta=a_*\widehat\beta$ and
$|R_\beta|\le C(E_{\rm a}\delta_Z+\mu_{\rm g}^2E_{\rm a}^3)$ from
Eq.~\eqref{tidal:nonlinear-remainder}, then gives $\ddot\beta(T)$:
\begin{equation}
 k_{\rm w}^2\beta(T)=\epsilon_{\rm a}+O(\epsilon_{\rm a}^3)+o(1),\quad
 \dot\beta(T)=O(\epsilon_{\rm a}/k_{\rm w}),\quad
 \ddot\beta(T)=a_*(T)\epsilon_{\rm a}+O(\epsilon_{\rm a}^3)+o(1)
 \label{tidal:endpoint}
\end{equation}

\subsection{Exact constraints and a common observation region}

The anisotropic solutions are so far defined on the homogeneous
core. To obtain developments of complete initial data, we cut off the
initial perturbation outside a fixed ball, solve the constraints exactly
across the cutoff, and join the interior solution to a short development
of the complete data. Before characteristic loss, geometric uniqueness
justifies this identification wherever the past lies in the unchanged
core. Afterwards the construction uses the particular exact interior
solution. It does not prove that every classical extension of the same
core data has that interior.

Write $r=|\boldsymbol x|$ for the initial Cartesian radius, equal to
the reference areal radius. Choose fixed constants with
\begin{equation}
 1<c_1<c_2<\bar c<\bar c_+<\frac{3}{2\sqrt2},\qquad
 \frac{\sqrt2\,\bar c_+}{3H_-}<R_g<\frac1{2H_-}
 \label{tidal:d39}
\end{equation}
The speeds $c_1,c_2$ will be those of the auxiliary metrics below,
$\bar c$ will bound every characteristic speed before characteristic loss, and
$1/\bar c_+$ will bound the slopes of the Cauchy surfaces. Both
choices are possible because $\sqrt2/3<1/2$ in
Eq.~\eqref{tidal:buffer}, so that $3/(2\sqrt2)>1$; the bound
$\bar c_+<3/(2\sqrt2)$ then makes the interval for $R_g$ nonempty.

Apply a fixed smooth cutoff to the metric perturbation in
Eq.~\eqref{tidal:seed}, leaving it unchanged for $r<R_g$ and zero
before $r=1/(2H_-)$. On the ball $r<1/(2H_-)$ set
$\mathcal K_{ij}=H_-\gamma_{ij}$, with
$\mathcal K=-\tfrac12\mathcal L_{e_0}\gamma$, and keep $\Psi$
spatially constant. Both exact data sets already have this form: the
reference data are the homogeneous data for $r\le1/(2H_-)$, and all
four rates $H_i$ in the initial data of Eq.~\eqref{tidal:seed} equal $-H_-$ at
$t_-$. The Codazzi tensor vanishes because $H_-$ is constant, and
the scalar momentum density because $\Psi$ is spatially constant, so
the momentum constraints hold exactly. The Hamiltonian constraint is
the one solved in Eq.~\eqref{area:scalar} of
Appendix~\ref{app:area-completion}, with the constant $c$ of that
appendix equal to $H_-$:
\begin{equation}
 3kG\,P^2=\mathcal C_{\rm H}[\gamma,H_-]
 =(1+4\alpha H_-^2)\,{}^{(4)}\mathcal R
 +\alpha\,\mathcal G[\gamma]+12H_-^2+24\alpha H_-^4,
 \qquad P=e_0\Psi
 \label{tidal:localized-constraint}
\end{equation}
where ${}^{(4)}\mathcal R$ and $\mathcal G[\gamma]$ are the scalar
curvature and Gauss--Bonnet density of the spatial metric, as in
Eq.~\eqref{area:gbshift}.

The right-hand side retains a fixed positive margin for large
$k_{\rm w}$. For the reference data it equals
$12H_-^2+24\alpha H_-^4$, since the reference spatial metric is flat
for $r\le1/(2H_-)$. The curvature terms involve at most two
derivatives of the metric, and the cutoff perturbation tends to zero
with every derivative, so for large $k_{\rm w}$ the right-hand side
stays above half of this value. Its square root with the reference
sign therefore solves the constraint. For $r<R_g$ it agrees with
the exact scalar data: there the spatial metric is that of
Eq.~\eqref{tidal:metric} at $t_-$, for which $\mathcal G[\gamma]=0$
and ${}^{(4)}\mathcal R=R_{\rm sp}
=-2k_{\rm w}^2\sinh^2(2\beta_{k_{\rm w}}(t_-))$, so
Eq.~\eqref{tidal:localized-constraint} reduces to
Eq.~\eqref{tidal:initial-scalar}.

Outside the cutoff the reference data are unchanged. The complete
constrained data therefore differ from the reference only in the fixed
ball $r<1/(2H_-)$, where they converge with every fixed number of
derivatives. Hence they converge also in the corresponding weighted
norms. Completeness and weak curvature persist, as do the positive
outgoing expansions of the coordinate spheres used to exclude trapped
surfaces.

Local existence in a strongly hyperbolic formulation
\cite{Kovacs2020} gives a short development of the complete data with
weak curvature. Since the data converge, the
same short time interval serves for the complete initial slices of all
sufficiently large $k_{\rm w}$; it includes a short past interval, so that the initial
slice lies in its interior. On the initial overlap, geometric uniqueness
identifies this development with the exact homogeneous solution
wherever the past stays inside $r<R_g$, where their data agree. On
fixed pre-transition intervals, Appendix~\ref{app:hyperbolic} supplies
uniqueness near the reference; after characteristic loss, the interior is supplied by
the exact reduced equations in Eq.~\eqref{tidal:acceleration-system}, with the bounds in Eqs.~\eqref{tidal:mean-bounds} and \eqref{tidal:bootstrap}.

Let $d_{\rm c}$ be the comoving distance traveled at unit speed since
$t_-$,
\begin{equation}
 d_{\rm c}(t)=\int_{t_-}^t\frac{\dd t'}{s(t')}
 \label{tidal:d40}
\end{equation}
In the coordinates $(d_{\rm c},\boldsymbol x)$ the reference core
metric is $s^2(-\dd d_{\rm c}^2+\dd\boldsymbol x^2)$, so there the
coordinate speed $|\dd\boldsymbol x/\dd d_{\rm c}|$ of a signal is its
proper speed relative to the comoving observers. We join the complete
short development to the homogeneous region
\begin{equation}
 r<R_g-\bar c\,d_{\rm c}(t)
 \label{tidal:d41}
\end{equation}
which is the same for all $k_{\rm w}$. Its boundary moves inward at
the speed $\bar c$, which exceeds every characteristic speed before
characteristic loss (see below), so the past of each point in this region before characteristic loss
stays inside $r<R_g$. Since Eq.~\eqref{tidal:buffer} gives
$d_{\rm c}(0)<\sqrt2/(3H_-)$, the lower bound on $R_g$ gives
$\bar c_+d_{\rm c}(0)<R_g$. We also take $T$ small enough that
$\bar c_+d_{\rm c}(T)<R_g$.

Metric light speeds are uniformly close to one in these coordinates,
and before characteristic loss the reference physical speeds are at most one, the
speed of the scalar, by Eq.~\eqref{tidal:frw-speed} with $1<K<\sqrt2$. Choose the
auxiliary inverse metrics
\begin{equation}
 g^{ab}+(1-c_i^{-2})\,e_0^ae_0^b,\qquad i=1,2
 \label{tidal:d42}
\end{equation}
with the speeds $c_1,c_2$ fixed above, which lie below $\bar c$. These
are the metrics of Eq.~\eqref{hyp:auxiliary-metrics} with
$\tilde a=c_1^{-2}-1$ and $\hat a=c_2^{-2}-1$, both in $(-1,0)$. They
are therefore Lorentzian, $e_0$ is timelike for both, and their speeds
$(1+\tilde a)^{-1/2}=c_1$ and $(1+\hat a)^{-1/2}=c_2$ exceed the
reference physical speeds. Their disjoint roots and common time
direction satisfy the requirements of Appendix~\ref{app:hyperbolic}.

These speed bounds persist for the perturbed solutions before characteristic loss.
The time covector is noncharacteristic on the reference: $K>0$ and
the matter and auxiliary time coefficients are nonzero. Compactness
of the fixed interval gives a positive lower bound on the smallest
singular value of the full time matrix $\mathsf P_2$ of
Eq.~\eqref{hyp:evolution-matrix}. Small metric and curvature
perturbations preserve a positive lower bound, so at each spatial
covector the characteristic frequencies are the roots of a polynomial
$\det\mathsf P(\omega)$ of fixed degree with nonvanishing leading
coefficient $\det\mathsf P_2$. They therefore depend continuously, as
an unordered set, on the principal-symbol coefficients. These depend
on curvature as well as the metric, as $K$ and $Y$ in
Eq.~\eqref{tidal:KY} do. The preceding curvature bounds therefore keep
all characteristic speeds before their degeneration below $\bar c$, and the physical ones below $c_1$,
for small $\epsilon_{\rm a}$; metric convergence in $C^1$ alone would
not suffice.

We foliate the glued development by complete Cauchy surfaces that
reach the time $T$ near the center. Choose $r_{\rm b}>0$ with
$r_{\rm b}+\bar c_+d_{\rm c}(T)<R_g$, and an early value $d_{\rm e}>0$ of $d_{\rm c}$ such that the early strip
$d_{\rm c}\le d_{\rm e}$ lies in that short development of the complete data. The cap
$d_{\rm cap}(r)$ is a smoothing of the radial profile
\begin{equation}
 \max\Bigl\{d_{\rm e},\ \min\Bigl\{d_{\rm c}(T),\
 d_{\rm c}(T)-\frac{r-r_{\rm b}}{\bar c_+}\Bigr\}\Bigr\}
 \label{tidal:d43}
\end{equation}
which is $d_{\rm c}(T)$ for $r\le r_{\rm b}$, decreases with slope
$-1/\bar c_+$ to $d_{\rm e}$, and stays constant outside. The
smoothing can keep
\begin{equation}
 \Bigl|\frac{\dd d_{\rm cap}}{\dd r}\Bigr|\le\frac1{\bar c_+},\qquad
 d_{\rm cap}(r)<\frac{R_g-r}{\bar c}\quad
 \text{where $d_{\rm cap}$ varies}
 \label{tidal:d44}
\end{equation}
Before smoothing, $(R_g-r)/\bar c$ minus the profile decreases along
the sloped part, because $\bar c<\bar c_+$, and at its outer end it
still exceeds $(\bar c_+/\bar c-1)d_{\rm e}>0$ by the choice of
$r_{\rm b}$. By the second condition, wherever $d_{\rm cap}$ varies,
the region between the initial slice and the graph of $d_{\rm cap}$
lies in the homogeneous region.

The leaves of the foliation are the graphs
\begin{equation}
 d_{\rm c}=\lambda_{\rm f}\,d_{\rm cap}(r),\qquad
 0\le\lambda_{\rm f}\le1
 \label{tidal:d45}
\end{equation}
Their normal covectors
$\dd d_{\rm c}-\lambda_{\rm f}\,\dd d_{\rm cap}$ have spatial part
at most $1/\bar c_+$ times their time part, while every characteristic
speed before characteristic loss is below $\bar c<\bar c_+$. The leaves are therefore
spacelike, with normals in every gravitational hyperbolicity cone before characteristic loss, and
outside a compact set they agree with complete early slices. Restrict
the glued development so that its past boundary lies below the leaf
$\lambda_{\rm f}=0$, which is the initial slice, and its future
boundary above the leaf $\lambda_{\rm f}=1$. Bounded coordinate causal
speeds prevent escape to infinity in finite time, so every
inextendible causal curve runs from the past to the future boundary
and crosses each leaf: the leaves are complete Cauchy surfaces for the
spacetime metric. The future boundary can lie above the top leaf
because each ODE solution extends beyond $T$; no common additional
lifetime is required.

The first characteristic loss of the homogeneous solution is also the
first characteristic loss on this foliation. On the leaf with parameter
$\lambda_{\rm f}$, the central region $r\le r_{\rm b}$ is at the
homogeneous time with $d_{\rm c}=\lambda_{\rm f}d_{\rm c}(T)$, the
latest time reached anywhere on that leaf. Where $d_{\rm cap}$ varies,
the leaf lies in the homogeneous region at earlier times; there the
geometry is homogeneous on each slice of constant $t$, so whether a
characteristic condition holds depends on $t$ alone.
Elsewhere the leaf lies in the early strip, where the curvature is
weak and the characteristic conditions hold. The first characteristic loss on this foliation therefore occurs in the
homogeneous region, on the leaf whose central time reaches the transition
of the homogeneous solution.

The times of first characteristic loss in the perturbed solutions tend
to zero. On each fixed closed interval in $[t_-,0)$ the reference satisfies the characteristic conditions
strictly, and the perturbed solutions converge to it smoothly, so no
characteristic loss occurs there for large $k_{\rm w}$. At each fixed $0<t<T$ the
convergence is also smooth, because
$E_{\rm a}(t)=\epsilon_{\rm a}\phi(t)/\phi(T)$ is exponentially small
in $k_{\rm w}$ by Eq.~\eqref{tidal:integrated-growth} applied between $t$ and $T$. Repeated
differentiation of the reduced equations introduces only
fixed powers of $k_{\rm w}$, as do spatial derivatives of the
homogeneous ansatz, so this smallness persists at every fixed
derivative order. At those positive times
the reference gravitational principal form
$-K\omega^2+Y|\boldsymbol\xi|^2$ is definite on real space--time
covectors, since $K>0>Y$, and small curvature perturbations preserve
this failure. Curvature remains bounded through these times, and
outside a compact set the leaves stay in the early strip, so the
construction does not require the exterior to be evolved to the time
reached near the center.

\subsection{Tidal curvature and continuous dependence}

We compute the tidal tensors of observers in the central region
$r\le r_{\rm b}$ from the scale factors and derive
Eq.~\eqref{tidal:separation}. Choose a compact ball $\mathscr B$ of
initial labels inside this region, which remains homogeneous up to the
time $T$ because $r_{\rm b}+\bar c\,d_{\rm c}(T)<R_g$. With $N=1$ and
zero shift, the constant-coordinate curves are the prescribed
unit-normal geodesics. The orthonormal frame dual to the coframe
$a_i\vartheta^i$ of Eq.~\eqref{tidal:metric} is parallel along them. The vectors dual to
the $\vartheta^i$ commute with $\partial_t$ and are stretched along
the geodesics at the rates $H_i$, because the
extrinsic curvature $\mathcal K_{ij}=-H_i\delta_{ij}$ is diagonal in
this frame, and the normalization factors $a_i^{-1}$ remove this
stretching.

Neighboring observers with fixed labels have a separation whose
components in this frame are $a_i$ times constants, so the
geodesic-deviation equation $J''=-(\mathcal E_{ij})J$ gives the
diagonal tidal tensor $\mathcal E_{ij}=-(\ddot a_i/a_i)\delta_{ij}$. With
$\log a_{1,2}=\ell_{\rm m}\pm\beta$,
\begin{equation}
 \mathcal E_{11}-\mathcal E_{22}
 =-2(\ddot\beta+2\dot\ell_{\rm m}\dot\beta)
 \label{tidal:split}
\end{equation}
The reference tidal tensor is isotropic,
$\mathcal E_{ij}[\mathcal U_*]=-(\ddot s/s)\delta_{ij}$, so the tidal
difference
$\mathcal E_{ij}[\mathcal U_{k_{\rm w}}]-\mathcal E_{ij}[\mathcal U_*]$
has the same eigenvalue spread, largest minus smallest eigenvalue, as
$\mathcal E_{ij}[\mathcal U_{k_{\rm w}}]$; by the diagonal form, this
spread is at least $|\mathcal E_{11}-\mathcal E_{22}|$. At $t=T$,
Eqs.~\eqref{tidal:endpoint} and \eqref{tidal:split} give, since
$\dot\ell_{\rm m}$ is bounded,
\begin{equation}
 \mathcal E_{11}-\mathcal E_{22}\big|_{t=T}
 =-2a_*(T)\,\epsilon_{\rm a}+O(\epsilon_{\rm a}^3)+o(1)
 \label{tidal:d46}
\end{equation}
where $a_*(T)>0$ because $Y<0$ after the transition. For small fixed
$\epsilon_{\rm a}$ the eigenvalue spread at $t=T$ therefore stays
bounded away from zero as $k_{\rm w}\to\infty$.

This frame-independent spread bounds a component of the tidal
difference in any prescribed orthonormal frame: every eigenvalue of a
symmetric $4\times4$ matrix is bounded in absolute value by four times
its largest entry in absolute value, so that entry is at least one
eighth of the spread. Choose the common measurement convention by
Gram--Schmidt on marked initial Cartesian vectors, followed by
parallel transport; carry the marking with the data under
diffeomorphisms. The spread at $t=T$ then gives
Eq.~\eqref{tidal:separation} on $\mathcal O=\mathscr B\times[t_-,T]$.
Homogeneity makes the eigenvalue spread independent of the observer
label. The tidal matrix therefore fails to converge at every fixed
label in $\mathscr B$ at the fixed time $T$, not merely at a moving
point where a supremum is attained. As proved above, the same sequence
converges smoothly at every fixed earlier time $t<T$. Another chosen
value of $T$ requires another normalization of the initial data.
A smaller lower bound also holds at the interior times
$T-c/k_{\rm w}$ for small fixed $c>0$, since the upper bound
$\phi'/\phi\le Ck_{\rm w}\mu_{\rm g}$ of Eq.~\eqref{tidal:growth} gives
$E_{\rm a}\ge e^{-Cc\mu_{\rm g}(T)}\epsilon_{\rm a}$ there.
The tidal discrepancy therefore persists at these times, with
$a_*(t)\to a_*(T)>0$. To obtain it, combine the equation for $\ddot\beta$
in the second line of Eq.~\eqref{tidal:coupled-errors} with the uniform
error bound in Eq.~\eqref{tidal:bootstrap}, then substitute the result in
Eq.~\eqref{tidal:split}. The components $\mathcal R_{ijkl}$, $\mathcal R_{0ijk}$ and
$\mathcal R_{0i0j}$ in the orthonormal frame are bounded using,
respectively, $\widehat\beta$, $k_{\rm w}\dot\beta$ and
$(\ddot\beta,\delta\ddot{\mathbf X})$, together with the bounded
reference curvature and the estimates for $\mathbf X$.
The anisotropic part of the metric perturbation is $O(k_{\rm w}^{-2})$
and its first derivatives are $O(k_{\rm w}^{-1})$; the errors in $\mathbf X$
obey Eq.~\eqref{tidal:mean-bounds}. Together with the pre-transition
estimates, this proves $C^1$ convergence on the full observation region.

For the conditional conclusion in
Corollary~\ref{cor:tidal-continuation}, suppose first that a proposed
continuation theory supplies classical developments for all nearby
data on a common observation region containing the marked central observers
from $t_-$ to some positive time after $t=0$. Choose a
sufficiently small $T>0$ inside that region, and only then normalize the
sequence by Eq.~\eqref{tidal:seed}. Each exact solution constructed here
is smooth on its own neighborhood of the observer geodesics through
$T$; no common additional lifetime after $T$ is needed. For each initial data set, compare the proposed classical development
with the corresponding exact solution, $\mathcal U_{k_{\rm w}}$ or
$\mathcal U_*$. Use overlapping regions that connect the initial slice
to the observation points. If local geometric uniqueness holds for
classical developments of the same initial data on these regions,
the two developments must agree there. Their measured tidal tensors
would then satisfy Eq.~\eqref{tidal:separation} despite convergence of
the data with every fixed number of derivatives. This contradicts
continuous dependence in the stated supremum norm.

Thus classical existence and the local geometric uniqueness assumed above
would force a failure of continuous dependence near $D_*$. The argument
does not decide which of the three requirements fails for general data.
Individual smooth extensions are compatible with this result: the
constructed solutions extend smoothly through characteristic loss because their
symmetry reduces the equations to ordinary differential equations.
The construction gives no genericity result or existence theorem for
arbitrary nonspherical data after characteristic loss.

The discrepancy is appreciable only within a time of order
$1/k_{\rm w}$ before $T$. For every finite $p\ge1$, the same estimates
give
\begin{equation}
 \int_{t_-}^T\sup_{\mathscr B}
 \max_{i,j}\bigl|\tilde\alpha\bigl(\mathcal E_{ij}[\mathcal U_{k_{\rm w}}]
                          -\mathcal E_{ij}[\mathcal U_*]\bigr)\bigr|^p\,\dd t
 \le C_p/k_{\rm w}+e^{-c_pk_{\rm w}+o(k_{\rm w})}
 \label{tidal:averages}
\end{equation}
Up to exponentially small terms, the tidal difference is at most a
constant times $E_{\rm a}(t)$, because it is controlled by
$\ddot\beta$ and $\delta\ddot{\mathbf X}$, which
Eqs.~\eqref{tidal:coupled-errors} and \eqref{tidal:mean-bounds} bound
in this way. Near $T$ the lower bound in Eq.~\eqref{tidal:growth}
gives exponential decay backward from $T$,
\begin{equation}
 E_{\rm a}(t)\le\epsilon_{\rm a}\,e^{-ck_{\rm w}(T-t)}
 \qquad(T/2\le t\le T)
 \label{tidal:d47}
\end{equation}
whose $p$th power integrates to at most $C_p/k_{\rm w}$; away from
$T$, $E_{\rm a}$ is exponentially small.

Equation~\eqref{tidal:averages} establishes convergence in
$L^p([t_-,T];L^\infty(\mathscr B))$ for each fixed finite $p\ge1$.
It implies convergence for fixed bounded time-weighting kernels
supported in $[t_-,T]$ and for averages over fixed positive-width windows
contained there. A window ending at $T$ is allowed; one extending past
$T$ requires further common estimates not proved here. The lower bound
near $T-c/k_{\rm w}$ also shows why convergence need not hold for
normalized windows of width $O(k_{\rm w}^{-1})$: the average magnitude of the tidal-tensor difference over such windows can remain nonzero.

For fixed initial separation $J(t_-)$ and relative velocity
$J'(t_-)$, the solutions $J$ of the geodesic-deviation equation and their
first derivatives converge uniformly on $[t_-,T]$. The difference of
the responses in $\mathcal U_{k_{\rm w}}$ and $\mathcal U_*$ vanishes
initially and obeys the deviation equation of $\mathcal U_{k_{\rm w}}$
with the added source
$-(\mathcal E_{ij}[\mathcal U_{k_{\rm w}}]-\mathcal E_{ij}[\mathcal U_*])$
applied to the reference response. The propagators are uniformly
bounded because the curvature components in these frames are bounded.
The response difference and its first derivative are therefore bounded
by a constant times the integral in Eq.~\eqref{tidal:averages} with
$p=1$. This argument concerns accumulated separation and velocity,
not instantaneous relative acceleration or every possible detector
protocol. It is a convergence result along this particular sequence,
not a well-posedness theorem for a neighborhood of arbitrary data or
an effective continuation of the evolution equations.
\bibliographystyle{unsrtnat}
\bibliography{ref}

@article{Deppe2012,
  author        = {Deppe, N. and Leonard, C. D. and Taves, T. and Kunstatter, G. and Mann, R. B.},
  title         = {Critical collapse in {Einstein--Gauss--Bonnet} gravity in five and six dimensions},
  journal       = {Phys. Rev. D},
  volume        = {86},
  pages         = {104011},
  year          = {2012},
  doi           = {10.1103/PhysRevD.86.104011},
  eprint        = {1208.5250},
  archivePrefix = {arXiv},
  primaryClass  = {gr-qc}
}

@article{Izumi2014,
  author        = {Izumi, Keisuke},
  title         = {Causal structures in {Gauss--Bonnet} gravity},
  journal       = {Phys. Rev. D},
  volume        = {90},
  pages         = {044037},
  year          = {2014},
  doi           = {10.1103/PhysRevD.90.044037},
  eprint        = {1406.0677},
  archivePrefix = {arXiv},
  primaryClass  = {gr-qc}
}

@article{Golod2012,
  author        = {Golod, Sveta and Piran, Tsvi},
  title         = {{Choptuik}'s critical phenomenon in {Einstein--Gauss--Bonnet} gravity},
  journal       = {Phys. Rev. D},
  volume        = {85},
  pages         = {104015},
  year          = {2012},
  doi           = {10.1103/PhysRevD.85.104015},
  eprint        = {1201.6384},
  archivePrefix = {arXiv},
  primaryClass  = {gr-qc}
}

@article{Papallo2017,
  author        = {Papallo, Giuseppe and Reall, Harvey S.},
  title         = {On the local well-posedness of {Lovelock} and {Horndeski} theories},
  journal       = {Phys. Rev. D},
  volume        = {96},
  pages         = {044019},
  year          = {2017},
  doi           = {10.1103/PhysRevD.96.044019},
  eprint        = {1705.04370},
  archivePrefix = {arXiv},
  primaryClass  = {gr-qc}
}

@article{Kovacs2020,
  author        = {Kov{\'a}cs, {\'A}ron D. and Reall, Harvey S.},
  title         = {Well-posed formulation of {Lovelock} and {Horndeski} theories},
  journal       = {Phys. Rev. D},
  volume        = {101},
  pages         = {124003},
  year          = {2020},
  doi           = {10.1103/PhysRevD.101.124003},
  eprint        = {2003.08398},
  archivePrefix = {arXiv},
  primaryClass  = {gr-qc}
}

@article{Maeda2008,
  author        = {Maeda, Hideki and Nozawa, Masato},
  title         = {Generalized {Misner--Sharp} quasilocal mass in {Einstein--Gauss--Bonnet} gravity},
  journal       = {Phys. Rev. D},
  volume        = {77},
  pages         = {064031},
  year          = {2008},
  doi           = {10.1103/PhysRevD.77.064031},
  eprint        = {0709.1199},
  archivePrefix = {arXiv},
  primaryClass  = {hep-th}
}

@article{Taves2012,
  author        = {Taves, T. and Leonard, C. D. and Kunstatter, G. and Mann, R. B.},
  title         = {Hamiltonian formulation of scalar field collapse in {Einstein--Gauss--Bonnet} gravity},
  journal       = {Class. Quantum Grav.},
  volume        = {29},
  pages         = {015012},
  year          = {2012},
  doi           = {10.1088/0264-9381/29/1/015012},
  eprint        = {1110.1154},
  archivePrefix = {arXiv},
  primaryClass  = {gr-qc}
}

@article{Boulware1985,
  author        = {Boulware, David G. and Deser, S.},
  title         = {String-generated gravity models},
  journal       = {Phys. Rev. Lett.},
  volume        = {55},
  pages         = {2656--2660},
  year          = {1985},
  doi           = {10.1103/PhysRevLett.55.2656}
}

@article{Reall2014,
  author = {Reall, Harvey S. and Tanahashi, Norihiro and Way, Benson},
  title = {Causality and hyperbolicity of {Lovelock} theories},
  journal = {Class. Quantum Grav.},
  volume = {31},
  pages = {205005},
  year = {2014},
  doi = {10.1088/0264-9381/31/20/205005},
  eprint = {1406.3379},
  archivePrefix = {arXiv},
  primaryClass = {hep-th}
}

@article{Cao2021,
  author = {Cao, Li-Ming and Wu, Liang-Bi},
  title = {Hyperbolicity and causality of {Einstein--Gauss--Bonnet} gravity in warped product spacetimes},
  journal = {Phys. Rev. D},
  volume = {103},
  pages = {064054},
  year = {2021},
  doi = {10.1103/PhysRevD.103.064054},
  eprint = {2101.02461},
  archivePrefix = {arXiv},
  primaryClass = {gr-qc}
}

@article{Beroiz2007,
  author = {Beroiz, Mart{\'i}n and Dotti, Gustavo and Gleiser, Reinaldo J.},
  title = {Gravitational instability of static spherically symmetric {Einstein--Gauss--Bonnet} black holes in five and six dimensions},
  journal = {Phys. Rev. D},
  volume = {76},
  pages = {024012},
  year = {2007},
  doi = {10.1103/PhysRevD.76.024012},
  eprint = {hep-th/0703074},
  archivePrefix = {arXiv}
}

@article{Konoplya2017,
  author = {Konoplya, R. A. and Zhidenko, A.},
  title = {Eikonal instability of {Gauss--Bonnet--(anti-)de Sitter} black holes},
  journal = {Phys. Rev. D},
  volume = {95},
  pages = {104005},
  year = {2017},
  doi = {10.1103/PhysRevD.95.104005},
  eprint = {1701.01652},
  archivePrefix = {arXiv},
  primaryClass = {hep-th}
}

@article{Nozawa2008,
  author = {Nozawa, Masato and Maeda, Hideki},
  title = {Dynamical black holes with symmetry in {Einstein--Gauss--Bonnet} gravity},
  journal = {Class. Quantum Grav.},
  volume = {25},
  pages = {055009},
  year = {2008},
  doi = {10.1088/0264-9381/25/5/055009},
  eprint = {0710.2709},
  archivePrefix = {arXiv},
  primaryClass = {gr-qc}
}

@article{Torii2008,
  author = {Torii, Takashi and Shinkai, Hisa-aki},
  title = {{$N+1$} formalism in {Einstein--Gauss--Bonnet} gravity},
  journal = {Phys. Rev. D},
  volume = {78},
  pages = {084037},
  year = {2008},
  doi = {10.1103/PhysRevD.78.084037},
  eprint = {0810.1790},
  archivePrefix = {arXiv},
  primaryClass = {gr-qc}
}

@article{CabreraMiao2018,
  author        = {Cabrera Pacheco, Armando J. and Miao, Pengzi},
  title         = {Higher dimensional black hole initial data with prescribed boundary metric},
  journal       = {Math. Res. Lett.},
  volume        = {25},
  number        = {3},
  pages         = {937--956},
  year          = {2018},
  doi           = {10.4310/MRL.2018.v25.n3.a10},
  eprint        = {1505.01800},
  archivePrefix = {arXiv},
  primaryClass  = {math.DG}
}

@article{Yoshino2011,
  author        = {Yoshino, Hirotaka},
  title         = {Black hole initial data in {Gauss--Bonnet} gravity: Momentarily static case},
  journal       = {Phys. Rev. D},
  volume        = {83},
  pages         = {104010},
  year          = {2011},
  doi           = {10.1103/PhysRevD.83.104010},
  eprint        = {1102.1810},
  archivePrefix = {arXiv},
  primaryClass  = {gr-qc}
}

@book{Olver1974,
  author    = {Olver, Frank W. J.},
  title     = {Asymptotics and Special Functions},
  publisher = {Academic Press},
  address   = {New York},
  year      = {1974},
  note      = {Reprinted by A K Peters, Wellesley, MA, 1997}
}

@article{TakahashiSoda2010,
  author        = {Takahashi, Tomohiro and Soda, Jiro},
  title         = {Master equations for gravitational perturbations of static {L}ovelock black holes in higher dimensions},
  journal       = {Prog. Theor. Phys.},
  volume        = {124},
  pages         = {911--924},
  year          = {2010},
  doi           = {10.1143/PTP.124.911},
  eprint        = {1008.1385},
  archivePrefix = {arXiv},
  primaryClass  = {gr-qc}
}

@article{Marques2012,
  author  = {Marques, Fernando Cod{\'a}},
  title   = {Deforming three-manifolds with positive scalar curvature},
  journal = {Ann. of Math.},
  volume  = {176},
  number  = {2},
  pages   = {815--863},
  year    = {2012},
  doi     = {10.4007/annals.2012.176.2.3},
  eprint  = {0907.2444},
  archivePrefix = {arXiv}
}

@article{MantoulidisSchoen2015,
  author  = {Mantoulidis, Christos and Schoen, Richard},
  title   = {On the {B}artnik mass of apparent horizons},
  journal = {Class. Quantum Grav.},
  volume  = {32},
  number  = {20},
  pages   = {205002},
  year    = {2015},
  doi     = {10.1088/0264-9381/32/20/205002},
  eprint  = {1412.0382},
  archivePrefix = {arXiv}
}

@article{GallowaySchoen2006,
  author  = {Galloway, Gregory J. and Schoen, Richard},
  title   = {A generalization of {H}awking's black hole topology theorem to higher dimensions},
  journal = {Commun. Math. Phys.},
  volume  = {266},
  number  = {2},
  pages   = {571--576},
  year    = {2006},
  doi     = {10.1007/s00220-006-0019-z},
  eprint  = {gr-qc/0509107},
  archivePrefix = {arXiv}
}

@article{AnderssonMarsSimon2008,
  author  = {Andersson, Lars and Mars, Marc and Simon, Walter},
  title   = {Stability of marginally outer trapped surfaces and existence of marginally outer trapped tubes},
  journal = {Adv. Theor. Math. Phys.},
  volume  = {12},
  number  = {4},
  pages   = {853--888},
  year    = {2008},
  doi     = {10.4310/ATMP.2008.v12.n4.a5},
  eprint  = {0704.2889},
  archivePrefix = {arXiv}
}

@unpublished{PaperII,
  author = {Suryaatmadja, Cendikiawan},
  title  = {Propagation and horizon formation in {Einstein--Gauss--Bonnet} scalar collapse. {II}. {N}arrow incoming pulses},
  note   = {In preparation},
  year   = {2026}
}

@incollection{Metivier2005,
  author        = {M{\'e}tivier, Guy},
  title         = {Remarks on the well-posedness of the nonlinear {Cauchy} problem},
  booktitle     = {Geometric Analysis of {PDE} and Several Complex Variables},
  series        = {Contemp. Math.},
  volume        = {368},
  pages         = {337--356},
  publisher     = {American Mathematical Society},
  address       = {Providence, RI},
  year          = {2005},
  doi           = {10.1090/conm/368/06790},
  eprint        = {math/0611441},
  archivePrefix = {arXiv}
}

@article{LernerNguyenTexier2018,
  author        = {Lerner, Nicolas and Nguyen, Toan T. and Texier, Benjamin},
  title         = {The onset of instability in first-order systems},
  journal       = {J. Eur. Math. Soc.},
  volume        = {20},
  number        = {6},
  pages         = {1303--1373},
  year          = {2018},
  doi           = {10.4171/JEMS/788},
  eprint        = {1504.04477},
  archivePrefix = {arXiv}
}

@article{Zwiebach1985,
  author        = {Zwiebach, Barton},
  title         = {Curvature squared terms and string theories},
  journal       = {Phys. Lett. B},
  volume        = {156},
  pages         = {315--317},
  year          = {1985},
  doi           = {10.1016/0370-2693(85)91616-8}
}

@article{MetsaevTseytlin1987,
  author        = {Metsaev, R. R. and Tseytlin, A. A.},
  title         = {Order {$\alpha'$} (two-loop) equivalence of the string equations of motion and the {$\sigma$}-model {Weyl} invariance conditions: Dependence on the dilaton and the antisymmetric tensor},
  journal       = {Nucl. Phys. B},
  volume        = {293},
  pages         = {385--419},
  year          = {1987},
  doi           = {10.1016/0550-3213(87)90077-0}
}

@article{Reall2021,
  author        = {Reall, Harvey S.},
  title         = {Causality in gravitational theories with second order equations of motion},
  journal       = {Phys. Rev. D},
  volume        = {103},
  pages         = {084027},
  year          = {2021},
  doi           = {10.1103/PhysRevD.103.084027},
  eprint        = {2101.11623},
  archivePrefix = {arXiv},
  primaryClass  = {gr-qc}
}

@article{RipleyPretorius2019,
  author        = {Ripley, Justin L. and Pretorius, Frans},
  title         = {Hyperbolicity in spherical gravitational collapse in a {Horndeski} theory},
  journal       = {Phys. Rev. D},
  volume        = {99},
  pages         = {084014},
  year          = {2019},
  doi           = {10.1103/PhysRevD.99.084014},
  eprint        = {1902.01468},
  archivePrefix = {arXiv},
  primaryClass  = {gr-qc}
}

@article{EastRipley2021,
  author        = {East, William E. and Ripley, Justin L.},
  title         = {Evolution of {Einstein--scalar--Gauss--Bonnet} gravity using a modified harmonic formulation},
  journal       = {Phys. Rev. D},
  volume        = {103},
  pages         = {044040},
  year          = {2021},
  doi           = {10.1103/PhysRevD.103.044040},
  eprint        = {2011.03547},
  archivePrefix = {arXiv},
  primaryClass  = {gr-qc}
}

@article{HegadeRipleyYunes2023,
  author        = {Hegade K. R., Abhishek and Ripley, Justin L. and Yunes, Nicol{\'a}s},
  title         = {Where and why does {Einstein--scalar--Gauss--Bonnet} theory break down?},
  journal       = {Phys. Rev. D},
  volume        = {107},
  pages         = {044044},
  year          = {2023},
  doi           = {10.1103/PhysRevD.107.044044},
  eprint        = {2211.08477},
  archivePrefix = {arXiv},
  primaryClass  = {gr-qc}
}

@article{Doneva2023,
  author        = {Doneva, Daniela D. and Arest{\'e} Sal{\'o}, Llibert and Clough, Katy and Figueras, Pau and Yazadjiev, Stoytcho S.},
  title         = {Testing the limits of {scalar--Gauss--Bonnet} gravity through nonlinear evolutions of spin-induced scalarization},
  journal       = {Phys. Rev. D},
  volume        = {108},
  pages         = {084017},
  year          = {2023},
  doi           = {10.1103/PhysRevD.108.084017},
  eprint        = {2307.06474},
  archivePrefix = {arXiv},
  primaryClass  = {gr-qc}
}

@article{Corman2024,
  author        = {Corman, Maxence and Lehner, Luis and East, William E. and Dideron, Guillaume},
  title         = {Nonlinear studies of modifications to general relativity: Comparing different approaches},
  journal       = {Phys. Rev. D},
  volume        = {110},
  pages         = {084048},
  year          = {2024},
  doi           = {10.1103/PhysRevD.110.084048},
  eprint        = {2405.15581},
  archivePrefix = {arXiv},
  primaryClass  = {gr-qc}
}

@misc{FiguerasKovacsYao2026EFT,
  author        = {Figueras, Pau and Kov{\'a}cs, {\'A}ron D. and Yao, Shunhui},
  title         = {Non-linear evolution of five-dimensional black strings in effective field theory},
  year          = {2026},
  eprint        = {2606.05319},
  archivePrefix = {arXiv},
  primaryClass  = {hep-th},
  note          = {arXiv:2606.05319 [hep-th]},
  url           = {https://arxiv.org/abs/2606.05319}
}

@misc{FiguerasKovacsYao2026EGB,
  author        = {Figueras, Pau and Kov{\'a}cs, {\'A}ron D. and Yao, Shunhui},
  title         = {The {Gregory--Laflamme} instability of five-dimensional black strings in {Einstein--Gauss--Bonnet} gravity},
  year          = {2026},
  eprint        = {2610.13612},
  archivePrefix = {arXiv},
  primaryClass  = {hep-th},
  note          = {arXiv:2610.13612 [hep-th]},
  url           = {https://arxiv.org/abs/2610.13612}
}
\end{document}